\documentclass{aa}

\usepackage{float}
\usepackage{graphicx}
\usepackage{txfonts}
\usepackage[colorlinks = true,
linkcolor = blue,
urlcolor  = blue,
citecolor = blue,
anchorcolor = blue]{hyperref}

\begin{document}
\title{LeMMINGs. V. Nuclear activity and bulge properties: a detailed multi-component decomposition of $e$-MERLIN
  Palomar galaxies with {\it HST}  }   
  \author{B.\ T.\ Dullo,\inst{1}\thanks{ E-mail: bdullo@ucm.es} 
  J.\ H.\ Knapen,\inst{2,3}
  R.\ J.\  Beswick,\inst{4}
  R.\ D.\ Baldi,\inst{5,6}
  D.\ R.\  A.\ Williams,\inst{4}
   I.\ M.\ McHardy,\inst{6}
   J.\ S.\ Gallagher,\inst{7}\\
   S.\ Aalto,\inst{8}
   M.\ K.\ Argo,\inst{9}
    A.\ Gil de Paz,\inst{1}
    H.-R.\   Kl\"ockner,\inst{10}
    J.\ M.\ Marcaide,\inst{11}
    C.\ G.\ Mundell,\inst{12}
    I.\ M.\ Mutie,\inst{13,4}
          \and
    P. Saikia \inst{14}}
\institute{\inst{1}Departamento de F\'isica de la Tierra y Astrof\'isica, IPARCOS (Instituto de F\'{\i}sica de Part\'{\i}culas y del Cosmos), Universidad Complutense de Madrid, E-28040, Madrid, Spain\\
\inst{2}Instituto de Astrof\'isica de Canarias, V\'ia
  L\'actea S/N, E-38205, La Laguna, Tenerife, Spain\\
\inst{3}Departamento de Astrof\'isica, Universidad de
  La Laguna, E-38206, La Laguna, Tenerife, Spain\\
\inst{4}Jodrell Bank Centre for Astrophysics, School of
  Physics and Astronomy, The University of Manchester, Alan Turing
  Building,\\ Oxford Road, Manchester, UK M13 9PL\\
\inst{5}Istituto di Radioastronomia - INAF, Via P. Gobetti 101, I-40129 Bologna, Italy\\
 \inst{6}School of Physics and Astronomy, University of Southampton, Southampton, SO17 1BJ, UK\\
\inst{7}Department of Astronomy, University of Wisconsin-Madison, Madison, Wisconsin, USA\\
 \inst{8}Department of Space, Earth and Environment, Chalmers University of Technology, 412 96 G\"oteborg,
Sweden\\
\inst{9}Jeremiah Horrocks Institute, University of Central Lancashire, Preston, Lancashire, PR1 2HE, UK\\
\inst{10}Max-Planck-Institut für Radioastronomie, Auf dem Hügel 69, 53121 Bonn, Germany\\
\inst{11}Real Academia de Ciencias, C/ Valverde 22, 28004, Madrid, Spain\\
\inst{12}Department of Physics, University of Bath, Claverton Down, Bath BA2 7AY, UK \\
\inst{13}Technical University of Kenya,  P.O. box 52428-00200, Nairobi, Kenya\\
\inst{14}Center for Astro, Particle and Planetary Physics (CAP$^3$), New York University Abu Dhabi, PO Box 129188, Abu Dhabi, UAE\\
  }
\date{\today}
\abstract
 {We use high-resolution {\it HST} imaging and $e$-MERLIN  1.5-GHz
  observations of galaxy cores from the LeMMINGs survey to investigate
  the relation between optical structural properties and nuclear radio
  emission for a large sample of galaxies.  We perform accurate,
  multi-component decompositions of new surface brightness profiles
  extracted from {\it HST} images for 163 LeMMINGs galaxies and fit up
  to six galaxy components (e.g., bulges, discs, AGN, bars, rings,
  spiral arms, and nuclear star clusters) simultaneously with S\'ersic
  and/or core-S\'ersic models.  By adding such decomposition data for 10
  LeMMINGs galaxies from our past work, the final sample of 173 nearby
  galaxies (102 Ss, 42 S0s, 23 Es plus 6 Irr) with bulge stellar mass
  (typically) $M_{*, \rm bulge}\sim 10^{6}-10^{12.5} M_{\sun}$,
  encompasses all optical spectral classes (LINERs, Seyferts, ALG and
  \mbox{H\,{\sc ii}}).  We show that the bulge mass can be
  significantly overestimated in many galaxies when components such as
  bars, rings and spirals are not included in the fits.  We
  additionally implement a Monte Carlo method to determine errors on
  bulge, disc and other fitted structural parameters.  Moving (in the opposite direction) across
  the Hubble sequence, i.e., from the irregular to elliptical galaxies, we
  confirm that bulges become larger, more prominent and round. Such
  bulge dominance is associated with a brighter radio core luminosity.
  We also find that the radio detection fraction increases with bulge
  mass. At $M_{*,\rm bulge} \ge 10^{11} M_{\sun}$, the radio detection
  fraction is 77\%, declining to 24\% for
  $M_{*,\rm bulge} < 10^{10} M_{\sun}$.  Furthermore, we observe
  core-S\'ersic bulges tend to be systematically round and to possess
  high radio core luminosities and boxy-distorted or pure elliptical
  isophotes. However, there is no evidence for the previously alleged
  strong tendency of galaxies' central structures (i.e., a sharp
  S\'ersic/core-S\'ersic dichotomy) with their radio loudness,
  isophote shape and flattening.}

\authorrunning{Dullo et al.}
\titlerunning {LeMMINGs. V.  Dissecting galaxies}
\maketitle

\section{Introduction}

In the standard structure formation model, elliptical galaxies and
massive bulges\footnote{The term `bulge' is traditionally associated
  with the spheroidal component of disc galaxies but it is used here
  to refer to the underlying host spheroid in case of elliptical
  galaxies and the bulge for S0, spiral and irregular galaxies.} of
lenticular galaxies (S0s) seen today are believed to have been
assembled, hierarchically, over cosmic time through mergers
\citep[e.g.,][]{1972ApJ...178..623T,1978MNRAS.183..341W,1984Natur.311..517B,1991ApJ...379...52W}. In
contrast, spiral galaxies may grow, within dark matter halos, through
the dissipational collapse of accreted gas
(\citealt{1980MNRAS.193..189F}) and subsequently experience
redistribution of their disc material through secular evolution
\citep[e.g.,][]{1982ApJ...257...75K,1993IAUS..153..209K}. The inflow
of gas into galaxy centres driven by galaxy mergers
\citep{1989Natur.340..687H,1991ApJ...370L..65B,1996ApJ...471..115B}
and instabilities in discs
\citep[e.g.,][]{1989Natur.338...45S,1995ApJ...454..623K,2000ApJ...528..677E,2009MNRAS.396..141D,2011MNRAS.415.1027H,2021ApJ...917...53A}
gives rise to star formation, growth of supermassive black holes
(SMBHs) and active galactic nuclei (AGN) activity. Feedback from an
accreting SMBH, which is often evident in the form of radio jets and outflows, particularly for the more common, lower luminosity AGN and LINERs which dominate the local universe, is believed to inject energy and momentum that expel
gas and regulate (the mass of the SMBH itself, e.g.,
\citealt{2010MNRAS.406L..55D}) and the star formation histories of the
host galaxy
\citep[e.g.,][]{1998A&A...331L...1S,2005ApJ...618..569M,2006MNRAS.366..417F,2006ApJ...645..986R,2006MNRAS.365...11C,2006MNRAS.370..645B,2014MNRAS.444.1518V,2017MNRAS.465.3291W}. How
such a process impacts the structural properties of the host galaxy
(e.g., stellar light distribution, stellar mass, concentration and
disc-to-total light ratio), remains a key unanswered question. The
empirical scaling relations between the SMBH mass and the properties
of its host bulge are understood as an indication of co-evolution of
SMBHs and their host bulges
\citep[e.g.,][]{1998AJ....115.2285M,2000ApJ...539L...9F,2000ApJ...539L..13G,2001ApJ...563L..11G,2002MNRAS.331..795M,2003ApJ...589L..21M,2004ApJ...604L..89H,2019ApJ...884..169G,2020ApJ...898...83D}.

The effect of AGN feedback has been invoked to explain the presence of
two types of structurally distinct elliptical galaxies: (i)
core-S\'ersic ellipticals with depleted cores whose (massive,
$n \ga 4$) bulges likely experienced `dry' (gas-poor) merger events
involving binary SMBHs and (ii) S\'ersic ellipticals which are
coreless, having less massive ($n \sim 3 \pm 1$) bulges thought to be
built in gas-rich processes
\citep[e.g.,][]{1997AJ....114.1771F,2008MNRAS.391..481S}. \citet{2009ApJS..182..216K}
argued that X-ray emitting, hot gas in core-S\'ersic ellipticals is
prevented from cooling and forming stars by radio-mode AGN feedback,
explaining the `dry' formation scenario. In contrast, S\'ersic
ellipticals undergo nuclear starburst events since the AGN feedback is
weaker and the galaxies' shallow potential well is incapable of
retaining hot gas \citep[e.g.,][]{2009Natur.460..213C}.  While the
influence of the AGN activity on the host galaxy structures is not yet
clear, a growing body of observational evidence points to heating by
(episodic) radio AGN from central galaxies to prevent the intracluster
medium from cooling in galaxy groups and clusters
\citep{2007MNRAS.380..877S,2008MNRAS.391..481S,2009ApJ...698..594M,2011ApJ...727...39M,2012ARA&A..50..455F}. High-resolution,
multi-wavelength data are crucial to investigate the central
structures of galaxies, AGN activities and jet structures.

The Legacy $e$-MERLIN Multi-band Imaging of Nearby Galaxies Survey
(LeMMINGs;
\citealt{2014evn..confE..10B,2018MNRAS.476.3478B,2021MNRAS.500.4749B,2021MNRAS.508.2019B})
is designed to reveal the physical origin of the observed relationships between different wavebands by combining high resolution observations from radio ($e$-MERLIN), through optical ({\it  Hubble Space Telescope, HST}) to X-ray ({\it Chandra}).
As part of LeMMINGs, we have recently
completed $e$-MERLIN 1.5 GHz observations of all 280 galaxies above
declination $\delta > +20^{\circ}$ from the Palomar bright
spectroscopic sample of nearby galaxies
\citep{1985ApJS...57..503F,1995ApJS...98..477H,1997ApJS..112..315H,1997ApJ...487..568H,1997ApJS..112..391H}.
Using our $e$-MERLIN 1.5 GHz data in \citet{2021MNRAS.508.2019B} we
reported different radio production mechanisms for the different
optical classes and SMBH masses: above
$M_{\rm BH} \sim 10^{6.5} M_{\sun}$ AGN are the main driver of the
nuclear radio emission in galaxies, whereas below this characteristic
mass stellar processes power the bulk of the radio
emission. Furthermore, our new 5 GHz $e$-MERLIN data for the full
LeMMINGs sample of galaxies have also been calibrated and are
currently in the process of being analysed (Williams et al.\, in prep.). 
To allow a proper understanding of the relationships between the radio properties revealed by $e$-MERLIN  and the structural properties of the host galaxies as revealed by optical observations, we require optical imaging with similar spatial resolution to the $e$-MERLIN observations. Only {\it HST} and {\it James Webb Space Telescope, JWST} can provide such  observations. The main aim of this paper is to provide high quality host galaxy optical structural parameters and then to investigate the relationship between those parameters and the radio parameters.

 To understand the  true nature of these correlations between  optical galaxy structural parameters and  radio core properties, it is necessary  first to
identify and model the
distinct photometric 
components of the galaxies such as bulges, discs and bars, rings, spiral arms, AGN,
nuclear star clusters (NSCs), depleted cores and haloes. Clues as to how bulges form and evolve are thought to have been
imprinted in the shapes of the galaxies' surface brightness profiles,
which exhibit a degree of diversity. Simulations have shown that
multiple major mergers would naturally produce a spheroidal structure
with a large S\'ersic index $n \ga 4.0$ (e.g.,
\citealt{2006MNRAS.370..681N,2006ApJ...636L..81N,2009ApJ...691.1424H,2009ApJS..181..486H}). This
is contrary to galaxies which underwent dissipative, unequal mass
mergers or have been shaped by secular processes, as they possess a
bulge with a low S\'ersic index ($n \la 3.0$). To measure
robust bulge structural properties, and investigate their connection
with sub-kpc radio core properties, the bulge and the remaining galaxy
components should be carefully separated by performing physically
motivated, multi-component decomposition
\citep[e.g.,][]{2005MNRAS.362.1319L,2010MNRAS.405.1089L,2015ApJS..219....4S,2016ApJ...818...47S,2017A&A...598A..32M,2019ApJ...876..155S,2019ApJ...873...85D,2014MNRAS.444.2700D,2015MNRAS.446.4039E,2016ApJS..222...10S,2016MNRAS.462.3800D,2017ApJS..232...21K,2017MNRAS.471.2321D,2019ApJ...871....9D,2019ApJ...886...80D,2021A&A...647A.100S}. When
fits are restricted to account only for the two main galaxy
components, i.e., a bulge-disc profile, as is commonly the case for fits
performed on a large sample of galaxies in an automated fashion (e.g.,
\citealt{2006MNRAS.371....2A,2011ApJS..196...11S}), the flux
contribution from prominent other components  can wrongly brighten
the bulge luminosity and give rise to erroneous central and global
structural parameters for the galaxy
\citep[e.g.,][]{2010MNRAS.405.1089L,2015ApJS..219....4S,2016MNRAS.462.3800D,2019ApJ...871....9D,2021A&A...647A.100S}. This
can lead to wrong conclusions about key correlations such as those
between bulges and SMBHs and radio core properties.  

Poor spatial resolution has also been shown as a key limitation in
deriving galaxy structural properties.  Structural studies that rely
on ground-based data
\citep[e.g.,][]{2005MNRAS.362.1319L,2010MNRAS.405.1089L,2017A&A...598A..32M}
and {\it Spitzer Space Telescope} data
\citep[e.g.,][]{2015ApJS..219....4S,2019ApJ...876..155S,2019ApJ...873...85D}
do not have the appropriate resolution to study sub-kpc structures,
such as depleted cores, AGN, NSCs, nuclear discs and inner bars, which
are commonly detected at {\it HST} resolution.

To date, {\it HST} studies of nearby galaxies have mainly focussed on
early-type galaxies
\citep{1995AJ....110.2622L,2007ApJ...662..808L,1997AJ....114.1771F,1997ApJ...481..710C,2001AJ....121.2431R,2001AJ....122..653R,2003AJ....125..478L,2003AJ....125.2951G,2004AJ....127.1917T,
  2006ApJS..164..334F,2009ApJS..182..216K,
  2011MNRAS.415.2158R,2012ApJS..203....5T,2012ApJ...755..163D,2013ApJ...768...36D,2014MNRAS.444.2700D,2015ApJ...798...55D,2021ApJ...908..134D,2013MNRAS.433.2812K,2013AJ....146..160R,2019ApJ...886...80D,2020A&A...635A.129K}. Multi-component
structural analysis of {\it HST} images for a large sample of
late-type galaxies is lacking.
\citet{1997AJ....114.2366C,1998AJ....116...68C} studied 40 spiral
galaxies using {\it HST} WFPC2 and NICMOS images.  These authors used
the inner (radially limited) $10\arcsec$ {\it HST} light profiles and
restricted their fits to describe only the bulge light distribution
using $R^{1/4}$, $R^{1/4}$+exponential, exponential or
double-exponential models. \citet{2008AJ....136..773F} modelled the
{\it HST} light profiles of 64 spiral galaxies using a two-component
S\'ersic bulge + exponential disc model. The robustness of these fits is limited as
innermost data points ($R \la 1\arcsec $), bars, rings and spiral arms
were excluded from the fits after being subjectively identified
through visual inspections.

Over the past two decades a few studies measured galaxy structural
properties with {\it HST} and explored the relation with radio-quiet
and radio-loud dichotomy
\citep[e.g.,][]{2005A&A...440...73C,2007A&A...469...75C,2006A&A...447...97B,2006A&A...451...35B,2005A&A...439..487D,2010ApJ...725.2426B,2011MNRAS.415.2158R}.
\citeauthor{2005A&A...440...73C} applied the Nuker model, which does
not provide an accurate description of galaxy surface brightness
profiles
\citep{2003AJ....125.2951G,2006ApJS..164..334F,2012ApJ...755..163D},
to radially limited $10\arcsec$ {\it HST} light profiles of early-type
galaxies, resulting in an approximate bulge structural classification
(`core' vs. `power-law'). Subsequent work by
\citet[][see also \citealt{2010ApJ...725.2426B}]{2011MNRAS.415.2158R} fit a S\'ersic, core-S\'ersic and
double-S\'ersic model to the {\it HST} surface brightness profiles of
110 galaxies with Hubble type earlier than Sbc, selected to avoid disc
dominated systems.  They reported that their fits were robust for
62/110 galaxies. However, there are limitations. Most of their fits
($>$ 75\%) rely on radially limited ($R \la 20\arcsec$) NICMOS light
profiles. Also the fits do not account for large-scale galaxy
components such as bars and rings which potentially contribute
significant light to the inner domain of the galaxy surface brightness
profiles.

In this paper we perform accurate multi-component decompositions of
the 1-dimensional {\it HST} surface brightness profiles of a
representative sample of 173 active and inactive galaxies from the
full sample of 280 LeMMINGs galaxies \citep[see
also][]{2016MNRAS.462.3800D,2018MNRAS.475.4670D,2019ApJ...871....9D,2019ApJ...886...80D}. Uniquely,
our sample contains a large number of 108 late-type galaxies. We fit
up to six galaxy components (e.g., bulges, discs, depleted core, AGN,
nuclear star clusters, bars, spiral arms, rings and stellar haloes)
simultaneously to the galaxy surface brightness profiles, which extend
out to large radii of $R \ga 80-100\arcsec$, adequate to accurately
quantify the shapes of the bulge and disc profiles along with other
large-scale galaxy components (e.g., bars, spiral arms, rings and
stelar haloes). Our work, which represents the largest, most detailed
structural analysis of nearby galaxies with {\it HST} to date, offers
the possibility to make a plausible connection between the optical
structural properties of galaxies and their nuclear optical and radio
activities in a homogenous manner and at a much higher resolution than
has hitherto been possible.  Dullo et al.\ (2022, MNRAS, LeMMINGsVI, submitted) investigate the connection between the radio core
luminosity and various properties of the host bulge.

The outline of this paper is as follows. In Section~\ref{Sec2}, we
discuss the LeMMINGs sample, radio and optical emission line data, our
{\it HST} imaging  data and data reduction
details.  Section~\ref{Models} presents an overview of the analytical
models used to describe galaxy components and then discusses the 1D  (one-dimensional) and 2D (two-dimensional)
multi-component decompositions of the galaxy stellar light distributions.    This
section also describes our colour calibration equations we derived to
transform the magnitudes obtained in various filters into $V$-band and
the stellar mass-to-light ratios employed to calculate the stellar
masses from the luminosities.  In Section~\ref{Secn4}, we study the
connection between bulge structural properties, optical emission-line
classes and nuclear radio activity. Finally, in Section~\ref{ConV} we summarise our main results and
conclude.  Notes on our decompositions and comparison with past fits in the 
literature  for 54 selected individual  galaxies are given in Appendix~\ref{NotesI}.
 The data tables, individual galaxy {\it HST} images, 1D  and 2D    decompositions
and our results are presented in Appendices~\ref{DataTables},
\ref{AppendB}, \ref{AppendC} and  \ref{Append2D}, respectively. The  Monte Carlo (MC)
method which we employ to estimate realistic errors on fitted
structural parameters and to test the robustness the 1D multi-component
decomposition of the galaxy light profiles, is described in
Appendix~\ref{AppendD}.

Throughout this paper, we use $H_{0}$
= 70 km s$^{-1}$ Mpc$^{-1}$, $\Omega_{m}$ = 0.3 and $\Omega_{\Lambda}$
= 0.7 \citep[e.g.,][]{2019ApJ...882...34F}, an average of the Planck 2018
Cosmology $H_{0}$ = 67.4 $\pm$ 0.5 km s$^{-1}$ Mpc$^{-1}$
\citep{2020A&A...641A...6P} and the LMC $H_{0}$ = 74.22 $\pm$ 1.82 km
s$^{-1}$ Mpc$^{-1}$\citep{2019ApJ...876...85R}. All magnitudes are in
the Vega system, unless specified otherwise.

\begin{table} 
\begin{center}
\setlength{\tabcolsep}{0.170in}
\begin {minipage}{90mm}
\caption{LeMMINGs {\it HST} data}
\label{Tab01}
\begin{tabular}{@{}llcccc@{}}
\hline
\hline
Galaxies&&&&Notes\\
 (1)&&&&(2)\\          
\hline       
173 (61.8\%)&&&&this work        \\
5 (1.8\%)&  &&  &dust extinction$^{\rm a}$     \\                     
 2 (0.7\%) & & &&merging/interacting$^{\rm  b}$   \\
 31 (11.1\%)& &&& complex morphology$^{\rm c}$      \\
16 (5.7\%)   & &&&data not suitable$^{\rm d}$     \\
 53 (18.9\%)   &&& & no {\it HST} observations$^{\rm e}$  &  \\
 \hline
 280 (100\%)  \\
\hline
\end{tabular} 
The LeMMINGs galaxies that are omitted are those (a) with centres
obscured by dust, (b) that belong to a pair of merging/interacting
galaxies, (c) with peculiar and/or edge-on morphology that render the
extraction and modelling of the galaxy light profiles difficult and
thus deferred to a future work, (d) with {\it HST} images that miss
the galaxy centre, or that were taken in narrow-band or mid-UV (F300W)
filters only and (e) with no {\it HST} observations.
 \end{minipage}
\end{center}
\end{table}

\begin{table} 
\begin{center}
\setlength{\tabcolsep}{0.00033in}
\begin {minipage}{90mm}
\caption{Imaging summary}
\label{Tab02}
\begin{tabular}{@{}llcccccc@{}}
\hline
\hline
&\multicolumn{3}{c}{{\it HST}  imaging from HLA}&\\

Instrument &Image
                                                scale&FOV
                                                       &Shape of &Number&\\
&(arcsec px$^{-1}$)&&the FOV&(our sample)&\\
(1)&(2)&(3)&(4)&(5)\\      
\hline                           
WFPC2/PC              &  0$\farcs$05             &$36\arcsec      \times 36\arcsec$&square&\\                       
WFPC2/WF              &  0$\farcs$10             &$160\arcsec       \times 160\arcsec$&`L'-shaped&104\\
ACS/WFC &  0$\farcs$05               &$202\arcsec\times 202\arcsec$&rhomboid&52\\
NICMOS/NIC2&               0$\farcs$05 & $19$\farcs$2 \times $19$\farcs$2& square&2\\
NICMOS/NIC3&               0$\farcs$10 & $51$\farcs$2 \times                             $51$\farcs$2& square&4\\
WFC3/IR              &  0$\farcs$09             &$123\arcsec                            \times                       136\arcsec$&rectangle&9\\
WFC3/UVIS              &  0$\farcs$04             &$162\arcsec       \times 162\arcsec$&rhomboid&2\\
&\multicolumn{2}{c}{Ground-based imaging}&\\
SDSS  &               0$\farcs$396 &--- & square&75\\
DSS/Schmidt&               1$\farcs$70 & 14$\farcm$9 $\times  $                           $14\farcm$9& square&4\\
 camera& & & \\
\hline
\end{tabular} 
Notes--- The FOV for the SDSS images used in this work is
\mbox{$\ge 9\farcm0$ $\times
  $$9\farcm$0}. The science images retrieved from the HLA are
(re)sampled and aligned north up using the DrizzlePac
\citep{2012drzp.book.....G} and MultiDrizzle
\citep{2003hstc.conf..337K} software packages. For the 104 sample
galaxies, we used the full WFPC2 mosaic images.
\end{minipage}
\end{center}
\end{table}

\begin{table} 
\begin{center}
\setlength{\tabcolsep}{0.00013518827in}
\begin {minipage}{89mm}
\caption{LeMMINGs optical and radio properties}
\label{Tab03}
\begin{tabular}{@{}llcccc@{}}
\hline
\hline
Galaxies&Number&Undetected&core-S\'ersic \\
&(ours/full sample)&(ours/full sample)&(our sample)\\
 (1)&(2)&(3)&(4)&\\          
\hline       
E    &23 (13.2\%)/26(9.3\%)& 52.2\%/50.0\% &65.2\%  \\
S0 & 42 (24.3\%)/55(19.6\%)&   40.5\%/41.8\%& 11.9\%\\                     
S  &102 (59.0\%)/189(67.1\%)&   54.0\%/60.3\% & 0\% \\
Irr  &6 (3.5\%)/10(3.6\%) & 66.7\%/70.0\%& 0\%  \\
\hline
Seyfert & 10 (5.8\%)/18 (6.4\%)& 20.0\%/27.8\% &  0\%    \\
ALG &  23 (13.3\%)/28 (10.0\%)   &78.3\%/75.0\%  &34.5\%  \\
LINER &71 (41\%)/94 (33.6\%)   &30.1\%/38.3\%& 15.5\%\\
{\sc h ii}  &69 (39.9\%)/140 (50.0\%) &     67.6\%/66.4\%& 1.7\%\\
\hline
Total&173 (100\%)/280 (100\%) &52.0\%/56.1\%&20/173(11.6\%)\\ 
\hline
\end{tabular}     
Notes--- The sample galaxies are first separated based on the galaxy
morphological and optical spectral classes (cols 1 - 2) and then
further divided based on their radio non-detection and core-S\'ersic
type central structure (cols 3 - 4). The term `full sample' refers to
the total LeMMINGs sample of 280 galaxies, whereas the term `our
sample' refers to the sub-sample of 173 LeMMINGs galaxies studied in
this paper.
 \end{minipage}
\end{center}
\end{table}

\section{The LeMMINGs sample and data}\label{Sec2}

The LeMMINGs
\citep{2014evn..confE..10B,2018MNRAS.475.4670D,2018MNRAS.476.3478B,2021MNRAS.500.4749B,2021MNRAS.508.2019B}
is the deepest, high-resolution radio survey of the local Universe and
constitutes radio continuum observations of 280 galaxies for a total
of 810 hrs with $e$-MERLIN at 1.25$-$1.75 GHz ($L$-band)  and $C$-band, centred at 5 GHz, with a 512 MHz band width, complemented
by synergies from high-resolution, multi-band (X-ray, optical and
near-IR) data. The survey sample is a subset of the magnitude-limited
($B_{T} \le 12.5$ mag and declinations $\delta > 0^{\circ}$) Palomar
spectroscopic sample of 486 bright, nearby galaxies
\citep{1995ApJS...98..477H,1997ApJS..112..315H}, which in turn were
drawn from the Revised Shapley-Ames Catalog of Bright Galaxies
\citep{1981rsac.book.....S} and the Second Reference Catalogue of
Bright Galaxies \citep{1976srcb.book.....D}. In order to discriminate
between star-formation and AGN activities in Palomar galaxies,
\citet{1995ApJS...98..477H,1997ApJS..112..315H} utilised optical
emission-line ratios and classified the emission-line nuclei into four
spectral classes: Seyferts, LINERs, \mbox{H\,{\sc ii}} and Transition
galaxies. For LeMMINGs galaxies,
\citet{2018MNRAS.476.3478B,2021MNRAS.500.4749B} used the diagnostic
line ratios mostly from
\citet{1985ApJS...57..503F,1995ApJS...98..477H,1997ApJS..112..315H,1997ApJ...487..568H,1997ApJS..112..391H}
plus new spectroscopic classifications from the literature and revised
the classification for the entire sample employing the emission line
diagnostic diagrams by \citet{2006MNRAS.372..961K} and
\citet{2010A&A...509A...6B}.  In this revised classification, used in
this paper, each object with emission lines was categorised as
\mbox{H\,{\sc ii}}, Seyfert or LINER after Transition galaxies
\citep{1995ApJS...98..477H} were identified either as LINER or
\mbox{H\,{\sc ii}}. There are also galaxies in the LeMMINGs sample
which lack emission lines and were classified as Absorption Line
Galaxies (ALGs), \citep{2018MNRAS.476.3478B,2021MNRAS.500.4749B}.

\begin{figure}
\hspace{-01.0271413cm}
\includegraphics[trim={-1.9cm -8.7cm 0 1.08703cm},clip,angle=0,scale=0.4207]{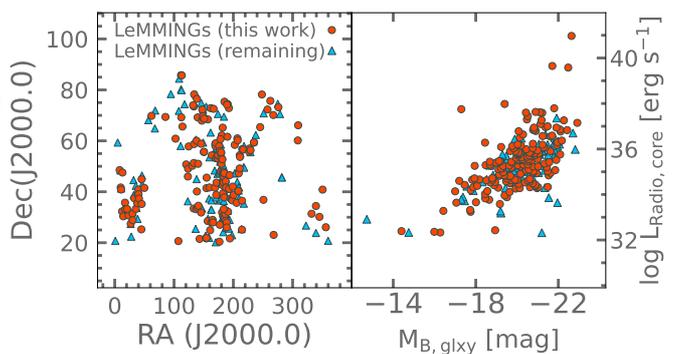}
\vspace{-4.291025cm}
\caption{The distribution of the full sample of 280 LeMMINGs galaxies.
  Left panel: shown in red are the 173 LeMMINGs galaxies with decent
  {\it HST} imaging studied in this paper, the remaining 107  (shown in blue) are galaxies either  (a) with centres
obscured by dust, (b) that belong to a pair of merging/interacting
galaxies, (c) with peculiar and/or edge-on morphology, (d) with {\it HST} images that miss
the galaxy centre, or that were taken in narrow-band or mid-UV (F300W)
filters only and (e) with no {\it HST} observations. Right panel: a plot of the radio core luminosity
  ($L_{\rm Radio,core}$) as a function of total galaxy $B$-band
  absolute magnitude from Hyperleda ($M_{B,\rm glxy}$). The sample studied in this work is representative of the  full LeMMINGs sample which is constructed  by selecting all the  galaxies  in Palomar
spectroscopic sample \citep{1995ApJS...98..477H,1997ApJS..112..315H} with $\delta > 20^{\circ}$.}
\label{Fig1}
\end{figure}

To
allow better radio visibility coverage for the $e$-MERLIN array, all
LeMMINGs galaxies were selected to have $\delta > +20^{\circ}$ (see
Fig.~\ref{Fig1}). An extensive description of this legacy survey
goals, radio data reduction technique, radio detection, maps and flux
measurements is presented in
\citet{2018MNRAS.476.3478B,2021MNRAS.500.4749B}. With an angular
resolution of $\sim 0\farcs15$ and sub-mJy sensitivity afforded by
e-MERLIN at 1.5 GHz, we were able to measure radio core luminosities
\mbox{$L_{\rm R,core} \sim 10^{34} - 10^{40}$ erg s$^{-1}$}.
Furthermore, the analysis of Chandra X-ray observations for 213
LeMMINGs galaxies is presented in \citet{2022MNRAS.510.4909W}.

\begin{figure}
\hspace{.347cm}
\includegraphics[angle=0,scale=0.25197342]{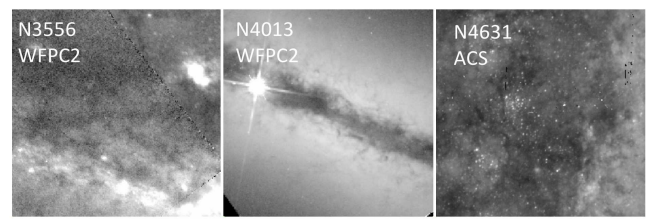}
\vspace{-.12cm}
\caption{Central $25\arcsec \times 25\arcsec$ regions
  of three LeMMINGs galaxies, taken with {\it HST}, which have their
  centres copiously obscured by dust (i.e., dark areas).
  Consequently, the galaxies were excluded from our analysis as the
  extraction of surface brightness profiles was not possible.  }
\label{FigDust}
\end{figure}

\begin{figure}
\hspace{-1.1837cm}
\includegraphics[trim={-2.9915cm -8cm -3.cm .8900cm},clip,angle=0,scale=0.3529]{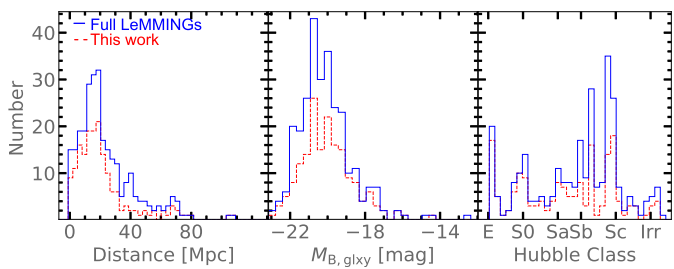}
\vspace{-3.4505263cm}
\caption{Histograms of distance (mostly from NED), absolute $B$-band
  galaxy magnitude and from HyperLeda ($M_{B, \rm glxy}$) and
  morphological T-type shown for the subset of 173 LeMMINGs galaxies
  in this work and the full, statistically complete sample of 280
  LeMMINGs galaxies.  The Hubble classes based on the numerical T-type
  and absolute $B$-band galaxy magnitudes are from HyperLEDA. }
\label{Fig2}
\end{figure}

In this work we perform accurate multi-component analysis of
the central and global structures for LeMMINGs/Palomar galaxies with
{\it HST} imaging data (Tables~\ref{Tab01}, \ref{Tab02}).  Our sample
comprises 173 galaxies ($\sim62$\%) from the full sample of 280
LeMMINGs galaxies (Tables~\ref{Tab01} and \ref{Tab03}).  As the best
balance between image availability and immunity to the obscuring
effect of dust, preference was given to {\it HST} images taken with
the redder optical band filter F814W. We did not use {\it HST} images
taken in narrow-band or mid-UV (F300W) filters. Of the 107 LeMMINGs
galaxies excluded in this work, 31 have peculiar and/or edge-on
morphologies that make the extraction and modelling of the galaxy
light profiles difficult and thus deferred to a future work, five
exhibit centres markedly obscured by dust (See~Fig.~\ref{FigDust}),
two belong to a pair of merging/interacting galaxies, 16 have {\it
  HST} images which miss the galaxy centre or were taken in
narrow-band or mid-UV (F300W) filters and 53 lack {\it HST}
observations (see Table~\ref{Tab01}). We rely on newly extracted {\it
  HST} surface brightness and associated geometrical profiles for
163/173 LeMMINGs galaxies, augmented by the {\it HST} structural data
from our previous work for the remaining 9 galaxies
(\citealt{2014MNRAS.444.2700D,2016MNRAS.462.3800D,2018MNRAS.475.4670D},
see Fig.~\ref{Fig2} and Appendix~\ref{DataTables}).

As one of the goals of this paper is to find out how radio luminosity
is related to optical properties, it is important to investigate the
parameter space coverage for our sample of 173 LeMMINGs galaxies and
ensure that the sample is representative of the full LeMMINGs
sample. As noted above the parent Palomar sample is statistically
complete and so is the full LeMMINGs sample as it represents all the
Palomar galaxies with $\delta > 20^{\circ}$. Fig.\ref{Fig1}~ shows the
distribution for our 173 galaxies and the remaining 107 LeMMINGs
galaxies. The two-dimensional Kolmogorov-Smirnov (KS) test
\citep{1983MNRAS.202..615P,1987MNRAS.225..155F} on these two
subsamples (Fig.\ref{Fig1}, left) yields a significance level
$P \sim 0.4$, thus we cannot reject the null hypothesis that the subsample are drawn from the same distribution.
In Fig.\ref{Fig1} (right) we also show a plot of
the radio core luminosity ($L_{\rm R,core}$) as a function of total
galaxy $B$-band absolute magnitude ($M_{B,\rm glxy}$) from Hyperleda
\citep{2014A&A...570A..13M}. It is evident that our sample provides
good coverage of the entire range of radio core luminosity and galaxy
luminosity afforded by the full LeMMINGs sample. We also employ the
2D KS test on the ($L_{\rm R,core}$, $M_{B,\rm glxy}$) dataset for the
subsample used in this work and the full LeMMINGs sample, finding a significance level of $P \sim 0.3$ which indicates    that 
 the null hypothesis, i.e.,  the ($L_{\rm R,core}$,
$M_{B,\rm glxy}$) data for the two samples are drawn from the same
distribution, cannot be rejected  and that our sample is representative (albeit
being statistically incomplete). Furthermore, Fig.~\ref{Fig2} also shows that our sample
covers a wide range in distance ($D\sim 0.7-100$ Mpc) and
morphological type between E and Im
(\citealt{1926ApJ....64..321H,1959HDP....53..275D}). These important
galaxy property considerations make the sample well suited for robust
characterisation of galaxy structural scaling relations.  The mean
distance for our sample is $\sim22$ Mpc, which translates to spatial
scales by {\it HST} of $5-10$ pc.  Details of our sample are
summarised in Tables~\ref{NewTabA1}$-$\ref{Tab4}.

 \subsection{Archival {\it HST}  and ground-based imaging data}\label{ArcIm}

 High-resolution {\it HST} images for the 163 target galaxies were
 retrieved from the Hubble Legacy Archive (HLA\footnote{HLA is the
   site for high-level {\it HST} archival products at \url
   {https://hla.stsci.edu.}}). These images were processed using the
 HLA pipeline which performs the standard data reduction procedures
 including bias subtraction, geometric distortion correction, dark
 current subtraction and flat fielding. The galaxy images were
 observed with the {\it HST} ACS, WFPC2, NICMOS or WFC3 cameras and
 taken in the following filters: $F547M$ (equivalent to the
 Johnson-Cousin $V$-band), $F555W$ ($\sim V$-band), $F606W$ (broad
 \mbox{$\sim V$-band}), $F702W$ (Cousins \mbox{$\sim R$-band}),
 $F814W$ (\mbox{$\sim I$-band}), $F850LP$ ($\sim $ SDSS-$z$ band),
 $F110W$ ($\sim J$-band) and $F160W$ ($\sim$ $H$-band).

 The fitted radial extent is of critical importance when decomposing a
 galaxy's surface brightness profile.  The recovery of accurate model
 parameters in particular for large-scale ($\sim 3.5 \pm 2.0$ kpc)
 galaxy components such as discs, rings and spiral arms is prone to
 how well the curvature of the light profile\footnote{The light
   profile of a galaxy is the radial distribution of its stellar
   light.}  that represents the outer most galaxy components is
 sampled by the data.  A related concern is the uncertainty in the
 shapes of the 1D light profiles of the outer galaxy components due to
 sky background errors.  {\it HST} images do not always probe a
 sufficiently large radial domain to perform a firm assessment of the
 background level and to sample the outer parts of the surface
 brightness profiles for nearby galaxies.  For 81/173 (46.8\%) of the
 sample galaxies their large-scale structures (e.g., discs, rings and
 spiral arms or low surface brightness stellar envelopes) extend
 beyond the FOV of the high-resolution {\it HST} images. Measuring
 accurate structural parameters for such galaxies may necessitate the
 high-resolution {\it HST} data at small radii to be matched and
 combined with radially extended ground-based data at large radii (see
 e.g.
 \citealt{2016MNRAS.462.3800D,2018MNRAS.475.4670D,2019ApJ...886...80D}). To
 extract 1D light profiles we therefore use ground-based
 \mbox{SDSS-$r,i,z$} and \mbox{DSS (103aE, i.e., optical red)} band
 images\footnote{\url{https://www.cadc-ccda.hia-iha.nrc-cnrc.gc.ca/en/dss/}}
 for the galaxies, which are downloaded from the SDSS
 \citep{2000AJ....120.1579Y} Data Release 13
 (DR13\footnote{\url{http://www.sdss.org/dr13}},
 \citealt{2017ApJS..233...25A}) and
 NED\footnote{\url{http://nedwww.ipac.caltech.edu}}, respectively (see
 Table~\ref{Tab02} and Section \ref{SB}).

A summary of the {\it HST} imaging, instrument and filter is presented in
Tables~\ref{Tab1}, \ref{Tab1Ring} and  \ref{Tab1cS}.

 The extraction of accurate surface photometry for the sample galaxies
 proceeds in several steps: sky background subtraction
 (Section~\ref{SkYB}), creation of mask files (Section~\ref{Msk}) and
 then fitting elliptical isophotes to the galaxy images using the IRAF
 task {\sc ellipse} and for some sample galaxies the creation of
 composite surface brightness, ellipticity, P.A. and $B_{4}$ profiles
 by matching {\it HST} and large FOV ground-based data
 (Section~\ref{SB}).

 \subsection{Sky background subtraction}\label{SkYB}

 As noted above, 81/173 sample galaxies extend beyond the FOV of the
 high-resolution {\it HST} images, as such background over-subtraction
 by the HLA pipeline is a potential concern. Indeed, we 
   find that,  because of sky over-subtraction by the HLA,  for most angularly extended galaxies in the sample the {\it HST}-derived brightness profiles tend to depart downward with respect to the SDSS profiles at radii of
   $R\ga$50-60$\arcsec$ (or $R\ga$10$\arcsec$ for NICMOS images. Therefore, for such
 galaxies we extract light profiles from the large FOV SDSS/DSS images
 and matched them with our inner {\it HST} data (as detailed in
 Section~\ref{SB}) to construct composite (i.e., inner {\it HST} plus
 outer ground-based) light profiles.  The SDSS images used for
   the 81 sample galaxies have $\ge 9\farcm$0 $\times$ 9$\farcm$0 FOV,
   considerably larger than the median major-axis diameter for our
   galaxies which is 2$\farcm$6 $\pm$ 2$\farcm$0. This allowed us to
   characterise the background level for our 1D and 2D decompositions
   more accurately. 
 
 For the remaining 92 sample galaxies, the quality of the HLA pipeline
 sky subtraction is examined by computing manually the median sky
 levels from several ($>5$) $20 \times 20$ pixel boxes at the edges of
 the {\it HST} CCDs. We find that the galaxy flux in the outermost
 parts is 10\% of the {\it HST} sky level, i.e., \mbox{$\sim$21.42
   $\pm$ 0.33 mag arcsec$^{-2}$} $I$-band, which converted here from a
 $V$-band value \citep{2005AJ....129.2138L} using $V-I$ = 1.08
 \citep{1995PASP..107..945F}.  We ensure that the average of the
 medians is near zero ($\sim 0.004 \pm 0.004$ e/s) when the galaxies
 are within the field of view of the {\it HST} CCDs.

 \subsection{Masking of images}\label{Msk}

 We inspected the galaxy images visually for dust structures, gaps
 between individual {\it HST} CCD detectors, image defects, bright
 foreground stars and background galaxies, which were then masked by
 taking extra care as discussed here. For each galaxy, an initial mask
 region file was first created using {\sc SExtractor}
 \citep{1996A&AS..117..393B}. To achieve this, we used a threshold
 background value, detection threshold of (typically) $2\sigma$ above
 the background and convolved the galaxy's image with a Gaussian
 filter {\sc gauss\_2.0\_3$\times$3} to generate catalogue sources
 with their image coordinates. This was first converted to a
 DS9\footnote{\url{ SAOImageDS9https://ds9.si.edu}} region file and
 then into a mask file using {\sc iraf}.  The initial mask was then
 added to a manual mask. In some cases we first removed a model galaxy
 image, defined in terms of the isophotal parameters derived for the
 science image, created by the {\sc iraf} task {\sc bmodel}. This was
 crucial to better identify dust-obscured regions, contaminating light
 sources and thus improve the initial mask.  Finally, we run the {\sc
   iraf} {\sc mskregions} task to create mask files with `.pl'
 extension, automatically recognisable by the {\sc iraf} task {\sc
   ellipse} (see Section~\ref{SB}).
   
 \subsection{Surface brightness profiles }\label{SB}

 The extraction of major-axis surface brightness profiles for our
 galaxies was carried out following data reduction steps similar to
 those described in
 \citet{2016MNRAS.462.3800D,2017MNRAS.471.2321D,2019ApJ...871....9D,2019ApJ...886...80D}.
 The {\sc iraf} task {\sc ellipse} was used to fit elliptical
 isophotes to the sky-subtracted galaxy images along logarithmically
 increasing semi-major axis \citep[see also
 \citealt{1983ApJ...266..562K}]{1987MNRAS.226..747J}, hence giving the
 most weight to the innermost portion of the galaxy.  We used
 $3\sigma$ clipping for flagging deviant pixels from contamination by
 cosmic rays and image defects.  {\sc ellipse} samples the 2D
 intensity distribution in the images as a function of an azimuthal
 angle $\phi$ starting from a first guess elliptical isophote defined
 by initial values of the isophote centre, position angle (P.A.) and
 ellipticity, $\epsilon = 1-b/a$, where $a$ and $b$ are the semi-major
 and semi-minor axes of the isophote, respectively. While $\epsilon$
 and P.A. were always set free to vary, the isophote centres were held
 fixed for most galaxies. When concentric isophotes are assumed for a
 galaxy that is actually well modelled by a nested isophotes and has a
 centre that wobble by few pixels it only slightly affects the galaxy
 surface brightness profile and the associated best-fitting structural
 parameters for the galaxy remain unchanged. For eight galaxies (NGC
 2770, NGC 2782, NGC 3034, NGC 3448, NGC 4605 and NGC 5474) with a low
 signal-to-noise ratio and complex morphology, the extraction of
 isophotal parameters and surface brightness profiles with {\sc
   ellipse} was difficult. To overcome the problematic nature of such
 galaxies we ran {\sc ellipse} multiple times using several sets of
 conditions: different initial semi-major axis lengths, occasionally
 fixing the ellipticity and P.A., increasing the logarithmic spacing
 between successive ellipses and in a few cases a residual image was
 created after a model image was built from {\sc ellipse} table
 (`.tab') file. The surface brightness profiles were then extracted
 when convergence is achieved with {\sc ellipse} and the residual
 images do not show strong galaxy light. We classify the fits to these
 eight galaxies' surface brightness profiles as Quality 2 (see
 Seciton~\ref{FitIN}).

 A Fourier expansion of the intensity distribution with an average
 intensity $ I_{0}$ can be written as

 \begin{equation}
 I(\phi) = I_{0} +    \sum_{j} [A_{j}~{\rm sin}~ (j\phi) + B_{j}~{\rm cos}~ (j\phi)].
  \label{Eq1}
 \end{equation}
 
 Higher order coefficients of the Fourier series ($j \ge 3$) carry
 crucial information as they quantify any deviations of the isophotes
 from perfect ellipses. Of particular importance is the coefficient of
 the cos($4\phi$) term namely the fourth-order moment $B_{4}$: a positive
$B_{4}$ value indicates the isophote is `discy', whereas negative $B_{4}$
 values signify that the isophotes are `boxy'.
 
 For edge-on disc galaxies, \citet{2015ApJ...810..120C} implemented
 modifications to {\sc iraf}/{\sc ellipse} task and introduced a new
 fitting routine dubbed {\sc isofit} to better represent the galaxy
 isophote shapes. While this latter task was not employed in this work
 to extract the light profiles for the 13 (7\%) sample galaxies with
 high inclination as determined from the galaxy axial ratio
 $i \ga 75^{\circ}$, our fit quality flag, which quantifies the
 reliability of the 1D light profile decompositions, reflects the
 effect of inclination (see Table~\ref{Tab1}). 
  Nonetheless, we find  a small dispersion on the best-fitting parameters, comparing, for each of  these 13 galaxies,  the decompositions of the original  1D light profile and its  ($\ge$100) simulations  (see Section~\ref{FitIN} and 
   Appendix~\ref{AppendD}).  Also, we perform 2D decompositions for six
   of the 13 high inclination galaxies, finding good agreement between
   the 1D and 2D modelling (see Section~\ref{Sec2V1} and 
   Appendices~\ref{AppendC} and \ref{Append2D}). As such, we decided to include all the 13
   galaxies in our analyses.  In the future, we plan to fit a 2D,
       edge-on exponential disc model using {\sc imfit}
       \citep{2015ApJ...799..226E} and remodel some of the high
       inclination sample galaxies to further explore the robustness
       of the galaxies' structural parameters. Likewise, for the 1D
       decompositions, {\sc isofit} will be used to extract new light
       profiles for the galaxies.

 In constructing composite surface brightness, ellipticity, P.A. and
 $B_{4}$ profiles for the 81 sample galaxies for which we used
 ground-based imaging to extend the {\it HST} FOV, see
 Section~\ref{ArcIm}, we manually shifted the ground-based profiles to
 match {\it HST} data by applying constant zero-point
 offsets. With the exception of two
 galaxies (NGC~3198 and NGC~3665), we could ensure that the {\it HST}
 and ground-based images from SDSS and DSS for the individual galaxies
 are taken with filters that are similar or close (e.g.,  {\it HST} $F814W$ and
  SDSS-$i$;   {\it HST} $F606W$ and SDSS-$r$), thus tracing
 similar stellar populations with no obvious colour gradients.  For
 NGC~3198 and NGC~3665, data from two different filters (NICMOS
 $F160W$ and SDSS-$z$) are combined. The matching procedure resulted 
   in excellent overlap between the two sets of
 data over a large radial range (i.e., typically
 $R\sim 2\arcsec-50\arcsec$), but the range is smaller when {\it HST}
 NICMOS and ground-based images are used ($R\sim
 2\arcsec-10\arcsec$). We find poor sky subtraction by the HLA 
 pipeline has negligible effect on  our  {\it HST}-derived light profiles
  at $R< 50\arcsec$ ($R< 10\arcsec$ for the NICMOS data).
 Note that the ground-based SDSS and DSS
 profiles depart downward with respect to the {\it HST} profiles
 inside $R\sim 1\arcsec$ where the blurring from seeing in the
 ground-based images becomes important.  
 
 Overall, the galaxy light profiles extend out to semi-major axis
 $R \ga 80-100\arcsec$, covering $\ga 2R_{\rm e,bulge}$ for 97\% of
 the sample.  The light profiles provide large ranges in surface
 brightness down to 26.0 mag arcsec$^{-2}$ in $I$-band with a $1 \sigma$
 error of 0.14 mag arcsec$^{-2}$.  Furthermore, we follow similar
 steps as in \citet{2006A&A...454..759P} and determine from SDSS-$i$
 images the limiting surface brightness ($\mu_{lim}$) as the surface
 brightness in which the galaxy profiles perturbed by
 $3\sigma_{\rm sky}$ deviate by 0.2 mag from the actual galaxy light
 profiles, where $\sigma_{\rm sky}$ is the uncertainty on the sky
 level measured using 20, 100 $\times$100 pixel boxes.  This yields
 \mbox{$\mu_{lim, I} \sim 24.0$ mag arcsec$^{-2}$}.  For the sample
 galaxies with SDSS-$i$ data, the median surface brightness value of
 the outermost data points in the light profiles is $\mu_{med,I}$
 \mbox{24.0 $\pm$ 0.5 mag arcsec$^{-2}$}. This suggests that the main
 bodies of the large-scale galaxy components such as bars, discs,
 rings and spiral arms in the light profiles are brighter than the
 surface brightness limit. That is, owing to the large radial extent
 of the light profiles, we are able to characterise and quantify the
 properties of bulges and constrain the shape of the stellar light
 distribution that describes large scale galaxy components. We quote
 all magnitudes in the VEGA magnitude system.
 
\begin{figure}
\hspace{-.6cm}
\includegraphics[trim={-1.42522cm -10.4cm -3.8cm -.cm},clip,angle=0,scale=0.375819027]{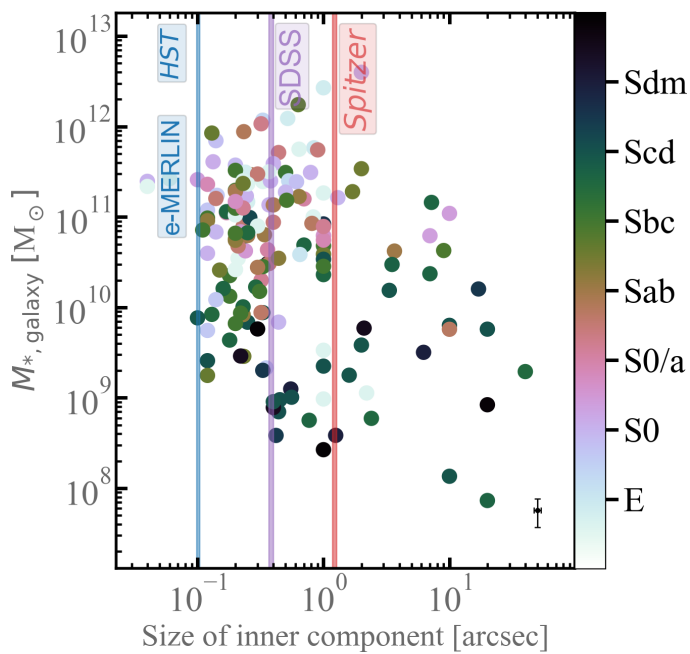}
\vspace{-4.3590cm}
\caption{Distribution of inner, sub-kpc stellar structures (point
  sources, nuclear star clusters, discs and rings, bars, and depleted
  stellar cores) detected in our photometric decompositions.  The size
  of the sub-kpc structure is defined to be the break radius for
  galaxies with a depleted core (i.e., core-S\'ersic galaxies),
  whereas for others the size determined by eye from the fits
  indicates the radius where the sub-kpc structure contributes
  significantly to the galaxy light. While a core-S\'ersic galaxy can
  be nucleated, here we only consider its depleted core as a sub-kpc
  structure.  The overwhelming majority (95\%) of our sample have sub-kpc
  stellar structures that are detected in our analyses of the {\it
    HST} imaging with a high spatial resolution $\le 0\farcs1$ (blue
  vertical line).  With high sensitivity $e$-MERLIN $L$-band (1.5 GHz)
  observations, we detected nuclear radio emission, at high resolution
  of $\le 0\farcs 2$, as such close comparison can be made with the
  analogous optical sub-kpc structure in the {\it HST} data. The red
  and magenta vertical lines delineate the spatial resolutions for
  SDSS and {\it Spitzer} images. While we show the pixel scale of
  SDSS, the actual resolution is set by the seeing, that is typically
  larger than $\sim 1\farcs{5}$ and varying across the sky.  Accounting
  for the effect of the PSF, only 15\% (22\%) of the kpc-scale
  structures from {\it HST} can be well resolved with {\it Spitzer}
  (SDSS) images.  Furthermore, we see no clear trend between the
  physical size of the inner, sub-kpc structures and total galaxy
  stellar mass ($M_{*,\rm glxy}$) or T-type.  The total galaxy mass
  ($M_{*,\rm glxy}$) is based on our light profile decompositions
  (Table~\ref{Tab4}) and the Hubble classes are based on the numerical
  T-type in HyperLEDA. A representative error bar is shown at the
  bottom. }
  \label{Innerkpc}
\end{figure}

\section{Decomposing the light profiles of early- and late-type galaxies} \label{Models}

\subsection{Light profile models} \label{Mods}

The analytic description of the stellar light distribution of galaxies
provides a means to quantify their photometric structures and to
examine how these structural properties change across galaxies,
wavelength and environment. A number of analytic functions have been
used to describe the stellar light distributions in early- and
late-type galaxies.  The three-parameter
\citet{1963BAAA....6...41S,1968adga.book.....S} $R^{1/n}$ model, a
generalisation of the two-parameter \citet{1948AnAp...11..247D}
$R^{1/4}$ has been shown to represent the stellar light profiles of
the main body of elliptical galaxies and the bulges of disc galaxies
very well over a large luminosity and radial range (e.g.,
\citealt{1993MNRAS.265.1013C,1994MNRAS.271..523D,1994MNRAS.268L..11Y,1995MNRAS.275..874A}). This
model can be defined as
 \begin{equation}
 I(R) = I_{e} \exp \left[ - b_{n}
 \left(\frac{R}{R_{e}}\right)^{1/n}-1\right],
  \label{EqIII8}
 \end{equation}
 where the S\'ersic index $n$ (a good proxy for galaxy light
 concentration; \citealt{2001ApJ...563L..11G}) describes the shape
 (curvature) of the radial brightness profile and $ I_{e}$ denotes the
 intensity at the half-light radius $R_{e}$. The variable $b_{n}$,
 defined to ensure that the half-light radius encloses half of the
 total luminosity, is determined by solving numerically
 $\Gamma(2n)=2\gamma(2n,b_{n})$, where $\Gamma(n)$ and $\gamma(n,x)$
 are the complete and incomplete gamma functions, respectively. For
 $1\la n\la 10$, $b_{n} \approx 2n- \frac{1}{3}$
 \citep{1993MNRAS.265.1013C}. When $n =$ 0.5 the S\'ersic model
 reduces to a two-parameter Gaussian function, whereas for $n=1$ it
 yields an exponential function. We use the S\'ersic model to describe
 the light profile of the bar component in disc galaxies.
 
 The luminosity for a S\'ersic model
 component within any radius $R$ can be determined as follows:

 \begin{equation}
 L_{\rm T,Ser}(<R) = I_{\rm e} R^{2}_{\rm e} 2 \pi n \frac{e^{b_{n}}}{(b_{n})^{2n}} \gamma (2n,x),
  \label{Eqq9}
 \end{equation}
 where $\gamma (2n,x)$ is the incomplete gamma function and $x=
 b_{n}(R/R_{\rm e})^{1/n}$ \citep{2005PASA...22..118G}. 

 In general, the radial intensity distribution of the discs of spiral
 and lenticular galaxies can be well modelled with an exponential
 function \citep{1970ApJ...160..811F}, given as
\begin{equation}
I(R)=I_{0,\rm d} \exp\left[\frac{-R}{h}\right],~~~~~~~~~~~~~~~~~~
\label{Eqq3}
\end{equation}
 where $I_{0,\rm d}$ and $h$ are the central intensity and scale length of
 the disc, respectively.

 We adopt the three-parameter Gaussian radial profile to describe the
 radial intensity distribution of the outer ring and spiral-arm
 components of the sample galaxies, given by
\begin{equation}
I(R)=I_{0,\rm r} \exp\left[\frac{-(R-R_{\rm r})^{2}}{2\sigma^{2}}\right],~~~~~~~~~~~~~~~~~~
\label{Eqq3}
\end{equation}
where $I_{0,\rm r}$ denotes the highest intensity value of the Gaussian
ring with a semi-major axis $R_{\rm r}$ and width $\sigma$. Setting
$R_{\rm r} =0$ yields a two-parameter Gaussian function, which is used
in this paper to model additional nuclear light components (i.e., AGN
and nuclear star clusters).

The bulge component of luminous early-type (core-S\'ersic) galaxies
have long been known to contain a depleted core (i.e., a luminosity
deficit), exhibited as a flattening of the inner stellar light
distribution relative to the inward extrapolation of the outer
S\'ersic profile
\citep[e.g.,][]{1978ApJ...222....1K,1978ApJ...221..721Y,1985ApJ...292..104L,2003AJ....125.2951G,2004AJ....127.1917T,2006ApJS..164..334F,2012ApJ...755..163D,2019ApJ...886...80D}. In
contrast, S\'ersic (elliptical, S0, spiral and irregular) galaxies do
not have such a luminosity deficit. Throughout this work, we regard
bulgeless spiral and irregular galaxies to be S\'ersic galaxies. For
core-S\'ersic galaxies, the radial intensity distributions are well
described by the core-S\'ersic model which is a blend of an inner
power-law and an outer S\'ersic model with a transition region
\citep{2003AJ....125.2951G}.  This model is defined as

 \begin{equation}
 I(R) =I' \left[1+\left(\frac{R_{\rm b}}{R}\right)^{\alpha}\right]^{\gamma /\alpha}
 \exp \left[-b\left(\frac{R^{\alpha}+R^{\alpha}_{\rm b}}{R_{\rm e}^{\alpha}}
 \right)^{1/(\alpha n)}\right], 
 \label{Eqq10}
  \end{equation}
 with 
 \begin{equation}
 I^{\prime} = I_{b}2^{-\gamma /\alpha} \exp 
 \left[b (2^{1/\alpha } R_{\rm b}/R_{\rm e})^{1/n}\right],
 \label{Eqq11}
 \end{equation}
 where $I_{b}$ is the intensity at the core break radius $ R_{b}$,
 $\gamma$ is the slope of the inner power-law profile, and $\alpha$
moderates  the sharpness of the transition between the inner power-law
 and the outer S\'ersic profile.  The half-light
 radius  of the outer S\'ersic model  represented by $R_{e}$ and the quantity $b$ has the same
 meaning as in the S\'ersic model (Eq.~\ref{EqIII8}). The total
  luminosity for the core-S\'ersic model  \citep[see e.g.,][their Eq.~A19]{2004AJ....127.1917T} can be written as  

 $L_{\rm T,cS}=2\pi I'n(R_{\rm e}/b^{n})^{2}\int\limits_{b(R_{\rm b}/R_{\rm e})^{1/n}}^{+\infty}e^{-x}x^{n(\gamma+\alpha)-1}$

 \begin{equation}
 ~~~~~~~~~~~~~~~~~~~~~~~~~~~~~~~~~\times\left[x^{n\alpha}-(b^{n}R_{\rm b}/R_{\rm e})^{\alpha}\right]^{(2-\gamma-\alpha)/\alpha}dx.~~~~~~~
 \label{Eqq37}
  \end{equation}
For $R_{\rm b}  \rightarrow  0$,  this expression becomes  the S\'ersic expression given by Eq.~\ref{Eqq9}.

 \begin{figure*}
\hspace{.9202912cm}
\includegraphics[trim={-1.47cm -11cm -8.26cm 2.47973cm},clip,angle=0,scale=0.39]{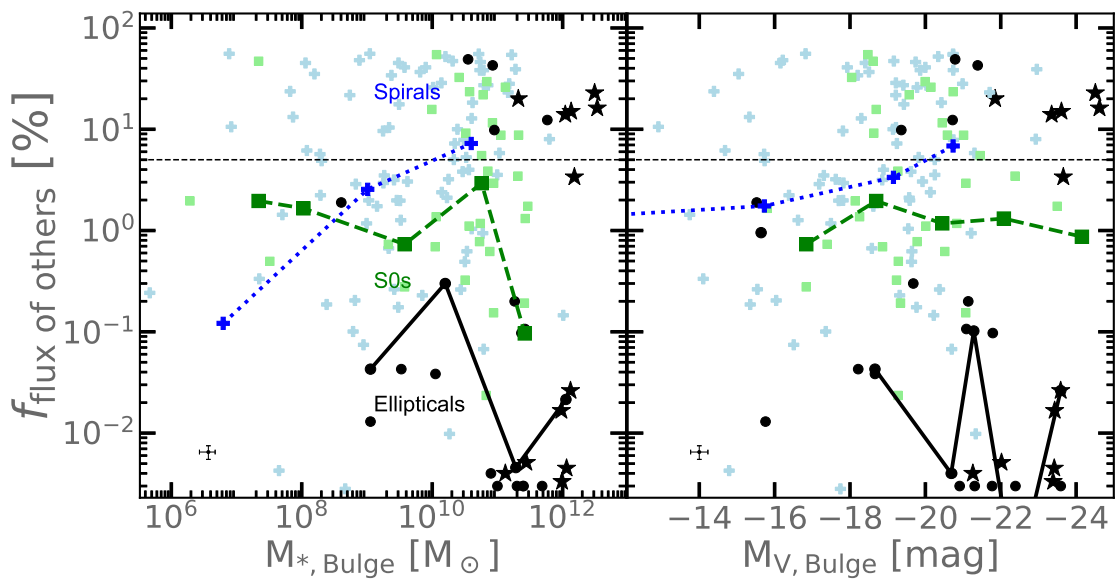}
\vspace{-4.9999027088560cm}
\caption{The fraction of galaxy light that is not ascribed to the
  bulge and disc components ($f_{\rm other}$) as a function of the
  bulge stellar mass and absolute $V$-band bulge magnitude corrected
  for Galactic and internal extinction. S0 and spiral galaxies are
  shown as green boxes and blue crosses, whereas the elliptical galaxies from
  this work are denoted by black circles.  We also add 13 massive elliptical
  galaxies (stars) from
  \citet{2014MNRAS.444.2700D,2019ApJ...886...80D}. Excluding irregular
  galaxies, the fraction of LeMMINGs galaxies having
  $f_{\rm other} \ge 20$\% is $\sim$26\% (43/167), this figure
  increases to $\sim$40\% (67/167) for $f_{\rm other} \ge 5$\% (i.e., above the dashed horizontal line).  The blue,
  green and black tracks trace the median trend for the binned spiral,
  S0 and elliptical galaxy data. Most (71\%) elliptical galaxies have
  $f_{\rm other} \sim 0$. A track is not shown for our irregular
  galaxies which are excluded due to their small number (6 objects).
  Representative error bars are shown at the bottom. }
  \label{Fracother}
\end{figure*}

\subsection{Fitting methodology and multi-component decomposition}\label{FitIN} 

Constraining how well galaxy optical properties such as mass, size and
stellar concentration correlate with each other, and also with the
radio, X-ray and optical emission line properties ultimately depends
on an accurate characterisation of the bulge and disc structural
properties.  High-resolution near-infrared and optical imaging and
IFU spectroscopy have revealed distinct morphological and/or
kinematical galaxy structures including bulges, discs, nuclear star
clusters, nuclear discs and rings, bars, and spirals
\citep[e.g.,][]{1997AJ....114.1771F,1998AJ....116...68C,2003AJ....125.2936G,2006ApJS..164..334F,2013MNRAS.433.2812K,2015MNRAS.446.4039E,2017ApJ...848...87C,2014MNRAS.444.2700D,2016MNRAS.462.3800D,2019ApJ...871....9D,2020A&A...635A.129K,2020MNRAS.495.2247J,2021MNRAS.500.4193J}.
 As noted in the Introduction, if these structures are not modelled
separately due to their coupling the derived bulge and disc structures
would be biased.  Multi-component decompositions are needed if we are
to understand properly the different mechanisms that build up these
very distinct photometric components. To this aim, we perform 1D    (one-dimensional) and 2D    (two-dimensional) multi-component decompositions. This section discusses our 1D and 2D decompositions and the reasons why we regard the  results from the 1D (multi-component) light profile modelling to be more appropriate for this study and thus base the analyses
throughout this paper on them.

 \subsubsection{1D Fitting Procedure}

 The 1D  multi-component fitting procedure utilised to decompose the
 one-dimensional galaxy light profiles with our personal Fortran 
 programming code was discussed in
 \citet{2016MNRAS.462.3800D,2017MNRAS.471.2321D,2019ApJ...871....9D,2019ApJ...886...80D}. The
 fitting code not only accounts for the PSF (see Section~\ref{PSF})
 but is also sophisticated, capable of fitting simultaneously up to
 six galaxy components, which are summed up to a full model with (up
 to) 16 free parameters.

 The cornerstone of our fitting strategy concerns remedying two
 commonly existing limitations of galaxy light profile
 parameterisation in the literature, which are summarised and
 addressed below:

 (1) {\it Excluding innermost structures. } Space-based NIR imaging
 data from {\it Spitzer} and most ground-based optical/NIR imaging do
 not have the necessary resolution to study reliably inner, sub-kpc
 stellar structures such as NSCs, discs and rings, bars, depleted
 stellar cores and non-thermal sources (AGN). Excluding the innermost
 portions of galaxies' light profiles has been a preferred strategy to
 overcome such resolution limitations.  In Fig~\ref{Innerkpc} we plot
 the distribution of sub-kpc optical/NIR structures detected in our
 analyses of the {\it HST} imaging with a high spatial resolution
 $\le 0\farcs1$.  For core-S\'ersic galaxies, the sub-kpc structure is
 defined to be the depleted core with size $R_{\rm b}$ ($\la$ 0.5 kpc,
 see Table~\ref{Tab1}). For S\'ersic galaxies, we inspect the results
 of the decompositions and measure the size of the sub-kpc structure
 by eye as the radius within which the innermost component contributes
 significantly to the galaxy light.  An overwhelming majority (95\%)
 of our sample have sub-kpc stellar structures detected with {\it
   HST}.  The high sensitivity $e$-MERLIN $L$ band (1.5 GHz)
 observations allow for analogous radio continuum detections at high
 resolution of $\le 0\farcs 2$.  Accounting for the effect of the PSF,
 we find only 15\% (22\%) of the kpc-scale, optical/NIR structures
 from {\it HST} can be well detected with {\it Spitzer} (SDSS) imaging
 data.

(2) {\it Restricting fits to bulge+disc profiles.}  It is common,
particularly for automated fits, to model elliptical galaxies with a
single S\'ersic model, and to consider Sersic+exponential bulge-disc
fits adequate for disc galaxies. The LeMMINGs sample covers a
morphology range that encompasses all Hubble types between E and
Im. Dissecting the galaxies into bulges, discs and bars, rings, spiral
arms, nuclear sources, cores and haloes, we derive the fraction of
galaxy light that is not ascribed to the bulge and disc components
($f_{\rm other}$) to highlight the need for fits beyond the two main
galaxy components, i.e., bulge-disc profile (see Fig~\ref{AppendC}).
Excluding irregular galaxies, we find that the fraction of LeMMINGs
galaxies with $f_{\rm other}>$ 20\% is $\sim$26\% (43/167), increasing
to $\sim$40\% (67/167) for $f_{\rm other}>$ 5\%
(Fig.~\ref{Fracother}).  Due to the assignment of $f_{\rm other}$ to
the bulge, one plausible outcome of a simple bulge+disc fit to disc
galaxies which contain bars, rings and spiral arms is an unreasonably
high $B/T$ ratio. In Fig.~\ref{Fracother} we show the dependence of
$f_{\rm other}$ on the bulge mass and luminosity.  We also plot green
and blue tracks which trace the median trend for the binned spiral, S0
and elliptical galaxy data.  For spirals, in general, the fraction
$f_{\rm other}$ tends to increase as the bulge mass and luminosity
increases, whereas S0s typically show $f_{\rm other} \sim$ 1.3\%. For
most (71\%) elliptical galaxies $f_{\rm other} \sim 0$, however at the
most massive end dominated by BCGs and cD galaxies the fractional
contribution from light components such stellar haloes and ICL
increases $f_{\rm other}$ typically to $\ga 5\%$. In summary, the
prevalence of galaxy components beyond bulge+disc and the
unprecedented level of detail required to decompose our galaxies
suggest that S\'ersic bulge (exponential disc) fits, especially when
forced in an automated fashion are incapable to reproduce our work  
and, in general, to derive accurate structural parameters of the galaxies' 
inner regions.
  
The best-fit model parameters which provide an optimal match to the
data are calculated iteratively using the Levenberg-Marquardt
minimisation algorithm (see
\citealt{2017MNRAS.471.2321D,2019ApJ...871....9D,2019ApJ...886...80D}). We
quantify the global goodness of the fits using the root-mean-square
(rms) residuals ($\Delta$). We let all fitting parameters free to
vary.  Our identification of a component in a galaxy is with guidance
from physical considerations such as the galaxy's morphological type,
velocity dispersion and structural features visible in the optical and
NIR images. We visually inspect the high resolution {\it HST} images
as well as the deeper ground-based (SDSS/DSS) data (when available)
before and after the fitting procedure.  We further check for any
systematic tendencies in radial $\epsilon$, P.A. and $B_{4}$
profiles\footnote{A caveat is the presence of strong dust structure in
  a galaxy can also cause patterns in the in $\epsilon$ and $B_{4}$
  profiles.}; galaxy components such bars and prominent rings are
typically accompanied by local minima/maxima in $\epsilon$, $B_{4}$
and a twist in P.A. For each galaxy, the fits were carefully examined
to determine the one that provides the best description. Fits that we
regard optimal are those with: (i) a minimum number of galaxy
components, (ii) low rms residual values and (iii) physical meaningful
best fitting model parameters that describe the full complexity of the
galaxies' stellar distributions well over the entire radial
range. Ideally, acceptable fits do not exhibit systematic
(`snake-like') residual patterns which commonly signify a mismatch
between the data and fitted model, admittedly however a few sample
galaxies are subject to modest residual patterns due to the presence
of dust, noise in the data or simply because of difficulties in
modelling the galaxy light profiles. In Section~\ref{FitAn}, we
measure the best fitting model component parameters, implementing an
MC approach, which are found to agree very well with the adopted
parametrisation from the multi-component decompositions discussed
here.

In addition to the rms residuals, each galaxy's goodness of fit is
given a ``quality flag'' designated by numbers (`1' or `2') and listed
in Tables~\ref{Tab1}, \ref{Tab1Ring} and  \ref{Tab1cS}  along with the
best-fitting model parameters form the decompositions.  Quality flag
`1' indicates fits with good or higher quality levels that meet all
our aforementioned criteria for an optimal fit, whereas those labeled
`2' are deemed questionable. The latter are mainly associated with
highly inclined ($i \ga 75^{\circ}$) and morphologically   complex  faint 
galaxies (see  Section~\ref{SB}) whose surface brightness profiles could 
not be accurately extracted and modelled, and in few cases the galaxy 
centre was affected by dust that results in unreliable decomposition 
(e.g., NGC~3665). Those with quality flag `2' make up $\sim 13$\% of
the sample.

\begin{figure*}
\hspace{-.780cm}
\includegraphics[trim={-2.808991cm -1.50cm .107cm -.59900cm},clip,angle=0,scale=0.2530950]{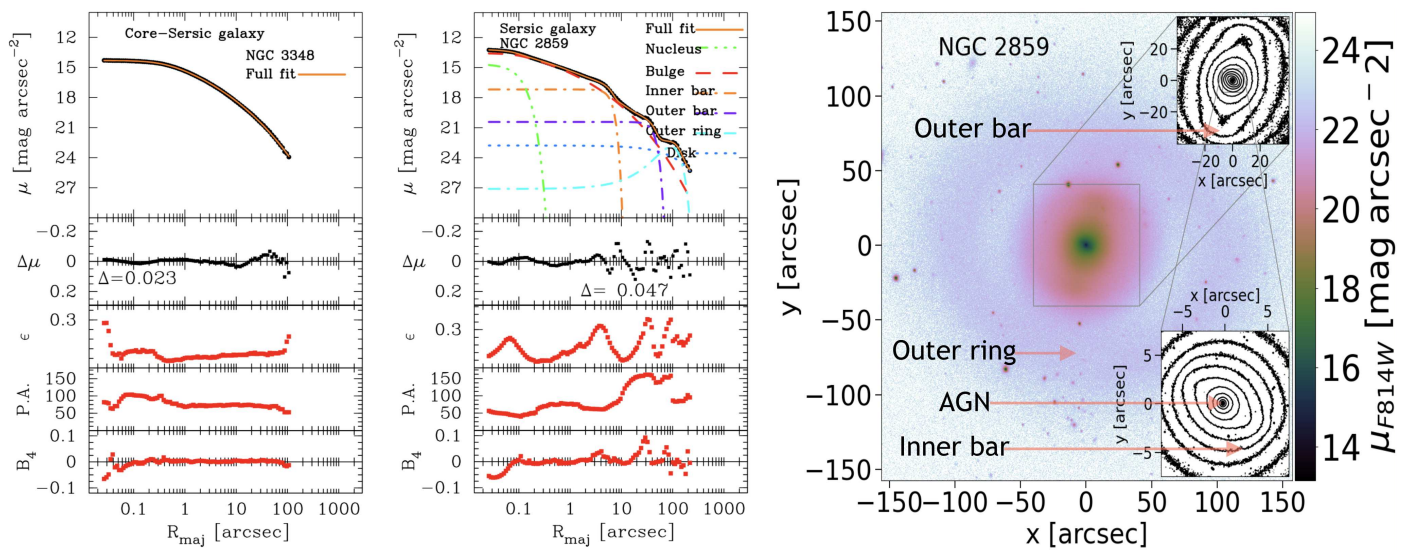}
\vspace{-.365700330cm}
\caption{Examples of  fits to the major-axis  surface
  brightness profiles of LeMMINGs galaxies. Left-hand panel:  the core-S\'ersic model fit to the major-axis  {\it
    HST} WFPC2 $F814W$-band  light profile (black dots) of the  elliptical LeMMINGs galaxy \mbox{NGC~3348} (orange solid curve).  The ellipticity
  ($\epsilon$), position angle (P.A.)  and B4 profiles, along with the  residual profile and rms residual ($\Delta \sim$ 0.023mag
  arcsec$^{-2}$) for this massive core-S\'ersic galaxy are also shown. Middle panel: composite ({\it
    HST} ACS+SDSS), $F814W$-band surface brightness,  
 $\epsilon$, P.A.  and B4 profiles of the
  doubled-barred lenticular LeMMINGs galaxy NGC 2859. The SDSS
  $i$-band data are zero-pointed to the {\it HST} $F814W$-band
  profile.  Six-component decomposition of the galaxy major-axis
  surface brightness profile (black dots) into a  nuclear component, a bulge, an
  inner bar and three outer galaxy components (bar+ring+disc) are
  denoted by various broken curves which are summed up to the final
  model (orange solid curve) that describes the galaxy. The stellar light distribution  of the nuclear
   component (which we tentatively identify as AGN, see \citealt{2018MNRAS.476.3478B,2021MNRAS.500.4749B}) 
   is modelled with a two-parameter
  Gaussian function and dominated by
  the S\'ersic ($n \sim3.6$) bulge. The inner and outer bars are described using two
  S\'ersic models each with $n \sim 0.2$.  The outer, extended
  exponential disc is the dotted blue curve.  The three-parameter
  Gaussian ring model (cyan dot-dashed curve) describes the outer
  ring. The two bars and the large-scale ring are accompanied by three
  local maxima in $\epsilon$ and $B_{4}$ and twists in P.A. Each model
  component is convolved with a Gaussian PSF.  The residual profile
  along with the small rms residual $\Delta \sim$ 0.047 mag
  arcsec$^{-2}$. We fit six model components which are summed up to a
  full model with 16 free parameters. Right-hand panel: SDSS image of
  NGC 2859. The top and bottom insets show the surface brightness
  contours of the galaxy's SDSS and {\it HST} ACS images,
  respectively. North is up, and east is to the left. }
   \label{MultD}
\end{figure*}

As an illustration of our fitting procedure, in Fig~\ref{MultD} (left)
we show a core-S\'ersic light profile and a six-S\'ersic-component
light profile: a model fit to the light profile of the giant
elliptical NGC~3348 and a six-component decomposition of a
doubled-barred lenticular LeMMINGs galaxy NGC 2859 into a nucleus, a
bulge, an inner bar plus three outer galaxy components
(bar+ring+disc). The bulge and bars are each described with a S\'ersic
model, while the disc is modelled with an exponential function.  We
fit a two-parameter Gaussian function to the nuclear source
tentatively identified here as AGN (see
\citealt{2018MNRAS.476.3478B,2021MNRAS.500.4749B}), whereas the outer
disc was well reproduced by a three-parameter Gaussian ring model.
Each galaxy model was convolved with a Gaussian PSF during every
fitting iteration. Inspection of the {\it HST} and SDSS images
(Fig~\ref{MultD}, right) strongly corroborates the results of the
decompositions.

 Figs.~\ref{FigSer1}, \ref{FigSerR} and \ref{FigCSer} display the
 decompositions of the major-axis surface brightness profiles of the
 galaxies listed in Tables~\ref{NewTabA1}$-$\ref{Tab1cS} and 
 the residual profiles from the fits. The root-mean-square (rms)
 residuals $\Delta$ are also shown. The median value of $\Delta$ for
 our fits is $\sim$ 0.065 mag arcsec$^{-2}$.

\subsubsection{PSF treatment}\label{PSF}
\begin{figure*}
\hspace{-4.038cm}
\includegraphics[trim={-10.940cm -6.5cm -14.598cm 1.80834cm},clip,angle=0,scale=0.352]{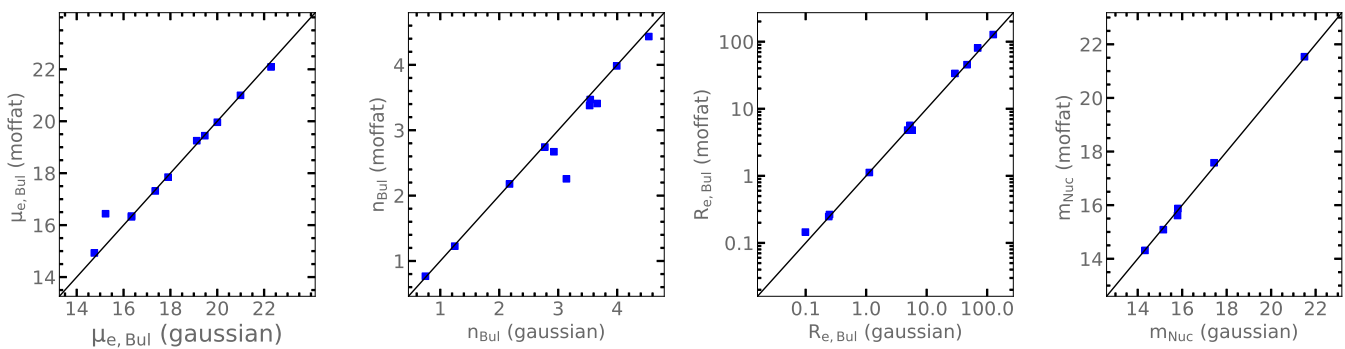}
\vspace{-2.819cm}
\caption{Modelling the {\it HST} PSF using a Moffat and a Gaussian
  function. A comparison of bulge S\'ersic indices, effective radii,
  effective surface brightnesses and point source magnitudes for a
  dozen LeMMINGs galaxies obtained from multiple S\'ersic model fits
  PSF-convolved using a Moffat and a Gaussian function. To explore the
  marked effects of the {\it HST} PSFs in the {\it HST} data, the
  galaxies were selected to have a steep inner light profile
  slope. The solid lines represent one-to-one relations.  The good
  agreement between the best-fitting structural parameters suggests
  that the light profile decomposition does not depend appreciably on
  selecting either of the two {\it HST} PSF modelling functions
  employed.}
  \label{Fig3}
\end{figure*}

When decomposing a galaxy light profile, at each fitting iteration,
the individual model components were convolved with the point-spread
function (PSF) in 2D.  The PSF implementation is discussed
\citet{2017MNRAS.471.2321D}, the mathematical expressions are
presented in \citet[][their Eq.\ 2]{2001MNRAS.321..269T}. For more
reference see \citet[][]{1981AJ.....86.1859P,1993MNRAS.264..961S}.
The PSF of a telescope + instrument, which scatters the stellar light
from the innermost concentrated regions of the imaged galaxy to more
extended regions, gives a measure of the image's spatial
resolution. Over the years, several analytic functions have been used
to describe the PSF.  \citet{2001MNRAS.328..977T} remarked that the
Moffat function is numerically more well behaved in modelling {\it
  HST} PSFs than a Gaussian function. While the design consideration
of our multi-component decomposition code is to perform PSF
convolution using either a Gaussian or Moffat function, we find that
modelling the {\it HST} PSF using the latter function is
computationally intensive, a striking contrast to the markedly faster
convolution with the former one.  Motivated by this difference we
devoted efforts to investigate the effect of our treatment of the PSF
on the best-fitting structural parameters.
  
As our intention  is to illuminate severe effects of the PSF, we select a dozen
LeMMINGs galaxies with a steep inner light profile slope and perform
decompositions with multiple S\'ersic models, PSF-convolved using a
Moffat and Gaussian function (Appendix~\ref{AppendC}).  The FWHMs of
the Gaussian PSFs as well as FWHMs and $\beta$ values for the Moffat
PSFs were determined from Gaussian and Moffat fits to the radial flux
profiles of several unsaturated stars present in the galaxy {\it HST}
images using the IRAF task {\sc imexamine}.  In Fig.~\ref{Fig3}, we
compare the S\'ersic indices, effective radii, effective surface
brightnesses of the bulges and point source magnitudes. The good
agreement between the structural parameters from the two PSF
convolution approaches suggests that no matter which of the two
modelling functions are selected to describe the PSF in the {\it HST}
data the light profile decompositions remain largely unaffected. We
therefore adopt the Gaussian function to convolve the {\it HST} PSFs
throughout this work.

\subsubsection{1D fitting analysis and galaxy components }\label{FitAn}

For luminous early-type galaxies with cores (i.e., core-S\'ersic
galaxies) we show that their bulges can be very well described with
the core-S\'ersic model, whereas for coreless (i.e., S\'ersic)
galaxies the underlying bulge light distributions are well fit by the
S\'ersic model (Appendix~\ref{AppendC}).  We identify 20 (5 S0s and 15
Es) core-S\'ersic galaxies out of the total sample of 173 galaxies
($\sim 11.6\%$) which are defined to have a deficit of light at the
centre with respect to the outer S\'ersic profiles. {\it HST's}
resolution implies we can detect depleted cores, as
measured by $R_{\rm b}$, as small as $\sim$10 pc. However, we caution
that the two small core galaxies NGC 5631 and NGC 6482 have
$R_{\rm b} \sim 0\farcs04$ ($\sim 10$ pc) and as these cores are from
{\it HST} imaging with a scale of $0\farcs05/{\rm px}$ and defined by
few innermost data points, their authenticity can be questioned.
While we tentatively classify two lenticular galaxies, NGC 1167 and
NGC 3665, as core-S\'ersic, they behave as intermediate objects, ergo
we also decomposed their light profiles fitting multiple S\'ersic
functions (see Appendix~\ref{AppendC}). For NGC 1167, both the S\'ersic and core-S\'ersic
approaches give good description to the galaxy light profile. Despite
the uncertainty which arises from the nuclear dust in the galaxy, the
core-S\'ersic fit of NGC~3665 is superior in quality than the S\'ersic
one. For the 5 core-S\'ersic S0s (NGC~0507, NGC~1167, NGC~2300, NGC~3665 and
NGC~5631)  the light profiles are modelled as
the sum of core-S\'ersic bulge and an exponential disc. Of the 15
core-S\'ersic elliptical galaxies in the full sample, six (NGC~0315, NGC~0410, NGC~2832,
NGC~3193, NGC~5322 and NGC~6482) have outer
stellar halo components modelled with an exponential function \citep{2019ApJ...886...80D,2021MNRAS.508.4786D}. We regard 
the outer bump (i.e., light excess) over the bulge profile of the core-S\'ersic elliptical
 (E6) galaxy (NGC~3613) to be a disc light that is well described using a S\'ersic $n\sim 0.81$
profile. Most ($80 \%$) core-S\'ersic bulges in the sample have
$n > 4$. On the other hand, an overwhelming majority (94\%) of
S\'ersic bulges have $n < 4$.  Furthermore, the bulges of 25 disc
galaxies, 80\% (68\%) of which are late-type (\mbox{H\,{\sc ii}})
galaxies, in the sample  follow a
near-exponential $n \sim 1.12 \pm 0.20$ light profile and have median
values of $M_{*,\rm bulge} \sim 9.2$, $\epsilon \sim 0.37$,
$B_{4} \sim 0.003$ and $ B/T \sim 0.17$. Given the exponential nature
coupled with the median properties these bulges are suspected to be
rotationally supported pseudo-bulge candidates.

 We identify 10 bulgeless galaxies (IC~2574, NGC~0672, NGC~0784,
NGC~1156, NGC~2537, NGC~3077, NGC~4183, NGC~4656, NGC~5112 and NGC~5907), which are all
late-types galaxies typically with low stellar masses
($M_{*,\rm glxy} < 10^{11} M_{\sun}$). Furthermore, for three low-mass
late-type galaxies (NGC~2541, NGC~3031 and NGC~4151) in our sample, the
bulges are very compact ($R_{\rm e} \sim 241-268$ pc) and with high S\'ersic indices
$(n\sim 3.9- 5.1$), which appear to be tell-tale signs of dynamically
hot classical bulges akin to those reported in
\citet{2015MNRAS.446.4039E}.

In general we find that the outer disc light distribution is well fit
with an exponential $n = 1$ model, for 17 late-type galaxies however
their outer disc light are better fitted with a near-exponential,
$n \sim 0.85 \pm 0.50$ S\'ersic model (see also \citealt
{2009MNRAS.396..121D,2015MNRAS.453.3729H}).  Of the 150 disc galaxies
(S, S0 and Irr) in the sample, our structural analyses have identified
bar structures in 47 (30 S, 15 S0, and 2 Irr).  The light
profiles of bars are modelled with a low-$n$ S\'ersic profile.
Detailed analysis of bar structures and the dependence on host galaxy
properties is deferred to a future work.

With spiral arms and rings in a galaxy disc typically resulting in
excess flux (`a bump') at large radii relative to the exponential disc
profile, they are generally well modelled using the three-parameter Gaussian
ring function.

In Appendix~\ref{NotesI}, we present notes on our decompositions and
literature comparison for 54 selected individual galaxies.

\begin{figure*}
\hspace{-1.0255080297593cm}
\includegraphics[trim={-3.940cm -9.5cm -10.598cm .32955cm},clip,angle=0,scale=0.349062]{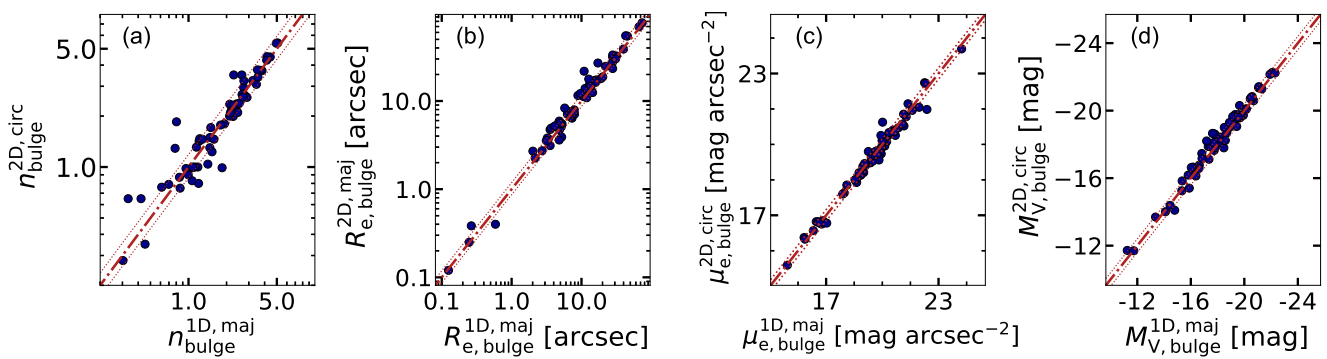}
\vspace{-3.7887905cm}
\caption{2D versus 1D bulge properties. Comparison of the
    S\'ersic indices ($n$), effective radii ($R_{\rm e}$), surface
    brightnesses at $R_{\rm e}$ ($\mu_{\rm e}$) and absolute, $V$-band
    magnitudes ($M_{V}$) from our 2D and 1D stellar light distribution
    analyses for 65 sample galaxies. The dashed-dotted lines show the
    one-to-one relations. The dotted lines show the 1$\sigma$
    uncertainties (see Section~\ref{ErrF} and Table~\ref{NewTabA1}).}

  \label{Fig2D_bul}
\end{figure*}

\begin{figure}
\hspace{-2.3797593cm}
\includegraphics[trim={-5.940cm -7.95cm -13.598cm -.115955cm},clip,angle=0,scale=0.5062]{ 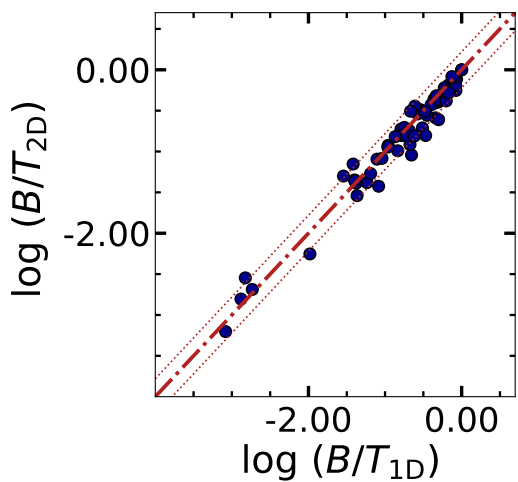}        
\vspace{-4.506575cm}
\caption{Comparison of the bulge-to-total ratios ($B/T$) from our
    2D and 1D decompositions for 65 sample galaxies. The dashed-dotted
    line shows the one-to-one relations. The dotted lines show the
    $\sigma$ uncertainty.  }
  \label{Fig2D_BTR}
\end{figure}

\subsubsection{2D fitting} 

Past studies (e.g., \citealt{2016PASA...33...62C,2016ApJS..222...10S}) highlighted  
the advantage of  2D decompositions of the full galaxy light distributions over 
  1D decompositions of  brightness profiles obtained from major-axis cuts (see Section~ \ref{Sec2V1}). 
  This is particularly true for  spiral galaxies which are known to have multiple components such as nuclei, bars, rings and  spiral arms.
In order to test the robustness of our 1D modelling, we perform 2D decompositions of the {\it HST} images of 65 sample
galaxies using {\sc
  imfit}\footnote{\url{https://www.mpe.mpg.de/~erwin/code/imfit/}} v1.8
\citep{2015ApJ...799..226E}.   The 65 galaxies with 2D decompositions, make up
$\sim$40\% of the 163 newly analysed galaxies in the
paper. Note that our 1D brightness  profiles for the sample galaxies are not derived from  major axis cuts (Section~\ref{SB}). 
Nonetheless,  in our 2D analysis  we  focus on  spiral galaxies  and perform 2D decompositions for nearly half of
the spiral galaxies in the paper (i.e.,
49/102).   These 49 spiral galaxies are representative   of  the full spiral galaxy sample  in terms of the morphological
  T-type, galaxy mass and AGN power as measured by the  O\,{\sc iii}  $\lambda$5007\,\AA\ line luminosity; we find that the  KS tests on the datasets from the 
  subsample and the full  sample give significance levels $P \sim$  0.52, 0.39 and 0.19 for the three quantities,   respectively. 
 
Our 2D fitting contains the same type and
number of galaxy structural components as the corresponding 1D
fitting. The only exception is NGC~2859: while we fit a six component
model to the galaxy's 1D, major-axis light profile (see
Fig.~\ref{MultD}), the 2D decomposition of the {\it HST} image reveal
a seventh component, i.e., a faint inner ring which encloses the inner
bar. In order to convolve the 2D model images, we created a Moffat PSF
using the {\sc imfit} task {\sc makeimage}.  The FWHM and $\beta$
values of the PSF were measured using stars in the {\it HST} images of
the galaxies. The 2D mask images were generated by converting the
`.pl' mask files from the {\sc ellipse} run into images using the {\sc
  iraf} task {\sc wfits}.  Appendix~\ref{Append2D} presents the best-fitting model
parameters from the 2D decompositions, whereas Appendix~\ref{NotesI}
provides a discussion by comparing our 2D decompositions with those in
literature for 23 galaxies.  Appendix~\ref{Append2D} shows the
residual images for a representative sample of 28/65 galaxies created
after subtracting the {\sc imfit} model images from the {\it HST}
images of the galaxies.

\subsubsection{Comparison of 1D and 2D decompositions}\label{Sec2V1}

We now compare our 1D and 2D decompositions
and explain why  the results and conclusions of this work are based on the 1D decompositions.
Fig.~\ref{Fig2D_bul} plots the 1D and 2D bulge
properties including the S\'ersic indices, $n$ (a), effective radii,
$R_{\rm e}$ (b), surface brightnesses at $R_{\rm e}$, $\mu_{\rm e}$,
(c) and absolute $V$-band magnitudes, $M_{V}$, (d) for the 65 sample
galaxies. We note that the 2D $R_{\rm e}$ values were converted into
major-axis values using the corresponding bulge ellipticities
(Fig.~\ref{Fig2D_bul}b).  The strong agreement between the 1D and 2D
measurements of $n$, $R_{\rm e}$, $\mu_{\rm e}$ and $M_{V}$ is such
that 85, 88, 92 and 86\% of the 1D versus 2D dataset pertaining to
$n$, $R_{\rm e}$, $\mu_{\rm e}$ and $M_{V}$, respectively, are within
1$\sigma$ (see Table~\ref{NewTabA1} and Section~\ref{ErrF}) of perfect
agreement. The 1D and 2D $n$, $R_{\rm e}$, $\mu_{\rm e}$ and $M_{V}$ values are within the 2$\sigma$ error ranges for 
91, 95, 97 and 97\% of the cases. Similarly, Fig.~\ref{Fig2D_BTR} reveals good agreement
between the 1D and 2D bulge-to-total ratios ($B/T$). Overall, we find
consistency between the 1D and 2D best fitting parameters when
obtained carefully through multi-component decompositions. Mild
discrepancies that we observe between the 1D and 2D measurements are expected to arise
from intrinsic differences between the two methods (see
Figs.~\ref{Fig2D_bul} and \ref{Fig2D_BTR}).

An advantage of performing a 2D decomposition over a 1D decomposition
is that the former makes use of all available information in a 2D
galaxy image to determine the best fit model image. As noted above, the 2D method of
galaxy decomposition is therefore considered to provide a better
description to non-axisymmetric stellar structures, which are common
in spiral galaxies. However, we note that non-axisymmetric structures
are also captured by azimuthally-averaged 1D light profiles
\citep[e.g.,][]{2016PASA...33...62C}.  2D decompositions have
limitations; each 2D model component, generated using e.g., a {\sc
  galfit} \citep{2010AJ....139.2097P}/{\sc imfit}
\citep{2015ApJ...799..226E}, is represented by a single ellipticity,
P.A., and boxy/discyness parameter constant over all scales. In
contrast, distinct galaxy components can have radial gradients in
ellipticity and P.A., as is the case for triaxial systems which are
seen in projection \citep[e.g.,][]{1981gask.book.....M}.  Determining
whether residual structures from a 2D fitting are caused by radial
variations in ellipticity and P.A. or are real galaxy features that
were unaccounted for in the 2D modelling can therefore be difficult
\citep[e.g.,][]{2002AJ....124..266P}.  This poses a problem when
determining the appropriate number of 2D model components that are
needed to fit a galaxy image.

 However, the radial tendencies of ellipticity,
P.A., and boxy/discyness profiles are well represented in 1D
brightness profiles determined from the azimuthal averaging of the
data in the galaxy image. As such 1D residual profiles are usually  informative 
 about distinct  galaxy components, as long as they are not  based on  major-axis cut fits.   This in part has motivated several authors
to advocate the 1D method (\citealt{2018MNRAS.477..845G}, see also \citealt{2016PASA...33...62C,2016ApJS..222...10S}).
These 1D methods can also be more beneficial than the 2D counterparts for
sky background determination since 1D brightness profiles can be
easily constructed by combining high-resolution space-based data and
deep ground-based data. To summarise, the strong agreement between our
1D and 2D decompositions, the capabilities of our 1D decompositions  to capture radial tendencies of ellipticity,
P.A., and boxy/discyness profiles of galaxies and the informative nature of the 1D residual profiles  are the  reasons  why we
adopte the 1D modelling throughout this paper.

\subsubsection{Ubiquity of nucleation }

The bulk of the sample galaxies (124/173) have central light excesses
(sub-kpc nuclear optical components), with respect to the inward
extrapolation of the S\'ersic or core-S\'ersic model fitted to the
bulge, detected in our analyses of the high spatial resolution {\it
  HST} imaging. This yields a high nucleation frequency of
$\sim 72\%$, but note that nuclei are less common ($\sim 10-20\%$) in
core-S\'ersic galaxies, which are among the most massive galaxies (see
also
\citealt{2006ApJS..165...57C,2012ApJ...755..163D,2012ApJS..203....5T,2014MNRAS.445.2385D,2017ApJ...849...55S,2019ApJ...886...80D}).
A two-parameter Gaussian function ($n=0.5$) provides a good
approximation to the light profile of 94/124 nuclei, whereas the
profiles for the remaining 30 nuclei are described fitting the
S\'ersic model with $0.4 \la n \la 2.5$ and a median
$n \sim 0.7 \pm 0.6$.

\begin{figure}
\hspace{1.797593cm}
\includegraphics[trim={-1.04395940cm .0095cm -6.40698cm -16.8115955cm},clip,angle=270,scale=0.427]{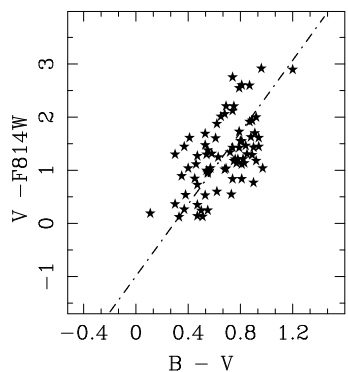}
\vspace{-2.59932805cm}
\caption{Transformation of {\it HST}  $F814W$ magnitudes into
  \mbox{$V$-band}. Each data point represents a galaxy.  The total $V$-band
  magnitudes and total $V-B$ colours are from HyperLEDA, whereas the
  total $F814W$  galaxy magnitudes are from this work.  The
  ordinary least squares (OLS) bisector fit to the data (dot-dashed
  line) is our transformation equation.}
  \label{Fig4}
\end{figure}

\begin{table} 
\begin{center}
\setlength{\tabcolsep}{.806in}
\begin {minipage}{90mm}
\caption{Transformation into $V$-band magnitude}
\label{Tab04}
\begin{tabular}{@{}llcccccc@{}}
\hline
\hline    
 \multicolumn{3}{c}{\bf$V=  F702W + 1.57 (B-V) -0.41$}\\
\multicolumn{3}{c}{\bf$V=  F814W + 3.31 (B-V) - 0.91$}\\
\multicolumn{3}{c}{\bf$V= F850LP + 1.71 (B-V) + 0.21$}\\
\multicolumn{3}{c}{\bf $V= FILTER + \Delta_{\rm mag,FILTER}$}\\
\hline
\end{tabular} 
For magnitudes in the {\it HST} filters $F547M$, $F555W$ and $F606W$
the applied, constant colour  term is $\Delta_{\rm mag,FILTER}$=0.0,
whereas for those in {\it HST} $F625W$ and $F160W$,
$\Delta_{\rm mag,F625W}$=+0.31 \citep{1995PASP..107..945F} and
$\Delta_{\rm mag,F160W}$=+2.95 \citep{2007ApJ...664..226L}. Our
adopted calibration from the {\it HST} $F110W$magnitudes is given by
$V= 0.98 (F110W)+ 2.89$
\citep{1995PASP..107..945F,2007ApJ...664..226L}. See the text for
details.
 \end{minipage}
\end{center}
\end{table} 

 \subsection{Magnitudes, $M/L$ and Stellar Masses }\label{Sec3.4}
  
 Apparent magnitudes of the galaxy components from the 1D fitting are determined using the
 best-fitting major-axis, S\'ersic and core-S\'ersic structural
 parameters along with the ellipticity values, which are integrated to
 $R = \infty$ (e.g.,
 \citealt{2014MNRAS.444.2700D,2019ApJ...886...80D}). For each fitted
 galaxy component, we use our ellipticity profiles
 (Appendix~\ref{AppendC}) to measure an average ellipticity value over
 the region  where the component contributes significant light to
 the galaxy light distribution. For the 2D galaxy components, the apparent magnitudes 
are measured by integrating the best-fitting structural parameters. Foreground Galactic extinction
 corrections of the magnitudes are based on the reddening values from
 \cite{2011ApJ...737..103S}. For disc (S0, spiral and irregular)
 galaxies, we additionally correct for the inclination ($i$)
 dependent, internal dust attenuation using \citet[their equations 1
 and 2 and Table 1]{2008ApJ...678L.101D}. These corrections differ
 between the bulge and disc components and can be expressed as:

 \begin{equation}
m^{\rm corr}_{\rm bulge,filter} =m^{\rm obs}_{\rm bulge,filter}
-b_{1}-b_{2}[1-{\rm cos} (i)]^{b_{3}}, \mbox{and}
\label{Eq1}
 \end{equation}
 
\begin{equation}
m^{\rm corr}_{\rm disc,filter} =m^{\rm obs}_{\rm disc,filter}
-d_{1}-d_{2}[1-{\rm cos} (i)]^{d_{3}}, 
\label{Eq2}
 \end{equation}
 where $m^{\rm corr}$ and $m^{\rm obs}$ denote the observed and
 corrected magnitudes, respectively, and the values of the
 coefficients $b_{1},b_{2},b_{3}, d_{1},d_{2}$ and $d_{3} $ vary with
 the filter type as tabulated in \citet[Table
 1]{2008ApJ...678L.101D}. To compute cos($i$) = $b/a$, the minor and
 major diameters of the galaxies were obtained from NED.  The same
 amount of internal dust correction applied to the bulge was applied to
 the nuclear and intermediate galaxy components, i.e., including nuclear discs. Analogously, the disc
 and other outer galaxy components were treated as the same for the
 internal dust correction.

 With component apparent magnitudes corrected for dust, they are then
 converted into stellar masses in a three-step procedure. First, we
 computed luminosities in solar units after transferring the apparent
 magnitudes into absolute magnitudes. To achieve this the galaxy
 distances from NED (3K CMB) and other sources
 (\citealt[][]{2004AJ....127.2031K} and
 \citealt[][]{2006ApJ...652..313B}) and absolute magnitude for the
 Sun from \citet{2018ApJS..236...47W} were used. Next, we calculated
 the stellar mass-to-light ratios ($M/L$) for the galaxies using the
 waveband dependent ($B-V$
 colour)$-$$(M/L)_{\lambda}$ relations from \citet[][their Table
 B1]{2009MNRAS.400.1181Z} appropriate for elliptical galaxies and
 those from \citet[][their Table 4]{2013MNRAS.430.2715I} for the disc
 galaxies assuming a Salpeter stellar initial mass function
 IMF\footnote{Using a \citet{2001MNRAS.322..231K} IMF reduces the
   Salpeter IMF-based disc galaxy
   $(M/L)_{\lambda}$ typically by a factor of 1.6  at optical wavelengths.}
 \citep{1955ApJ...121..161S}. The total
 $B-V$ colours are from HyperLEDA and for each sample galaxy 
 all the fitted  components are assumed to have the same $M/L$.  Finally, the luminosities coupled
 with the estimates of stellar mass-to-light ratios in the
 corresponding bands yield the stellar masses. To calculate the total
 stellar mass of a galaxy we add the stellar masses from all its
 components (see Table~\ref{Tab4}).

 In order make a direct comparison with previous work and among the
 sample galaxies, as the data are measured in various {\it HST}
 photometric bands (see Table~\ref{Tab1}), the component absolute
 magnitudes are transformed into
 $V$-band. To do so, the galaxies' total $V$ magnitudes and
 $B-V$ colours were taken from HyperLEDA.  The bulk (61\%) of the
 sample galaxies have their magnitudes measured in {\it HST}
 $F814W$ ($\sim I$-band), which we calibrated into
 $V$-band by comparing their total $V-F814W$ and
 $B-V$ colours. Fig.~\ref{Fig4} shows the correlation between the two
 colours and the resulting ordinary least squares (OLS) bisector fit,
 which is the adopted transformation equation.  Applying the same
 prescription, transformation equations were also derived for the {\it
   HST} $F702W$ and
 $F850LP$ filters. For the remaining {\it HST} filters when necessary
 the calibration was performed using \citet[][their
 Table~3]{1995PASP..107..945F} and \citet{2007ApJ...664..226L}. The
 absolute magnitudes, stellar masses,
 $M/L$ for the sample galaxies are reported in Table~\ref{Tab4}.

\begin{figure}
\hspace*{-0.558160910cm} 
\includegraphics[trim={-1.9cm -10.60cm -2cm 1.404cm},clip,angle=0,scale=0.39]{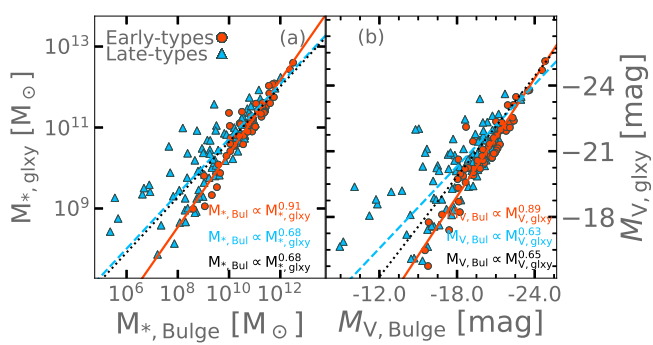}
\vspace{-4.62cm}
\caption{Left panel: correlation between the galaxy stellar mass
  ($M_{*,\rm glxy}$) and bulge stellar mass ($M_{*,\rm bulge}$) for
  our sample.  Right panel: correlation between absolute $V$-band
  galaxy magnitude ($M_{\rm V,glxy}$) and absolute $V$-band bulge
  magnitude ($M_{V, \rm bulge}$, see Table~\ref{Tab4}). Bulgeless
  galaxies have been excluded from our analyses.  The OLS bisector
  fits to the early-type data (solid line), late-type data (dashed
  line) and full galaxy sample (dotted line). }
 \label{Total_vs_Bulge}
\end{figure}

 Fig.~\ref{Total_vs_Bulge}(a) plots the total stellar mass of a galaxy
 ($M_{*,\rm glxy}$) as a function of its bulge stellar mass
 ($M_{*,\rm bulge}$). We also show the correlations between the
 absolute $V$-band galaxy magnitude ($M_{\rm V,glxy}$) and absolute
 $V$-band bulge magnitude ($M_{V, \rm bulge}$) for the sample galaxies
 (Fig.~\ref{Total_vs_Bulge}b). We find strong correlations between
 $M_{*,\rm glxy}$ and $M_{*,\rm bulge}$ for our early-type, late-type
 and full sample of galaxies (Pearson correlation coefficients
 $r \sim 0.77 - 0.94$, see Table~\ref{Tabnew5}). Akin to the
 $M_{*,\rm glxy}-M_{*,\rm bulge}$ relations, we observe tight
 correlations between $M_{\rm V,glxy}$ and $M_{V, \rm bulge}$
 ($r \sim 0.72 - 0.94$, Table~\ref{Tabnew5}).  After excluding
 bulgeless galaxies and applying the ordinary least squares (OLS)
 bisector regressions \citep{1992ApJ...397...55F}, our fits to the
 ($M_{*,\rm glxy}, M_{*, \rm bulge}$) and
 ($M_{\rm V,glxy}, M_{V, \rm bulge}$ ) data are listed in
 Table~\ref{Tabnew5}. For late-type galaxies, we find that the scatter
 in the $M_{*,\rm glxy}-M_{*,\rm bulge}$ and
 $M_{\rm V,glxy}-M_{V, \rm bulge}$ relations increases as the
 galaxy/bulge stellar mass and luminosity decrease. This suggests that
 SMBH scaling relations which are based on $M_{*,\rm glxy}$ and
 $L_{\rm glxy}$ do not have the same predictive power as those
 constructed using $M_ {*,\rm bulge}$ and $L_{\rm bulge}$ for a galaxy
 sample containing late-type galaxies \citep[e.g.,][]{2016ApJ...817...21S}. Given that the masses of SMBHs
 are known to correlate well with the properties of the bulges
 \citep[e.g.,][]{2013ARA&A..51..511K,2016ASSL..418..263G},
 Fig.~\ref{Total_vs_Bulge} underscores the importance of separating
 the bulge component from the rest of the galaxy through
 decomposition, especially for late-type galaxies.

\setlength{\tabcolsep}{0.019650390in}
\begin{table}
\begin {minipage}{90mm}
\caption{Correlations between  galaxy  and   bulge properties.} 
\label{Tabnew5}
\begin{tabular}{@{}lllccc@{}}
\hline
%\Xhline{2\arrayrulewidth}
OLS bisector fit &$r_{\rm
                                              p}$&Note\\
\hline
%\Xhline{2\arrayrulewidth}
        $\mbox{$\log$}\left(\frac{M_{\rm *,glxy}}{  \mbox{$M_{\sun}$}}\right)= (0.91\pm
           0.04) \mbox{log}\left(\frac{M_{\rm *,bul}}{\mbox{$M_{\sun}$}}\right)$ +~($1.27 ~ \pm  0.50$)
 &0.94&ET\\
   $\mbox{$\log$}\left(\frac{M_{\rm *,glxy}}{  \mbox{$M_{\sun}$}}\right)= (0.68\pm
           0.05) \mbox{log}\left(\frac{M_{\rm *,bul}}{\mbox{$M_{\sun}$}}\right)$ +~($3.94 ~ \pm  0.53$)
 &0.82&LT\\
    $\mbox{$\log$}\left(\frac{M_{\rm *,glxy}}{  \mbox{$M_{\sun}$}}\right)= (0.68\pm
           0.04) \mbox{log}\left(\frac{M_{\rm *,bul}}{\mbox{$M_{\sun}$}}\right)$ +~($3.81 ~ \pm  0.43$)
 & 0.77&all\\
 
$\mbox{log}\left(M_{V,\rm glxy}\right)= (0.89   \pm
            0.04)\left( M_{V,\rm bul}\right)$ +~($ -2.86 ~ \pm   0.89$)
             &0.94&ET \\
            $\mbox{log}\left(M_{V,\rm glxy}\right)= (0.63   \pm
            0.06)\left( M_{V,\rm bul}\right)$ +~($ -8.94 ~ \pm   1.10$)
             & 0.72&LT \\
$\mbox{log}\left(M_{V,\rm glxy}\right)= ( 0.65   \pm
              0.05)\left(M_{V,\rm bul}\right)$ +~($  -8.25~ \pm  0.94$)
             & 0.77&all \\
\hline
\end{tabular} 
Note: Galaxy stellar mass ($M_{\rm *,glxy}$) and absolute $V$-band
galaxy magnitude ($M_{\rm V,glxy}$) as a function of bulge stellar
mass ($M_{\rm *,bulge}$) and absolute $V$-band galaxy magnitude
($M_{V, \rm bulge}$) for early-type galaxies (ET), late-type galaxies
(LT) and combination of both. The different columns represent the OLS
bisector fits to the data and the Pearson correlation coefficients
($r_{\rm p}$).
\end{minipage}
\end{table}

\begin{figure}
\hspace*{-2.0950cm}
\includegraphics[trim={-4.49cm -5cm -2cm .1258cm},clip,angle=0,scale=0.47]{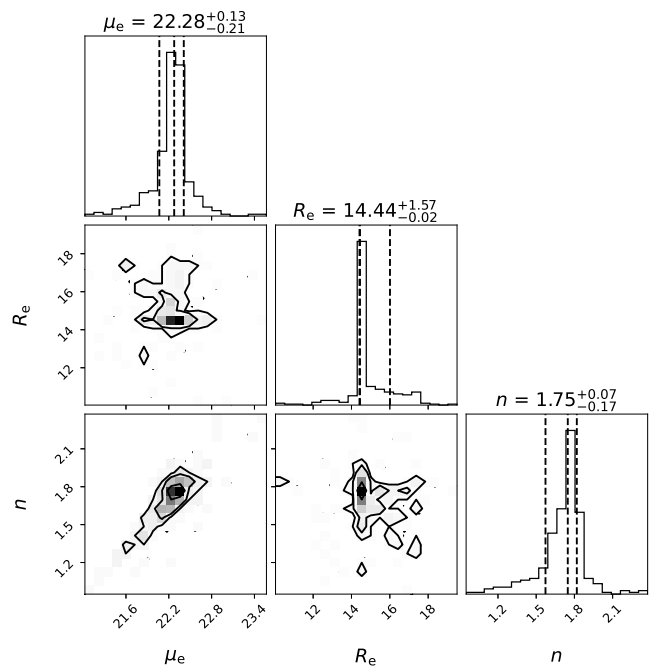}
\vspace{-3.02608cm}
\caption{ To illustrate the parameter coupling and errors associated
  with the fitted model parameters, we show a corner plot for the
  bulge of one LeMMINGs galaxy NGC 959. We fitted each of the 300
  realisations of the galaxy profiles with a four-component
  (bulge+disc+spiral-arm+nucleus) model as done for the
  actual galaxy light profile (see Fig.~\ref{FigSer1}).  Panels show
  best-fit parameters for the bulge (grey boxes) and the pertaining
  histograms and $1\sigma$ errors.}
 \label{Cornpt}
\end{figure}

\subsection{Error Analysis}\label{ErrF}

We determine realistic errors for the 1D best-fitting structural
parameters, the bulge magnitudes and stellar masses after performing
multi-component decomposition of simulated galaxy light
profiles.  We generated over 100 realisations of the surface brightness  profile  for each galaxy by running a series of MC
simulations. This prescription, discussed in detail in Appendices
\ref{AppendB} and \ref{AppendC}, was also used to test the robustness
of the multi-component decomposition of the galaxy light profiles.  As an
illustration of this exercise of measurements of the errors, in
Fig~\ref{Cornpt} we show a corner plot for the bulge of the LeMMINGs
galaxy NGC 959 fitting each of the 300 realisations of the galaxy with
a four-component (bulge+disc+spiral-arm+nucleus) model, as
done for the original  galaxy light profile (see Fig.~\ref{Fig2}).

\section{Bulge structural properties and connection with 
  emission-line class and radio core emission }\label{Secn4}

In this section we study sets of bulge structural properties over
large stellar mass and morphology ranges obtained from detailed modelling of {\it HST}
surface brightness profiles for a sample of 173 LeMMINGs galaxies. We
also examine how the bulge properties vary as a function of host
optical emission class, and radio morphological structures from
$e$-MERLIN  (\citealt{2018MNRAS.476.3478B,2021MNRAS.500.4749B,2021MNRAS.508.2019B}). In the analysis of the
  galaxies properties, we divided the sample into two optical morphological
  classes, early-type galaxies (Es and S0s) and late-type galaxies (Ss
  and Irrs), \citealt{1926ApJ....64..321H,1961hag..book.....S}. While
the study of the dependency of bulge structural properties  on galaxy properties such
as stellar mass and morphology is not new
\citep[e.g.,][]{2012ApJS..198....2K,2014ARA&A..52..291C}, robust characterisation of 
galaxy structures using homogeneously measured high-resolution optical
and radio data for a large sample of galaxies has not been possible to
date. In particular, our analysis focuses on the bulge component as it
is known to correlate better with the mass of the SMBH and other
central galaxy properties than the galaxy disc at large radius.
 
\subsection{Inner light profile slope of the bulge }\label{InnerS}
\begin{figure}
\hspace{-2.20188cm}
\includegraphics[trim={-7.29cm -5cm -9cm 2.91964cm},clip,angle=0,scale=0.341]{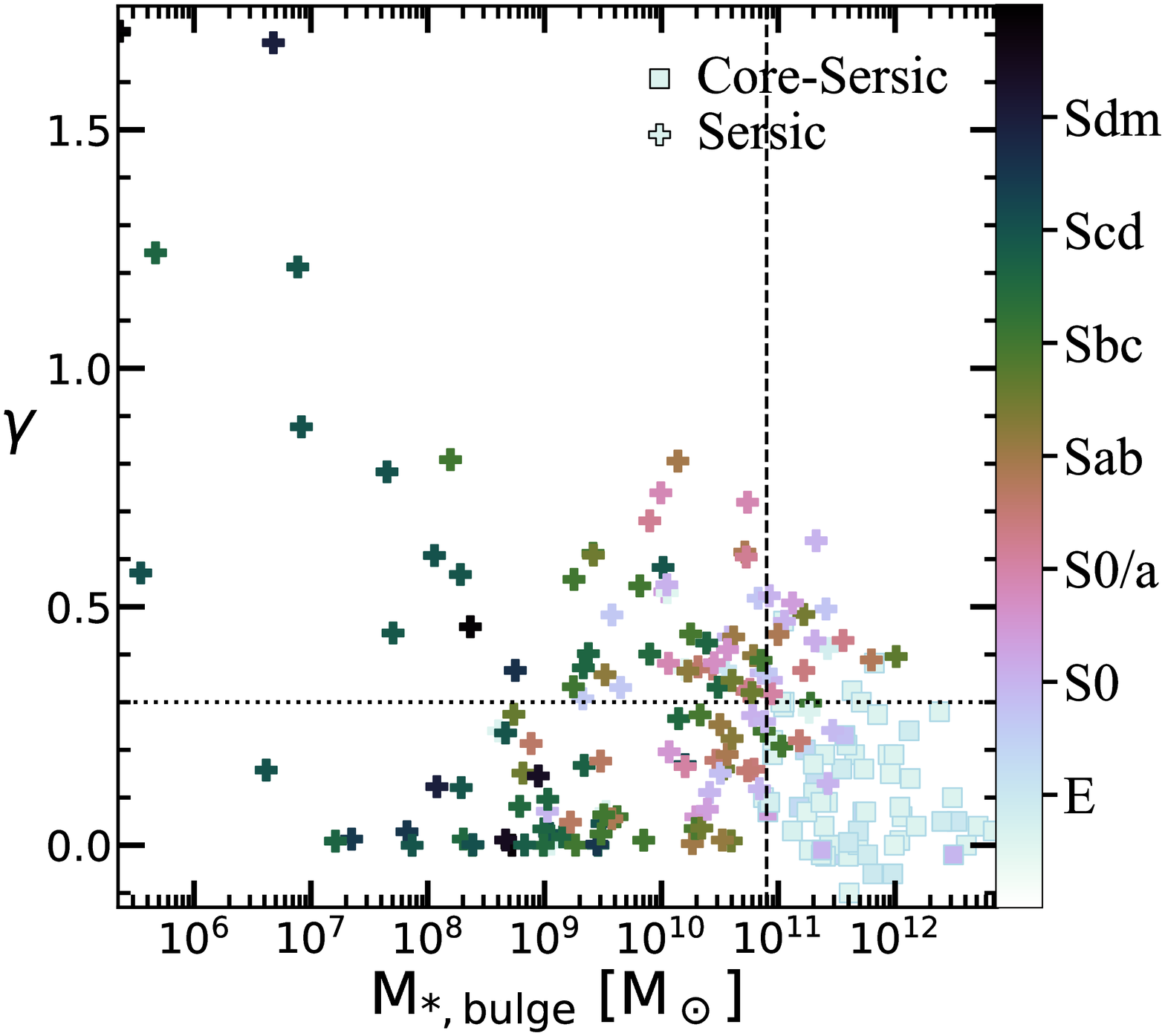}
\vspace{-2.8cm}
\caption{Negative, inner logarithmic slope of the bulge light profile
  ($\gamma$) against the bulge stellar mass ($M_{*, \rm bulge}$) for
  211 (66 core-S\'ersic and 145 S\'ersic) galaxies. These bulges
  modelled with the S\'ersic or core-S\'ersic model.  Of the 211 galaxies, 173  are from this work, whereas the remaining  38 are non-LeMMINGs core-S\'ersic galaxies  from \citet{2014MNRAS.444.2700D,2019ApJ...886...80D,2021ApJ...908..134D}. 
  For S\'ersic galaxies $\gamma$ is measured at $R = 0\farcs1$ (see the
  text for details). The Hubble classes are based on the numerical
  T-type in HyperLEDA. All core-S\'ersic galaxies have
 $M_{*, \rm bulge} \ge 8 \times 10^{10} M_{\sun}$ and low values of
 $\gamma \le 0.35$, except for NGC~584 having $\gamma \sim 0.47$.  We note that galaxies with  low values of  $\gamma$ have shallow inner light profiles, while those with high values of  $\gamma$  have light profiles that are steep all the way into the galaxy centre. }
  \label{GamMas}
\end{figure}

\begin{figure}
\hspace{-1.940239348cm}
\includegraphics[trim={-7.649cm -.822cm -9cm 2.989cm},clip,angle=0,scale=0.451]{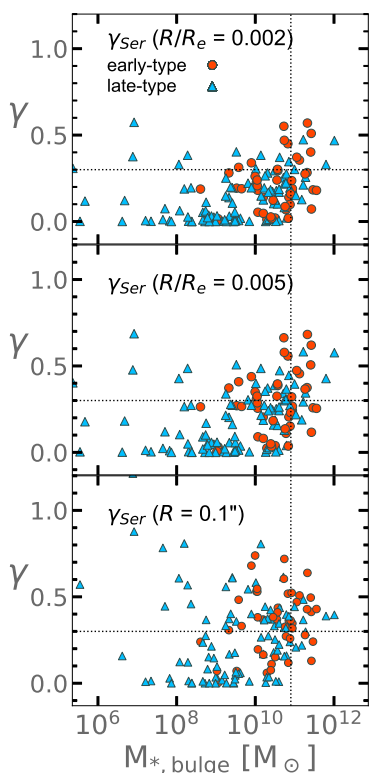}\\
\vspace{-2.2484cm}
\caption{Local logarithmic slopes for the
 bulges of our  S\'ersic galaxies calculated at constant $R/R_{\rm e}$ values of 0.002 (top panel) and  0.005 (middle panel) and at $R = 0\farcs1$ (bottom panel).}
  \label{Gamall_M}
\end{figure}

Inner slopes of the surface brightness profiles of early-type galaxies
have been extensively studied as a diagnostic of galaxy formation
mechanisms. The advent of high resolution data afforded by {\it HST}
prompted the detection of a bimodal distribution of the negative,
logarithmic, inner profile  slopes ($\gamma$) for early-type
galaxies
\citep[e.g.,][]{1994AJ....108.1598F,1995AJ....110.2622L,1996AJ....112..105G,1996AJ....111.1889B}.
With their bulges preferentially pressure-supported and built amid
major dry mergers, bright galaxies with depleted cores display shallow
inner profiles with power-law indices  $\gamma \la 0.3$
\citep[e.g.,][]{2003AJ....125.2951G,2006ApJS..164..334F,2012ApJ...755..163D,2019ApJ...886...80D}. In
contrast, early studies have shown that low- and
intermediate-luminosity early-type bulges---which are likely a
consequence of gas-rich processes--- are consistent with being
rotationally supported and possess profiles that are steep
($\gamma \ga 0.5$) all the way into the centre. The evidence for the
bimodal distribution of $\gamma$ however remains a matter of hot
debate. Some studies identified a few galaxies having intermediate
slopes $0.3 \la \gamma \la 0.5$ and cautioned that the bimodality is
weak \citep{2001AJ....121.2431R,2001AJ....122..653R} while others
later argue the bimodality is entirely non-existent after finding more
objects with intermediate slopes. Another key unknown is the
distribution of $\gamma$ for late-type bulges and how it compares with
those of early-type bulges. As discussed extensively in
\citet{2012ApJ...755..163D} , see also Section~\ref{Mods}, our core identification is not based on
$\gamma$ but instead a `core-S\'ersic' galaxy is defined to have a
central stellar light deficit relative to the inward extrapolation of
the bulge's outer S\'ersic profile. In contrast, a S\'ersic galaxy has
no such central light deficit (Sections~\ref{Mods}
and. \ref{FitIN}). Nonetheless it is beneficial to explore any
systematic trend of $\gamma$ with bulge stellar mass for our S\'ersic
and core-S\'ersic galaxies.

In Fig.~\ref{GamMas} we plot the negative, logarithmic, inner profile
slope of the bulge, unaffected by PSF convolution and not biased by
the nuclear light component, against its stellar mass
($M_{*,\rm bulge}$) for 211 (66 core-S\'ersic plus 145 S\'ersic)
galaxies.    Of the 211 galaxies, 173  are from this work, whereas the remaining  38 are non-LeMMINGs core-S\'ersic galaxies  from \citet{2014MNRAS.444.2700D,2019ApJ...886...80D,2021ApJ...908..134D}. 
For a core-S\'ersic bulge, $\gamma$ is the slope of the inner power-law
profile of the  fitted  core-S\'ersic model given by Eq.~\ref{Eqq10}, while for a S\'ersic bulge $\gamma$ determined  from the fitted S\'ersic model
at any radius $R$ can be written as:

 \begin{equation}
\gamma_{\rm Ser} (R) = -\frac{d \log I (R)}{d \log R} = \frac{b}{n}
\left(\frac{R}{R_{\rm e}}\right)^{1/n}
 \end{equation}

 We use local slopes measured at a representative {\it HST} resolution
 limit $R = 0\farcs1$, i.e., $\gamma_{\rm Ser}$($R = 0\farcs1$) for
 S\'ersic bulges. As
 \citet{2006ApJS..164..334F,2007ApJ...671.1456C,2011ApJ...726...31G},
 we find no evidence for a bimodal $\gamma$ distribution for
 early-type galaxies in the $\gamma-M_{*, \rm bulge}$ plane, instead
 the distribution appears continuous. All core-S\'ersic galaxies have
 $M_{*, \rm bulge} \ga 8 \times 10^{10} M_{\sun}$ and low values of
 $\gamma \la 0.35$, except for NGC~584 with $\gamma \sim 0.47$,
 whereas S\'ersic early-type galaxies
 ($M_{*, \rm bulge} \sim 4.0 \times 10^{8}- 3.6 \times 10^{10}
 M_{\sun}$) are spread over a large $\gamma$ range covering from 0.01
 to 0.70 and only 11/46 (24\%) of them have $\gamma > 0.5$.  For
 early-type galaxies, a S\'ersic to core-S\'ersic regime transition
 occurs across a mass range
 $M_{*, \rm bulge} \sim 6 \times 10^{10} - 3 \times 10^{11} M_{\sun}$
 where 13 early-type S\'ersic galaxies (10/13 with
 $0.3 \la \gamma \la 0.5$) and 27 core-S\'ersic galaxies coexist.

 For the first time, we examine the inner logarithmic slope
 distribution for the bulges of late-type galaxies as a function of
 stellar mass. Akin to early-type S\'ersic galaxies, late-type
 S\'ersic galaxies exhibit a wide range of $\gamma $, from shallow
 slope values ($\gamma \sim 0$) to very steep ones
 ($\gamma \sim 1.71$).  For late-types, we also find a mild tendency
 for the slopes to be systematically steepened as the bulge masses
 decreased. 
   
 A caveat for our inner slope analysis of S\'ersic galaxies is the
 effect of distance. The local S\'ersic slope $\gamma$($R = 0\farcs1$)
 is calculated at an angular radius of $R = 0\farcs 1$ instead of using a
 fixed physical radius, thus identical S\'ersic galaxies at different
 distances will have different $\gamma$($R = 0\farcs1$)
 values. Concerned by this, we measure local slopes for the S\'ersic
 bulges at constant $ R/R_{\rm e}$ values of 0.002 and 0.005. Fig.~\ref{Gamall_M} shows local slopes   $\gamma$($R/R_{\rm e} = 0.002 $), $\gamma$($R/R_{\rm e} = 0.005$) and $\gamma$($R = 0\farcs1$) as a function of bulge mass for our S\'ersic galaxies. The plot reveals that the 
  $\gamma$ distributions for S\'ersic galaxies obtained from the three methods are largely 
 consistent. That is there is no strong impact of  the galaxy distance  on the $\gamma$($R = 0\farcs1$) distribution discussed above. This is expected given the bulk (60\%) of our
 galaxies have distances $\sim 10-30$ Mpc. 
 
 To summarise,  our results demonstrate  that  $\gamma $ alone 
 cannot uniquely separate  core-S\'ersic and S\'ersic  galaxies.

\subsection{Global bulge properties, spectral emission class and nuclear radio
emission}

\subsubsection{Hubble type versus bulge properties
}

\begin{figure}
\hspace{-.024982cm}
\includegraphics[trim={-.44cm -2.6cm -13cm 2.436936518563cm},clip,angle=0,scale=0.28538]{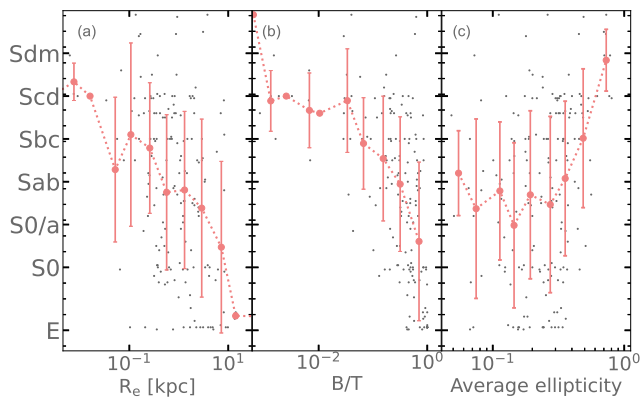}
\vspace{-1.40625320cm}
\caption{Hubble type as a function of (a) the size of the bulge as
  measured by its effective (half-light) radius $R_{\rm e}$, (b) the
  bulge-to-total light ratio (B/T) and (c) the average ellipticity of
  the galaxy within $R_{\rm e}$ calculated excluding the most-PSF
  affected data points inside $R \sim 0\farcs1-0\farcs2$. The mean
  trends together with the standard deviation within each bin
  $1\sigma$ are shown. Going across the Hubble sequence from the
  irregular types to elliptical galaxies, the bulge becomes larger,
  dominant and round. }
  \label{HubPro}
\end{figure}

Having carefully separated the bulge component via detailed
decompositions, in Fig.~\ref{HubPro} we plot the galaxy morphological
type as a function of the bulge effective (half-light) radius
($R_{e}$), dust-corrected bulge-to-total light ratio ($B/T$) and average
bulge  ellipticity value within $R_{\rm e}$ \citep[e.g.,][]{2020ApJ...890..128F} calculated excluding the most-PSF
  affected data points inside $R \sim 0\farcs1-0\farcs2$ for the sample of 173
galaxies, consisting of 23 Es, 42 S0s, 102 Ss and 6 Irrs. When
calculating the average galaxy ellipticities, we excluded the most
PSF-affected data.  The mean trends for the different correlations
together with the standard deviation within each bin are shown.  A
clear trend emerges as one moves across the Hubble sequence from the
irregular types to elliptical galaxies, the bulges are generally
larger, more prominent and round.  Our results confirm previous work
\citep{2008MNRAS.388.1708G,2010MNRAS.405.1089L,2014MNRAS.439.1245K},
but ours are based on detailed  decompositions of high-resolution {\it
  HST} data for a large sample of galaxies. For spiral, S0 and
elliptical galaxies, we find median values of $B/T \sim$ 0.21, 0.46
and 0.86, respectively.  While most Sab galaxies in the sample have
flatter ellipticities than S0s and Es, four (IC 520, NGC~278, NGC~3344
and NGC~7217) have roughly the same round ellipticities as S0s. Most
massive elliptical galaxies posses a single spheroidal component,
therefore their $B/T \sim 1$. All the 10 bulgeless (i.e., $B/T = 0$)
galaxies in our sample, which are not shown in Fig.~\ref{HubPro}, have
morphological type later than Sbc (i.e., Hubble stage
$T \sim 5.2-9.8$).

 \subsubsection{The relations between optical emission-line class and
   bulge properties}

\begin{figure}
\hspace{-.791312cm}
\includegraphics[trim={-2.087cm -8cm -13cm 1.709209903cm},clip,angle=0,scale=0.48802]{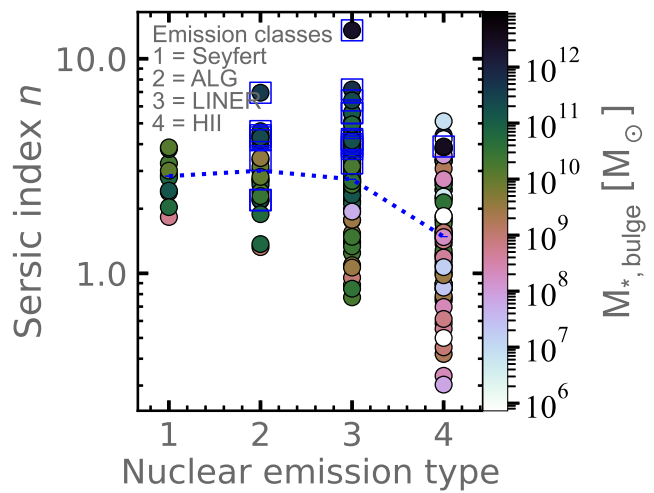}
\vspace{-4.529180cm}
\caption{Bulge S\'ersic index ($n$) plotted against nuclear emission
  type and bulge stellar mass. Core-S\'ersic galaxies (enclosed in
  boxes) are massive and they harbor either a LINER or an ALG
  nucleus. The only core-S\'ersic galaxy with nuclear emission
  associated with star-formation ({\sc h ii}) is the massive
  ($M_{\rm *,bulge} \sim 3.2 \times 10^{12} M_{\sun}$) S0 \mbox{NGC
    3665}. The blue curve connects the median S\'ersic indices for the
  four optical emission-line classes. }
  \label{Spec}
\end{figure}

In Fig.~\ref{Spec}, we plot the relation between the bulge S\'ersic
index ($n$) and nuclear emission-type colour-coded on the basis of the
bulge stellar mass ($M_{*,\rm bulge}$). Seyferts, ALGs, LINERs and
\mbox{H\,{\sc ii}} galaxies constitute 5.8\%, 13.3\%, 41.0\%, 39.9\%
(6.4\%, 10.0\%, 33.6\% and 50.0\%) of our (the full LeMMINGs) sample,
respectively (Table~\ref{Tab03}). Regarding morphological type nearly half (52.0\%) of the elliptical
galaxies in the sample are ALGs and the remaining are LINERs
(48.0\%). We find that 65.0\% of elliptical galaxies have bulge mass
$M_{*,\rm bulge} \ga 10^{11}M_{\sun}$ and that all those with
$M_{*,\rm bulge} < 10^{11}M_{\sun}$ are classified as ALGs. For S0s
the emission-line type breakdown is 7.1\% Seyfert, 57.2\% LINER, 26.2\%
ALG and 9.5\% \mbox{H\,{\sc ii}} galaxies.  An overwhelming majority
(94.0\%) of \mbox{H\,{\sc ii}} galaxies are spiral galaxies, the
remaining 6.0\% are S0s.  Furthermore, all irregular type galaxies in
the sample are \mbox{H\,{\sc ii}} galaxies.

\begin{figure}
\hspace{-4.09581312cm}
\includegraphics[trim={-7.47cm -6.4cm -9cm 1.573cm},clip,angle=0,scale=0.557]{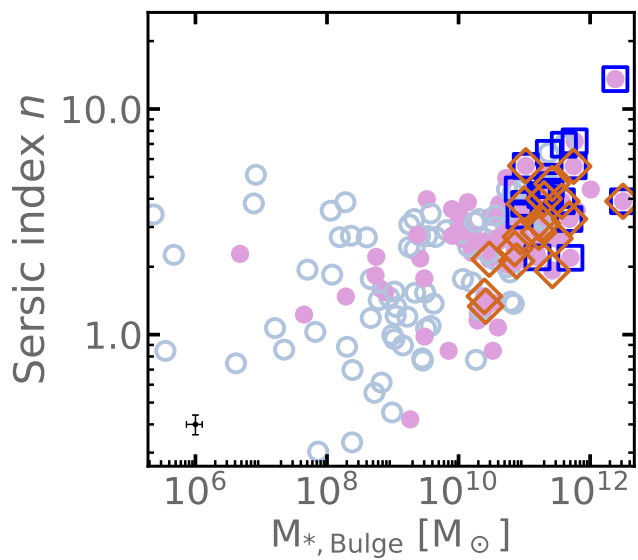}
\vspace{-4.135290cm}
\caption{Dependence of radio detection (Table~\ref{Tab4}) on the
  S\'ersic index ($n$) and stellar mass of the bulge
  ($M_{*,\rm bulge}$), Table~\ref{Tab1}. Filled circles show the
  galaxies in our sample that are radio detected with $e$-MERLIN at 1.5
  GHz, whereas open circles are for undetected galaxies.
  Core-S\'ersic galaxies are enclosed in blue boxes and 7/20 (35\%) of
  them are undetected with $e$-MERLIN at 1.5 GHz. For comparison, the
  radio detection rate with $e$-MERLIN at 1.5 GHz for the full LeMMINGs
  sample of 280 galaxies is 44\%.  Radio-loud galaxies are enclosed in
orange  diamonds.  A representative error bar is shown at the
  bottom. }
  \label{RadA}
\end{figure}

\begin{figure}
\hspace*{-1.9458cm}
\includegraphics[trim={-7.7cm -1.92cm -10.cm 2.7212264803cm},clip,angle=0,scale=0.250]{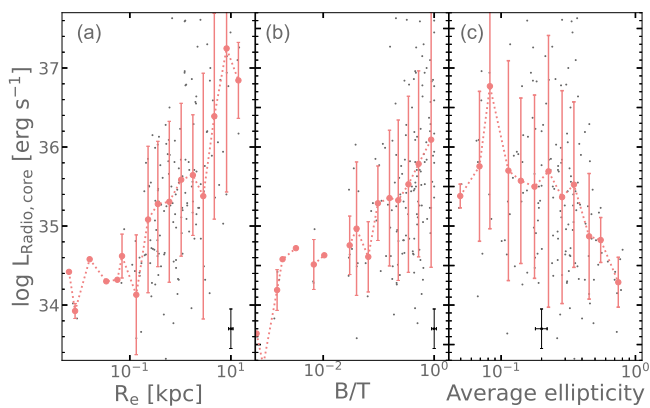} 
\vspace{-1.1327cm}
\caption{The radio core luminosity ($L_{\rm Radio,core}$) against (a)
  the size of the bulge as measured by its effective (half-light)
  radius $R_{\rm e}$, (b) the bulge-to-total light ratio ($B/T$) and
  (c) the average ellipticity within $R_{\rm e}$ calculated excluding
  the most-PSF affected data points inside $R \sim 0\farcs1-0\farcs2$.
  The dotted curves trace the mean trend. Apparent is the tendency for
  galaxies with brighter radio core luminosities to be bulge
  prominent, round and with large bulge sizes.  A representative error
  bar is shown at the bottom of each panel.}
  \label{Lco_Tr}
\end{figure}

We find that $n$ is closely correlated with $M_{*,\rm bulge}$, which
is expected as high-$n$ galaxies are shown to be brighter (see
Section~\ref{FitIN}). Galaxies with large values of $n$ are massive,
except for the low mass tiny classical bulges having large values of
$n$ (see also Section~\ref{FitAn}). We find that most ($\sim74.0\%$) disc
galaxies have a bulge S\'ersic index $n \la 3$. LINERs, ALGs and
\mbox{H\,{\sc ii}} span a broad range in $n$, unlike Seyferts which
have a narrow $n$ distribution: $ n_{\rm ALG}  \sim 1.3-6.9$,
$n_{\rm LINER} \sim 0.8-13.6$, $n_{ \mbox{H\,{\sc ii}}} \sim 0.4-5.0$,
and $n_{\rm Seyfert} \sim 1.8-3.7$.  As can be seen, LINERs and ALGs
exhibit similar distributions in the $n-M_{*, \rm bulge}$ diagram,
they are among the most massive galaxies with high $n$ values in the
sample. Conversely, \mbox{H\,{\sc ii}} galaxies are almost exclusively
associated with low-mass galaxies typically with $n \la 2.5$. At fixed
$n$, LINERs, ALGs and Seyferts have higher bulge masses than
\mbox{H\,{\sc ii}} galaxies.  Of the 69 \mbox{H\,{\sc ii}} galaxies in
the sample 61 (88\%) have
$M_{*, \rm bulge} \la 3 \times 10^{10} M_{\sun}$. Only 2/69 (2\%) have
$n > 3$ and $M_{*, \rm bulge} > 8 \times 10^{10} M_{\sun}$, as such
this domain in $n$ and $M_{*, \rm bulge}$ can be coupled and used as a
reasonably reliable diagnostic to rule out \mbox{H\,{\sc ii}}
galaxies.  Seyferts are observed across a range of $n$ and $M_{*}$
that is typically intermediate between those of the LINERs and
\mbox{H\,{\sc ii}} galaxies. The median values of the S\'ersic indices
for the bulges Seyfert, ALG, LINER and \mbox{H\,{\sc ii}} hosts are
$n\sim 2.80 \pm 0.64$, $\sim 3.01\pm1.25$, $\sim 2.74\pm 1.77$ and
$\sim 1.48\pm1.13$, respectively. As for the median $B/T$, the
Seyfert, ALG, LINER and \mbox{H\,{\sc ii}} hosts have values of
$\sim 0.29 \pm 0.19$, $\sim 0.56 \pm 0.28$, $\sim 0.44\pm 0.30$ and
$\sim 0.16\pm 0.11$. The majority (83\%) of AGN (i.e., Seyfert +
LINER) hosts have $M_{\rm bulge}$ $\ga 10^{10} M_{\sun}$
($M_{\rm glxy} \ga 10^{10.5} M_{\sun}$).  For low mass galaxies
($M_{*, \rm glxy} \la 8 \times 10^{9} M_{\sun}$), we find an AGN
fraction of 6.2\%.

 \begin{figure*}
\hspace{.29cm}
\includegraphics[trim={-3.39cm -4cm -6.cm 2.5993cm},clip,angle=0,scale=0.51]{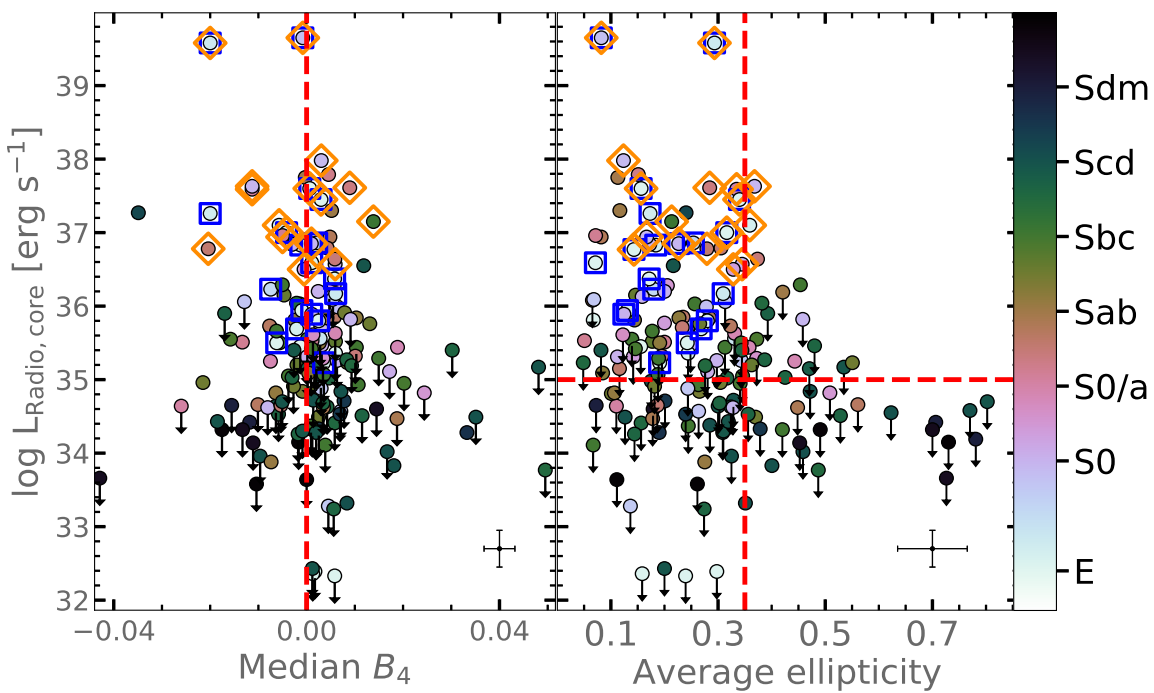}\\
\vspace{-3.098308cm}
\caption{The radio core luminosity against the median of the isophote
  shape parameter of the bulge inside $R_{\rm e}$ ($B_{4}$) and
  average bulge ellipticity within $R_{\rm e}$ both derived after
  excluding the most-PSF affected data points inside
  $R \sim 0\farcs1-0\farcs2$.  Downward arrows denote upper limits for undetected
  sources. Boxy and discy isophotes have $B_{4} < 0$ and $B_{4} > 0$,
  respectively. Core-S\'ersic galaxies (enclosed in boxes) show a
  tendency to possess brighter radio core luminosities, they are
  also systematically rounder (see Section~\ref{IsFKR}). However, only 7/20 (35\%) of our
  core-S\'ersic galaxies are radio-loud.  Radio-loud galaxies are
  enclosed in orange diamonds. In the left panel, the red-dashed line separates boxy and discy galaxies, whereas in the right panel such  lines are shown to demarcate  the location of  core-S\'ersic galaxies.    Nuclear radio emission is more
  prevalent in round, boxy sources than in flat, discy ones. A
  representative error bar is shown at the bottom of each panel. }
\label{RadBE}
\end{figure*}

While not shown in Fig.~\ref{Spec}, all the 10 bulgeless galaxies in
our sample are \mbox{H\,{\sc ii}} galaxies, pointing to the low
incidence of AGN for bulgeless galaxies, some of which are known to
house a SMBH but rarely an AGN (See
\citealt{2003MNRAS.341...54K,2008ApJ...688..159G,2013MNRAS.429.2199S}).

At $M_{\rm *,bulge} \ga 2 \times 10^{11} M_{\sun}$, core-S\'ersic
galaxies (enclosed in boxes, Fig.~\ref{Spec}) make up $61\%$ of the
galaxy population, although a core-S\'ersic bulge can possess a
stellar mass as low as
$M_{\rm *,bulge} \sim 8 \times 10^{10} M_{\sun}$
(Fig.~\ref{Spec}). They have $B/T \ga 0.5$, $M_{V} \la -20.7$ mag and
harbour either a LINER or an ALG nucleus. The only exception is the
most massive \mbox{H\,{\sc ii}} galaxy \mbox{NGC 3665}
($M_{\rm *,bulge} \sim 4 \times 10^{12} M_{\sun}$), a jetted S0 galaxy \citep{2021MNRAS.500.4749B} which
we tentatively classified as a core-S\'ersic galaxy
(Section~\ref{FitAn}).  On the other hand, the bulk ($\sim 90$\%) of
S\'ersic bulges have $M_{*, \rm bulge} \la 10^{11} M_{\sun}$ and
$n \la 4$.  For our core-S\'ersic galaxies, the median values of
$M_{V, \rm bulge}$, $M_{*, \rm bulge} $ and $n$ are $-$21.40 mag,
$ 2.4 \times 10^{11} M_{\sun}$ and 4.3,  respectively. For S\'ersic galaxies, the
median values of $M_{V, \rm bulge}$, $M_{*, \rm bulge} $ and $n$ are
$-19.0$ mag, $10^{10} M_{\sun}$ and 2.4,  respectively.

\subsubsection{ The relations between radio detection and bulge properties}\label{Sec4.2.3}

Fig.~\ref{RadA} shows the $n-M_{*,\rm bulge}$ relation colour-coded by
galaxy radio detection.  Of the 173 sample galaxies, we detect radio
emission $\ga 0.2~\rm {mJy}$ from 83 with $e$-MERLIN at 1.5 GHz (filled
circles). This gives a detection rate of 48\%, in fair agreement with
that from the full sample (44\%). Of the 83 radio-detected galaxies,
five (NGC~3034, NGC~3838, NGC~4242, NGC~5273 and NGC~5907) are
radio-detected but `core-unidentified'\footnote{We use the term  `core-unidentified' when referring  to radio-detected galaxies that do not have clearly  identified radio cores \citep{2018MNRAS.476.3478B,2021MNRAS.500.4749B}.} (see Table~\ref{Tab4}); they have low S\'ersic indices
($n \le 3$), faint bulge magnitudes ($M_{V,\rm bulge} \ga -18.9$ mag)
and low stellar masses ($M_{*,\rm bulge} \la 10^{10} M_{\sun}$).  For
core-S\'ersic galaxies the detection rate is 65\%.  The remaining 90
undetected sources are denoted by open circles.  The radio detection
fraction increases with bulge mass, $n$ and $B/T$. At
$M_{*,\rm bulge} \ga 10^{11} M_{\sun}$ ($n \sim 4.00 \pm 1.50$,
$B/T \sim 0.78 \pm 0.24$), this fraction is 77\%, but it plummets to
24\% for $M_{*,\rm bulge} < 10^{10} M_{\sun}$ ($n \sim 1.55 \pm 1.06$,
$B/T \sim 0.10 \pm 0.10$).  Large $M_{*,\rm bulge}$ and high $n$
values, however, are not strictly associated with radio detection at
$e$-MERLIN sensitivity. Most massive undetected sources are ALGs, the
second most common being LINERs.

\subsubsection{Radio loudness }\label{Sec4.2.3}

\citet[][see Section 3]{2021MNRAS.508.2019B} used
$L_{\rm core}/L_{\rm {[OIII]}}$ together with SMBH masses for
discriminating `radio-loud' galaxies from `radio-quiet' ones in the
LeMMINGs sample. They classified as radio-loud those galaxies having
log ($L_{\rm core}/L_{\rm {[OIII]}}$) > $-$2 and log
($M_{\rm BH}/M_{\sun}$) > 7.7. Using the
\citet[][]{2021MNRAS.508.2019B} classification, we find 18 radio loud
galaxies out of the sample of 173 galaxies, 17 of which are LINERs and
one is a (jetted) \mbox{H\,{\sc ii}} galaxy (see
Table~\ref{Tab4}). For radio loud galaxies, we find that the median
\mbox{log ($M_{*,\rm bulge}/M_{\sun}$)}, $n$ and $B/T$ are 11.3 $\pm$
0.5, 3.0 $\pm$ 1.2 and 0.67 $\pm$ 0.24, respectively (see
Fig.~\ref{RadA}). On the other hand, radio-quiet galaxies have median
\mbox{log ($M_{*,\rm bulge}/M_{\sun}$)}, $n$ and $B/T$ values of 10.0
$\pm$ 1.7, 2.5 $\pm$ 1.4 and 0.29 $\pm$ 0.19, respectively.

\subsection{Correlations between radio luminosity and bulge size,
  $B/T$ and ellipticity } 
 
Here we compare the global optical properties of our galaxies with the
radio core luminosity. Fig.~\ref{Lco_Tr} shows a plot of
$L_{\rm Radio,core}$ as a function of the bulge effective (half-light)
radius ($R_{e}$), dust corrected bulge-to-total light ratio ($B/T$)
and average galaxy ellipticity value within $R_{\rm e}$ (i.e., $\epsilon$).
$L_{\rm Radio,core}$ very nicely correlates with $R_{e}$ (Pearson
 coefficients $r_{p} \sim 0.53$ and Pearson  probability $P \sim 6 \times 10^{-14}$) and
$B/T$ ($r_{p} \sim 0.33$, $P \sim 8 \times 10^{-6}$). $L_{\rm Radio,core}$ also correlates reasonably well with
$\epsilon$ ($r_{p} \sim -0.24$, $P \sim 1 \times 10^{-3}$), but the scatter in this relation is large
(see Fig.~\ref{Lco_Tr}c). The mean trends observed here are reminiscent of
those in Fig.~\ref{HubPro}: moving towards brighter radio core
luminosities the bulge becomes progressively dominant, round and
larger. Although not shown in Fig.~\ref{Lco_Tr},  we find that the
galaxies' radio core luminosities correlate well with their Hubble types  ($r_{p} \sim -0.46$, $P \sim 3 \times 10^{-10}$, see Fig.~\ref{RadBE}).

\subsection{The S\'ersic/core-S\'ersic dichotomy: isophote shape,
  flattening, kinematics and radio loudness}\label{IsFKR}

The ``core''/``power-law'' structural dichotomy based on the inner
slope of the Nuker model
(\citealt{1995AJ....110.2622L,2005AJ....129.2138L}, see
Section~\ref{InnerS}) was reported to be associated with various
properties of early-type galaxies such as radio loudness
\citep{2005A&A...440...73C,2006A&A...447...97B,2011MNRAS.415.2158R},
X-ray luminosity \citep{2005MNRAS.364..169P,2009Natur.460..213C}, a
measure of departures of isophotes from pure elliptical shape
($B_{4}$,
\citealt{1989A&A...215..266N,1991A&A...244L..37N,2009ApJS..181..135H}),
ellipticity ($\epsilon$,
\citealt{1996AJ....111.2243T,1994AJ....108.1598F,1997AJ....114.1771F,2011MNRAS.414..888E})
and kinematics
(e.g. \citealt{1983ApJ...266...41D,1988A&A...193L...7B,1994AJ....108.1567J,2011MNRAS.414..888E}).
However, a key issue with some of these earlier studies was the
identification of depleted cores by the Nuker model.  Extensive
investigations by \citet{2003AJ....125.2951G,2004AJ....127.1917T,2012ApJ...755..163D,2013ApJ...768...36D, 2014MNRAS.444.2700D, 2019ApJ...886...80D,2021ApJ...908..134D} have shown that
application of the Nuker model could wrongly identify cores in genuine
coreless (i.e., S\'ersic) galaxies which do not have depleted cores
relative to the outer S\'ersic profile. The S\'ersic and core-S\'ersic
models provide robust means to identify depleted cores of
core-S\'ersic galaxies and to parameterise the central and global
structural properties of S\'ersic and core-S\'ersic galaxies
(Sections~\ref{Mods} and. \ref{FitIN}).

As noted in Section~\ref{SB}, $B_{4}$ provides a measure of the
deviations of isophotes from perfect ellipses. Notably, boxy
($B_{4} < 0$) and pure elliptical ($B_{4} = 0$) isophotes of massive
early-type galaxies are widely regarded as consequences of `dry',
violent relaxation of their stars
\citep[e.g.,][]{1982MNRAS.201..939V,1984ApJ...281...13M,2006MNRAS.373.1013C,2006ApJ...636L..81N,2008ApJ...685..897B,2009ApJS..181..486H}. On
the other hand, discy ($B_{4} > 0$) isophotes commonly found in less massive
galaxies form in dissipative (gas rich) processes
\citep[e.g.,][]{1993ApJ...416..415H,1994ApJ...437L..47M,2006ApJ...645..986R,1997AJ....114.1771F,2009ApJS..181..135H}.
 
By separating the sample into S\'ersic and core-S\'ersic galaxies, we
can identify any trends between the radio core luminosity
($L_{R,\rm core}$) and the median of the isophote shape parameter of
the bulge inside $R_{\rm e}$ ($B_{4}$) and average bulge ellipticity
within $R_{\rm e}$ (Fig.~\ref{RadBE}).  For $B_{4}$, we opt for the median as the mean $B_{4}$ values can  easily be biased negative or positive (away from the central tendency) by a few extremely  high and low $B_{4}$ values in the dataset. 
We witness a trend for
core-S\'ersic galaxies (enclosed in boxes in the figure) to be
systematically round. They also show a tendency to possess high radio
core luminosities and boxy-distorted or pure elliptical (i.e.,
neutral) isophotes, in broad agreement with past work
\citep[e.g.,][]{1983A&A...127..205H,1997AJ....114.1771F,1984MNRAS.206..899D,1991AJ....101..148W,1965ApJ...141.1560S,1995MNRAS.276.1373S,2005A&A...440...73C,2006A&A...447...97B,2011MNRAS.415.2158R}. Note that we consider  $|B_{4}|$ values < 0.001 as neutral. 
In general, S\'ersic galaxies tend to have discy isophotes, high
ellipticities and lower radio core luminosities. However, we did not
find evidence for the previously alleged strong tendency of the
central structures to correlate with radio core luminosity,
radio-loudness, isophote shape and bulge ellipticity
\citep[e.g.,][]{2012ApJ...759...64L,2009ApJS..182..216K,2013ARA&A..51..511K}
as revealed by the fact that roughly half (11/20) of the core-S\'ersic
galaxies show boxy/neutral isophotes, while the remaining half exhibit
discy isophotes. 

 Furthermore, all core-S\'ersic galaxies fall inside
the region defined by $L_{R,\rm core} \ga 10^{35}$ erg s$^{-1}$ and
$\epsilon \la 0.3$ but they are cospatial with early- and late-type
S\'ersic galaxies which cover a wide range of $L_{R,\rm core}$,
$B_{4}$, and $\epsilon$. For core-S\'ersic, S\'ersic, early-type and
late-type galaxies in our sample, we measure median $B_{4}$
values \mbox{$\sim$($-0.007 \pm 0.08)\times10^{-2}$}, \mbox{($0.202 \pm 0.90) \times10^{-2}$},  \mbox{(0.174  $\pm$ 0.80)$\times10^{-2}$},
and  \mbox{($0.162 \pm 0.90)\times10^{-2}$} and average $\epsilon$ values $\sim$ 0.21 $\pm$
0.08, 0.30 $\pm$ 0.15, 0.24 $\pm$ 0.10 and 0.32 $\pm$ 0.16.
For the sample disc galaxies, we find that the bulge ellipticity differs (by more than 15\%) from the average galaxy ellipticity exterior to $R_{\rm e}$ in 77\% of the cases, indicating the ellipticities in these systems are not due to projection. As for the
radio-loudness of our core-S\'ersic galaxies, we show that only 7/20 (35\%)
are radio-loud. However, as in the case of core-S\'ersic galaxies the
hosts of most radio-loud sources are massive bulges with low
ellipticities and boxy-distorted isophotes. The fraction of radio-loud
sources with $M_{*,\rm bulge} > 10^{11} M_{\sun}$, $B_{4} < 0$ and
$\epsilon < 0.32$ are 72\%, 61\% and 67\%, respectively. We determine
the typical uncertainties associated with $\epsilon$ and $B_{4}$ to be
$\sim 20\%$ and $\sim 35\%$. In summary, S\'ersic and core-S\'ersic
bulges cannot be distinctively distinguished by their radio-loudness,
$L_{R,\rm core}$, $B_{4}$ or $\epsilon$, in clear departure from past
conclusions. 

In their IFU stellar kinematic study of the ATLAS$^{\rm 3D}$ galaxy
sample of 260 early-type galaxies,
\citet{2007MNRAS.379..401E,2011MNRAS.414..888E,2013MNRAS.433.2812K}
revealed that most slow rotators (SR) which are relatively round
systems with $< 0.3$ are core-S\'ersic galaxies, further reinforcing
the dry merger scenario, while most fast rotators (FR) which span a
large range in $\epsilon \sim 0.05-0.6$ are coreless
galaxies (see also \citealt{2014MNRAS.444.3357N}). \citet{2013MNRAS.433.2812K} also went on to remark that the
correspondence between these galaxies' central structure and
kinematics is not one-to-one. There are 30 early-type galaxies in
common between our sample and the ATLAS$^{\rm 3D}$ galaxy sample
\citep{2011MNRAS.414..888E} which encompass 4 SRs and 26 FRs. We
confirm the tendency for SRs (FRs) to be core-S\'ersic (S\'ersic)
galaxies: 3/4 SRs are core-S\'ersic galaxies, while 22/26 FRs are
S\'ersic galaxies. While the radio
detection fraction that we find for FRs (12/26)  is different from that of the SRs (1/4), conclusions, about the
connection between the bulge kinematics and nuclear radio emission,
cannot be drawn due to the very low number of SRs in the subsample.

\section{Conclusions}\label{ConV} 

We present an accurate structural analysis of high-resolution {\it
  HST} imaging for a representative subsample of 173 (23 Es, 43 S0s,
102 Ss and 6 Irrs) galaxies drawn from the full sample of 280 nearby
galaxies in the $e$-MERLIN legacy survey (LeMMINGs,
\citealt{2014evn..confE..10B}). The aim is to investigate the nuclear
activity, optical and radio properties at sub-arcsec resolution using
{\it HST} and 1.5 GHz $e$-MERLIN radio observations.  This work focuses
on the results from our {\it HST} imaging analyses, which is coupled
in more detail with our 1.5 GHz $e$-MERLIN radio data in Dullo et al.\
(2022, submitted).  Using {\it HST} (ACS, WFPC2, WFC3 and NICMOS)
images, we have extracted new, 1D major-axis surface brightness,
$B_{4}$, P.A. and ellipticity profiles for a sample 163 LeMMINGs
galaxies and these are combined with data for an additional 10
LeMMINGs galaxies from our previous work.  We perform accurate
multi-component decompositions of the surface brightness profiles,
which extend out to $R \ga 80-100\arcsec$ and cover
$\ga 2R_{\rm e,bulge}$ for 97\% of the sample, fitting simultaneously
up to six galaxy components (i.e., bulge, disc, depleted core, AGN,
NSC, bar, spiral arm, ring and stellar halo), which are summed up to a
full model with (up to) 16 free parameters. The median rms residual
for our fits is $ \Delta \sim$ 0.065 mag arcsec$^{-2}$. Galaxy components were carefully identified before models were fit
  to the galaxy profiles.  We also perform 2D decompositions of the
  {\it HST} images 65 sample galaxies, including nearly half of the
  spiral galaxies in the paper (i.e., 49/102). We find that,
  regardless of the galaxy morphology, careful 1D and 2D galaxy
  decompositions result in strong agreements.  This strong agreement,
   the capabilities of our 1D decompositions (unlike 2D decompositions)  
   to capture radial tendencies of ellipticity,
P.A., and boxy/discyness profiles of galaxies and the informative nature
 of 1D residual profiles motivate us to adopt the results
  from the 1D decompositions throughout the this work.  

The LeMMINGs {\it HST} sample encompasses all morphological types and spans over six
orders of magnitude in stellar mass
($6 \la \log M_{*, \rm bulge} \la 12.5$).  Our work represents the
largest, most detailed structural analysis of nearby galaxies with {\it HST} to date,
providing accurate structural parameters for 173 galaxies, a major
improvement over past studies especially given the large number of 108
late-type galaxies (Tables~ \ref{Tab1}, \ref{Tab1Ring} and \ref{Tab4}).
Having carefully isolated galaxy components, we derived luminosities
and stellar masses for the bulges, nuclear components and host
galaxies (Table~\ref{Tab4}).  We have also implemented an innovative
method to estimate realistic uncertainties on the fit parameters after
creating over 100 realisations of each galaxy's light profile that
were later decomposed akin to the original galaxy profiles.

The main results  are as follows:

(1) We have highlighted  the need for performing fits beyond the two main
galaxy components (i.e., bulge-disc profile), and calculated, from
the detailed decompositions, the fraction of galaxy light outside the
bulge+disc component ($f_{\rm other}$): $f_{\rm other} \ga$ 20\% for
43/167 of sample galaxies and $f_{\rm other}>$ 5\% for 67/167. To do
this, we excluded irregular galaxies.  Fitting a bulge-disc model to
light profiles of disc galaxies which have components such as strong
bars, rings and spiral arms could thus overestimate the actual
bulge/disc mass by $\ga$ 25\% for $\sim$26\% of the cases.  For
spirals, $f_{\rm other}$ tend to increase as the bulge mass and
luminosity increases, whereas S0s typically show $f_{\rm other} \sim$
1.3\%. For most (71\%) elliptical galaxies $f_{\rm other} \sim 0$,
although for the most massive galaxies (such as BCGs and cD galaxies)
$f_{\rm other}$ is typically $\ga 5\%$. For  nearby galaxies, we suggest  using imaging data with a resolution of 
$\la 0\farcs2$, particularly  in the inner regions, for accurate  multi-component decompositions and to reproduce  
  the level of detail achieved in our galaxy decomposition analysis.

(2) An overwhelming majority (95\%) of our sample galaxies have
sub-kpc stellar structures (including depleted cores) detected with
the analysis of the {\it HST} imaging.  Correspondingly, we find that
our high-sensitivity $e$-MERLIN $L$-band observations permit for
analogous radio continuum detections at a comparable resolution to
that of the {\it HST}, however, only $\sim$15\% ($\sim$22\%) of the
sub-kpc structures can be well studied with {\it Spitzer} or
ground-based imaging data (e.g., SDSS).

(3) Our analysis of the high spatial resolution {\it HST} imaging
reveals that nuclei, presented as central light excesses, are
ubiquitous in nearby galaxies with a frequency of $\sim 72\%$
(124/173).

(4) All 20 of our core-S\'ersic galaxies have
$M_{*, \rm bulge} \ga 8 \times 10^{10} M_{\sun}$ and most (80\%) of
them have $n \ga 4$.  In contrast, the vast majority ($\sim 90$\%) of
S\'ersic bulges have $M_{*, \rm bulge} \la 10^{11} M_{\sun}$ and
$n \la 4$.  For our core-S\'ersic galaxies, the median values of
$M_{V, \rm bulge}$, $M_{*, \rm bulge} $ and $n$ are $-$21.40 mag,
$ 2.4 \times 10^{11} M_{\sun}$ and 4.3. For S\'ersic galaxies, the
median values of $M_{V, \rm bulge}$, $M_{*, \rm bulge} $ and $n$ are
-19.0 mag, $10^{10} M_{\sun}$ and 2.4.
 
(5) We find no evidence for a previously reported bimodal distribution
of the inner logarithmic light profile slope $\gamma$
\citep[e.g.,][]{2007ApJ...664..226L} in the $\gamma-M_{*,\rm bulge}$
plane due to S\'ersic and core-S\'ersic galaxies, that could be
interpreted as a diagnostic of galaxy formation mechanisms. Instead,
the distribution appears  continuous, in good agreement with e.g.,
\citet{2007ApJ...671.1456C}. Overall, core-S\'ersic bulges have
$\gamma \la 0.35$, whereas S\'ersic early-type bulges exhibit a large
$\gamma$ range from 0.01 to 0.70.  For the first time, we show that
late-type S\'ersic galaxies exhibit a wide range of slopes
($\gamma \sim 0.00- 1.71$), analogous to early-type S\'ersic galaxies.

(6) There are strong trends between galaxy Hubble type and bulge size
($R_{e}$), bulge-to-total light ratio ($B/T$) and ellipticity
($\epsilon$). Moving from the irregular types to elliptical galaxies,
bulges on average are larger, more prominent and round, confirming
past work.  Similarly, tight correlations between the radio core
luminosity ($L_{\rm R,core}$) and $R_{e}$, $B/T$ and $\epsilon$ are
such that bulges with brighter $L_{\rm R,core}$ are dominant, round
and larger. Unsurprisingly, all the 10 bulgeless galaxies (5.8\%) in
our sample have a morphological type later than Sbc.  We find median
values of $B/T \sim$ 0.21, 0.46 and 0.86 for spiral, S0 and elliptical
galaxies, respectively. The median values of bulge S\'ersic indices
($n$) and $B/T$ for Seyfert, ALG, LINER and \mbox{H\,{\sc ii}} nuclei
are $n\sim 2.69 $, $3.01$, $2.84$ and $1.48$, and $B/T \sim 0.29 $,
$ 0.57 $, $0.44$ and $ 0.11$.

(7) We find that the fraction of galaxies harbouring emission-line
AGN is a strong function of $M_{*,\rm bulge}$ and
$M_{V,\rm bulge}$.  The majority of AGN (83\%) hosts
have $M_{\rm bulge}$ $\ga 10^{10} M_{\sun}$
($M_{\rm glxy} \ga 10^{10.5} M_{\sun}$).  For low mass galaxies
($M_{*, \rm glxy} \la 8 \times 10^{9} M_{\sun}$), we find an AGN
fraction of 6.2\%.   All the 10 bulgeless galaxies in
our sample are \mbox{H\,{\sc ii}} galaxies, confirming the low
incidence of AGN for bulgeless galaxies, some of which are known to
house a SMBH but rarely an AGN (See \citealt{2003MNRAS.341...54K,2008ApJ...688..159G,2013MNRAS.429.2199S}).

(8) We find that the radio detection fraction increases with bulge
mass $M_{*,\rm bulge}$, $n$ and $B/T$. At
$M_{*,\rm bulge} \ga 10^{11} M_{\sun}$, the radio detection fraction
is 77\%, declining to 24\% for $M_{*,\rm bulge} < 10^{10} M_{\sun}$.
Furthermore, we report a tendency for core-S\'ersic galaxies to be
systematically round and to possess high radio-core luminosities and
boxy-distorted/pure elliptical isophotes but there is no evidence for
the previously alleged strong correlation of the central structures
(i.e., a sharp S\'ersic/core-S\'ersic dichotomy) with radio-loudness, $B_{4}$,
$L_{R,\rm core}$ and $\epsilon$ (e.g.,
\citealt{2009ApJS..182..216K,2012ApJ...759...64L}).  Of the 20
core-S\'ersic galaxies in the sample, only 7/20 (35\%) are
radio-loud. Also, all core-S\'ersic galaxies are confined to the
region defined by $L_{R,\rm core} \ga 10^{35}$ erg s$^{-1}$ and
$\epsilon \la 0.3$ but they are cospatial with S\'ersic galaxies, the
latter cover a large range of $L_{R,\rm core}$, $B_{4}$, and
$\epsilon$.  Nonetheless, our results, are overall in accordance with
cosmological models which predict that the most massive early-type
galaxies are more round, and have boxy isophotes, compatible with their
formation and evolution scenario as a more evolved object which have
undergone several major (dry) mergers. This formation process then
results in a dominant bulge housing a massive black hole, a crucial
precondition for supporting the launch of jets and outflows in the
radio band \citep[e.g.,][]{2014ARA&A..52..589H,2021MNRAS.508.2019B}.

We will investigate the relations between the radio core luminosity
and the host bulge properties in an upcoming paper. A multi-wavelength
view of the origin and formation mechanisms of nuclei (NSCs/AGN) and
the AGN triggering processes and their relation with host galaxy
environments will be presented in future papers. Further observations
and analysis are in progress to exploit synergies from a large sample
of multi-wavelength ({\it HST} optical/near-infrared, $e$-MERLIN radio
and Chandra X-ray) data.

\section{ACKNOWLEDGMENTS}
We thank the  anonymous referee for their careful reading of the manuscript 
and for  their suggestions.  We would like to thank 
Alex Rosenthal for the help with the 2D {\sc imfit} decompositions of the {\it HST} images. 
B.T.D acknowledges support from grant `Ayudas para la realizaci\'on de
proyectos de I+D para j\'ovenes doctores 2019.' for the HiMAGC
(High-resolution, Multi-band Analysis of Galaxy Centres) project
funded by Comunidad de Madrid and Universidad Complutense de Madrid
under grant number PR65/19-22417. This work has been supported by  the AEI-MCINN grant RTI2018-096188-B-I00.
J.H.K. acknowledges financial
support from the State Research Agency (AEI-MCINN) of the Spanish
Ministry of Science and Innovation under the grant `The structure and
evolution of galaxies and their central regions' with reference
PID2019-105602GB-I00/10.13039/501100011033, from the ACIISI,
Consejer\'{i}a de Econom\'{i}a, Conocimiento y Empleo del Gobierno de
Canarias and the European Regional Development Fund (ERDF) under grant
with reference PROID2021010044, and from IAC project P/300724,
financed by the Ministry of Science and Innovation, through the State
Budget and by the Canary Islands Department of Economy, Knowledge and
Employment, through the Regional Budget of the Autonomous
Community. JSG thanks the University of Wisconsin-Madison for partial
support of this research.  CGM acknowledges financial support from Jim
and Hiroko Sherwin.  We would like to acknowledge the support the
$e$-MERLIN Legacy project `LeMMINGs', upon which this study is based.
$e$-MERLIN, and formerly, MERLIN, is a National Facility operated by
the University of Manchester at Jodrell Bank Observatory on behalf of
the STFC. We acknowledge Jodrell Bank Centre for Astrophysics, which
is funded by the STFC.

This work has made use of {\sc numpy} \citep{2011CSE....13b..22V},
{\sc matplotlib} \citep{Hunter:2007} and {\sc corner}
\citep{2016JOSS....1...24F} and {\sc astropy}, a community-developed
core {\sc python} package for Astronomy
\citep{2013A&A...558A..33A,2018AJ....156..123A}, and of {\sc topcat}
(i.e., `Tool for Operations on Catalogues And Tables',
\citealt{2005ASPC..347...29T}).
 
\section{Data availability}

The data underlying this article are available in the article and in
its online supplementary material.

\bibliographystyle{aa}
\bibliography{Bil_Paps_biblo}

\newpage
\newpage

\setcounter{section}{0}
\renewcommand{\thesection}{A\arabic{section}}

\setcounter{figure}{0}
\renewcommand{\thefigure}{A\arabic{figure}}

\section{ Notes on Selected Individual Galaxies}\label{NotesI}

Direct comparison of our multi-component structural decomposition with
past fits in the literature is not straightforward as detailed
decompositions of {\it HST} data for a large sample of (nearby)
late-type galaxies are not available. Here, we comment on our fits for
54/173 galaxies and compare them, primarily, with those in
\citet{2015ApJS..219....4S} and \citet{2019ApJ...873...85D} and, in a
few cases, with those from other studies in the literature.
\citet{2015ApJS..219....4S} provided two-dimensional, multi-component
decompositions of 3.6 $\mu$m images for 2352 nearby galaxies, while
\citet{2019ApJ...873...85D} presented one-dimensional, multi-component
decompositions for 43 nearby spiral galaxies with measured SMBHs
relying mostly on 3.6 $\mu$m {\it Spitzer} data.  We include all the
sample galaxies in common with \citet{2019ApJ...873...85D}. Of the 101
galaxies in common between us and \citet{2015ApJS..219....4S}, here we
discuss 38. In selecting these 38 galaxies we try to create a
representative subsample (in terms of the morphology  and  number of fitted galaxy  components) of the 101 overlapping galaxies and to
include all galaxies when there are notable disagreements with
\citet{2015ApJS..219....4S}.

 {\it IC~2574.}  The 1D {\it HST} surface brightness  profile is
 well described by the exponential disc + S\'ersic bar + two-parameter
 Gaussian nucleus model (Fig.~\ref{FigSer1}). \citet{2015ApJS..219....4S} fitted a 2D
 exponential disc model to their 3.6 $\mu$m  {\it Spitzer} data.

 {\it NGC~2273.}  We decompose the  1D {\it HST} surface
 brightness profile into five components (bulge, disc, bar, ring and
 nucleus), which are best fitted with four S\'ersic models and a
 Gaussian function. The {\it HST} surface brightness profile
 decomposition of this galaxy by \citet{2019ApJ...873...85D} shows a
 bulge, a disc, a bar plus six, 3-parameter Gaussian components. They
 did not detect the nuclear component.

 {\it NGC~2634.} This galaxy is classified as an elliptical galaxy in
 the Third Reference Catalogue, RC3
 \citep{1991rc3..book.....D}. \citet{2010ApJS..190..147B,2012ApJ...753...43K}
 showed the presence of multiple outer shell structures in the galaxy,
 which may indicate of recent gravitational interaction or merger.   We
 therefore fit 1D and 2D S\'ersic bulge  +  outer exponential shell  + Gaussian nucleus model to the galaxy data.

  {\it NGC~2655. }  We fit the {\it HST} data with a S\'ersic bulge + an outer exponential disc + a
 Gaussian nucleus model.  \citet{2015ApJS..219....4S} fitted a
 3-component bulge+bar+disc 2D model to their 3.6 $\mu$m  Spitzer image.
 The best-fitting S\'ersic indices from our  1D and 2D S\'ersic  bulge models are $n \sim 2.7$, smaller than
  that from   \citet[][$n \sim 4.4$]{2015ApJS..219....4S}. The latter value is high for a S\'ersic 
galaxy.

 {\it NGC~2748.}   The 2D and 1D {\it HST} data  is well fitted by a ($n  \sim 2.2$)  S\'ersic
 bulge, an outer exponential disc plus a Gaussian nucleus model.
 \citet{2019ApJ...873...85D} fitted their major-axis {\it HST} surface brightness profile for the galaxy with a 
 ($n  \sim 1.59$) S\'ersic bulge+ Edge-on disc model and two, faint, Gaussian
 components. \citet{2015ApJS..219....4S} performed 2D decomposition of
 the 3.6 $\mu$m  {\it Spitzer} galaxy image into an exponential disc
 plus a Gaussian nucleus.

 {\it NGC~2768.}  This galaxy is classified as an E6 in the RC3
 \citep{1991rc3..book.....D}, but it was classified as an S0 by
 \citet{1981rsac.book.....S}. We fit a S\'ersic bulge plus an outer
 exponential disc plus a Gaussian nucleus model to our composite {\it
   HST}+ground-based surface brightness profile.  \citet{2015ApJS..219....4S}
 fit a 2D S\'ersic bulge + outer exponential disc model to their 3.6 $\mu$m 
 {\it Spitzer} image.

 {\it NGC~2770.} The galaxy is classified as an SA(s)c in the RC3
 \citep{1991rc3..book.....D}. We therefore fit a S\'ersic bulge, an
 outer exponential disc, and a Gaussian nucleus model, whereas
 \citet{2015ApJS..219....4S} reported  the galaxy has a bar. They fitted
 the bulge component with an exponential model, in good agreement with
 our S\'ersic ($n \sim 0.9$) bulge model.

 {\it NGC~2787.} We decompose our 1D {\it HST} surface brightness
 profile in to bulge, a disc, a bar, an inner disc and a nucleus,
 whereas the 2D bulge+disc+bar model fitted to the {\it
   Spitzer} image by \citet{2015ApJS..219....4S} did not include a nucleus
 and an inner disc. Our best-fitting S\'ersic bulge model  ($n \sim 1.3$),  compare to
  that from   \citet[][$n \sim 2.9$]{2015ApJS..219....4S}.

 {\it NGC~2859.}   The 1D {\it HST}+ground-based light profile is well decomposed into six components:  a bulge, a disc, two bars, a ring and
 a nucleus (Figs.~\ref{MultD} and \ref{FigSerR}). Our 2D decomposition  of the galaxy's {\it HST} image identified a seventh component, i.e., an inner ring which encloses  the inner bar.  In their 2D decomposition of the galaxy's {\it Spitzer}
 image, \citet{2015ApJS..219....4S} fitted a bulge, an outer
 exponential disc, a secondary disc fitted by a Ferrers function and a
 bar.

 {\it NGC~2964.}  Our 1D and 2D decompositions  show that the galaxy has a bulge,
 a disc, a spiral-arm component and a nucleus. We fit
 the spiral-arm component with a 3-parameter Gaussian
 function. \citet{2015ApJS..219....4S} fitted a 2D bulge+disc model to
 their {\it Spitzer} data. The bulge+disc model was also used by
 \citet{2009MNRAS.395.1669G} to fit the 1D {\it HST}+ground based data
 for the galaxy.

 {\it NGC~3031.}  We fit the 1D {\it HST}+SDSS light profile
 with a ($n \sim 4$) S\'ersic bulge, an exponential disc, a weak
 S\'ersic Bar and a Gaussian nucleus. The 2D decomposition of
 the {\it Spitzer} 3.6 $\mu$m stellar distribution of the galaxy by
 \citet{2015ApJS..219....4S} show a ($n \sim 3.6$) S\'ersic bulge plus
 an outer, exponential disc components. In contrast,
 \citet{2019ApJ...873...85D} modelled the 1D {\it Spitzer} 3.6 $\mu$m
 major-axis surface brightness profiles of NGC~3031 with a
 core-S\'ersic bulge + exponential disc model, they also added four,
 relatively faint Gaussian components.

 {\it NGC~3073.}  We decompose the 1D and 2D  {\it HST} light distributions into  four components (bulge-disc-bar-nucleus),
 fitting three S\'ersic models and a Gaussian function.
 \citet{2015ApJS..219....4S} modelled the 2D {\it Spitzer} 3.6 $\mu$m
 stellar distribution of the galaxy with a S\'ersic bulge and an outer
 exponential disc model.

 {\it NGC~3077.}  This galaxy is classified as an I0 in the RC3
 \citep{1991rc3..book.....D}.  We classify NGC~3077
 as `bulgeless' and fit its  1D and 2D  {\it HST}(+SDSS)
data with an outer ($n \sim 1.8$) S\'ersic disc, a
 S\'ersic inner disc and a Gaussian nucleus model.  \citet{2015ApJS..219....4S} modelled their 2D
 {\it Spitzer} 3.6 $\mu$m stellar distribution with a single, 
 ($n \sim 3.2$) S\'ersic bulge model.

 {\it NGC~3079.} We decompose the {\it HST}+SDSS surface
 brightness profile into a ($n \sim 2.3$ S\'ersic
 bulge)-disc-bar-(spiral-arm)-nucleus model profile. The 1D
 decomposition of the {\it Spitzer} 3.6 $\mu$m major-axis data by
 \citet{2019ApJ...873...85D} was performed using a ($n \sim 0.5$)
 S\'ersic bulge, an edge-on disc, a Ferrers bar plus three,
 additional, Gaussian components.  \citet{2015ApJS..219....4S}
 modelled the 2D { \it Spitzer} 3.6 $\mu$m stellar distribution of the
 galaxy with a 2D ($n \sim 1.7$) S\'ersic disc plus an edge-on
 exponential disc model.

 {\it NGC~3184.}  We model the 1D {\it HST}+SDSS surface brightness
 profile for this galaxy with a ($n \sim 2.3$) S\'ersic bulge, an
 outer disc and a nucleus. Similarly,  \citet{2015ApJS..219....4S} modelled their 2D
 {\it Spitzer} 3.6 $\mu$m stellar distribution with a ($n \sim 1.4$)
 S\'ersic, an outer  disc and a nuclear component.  
 
 {\it NGC~3198.}  This galaxy is a barred spiral galaxy
 \citep{1991rc3..book.....D}. We find the galaxy's 1D {\it HST}+SDSS
 light profile is well fitted with a five-component bulge
 ($n \sim 3.4$)+disc+bar+spiral-arm+nucleus model. In contrast,
 \citet{2015ApJS..219....4S} modelled the 2D {\it Spitzer} 3.6 $\mu$m
 stellar distribution with a bulge+disc 2D model and the resulting
 best fitting bulge S\'ersic index ( $n \sim10.0$) is unusually large
 for a S\'ersic galaxy.  

 {\it NGC~3486.}  This galaxy is well modelled by a four-component
 bulge+disc+ spiral-arm+nucleus decomposition
 model.  \citet{2015ApJS..219....4S} fitted the 2D {\it Spitzer} 3.6
 $\mu$m stellar distribution of the galaxy with a bulge and two
exponential disc components.  The S\'ersic index
 of the bulge from our 1D and 2D decompositions is $n \sim 2.4$, in close agreement with that from
 \citet[][$ n \sim 2.0$]{2015ApJS..219....4S}.

 {\it NGC~3504.} The surface brightness profile of this galaxy is well
 described by the bulge+disc+ring+bar+spiral-arm+nucleus decomposition
 model.  \citet{2015ApJS..219....4S} fitted a 2D, three-component
 bulge+disc+bar decomposition model to their {\it Spitzer} 3.6 $\mu$m
 data. 
 
 {\it NGC~3600.}  We fit the 1D and 2D {\it HST} data
 with a ($n \sim 0.8-1.0$) S\'ersic bulge and an exponential disc
 model. \citet{2015ApJS..219....4S} modelled their 2D {\it Spitzer}
 3.6 $\mu$m data with a 2D  ($n \sim 2.0$) S\'ersic bulge plus + edge-on
 disc model.

 {\it NGC~3610.}  This galaxy is an elliptical galaxy
 \citep{1991rc3..book.....D} which hosts an inner disc
 \citep{1990A&A...235...49S,1997AJ....114.1797W}. The galaxy's {\it
   HST} light profile is well described by the S\'ersic bulge + inner
 ($n \sim 0.4$) S\'ersic disc + Gaussian nucleus decomposition model.

 {\it NGC~3613. } This galaxy is classified as an E6 in the RC3
 \citep{1991rc3..book.....D}. However our decomposition of the 1D {\it
   HST} +SDSS surface brightness profile for the galaxy reveals a
 core-S\'ersic bulge and an outer ($n \sim 0.8$) S\'ersic disc.  
 \citet{2004AJ....127.1917T} also modelled their 1D {\it HST} brightness
 profile and identified a `possible core' in the galaxy.

 {\it NGC~3665.} While the {\it HST} NICMOS+SDSS surface
 brightness profile is well fitted by a ($n \sim 3.9$) core-S\'ersic
 bulge and an outer exponential disc model, the nuclear dust in the
 galaxy implies our core identification is tentative.
 \citet{2010MNRAS.405.1089L} performed a 2D decomposition of to their
 ground-based data for the galaxy into a ($n = 2.7$) S\'ersic bulge
 plus an outer exponential disc model.

 {\it NGC~3718.}  Our 1D and 2D  five-component (bulge + outer disc + inner
 disc + spiral-arm + nucleus) decompositions of the {\it HST} data
 yield a bulge S\'ersic index of $n \sim 2$. In contrast,
 the 2D, three-component (S\'ersic bulge + exponential disc + Ferrer
 disc) decomposition of the {\it Spitzer} 3.6 $\mu$m data by
 \citet{2015ApJS..219....4S} yielded a bulge S\'ersic
 index of $n \sim 9.0$, which is unusually high for a S\'ersic galaxy.

 {\it NGC~3884.} This galaxy is classified as an SA(r)0/a in the RC3
 \citep{1991rc3..book.....D}, we therefore fit a ($n \sim 2.7 $)
 bulge + disc + ring + nucleus model to the 1D {\it HST} light
 profile and 2D {\it HST} image. Because \citet{2006AJ....131.1236D} modelled their 2MASS
 data in 2D with a single, S\'ersic bulge model and did not account
 for the disc and ring components, the resulting best-fitting bulge
 $n$ is biased high (i.e., $ \sim 3.8 $).

 {\it NGC~3898.}   Decomposing the 1D {\it HST}+SDSS light profile of NGC~3898 into 
 a bulge, a disc, a spiral-arm component  and  a nucleus, we find $n \sim 3.4$ for the bulge. In contrast, 
 \citet{2015ApJS..219....4S} fitted
 a, 2D, ($n \sim 5.2$) S\'ersic plus an exponential disc model to the {\it
   Spitzer} 3.6 $\mu$m data.

 {\it NGC~3949.}  We describe the 1D {\it HST} surface brightness profile and the 2D {\it HST} image for the galaxy as the sum
 of a ($n \sim 1.0-1.3$) S\'ersic bulge, an exponential disc, a
 Gaussian spiral-arm and a Gaussian nucleus. Our decomposition can be compared to
  \citet{2009MNRAS.395.1669G} who fitted a
 ($n \sim 1.1$) bulge plus an exponential disc model to their 1D {\it
   HST}+ground-based light profile.

 {\it NGC~3982.}  This galaxy has a weak bar component. In our
 (bulge + disc + bar + spiral-arm + nucleus) decomposition, the bar was fitted
 with a ($n \sim 0.2$) S\'ersic model. In contrast,  \citet{2015ApJS..219....4S}
  fitted a  two-component  (exponential  disc + Gaussian nucleus) decomposition
   model to their  {\it Spitzer} 3.6 $\mu$m data. 

 {\it NGC~3998.}  This galaxy has a ($n \sim 2.4$) S\'ersic bulge, an
 exponential disc, a ($n \sim 0.3-0.7$) S\'ersic lens and ($n \sim 0.3-0.9$)
 S\'ersic nucleus (Fig.~\ref{FigSer1}).  In their 2D decomposition of the {\it Spitzer} 3.6
 $\mu$m image, \citet{2015ApJS..219....4S} fitted a ($n \sim 1.5$)
 S\'ersic bulge, an exponential disc, a Gaussian nucleus.

 {\it NGC~4026.} This galaxy has an inner stellar disc
 \citep{2009ApJ...695.1577G}. In our decomposition, we fitted the {\it
   HST} surface brightness profile as the sum of a ($n \sim 1.3$)
 S\'ersic bulge, an outer exponential disc, two ($n \sim 0.3$)
 S\'ersic discy components and a Gaussian nucleus.

 {\it NGC~4036.}  We find that the three-component bulge-disc-nucleus
 light profile of NGC~4036 is well fitted by the ($n \sim 1.4$)
 S\'ersic bulge + exponential disc + ($n \sim 0.4$) S\'ersic nucleus
 decomposition model. Similarly, in their 2D decomposition
 of the galaxy's $i$-band SDSS data, \citet{2012MNRAS.419.2497B}
 fitted a ($n \sim 1.7$) S\'ersic bulge plus an outer exponential disc
 model.

 {\it NGC~4041.}   Our 1D and 2D decompositions reveal that this galaxy has a
 ($n \sim 0.9$) S\'ersic bulge, an outer exponential disc and a
 ($n \sim 0.8$) S\'ersic nucleus.  Our structural analysis can be compared 
 to the three-component, 2D fit by \citet{2015ApJS..219....4S}: a
 ($n \sim 0.9$) S\'ersic disc, an outer exponential disc and a
 Gaussian nucleus.

 {\it NGC~4062.}   We describe the 1D  {\it HST}  light profile and 2D {\it HST}  stellar light distribution of NGC~4062 as the sum
 of a ($n \sim 0.4 -0.7$) S\'ersic bulge, an exponential disc and a
 Gaussian nucleus.  \citet{2015ApJS..219....4S} modelled the galaxy's
 2D, 3.6 $\mu$m galaxy stellar light distribution with a
 ($n \sim 0.7$) S\'ersic bulge plus an outer exponential disc.

 {\it NGC~4096.} This galaxy is classified as an S0 in the RC3. We did
 not identify a bar, thus we fit a 1D/2D bulge-disc-(spiral-arm) model to our
 {\it HST} light profile/image. In contrast, \citet{2015ApJS..219....4S}
 fitted the 3.6 $\mu$m galaxy light distribution with a outer disc, a bar
 and a nucleus.

 {\it NGC~4144.} We fit this galaxy's {\it HST} surface brightness
 profile with a ($n \sim 0.9$) S\'ersic bulge and an exponential disc decomposition  model.
 \citet{2015ApJS..219....4S} identified our bulge component as a bar
 and modelled the 2D, 3.6 $\mu$m stellar light distribution for the
 galaxy with a Ferrers bar plus an outer exponential disc.

 {\it NGC~4151.}  
 The galaxy has a bright point like Seyfert nucleus 
\citep{2017MNRAS.472.3842W,2020MNRAS.495.3079W}, and a large-scale oval bar, 
an inner disc-like  feature seen in molecular gas and optical 
 dust maps \citep{2010ApJ...721..911D}.  The inner  disc-like component  is likely due to the 
gaseous response to the primary bar  \citep{1999MNRAS.304..481M,1999MNRAS.304..475M,2005MNRAS.359..408A}.
Inside this inner gaseous feature is an outflowing ionised gas from the AGN \citep{2022ApJ...940...28K}. \citet{2010ApJ...719L.208W} reported that the galaxy might  have experienced an Eddington-rate outburst $\sim$ 10,000 yrs ago.

We decompose the 1D {\it HST}+SDSS brightness
 profile of NGC~4151 into a ($n \sim 3.9$) S\'ersic bulge, an outer
 exponential disc plus a bar, an inner disc, a spiral-arm component
 and a nucleus. In their five-component decomposition of the 1D {\it
   Spitzer} 3.6 $\mu$m light profile, \citet{2019ApJ...873...85D}
 fitted a ($n \sim 2.2$) bulge, a disc, a bar plus two Gaussian
 components. On the other hand, \citet{2015ApJS..219....4S} decomposed
 the { \it Spitzer} 3.6 $\mu$m galaxy light distribution in 2D with a
 ($n \sim 1.8$) bulge and an outer exponential disc.

 {\it NGC~4203.} This galaxy is classified as an SAB0 in the
 RC3. While \citet{2015ApJS..219....4S} fitted a bar component in
 their 2D decomposition, it was too weak to be included in our
 decomposition. We therefore describe the {\it HST} brightness profile
 as the sum of a S\'ersic bulge, an exponential disc and a Gaussian
 nucleus, while \citet{2015ApJS..219....4S} fitted a 2D (S\'ersic
 bulge)-(exponential disc)-(Ferrers bar) model to their {\it Spitzer}
 3.6 $\mu$m data.  Our best-fitting bulge S\'ersic index
 ($n \sim 2.5$) is in fair agreement (within the quoted 12\% error,
 Table~\ref{Tab1}) with that from \citet[][$n \sim 3.0$]{2015ApJS..219....4S}.

 {\it NGC~4245.} The five-component light profile for this galaxy is
 well described by the ($n \sim 1.7$) S\'ersic bulge + exponential
 disc + S\'ersic bar + Gaussian ring + S\'ersic nucleus decomposition
 model. Our decomposition can be compared to the 2D
 decomposition by \citet{2015ApJS..219....4S} who fit a
 three-component decomposition model: ($n \sim 1.1$) S\'ersic bulge +
 exponential disc + Ferrers bar.

 {\it NGC~4258.} We describe our 1D {\it HST} brightness profile for
 this galaxy as the sum of a ($n \sim 3.6$) S\'ersic bulge, an
 exponential disc plus a bar \citep{2009ApJ...693..946S}, a spiral-arm
 and a nuclear component. Similarly, the decomposition of the
 galaxy's {\it Spitzer} 3.6 $\mu$m data by \citet{2019ApJ...873...85D}
 shows a ($n \sim 3.2$) S\'ersic bulge, an exponential disc plus a
 Ferrers bar and Gaussian nucleus, but they did not account for the
 spiral-arm component.

 {\it NGC 4448.}   In our 1D and 2D five-component
 (bulge+disc+bar+spiral-arm+nucleus) decomposition of this galaxy's 
 stellar light distributions, the bulge is described by a
 ($n \sim 2$) S\'ersic model, while in the  bulge+disc+bar
 decomposition by \citet{2015ApJS..219....4S} the bulge was
 fitted with a ($n \sim 0.9$) S\'ersic model. 

 {\it NGC~4449.}  Our 1D and 2D bulge+disc+bar+nucleus decomposition of the
 {\it HST} data for this galaxy reveals a small ($R_{\rm e} \sim 0.13 -0.16$
 kpc) bulge which is well fitted by a ($n \sim 2.1-3.2$) S\'ersic
 model. This is contrary to the fit in \citet{2015ApJS..219....4S} who
 modelled the 2D {\it Spitzer} 3.6 $\mu$m stellar distribution of the
 galaxy with the exponential disc + Ferrers bar + Gaussian nucleus
 decomposition model.

 {\it NGC~4565.}  Our multi-component decomposition of the {\it HST}
 data for this galaxy shows a bulge, an outer disc, an inner disc, a
 boxy bulge and a nucleus. Our fit agrees well with the 1D {\it HST}
 surface brightness decomposition in \citet{2010ApJ...715L.176K},
 although they did not account for the faint, inner disc and nucleus
 (see also \citealt{2016ASSL..418...77L}).

 {\it NGC 4605.}  We model the 1D and 2D light distributions  of the galaxy
 with a ($n \sim 0.3-0.5$) S\'ersic bulge, a nucleus and  an outer exponential disc
 model. \citet{2015ApJS..219....4S} fitted a ($n \sim 0.7$) S\'ersic
 disc plus an outer exponential disc decomposition  model to their 3.6 $\mu$m {\it
   Spitzer} data.

 {\it NGC 4648.}  This galaxy is classified as an E3 in the RC3
 \citep{1991rc3..book.....D}, therefore we describe the 1D and 2D {\it HST}
data as a sum of a ($n \sim 2.6-2.8$) S\'ersic bulge, an outer
 exponential halo, a S\'ersic ring and a Gaussian
 nucleus. \citet{2014ApJ...791..133B} modelled the two-dimensional
 {\it HST} light distribution of the galaxy with an inner S\'ersic
 component plus an outer exponential disc.

 {\it NGC~4736.} In our decomposition, we fit a near-exponential ($n \sim 0.9$)
 S\'ersic bulge, an outer exponential disc plus an inner (S\'ersic)
 ring, a (3-parameter) Gaussian spiral-arm and a (2-parameter)
 Gaussian nucleus to this galaxy's 1D {\it
   HST}+SDSS light profile.  \citet{2015ApJS..219....4S} performed 2D
 decomposition of the {\it Spitzer} 3.6 $\mu$m data of the galaxy into
 a ($n \sim 1.8$) S\'ersic bulge and two exponential discs. On the
 other hand \citet{2019ApJ...873...85D} described the galaxy's 1D
 (major-axis) {\it Spitzer} 3.6 $\mu$m light profile with
 ($n \sim 0.9$) S\'ersic bulge, an outer exponential disc plus four
 (3-parameter) Gaussian components.

 {\it NGC~4800.} This galaxy show a near-exponential (i.e.,
 $n \sim 1.3$) bulge, an outer exponential disc, a circumnuclear ring
 \citep{2008A&A...478..403C}, a spiral-arm component and a Gaussian
 nucleus.  \citet{2015ApJS..219....4S} fitted a two exponential discs
 plus a Gaussian nucleus.

 {\it NGC 4826.} The 1D {\it HST}+SDSS light profile is well
 described as a sum of a S\'ersic bulge, an exponential disc, a
 Gaussian ring and a Gaussian nucleus.  The S\'ersic index of the
 bulge is $n \sim $2.3.  \citet{2015ApJS..219....4S} modelled their 3.6 $\mu$m 
 {\it Spitzer} data for the galaxy with a ($n \sim 4.2$) S\'ersic
 bulge and two exponential discs, whereas \citet{2019ApJ...873...85D}
 modelled their 1D
 (major-axis) 3.6 $\mu$m  {\it Spitzer} light profile  with a
 ($n \sim 0.7$) S\'ersic bulge, an outer exponential disc plus five
 Gaussian components.

 {\it NGC~5005.}  While this SAB(rs)bc galaxy
 \citep{1991rc3..book.....D} is reported to have a bar by
 \citet{2015ApJS..219....4S}, it is not visible in our structural
 analysis.  Also, the multi-wavelength structural analysis by
 \citet{2015MNRAS.449.3981R} did not detect a bar component in the
 galaxy.  We fitted the 1D {\it HST}+SDSS light profile as the
  sum of  a bulge, an outer disc, an inner clumpy disc, a
 spiral arm and a nucleus.

 {\it NGC~5055.}  In our decomposition of this galaxy's 1D and 2D {\it
   HST}+SDSS data, we fit a ($n \sim 0.8-1.3$) S\'ersic bulge, an
 outer exponential disc plus  a spiral-arm component and a nucleus   which
 are described by two Gaussian functions.   In their 2D decomposition
 of the galaxy's {\it Spitzer} 3.6 $\mu$m data,
 \citet{2015ApJS..219....4S} fitted two exponential discs plus a
 Gaussian nucleus.  On the other hand, \citet{2019ApJ...873...85D}
 modelled the 1D major-axis {\it Spitzer} 3.6 $\mu$m data with a
 ($n \sim 2.2$) S\'ersic bulge, an exponential disc and a Gaussian
 nucleus.

 {\it NGC~5308.} \citet{1996A&A...310...75S} reported the presence of
 a rapidly rotating inner disc component in this galaxy. In addition,
 the kinematic maps by \citet{2004MNRAS.352..721E} reveal that the
 central dispersion structure in the galaxy is box-shaped. Our
 decomposition shows a near-exponential (i.e., $n \sim 1.4$ S\'ersic)
 bulge and a ($n \sim 0.9$) S\'ersic nuclear disc.  We also fit
 an intermediate-scale component which we refer to as a `boxy bulge'.

 {\it NGC~5353.} The 1D {\it HST}+SDSS surface brightness
 profile is well described as the sum of a ($n \sim 1.3$) S\'ersic
 bulge, an exponential disc and an X-shaped S\'ersic bar (e.g.,
 \citealt{2011MNRAS.418.1452L}).

 {\it NGC~5377.} Our six-component decomposition of this galaxy's 1D
 {\it HST}+SDSS light profile reveals a ($n \sim 2.2$) S\'ersic plus
 outer disc, an inner disc, a bar, a spiral-arm component and a
 nucleus. Decomposing the galaxy's 2D {\it Spitzer} 3.6 $\mu$m stellar
 distribution, \citet{2015ApJS..219....4S} reported a
 ($n \sim 2.2$) S\'ersic bulge, an exponential disc and a Ferrers bar.

 {\it NGC~5585.}  The 1D and 2D decomposition of our {\it HST} data for
 this galaxy reveals a ($n \sim 0.8-0.9$) S\'ersic bulge, an outer,
 exponential disc and a ($n \sim 1-3$) S\'ersic nucleus.
 \citet{2009MNRAS.395.1669G} fitted their {\it HST}+ground light
 profile with a ($n \sim 1.0$) S\'ersic bulge and an outer
 exponential disc model.

 {\it NGC~7640.}   This galaxy has a  ($n \sim 0.8- 1.1$) S\'ersic bulge, a box-shaped component 
 \citep{2000A&A...362..435L} and an outer exponential disc. 
 
 {\it NGC~7741.}   This galaxy is classified as a barred S(s)cd in
 RC3. The bulge component in our 1D and 2D three-component (bulge-disc-spiral-arm) decomposition may be a bar.
 \citet{2015ApJS..219....4S} modelled their {\it Spitzer} 3.6 $\mu$m
 stellar light distribution of the galaxy in 2D with a Ferrers bar
 plus and exponential disc decomposition model.

\section{Data Tables}\label{DataTables}

Table~\ref{NewTabA1} lists the tables, which present the best-fitting
structural parameters for our sample of 163 newly analysed LeMMINGs
galaxies, and the corresponding figures which show the
decompositions. Tables~\ref{Tab1}$-$\ref{Tab1cS} list the best-fitting
structural parameters from the surface brightness profile
decompositions. Table \ref{Tab4} provides global and central
properties of sample galaxies including distance, morphological
classification, velocity dispersion, bulge and galaxy stellar masses,
optical and radio luminosities, ellipticity, isophote shape parameter
and inner logarithmic slope of the galaxies' inner light profiles.

\setcounter{table}{0}
\renewcommand{\thetable}{A\arabic{table}} 

\begin{table} 
\setcounter{table}{0} 
\setlength{\tabcolsep}{.01853in}
\begin {minipage}{90mm}
\caption{Newly  analysed LeMMINGs galaxies.}
\label{NewTabA1}
\begin{tabular}{@{}lclclclclclccc@{}}
  \hline
  \hline
Galaxy&&Table/Figure &&Galaxy&&Table/Figure&&Galaxy&&Table/Figure & \\                                       
(1)& &(2)& &(1)& &(2)&&(1)& &(2) \\
 \hline                                                                                                                                                        
I0239     &  &A2 & &       N3034   &    & A2                              & & N4258     &  &A3 \\
I0356       &  &A2 & &      N3073    &  &A2                                 & & N4274       &  &A3 \\
I0520      &  &A2& &           N3077   &  &A2                           & & N4314      &  &A3 \\
I2574       &  &A2 & &         N3079   &  &A3                                  & &N4414&  &A3 \\
N0147       &  &A2 & &         N3184     &  &A2                               & &N4448      &  &A3\\
N0205     &  &A2 & &           N3185     &  &A3                          & &N4449    &  &A2\\
N0221    &  &A2& &             N3190       &  &A2                          & &N4485      &  &A2 \\
N0266       &  &A3 & &          N3193   &  &A4                      & &  N4490      &  &A2\\
N0278      &  &A3 & &           N3198     &  &A3                           & & N4559      &  &A3\\
N0315       &  &A4 & &       N3245        &  &A2                                      & &  N4565      &  &A2  \\
N0404        &  &A2 & &      N3319     &  &A3                    & & N4605     &  &A2  \\
N0410      &  &A2 & &          N3344     &  &A2                  & &N4648    &  &A2  \\
N0598         &  &A3 & &      N3348       &  &A4                    & &N4656        &  &A2  \\
N0672        &  &A2 & &       N3414      &  &A2                    & & N4736      &  &A3   \\
N0777        &  &A2 & &       N3448      &  &A2                & &N4750        &  &A3  \\
N0784     &  &A2& &             N3486    &  &A3                  & & N4800      &  &A3  \\
N0841      &  &A3& &          N3504     &  &A3                   & &N4826      &  &A3   \\
N0890      &  &A2 & &          N3516        &  &A2                    & &N4914     &  &A4   \\
N0959       &  &A3 & &       N3600     &  &A2                      & &N5005       &  &A3   \\
N1003       &  &A2 & &       N3610  &  &A2                     & &N5033     &  &A3   \\
N1023        &  &A2 & &      N3613       &  &A4                    & &N5055        &  &A3  \\
N1058       &  &A3 & &      N3631      &  &A3                     & &N5112  &  &A3   \\
N1156    &  &A2 & &       N3642     &  &A3                        & & N5204     &  &A2  \\
N1161     &  &A2 & &      N3665      &  &A2/A2(A4)                        & &N5273      &  &A3  \\
N1167      &  &A2/A2(A4) & &   N3675        &  &A3                      & &N5308       &  &A2   \\
N1275       &  &A2 & &      N3718       &  &A3                            & & N5353       &  &A2   \\
N1961         &  &A2 & & 	N3729       &  &A3 			 & & N5354       &  &A2   \\
N2273        &  &A2& &   	N3756     &  &A3 			& &N5377      &  &A3  \\
N2276           &  &A3 & &  	N3838     &  &A2 			& & N5448    &  &A2  \\
N2342         &  &A3 & &  	N3884        &  &A3 			& & N5457   &  &A3    \\
N2366      &  &A3 & &   	N3898     &  &A3 			& & N5474       &  &A3   \\
N2403      &  &A2 & &     	N3900   &  &A3			& & N5548       &  &A3   \\
N2500     &    & A2  & &	N3949      &  &A3 				  & &N5585      &  &A2 \\
N2537     &  &A2 & &    	N3982      &  &A3 					& & N5631       &  &A4  \\
N2541    &  &A2 & &     	N3992     &  &A3 					& &N5866    &  &A2 \\
N2549     &  &A2 & &    	N3998         &  &A2 					& & N5879        &  &A2 \\
N2634     &  &A2 & &    	N4026        &  &A2					& &N5907       &  &A2\\
N2639     &  &A2 & &       N4036       &  &A2                        		 	& & N5985       &  &A3\\
N2655     &  &A2 & &   	N4041      &  &A2 					& &N6140    &  &A3\\
N2683     &  &A2 & &        N4062     &  &A2      					& & N6207    &  &A2\\
N2685	    &  &A2 & &	N4096     &  &A3 			  & &N6217    &  &A3\\
N2748   &    & A2  & &  	 N4102       &  &A3			& &  N6340       &  &A3 \\
N2768     &  &A2 & &    	N4125      &  &A2			 & &  N6412    &  &A3\\
N2770   &  &A2& &     	N4138     &  &A 3			& &   N6482       &  &A4\\
N2782     &  &A2 & &  	N4143     &  &A2 			& &  N6503    &  &A3\\
N2787     &  &A2 & &		N4144      &  &A2 		    & & N6654      &  &A3 \\
N2832   &  &A4& &         N4150        &  &A2 		                      & &    N6946       &  &A2 \\
N2841    &  &A2 & &        N4151     &  &A3 			    & &  N6951       &  &A3\\
N2859    &  &A3 & &       N4183       &  &A2    		     & &  N7217      &  &A3\\
N2903    &  &A2 & &       N4203      &  &A2      		  & &N7331     &  &A3 \\
N2950     &  &A2 & &    N4217     &  &A2                         & &N7457       &  &A2\\
N2964    &  &A3 & &    N4220    &  &A3                          & & N7640   &  &A2\\
N2976    &  &A2 & &    N4242     &  &A2                          & &  N7741       &  &A3\\
N2985    &  &A2 & &    N4244       &  &A2                            & &    &  &\\
N3031   &  &A2 & &     N4245     &  &A3                            & & &  &\\
 \hline        
 \hline      
\end{tabular}                                                                                                                 
Notes.---Col. (1) galaxy name. Col. (2) indicates the tables, which  list the galaxy structural parameters for our sample of 163 newly  analysed LeMMINGs galaxies, and the associated figures where the fits are shown. 
\end {minipage}
\end{table}

\begin{center}
\begin{table*} 
\setlength{\tabcolsep}{0.000790in}
\begin {minipage}{186mm}
\caption{Structural parameters for S\'ersic LeMMINGs galaxies.}
\label{Tab1}
\begin{tabular}{@{}llcccccccccccccccccccccc@{}}
\hline
\hline      
I0239 (1D)   &{\it HST} Instrument = WFPC2 &Filter = F814W  &&$\#$ of comp = 4 & &Quality = 1  \\
&Bulge &S\'ersic& 21.37 $\pm$ 0.23 &  17.2 $\pm$ 1.4 &  0.83 $\pm$ 0.07  &   1.21 $\pm$  0.05\\
&Disc&exponential  &22.95 &  80.8  \\
&Spiral Arm&S\'ersic& 22.77 &  67.6 &   0.7 &\\
&Nucleus & 2-parameter Gaussian function &16.3&19.9  \\\\

I0239 (2D)   &{\it HST} Instrument = WFPC2 &Filter = F814W  &&\# of comp = 4 & &Quality = 1  \\
&Bulge &S\'ersic& 21.72 &  25.0  & 1.20 & 1.42&\\
&Disc&exponential  &21.78 &  95.0  \\
&Spiral Arm&S\'ersic& 25.73 &  35.5 &   0.43 &\\
&Nucleus & 2-parameter Gaussian function & 14.66 & 19.2& \\
&$\epsilon$/P.A./c$_{0}$&Bul:   0.14, 125, 0.08&Dis: 0.26, 40, 0.90& SAr: 0.00, 68, 5.0 &\\
&&Nuc: 0.41, 144, ---&\\\\

I0356  (1D)   &{\it HST} Instrument = ACS&Filter = F814W   &&\# of comp= 4 && Quality = 1  \\
&Bulge &S\'ersic&19.36 $\pm$ 0.21& 7.89  $\pm$  $0.50$  & 0.44  $\pm$  $0.27$  &   3.8  $\pm$  $0.44$ &\\
&Disc&exponential  & 18.55 &  34.5 &\\
&Innerdisc&S\'ersic& 19.58&  10.8&   0.6&\\ 
&Nucleus & 2-parameter Gaussian function &  11.7  &19.3 \\\\

 I0356   (2D)  &{\it HST} Instrument = ACS&Filter = F814W   &&\# of comp = 4 && Quality = 2  \\
&Bulge &S\'ersic& 19.22 & 7.2  &  0.40  &   3.71 &\\
&Disc&exponential  & 18.48 & 30.5 &\\
&Innerdisc&S\'ersic& 19.62& 11.03&   0.5&\\
&Nucleus & 2-parameter Gaussian function &  11.8& 19.9& \\
&$\epsilon$/P.A.&Bul: 0.18, 92&Dis: 0.17, 97& IDi: 0.37, 97 &Nuc: 0.57, 180&\\\\

I0520   &{\it HST} Instrument = ACS&Filter = F606W   & &\# of comp = 4 && Quality = 1  \\  
&Bulge &S\'ersic&19.24$\pm$  0.21 &   3.2 $\pm$ 0.6 &  0.78 $\pm$  0.15  &   3.09 $\pm$ 0.22 \\
&Disc&exponential  &  21.77 &  29.4 &\\
&Spiral Arm&S\'ersic& 21.77 &  24.1 &   1.1 &\\
&Nucleus & 2-parameter Gaussian function  &15.3  & 18.7& \\\\

I2574$^{\dagger}$   &{\it HST} Instrument = ACS & Filter =F814W && \# of comp = 3 && Quality = 1  \\  				
&Disc&exponential  & 20.74 &  380.7&   \\
&Bar&S\'ersic& 24.71 &  2.2 &  0.4&\\
&Nucleus & 2-parameter Gaussian function &20.1&21.5 \\\\

N0147   &{\it HST} Instrument = WFPC2&Filter = F814W  &Outer Data =  DSS&\# of comp = 2 && Quality = 1  \\  
&Bulge &S\'ersic& 23.63$\pm$  $0.27$ &  680.8 $\pm$  24.3 & 2.73 $\pm$  0.07   &   1.33 $\pm$  0.11\\
&Nucleus &S\'ersic& 20.04 &  1.4  & 1.6 & \\\\

N0205  (1D)  &{\it HST} Instrument = WFPC2&Filter = F555W  &Outer Data =  SDSS~&\# of comp = 2 && Quality = 1  \\  
&Bulge &S\'ersic&24.23 $\pm$  $0.25$ &  777.5  $\pm$ 22.2 &  3.11$\pm$ 0.07   &   2.73 $\pm$ 0.06 \\
&Nucleus &  S\'ersic &15.88 &   0.4 &  2.5 \\

\hline
\hline
\end{tabular} 
 Notes.--- Best-fitting model parameters for LeMMINGs galaxies  from our 1D (major-axis) and 2D  decompositions (see
Figs.~\ref{FigSer1} and \ref{2Dfits}). Unless labelled otherwise, we present  1D  best-fitting  parameters.  Shown are the galaxy name, the {\it HST}
instrument and filter, the goodness of the fit, the number and nature
of the fitted components and the ground-based data at large radii for
galaxies with composite profiles, i.e., inner {\it HST} plus outer
ground-based data. Bulgeless galaxies are indicated by the superscript
`$\dagger$'.  Quality flag `1' indicates fits with good or higher
quality levels whereas those labeled `2' are questionable primarily as
a result of the difficulty in modelling highly inclined
($ i \ga 75^{\circ}$) galaxies, see the text for further detail. For each 2D galaxy component, we present  the best-fitting  {\sc imfit}  ellipticity ($\epsilon$) and  position angle (P.A.). The {\sc imfit } 'boxy'/'discy' parameter c$_{0}$ is shown for 2D fits that were performed using generalised ellipses \citep{2015ApJ...799..226E}. We do not show the errors on the best-fitting 2D parameters from {\sc imfit} as they are unrealistically low, see \citet{2015ApJ...799..226E}. 
The fitted models (see Section~\ref{Mods}) are: {\it S\'ersic Bulge}
($\mu_{\rm e}$, $R_{\rm e}$ [arcsec], $R_{\rm e}$ [\mbox{kpc}], $n$),
{\it S\'ersic model for galaxy components other than the bulge}
($\mu_{\rm e}$, $R_{\rm e}$ [arcsec], $n$), {\it exponential model}
($\mu_{\rm disc,0}$, $h$ [arcsec]) and {\it 2-parameter Gaussian
  function} ($\mu_{\rm Gauss,0}$, $m_{\rm Gauss}$).
$\mu_{\rm disc,0}$ and $h$ are central surface brightness and scale
length of the large-scale exponential disc. $\mu_{\rm Gauss,0}$ and
$m_{\rm Gauss}$ are central surface brightness and apparent magnitude
of nuclear components modelled with a 2-parameter Gaussian function.
The surface brightnesses $(\mu)$ are in units of mag arcsec$^{-2}$,
where the magnitudes are given in the Vega magnitude system.  Unless specified otherwise, the errors associated with
$\mu_{\rm e}$, $R_{\rm e}$, $n$, $\mu_{\rm disc,0}$, $h$,
$\mu_{\rm Gauss,0}$ and $m_{\rm Gauss}$ are 0.24, 15\%, 15\%, 0.28,
20\%, 0.2 and 0.3, respectively. Sources: (a)
\citet{2016MNRAS.462.3800D}; (b) \citet{2014MNRAS.444.2700D}; (c)
\citet{2018MNRAS.475.4670D}.
\end {minipage}
\end{table*}
\end{center}

\setcounter{table}{1}
\renewcommand{\thetable}{A\arabic{table}} 
\begin{center}
\begin{table*} 
\setlength{\tabcolsep}{0.0007900in}
\begin {minipage}{184mm}
\caption{(\it continued)}
\label{Tab1}
\begin{tabular}{@{}llcccccccccccccccccccccc@{}}
\hline           
\hline
N0205 (2D) &{\it HST} Instrument = WFPC2&Filter = F555W  &&\# of comp = 2 && Quality = 1  \\  
&Bulge &S\'ersic& 24.04 &  652.2 & 2.61  &   2.95& \\
&Nucleus &  S\'ersic & 15.70 &  0.3 & 2.67 &\\
&$\epsilon$/P.A.&Bul: 0.33, 164 &Nuc: 0.08, 0&\\\\

N0221  &{\it HST} Instrument = WFPC2&Filter = F814W  &Outer Data =  SDSS&\# of comp = 3 && Quality = 1  \\  
&Bulge &S\'ersic& 17.32 $\pm$ 0.13 &  25.8 $\pm$ 1.2 &  0.10 $\pm$ 0.01   &   2.70 $\pm$ 0.11\\
&Disc&exponential  & 21.09&   295.2 & \\
\hline
\hline
\end{tabular} 

\end {minipage}
\end{table*}
\end{center}

\setcounter{table}{1}
\renewcommand{\thetable}{A\arabic{table}} 
\begin{center}
\begin{table*} 
\setlength{\tabcolsep}{0.000790in}
\begin {minipage}{184mm}
\caption{(\it continued)}
\label{Tab1}
\begin{tabular}{@{}llcccccccccccccccccccccc@{}}
\hline       
\hline   
N1961  &{\it HST} Instrument = WFPC2&Filter = F606W  &&\# of comp = 3 && Quality = 1  \\  
&Bulge &S\'ersic&  22.33 $\pm$ 0.27 &  31.3 $\pm$ 1.7 &  8.32 $\pm$ 0.45  & 3.11 $\pm$ 0.30 &\\
&Disc&S\'ersic& 22.26 &  44.7 &   0.6&\\
&Nucleus & 2-parameter Gaussian function&  17.0 &	17.9 &  \\\\

N2273  &{\it HST} Instrument = WFPC2&Filter = F606W  &&\# of comp = 5 && Quality = 1  \\  
&Bulge &S\'ersic&  21.14 $\pm$ 0.14 &  16.0 $\pm$ 0.9 &  2.07 $\pm$ 0.12  & 2.88 $\pm$ 0.27 &\\
&Disc&S\'ersic& 22.95 &  45.8 &  0.6&\\
\hline
\hline
\end{tabular} 
\end {minipage}
\end{table*}
\end{center}

\setcounter{table}{1}
\renewcommand{\thetable}{A\arabic{table}} 
\begin{center}
\begin{table*} [hbt!]
\setlength{\tabcolsep}{0.000790in}
\begin {minipage}{184mm}
\caption{(\it continued)}
\label{Tab1}
\begin{tabular}{@{}llcccccccccccccccccccccc@{}}
\hline     
\hline
 N2655 (2D)&{\it HST} Instrument = WFPC2 &Filter = F547W  &&\# of comp = 3 && Quality = 1 \\  
&Bulge &S\'ersic&  19.51 &  11.3 &  1.10   &    2.71&\\
&Disc&S\'ersic & 20.80 & 24.6 &  1.04 &\\
&Nucleus & 2-parameter Gaussian function&  15.2 &	14.5 &  \\
&$\epsilon$/P.A./c$_{0}$&Bul: 0.28, 89, -0.08&Dis: 0.30, 82, -0.08 &Nuc: 0.62, 78, ---&\\\\

N2681$^{\mbox{a}}$   &{\it HST} Instrument = WFPC2&Filter = F814W  &Outer Data = SDSS&\# of comp = 6 && Quality = 1  \\  
&Bulge &S\'ersic&  16.90 $\pm$ 0.18 &  4.1 $\pm$ 0.3 &  0.24 $\pm$ 0.02  & 3.61 $\pm$ 0.5 &\\
\hline
\hline
\end{tabular} 
\end {minipage}
\end{table*}
\end{center}

\setcounter{table}{1}
\renewcommand{\thetable}{A\arabic{table}} 
\begin{center}
\begin{table*}  [hbt!]
\setlength{\tabcolsep}{0.000790in}
\begin {minipage}{184mm}
\caption{(\it continued)}
\label{Tab1}
\begin{tabular}{@{}llcccccccccccccccccccccc@{}}
\hline     
\hline

N2787  &{\it HST} Instrument = WFPC2&Filter = F814W  &&\# of comp = 5 && Quality = 1  \\  
&Bulge &S\'ersic&  17.50 $\pm$ 0.13 &  6.7 $\pm$ 1.6 &  0.35 $\pm$ 0.09  & 1.25 $\pm$ 0.21 &\\
&Disc&exponential  & 18.34 &  24.9&   \\
&inner Disc?&S\'ersic& 15.47&  0.6 &  0.7&\\
&Bar&S\'ersic& 19.63 &  13.5 & 0.4&\\     
&Nucleus & 2-parameter Gaussian function&  13.0 &	16.4 &  \\\\

N2841 (1D)  &{\it HST} Instrument = ACS  &Filter = F814W  &Outer Data = SDSS&\# of comp = 2 && Quality = 1  \\
\hline
\hline
\end{tabular} 

\end {minipage}
\end{table*}
\end{center}

\setcounter{table}{1}
\renewcommand{\thetable}{A\arabic{table}} 

\begin{center}
\begin{table*}  [hbt!]
\setlength{\tabcolsep}{0.000790in}
\begin {minipage}{184mm}
\caption{(\it continued)}
\label{Tab1}
\begin{tabular}{@{}llcccccccccccccccccccccc@{}}
\hline 
\hline
N3073  (1D)&{\it HST} Instrument = ACS &Filter = F850LP  &Outer Data = SDSS&\# of comp= 4 && Quality = 1  \\  
&Bulge &S\'ersic& 20.6 $\pm$ 0.09 &  11.0 $\pm$ 0.5 &  1.02 $\pm$ 0.05  & 	2.71 $\pm$ 0.08 &\\
&Disc&exponential & 23.34 &63.4&\\
&Bar&S\'ersic& 17.39 &  1.0 &  0.4&\\
&Nucleus & 2-parameter Gaussian function&  8.1  &  20.4 &  \\\\

 N3073 (2D)  &{\it HST} Instrument = ACS &Filter = F850LP  && \# of comp = 4 && Quality = 1  \\  
&Bulge &S\'ersic& 20.49 &  10.8 &  0.99  & 	2.64&\\
\hline
\hline
\end{tabular} 
\end {minipage}
\end{table*}
\end{center}

\setcounter{table}{1}
\renewcommand{\thetable}{A\arabic{table}}

\begin{center}
\begin{table*}  [hbt!]
\setlength{\tabcolsep}{0.000790in}
\begin {minipage}{184mm}
\caption{(\it continued)}
\label{Tab1}
\begin{tabular}{@{}llcccccccccccccccccccccc@{}}
\hline
\hline
\multicolumn{1}{c}{}      
N3448 (1D) &{\it HST} Instrument = NICMOS NIC3 &Filter = F160W  &&\# of comp  = 2 && Quality = 2  \\  
&Bulge &S\'ersic&  15.82 $\pm$ 0.16 &  8.9 $\pm$ 0.9 &  0.93 $\pm$ 0.09  & 1.15 $\pm$ 0.05 &\\
&Disc&S\'ersic& 16.99&  20.9 &  0.5&\\\\

 N3448  (2D)&{\it HST} Instrument = NICMOS NIC3 &Filter = F160W  &&\# of comp = 2 && Quality = 2  \\  
&Bulge &S\'ersic&  16.06 &  8.8 &  0.92  & 1.31&\\
&Disc&S\'ersic&  17.03  &  25.9 &  0.80 &\\
&$\epsilon$/P.A.&Bul: 0.42, 60&Dis: 0.60, 66&\\
\hline
\hline
\end{tabular} 
\end {minipage}
\end{table*}
\end{center}

\setcounter{table}{1}
\renewcommand{\thetable}{A\arabic{table}} 
\begin{center}
\begin{table*}  [hbt!]
\setlength{\tabcolsep}{0.000790in}
\begin {minipage}{184mm}
\caption{(\it continued)}
\label{Tab1}
\begin{tabular}{@{}llcccccccccccccccccccccc@{}}
\hline
\hline
N4026 &{\it HST} Instrument =WFPC2 & Filter = F814W  &&\# of comp = 5 && Quality = 2  \\  
&Bulge &S\'ersic&  18.53 $\pm$ 0.21 &  12.9 $\pm$ 1.6 &  1.05 $\pm$ 0.13  & 3.27 $\pm$ 0.30 &\\
&Disc (outer)&exponential  &18.31 & 28.9&   \\
&inner Discy-comp&S\'ersic& 14.41 &  0.3 &0.3 &\\
&intermediate Discy-comp&S\'ersic& 18.04&  3.7 & 0.3&\\
&Nucleus & 2-parameter Gaussian function &10.6  &16.4 &  \\\\

N4036&{\it HST} Instrument = WFC2 & Filter = F814W &&\# of comp = 3 && Quality = 1  \\
\hline
\hline
\end{tabular} 
\end {minipage}
\end{table*}
\end{center}

\setcounter{table}{1}
\renewcommand{\thetable}{A\arabic{table}}

\begin{center}
\begin{table*}  [hbt!]
\setlength{\tabcolsep}{0.000790in}
\begin {minipage}{184mm}
\caption{(\it continued)}
\label{Tab1}
\begin{tabular}{@{}llcccccccccccccccccccccc@{}}
\hline           
\hline   
 N4150 (2D)&{\it HST} Instrument = WFPC2& Filter = F814W &Outer Data = SDSS& \# of comp = 3 && Quality = 1  \\  
&Bulge &S\'ersic&  16.73 &  2.4 &  0.08  & 0.98\\
&Disc&S\'ersic& 19.98 & 23.4 & 1.25 &\\
&Nucleus &2-parameter Gaussian function& 14.9  & 14.8 &  \\
&$\epsilon$/P.A.&Bul: 0.21, 146&Dis: 0.35, 147 &Nuc: 0.52, 75&\\\\

N4183$^{\dagger}$   &{\it HST} Instrument = ACS& Filter = F814W &&\# of comp = 1 && Quality = 2  \\  
&Disc &S\'ersic&  20.75 &  21.8 &   1.11 &\\
\hline
\hline
\end{tabular} 
\end {minipage}
\end{table*}
\end{center}

\setcounter{table}{1}
\renewcommand{\thetable}{A\arabic{table}}

\begin{center}
\begin{table*}  [hbt!]
\setlength{\tabcolsep}{0.000790in}
\begin {minipage}{184mm}
\caption{(\it continued)}
\label{Tab1}
\begin{tabular}{@{}llcccccccccccccccccccccc@{}}
\hline           
\hline
N4490&{\it HST} Instrument = WFPC2& Filter = F814W && \# of comp = 2 && Quality = 1  \\  
&Bulge &S\'ersic&  19.33 $\pm$ 0.05 &  17.0 $\pm$ 1.0 &  0.9$\pm$ 0.05  & 0.79 $\pm$ 0.03 &\\
&Disc&S\'ersic&19.93 & 55.1  & 0.7&\\\\

N4565&{\it HST} Instrument = WFPC2& Filter = F814W && \# of comp  = 5 && Quality = 2  \\  
&Bulge &S\'ersic&  18.48 $\pm$ 0.18 &  11.1 $\pm$ 2.2 &  1.11 $\pm$ 0.22 & 2.43 $\pm$ 0.14 &\\
&Disc&exponential  & 19.82 & 90.3&   \\
&inner Disc&S\'ersic& 18.15& 1.9 & 0.4&\\
\hline
\hline
\end{tabular} 
\end {minipage}
\end{table*}
\end{center}

\setcounter{table}{1}
\renewcommand{\thetable}{A\arabic{table}}

\begin{center}
\begin{table*}  [hbt!]
\setlength{\tabcolsep}{0.000790in}
\begin {minipage}{184mm}
\caption{(\it continued)}
\label{Tab1}
\begin{tabular}{@{}llcccccccccccccccccccccc@{}}
\hline           
\hline   
N5353&{\it HST} Instrument = WFC3 IR& Filter = F110W && \# of comp = 3 && Quality = 1  \\  
&Bulge &S\'ersic&  16.13 $\pm$ 0.08 &  4.0 $\pm$ 0.4 & 0.69 $\pm$ 0.07  & 1.34 $\pm$ 0.05 &\\
&Disc&exponential  &17.02& 15.9&   \\
&X-shaped Bar&S\'ersic&17.26 & 12.0 &0.3 &\\\\

N5354 (1D)&{\it HST} Instrument = WFC3 IR& Filter = F160W && \# of comp = 3 && Quality = 1  \\  
&Bulge &S\'ersic&  16.58 $\pm$ 0.11 &  6.4 $\pm$ 0.4 & 1.21 $\pm$ 0.47  & 2.12 $\pm$ 0.08 &\\
&Disc&exponential  &17.49& 18.0&   \\
\hline
\hline
\end{tabular} 
\end {minipage}
\end{table*}
\end{center}

\setcounter{table}{1}
\renewcommand{\thetable}{A\arabic{table}}

\begin{center}
\begin{table*}  [hbt!]
\setlength{\tabcolsep}{0.000790in}
\begin {minipage}{184mm}
\caption{(\it continued)}
\label{Tab1}
\begin{tabular}{@{}llcccccccccccccccccccccc@{}}
\hline 
N6207 (1D)&{\it HST} Instrument = ACS& Filter = F814W && \# of comp  = 3 && Quality = 1  \\  
&Bulge &S\'ersic&  20.70 $\pm$ 0.40 &  7.9 $\pm$ 1.3 & 0.48 $\pm$ 0.08  & 1.42 $\pm$ 0.18 &\\
&Disc&S\'ersic& 19.88& 27.7 & 0.6 &\\
&Nucleus & 2-parameter Gaussian function& 16.8 & 21.1 &  \\\\

N6207 (2D)&{\it HST} Instrument = ACS& Filter = F814W &&\# of comp = 3 && Quality = 1  \\  
&Bulge &S\'ersic& 20.38 & 5.7 & 0.34 & 1.04 &\\
&Disc&S\'ersic& 19.97 & 29.3 & 0.75 &\\
\hline
\hline
\end{tabular} 
\end {minipage}
\end{table*}
\end{center}

\begin{center}
\begin{table*} 
\setlength{\tabcolsep}{0.000790in}
\begin {minipage}{184mm}
\caption{Structural parameters for S\'ersic LeMMINGs galaxies with a large-scale Gaussian component.}
\label{Tab1Ring}
\begin{tabular}{@{}llcccccccccccccccccccccc@{}}
\hline        
\hline   
N0266&{\it HST} Instrument = WFPC2& Filter = F547M &Outer Data = SDSS& \# of comp  = 6 && Quality = 1  \\  
&Bulge &S\'ersic&  22.33 $\pm$ 0.27&  28.0 $\pm$ 1.3 & 8.29 $\pm$ 0.42  & 3.54 $\pm$ 0.32 &\\
&Disc&exponential  & 23.27& 63.1&   \\
&inner Spiral/Disc&S\'ersic& 18.88& 1.5 & 0.6 &\\
&Bar&S\'ersic& 24.07&	37.3&	0.6\\
&Spiral Arm&3-parameter Gaussian ring model& 	23.36 & 14.8 & 40.2 &\\
&Nucleus & 2-parameter Gaussian function& 16.1 & 19.8 &  \\\\
\hline
\hline
\end{tabular} 
Notes.---Similar to Table~\ref{Tab1} but here showing S\'ersic
galaxies with a large-scale Gaussian component (Figs.~\ref{FigSerR} and \ref{2Dfits}).
The fitted 3-parameter Gaussian ring model parameters (see
Section~\ref{Mods}) are $\mu_{r}$, $\sigma$, $R_{r}$, where
$\mu_{\rm r}$ denotes the brightest value of the Gaussian ring with a
semi-major axis $R_{r}$ and a full width at half maximum of FWHM = 2
$\sqrt {\rm 2ln 2}$ $\sigma$. The errors associated with the
best-fitting parameters $\mu_{r}$, $\sigma$ and $R_{r}$ are 0.28, 15\%
and 12\%, respectively.
\end {minipage}
\end{table*}
\end{center}

\setcounter{table}{2}
\begin{center}
\begin{table*} 
\setlength{\tabcolsep}{0.000790in}
\begin {minipage}{184mm}
\caption{(\it continued)}
\label{Tab1Ring}
\begin{tabular}{@{}llcccccccccccccccccccccc@{}}
\hline        
\hline   
N2276  (1D)&{\it HST} Instrument = WFPC2 &Filter = F814W  &Outer Data =  DSS&\# of comp = 3 && Quality = 1  \\  
&Bulge &S\'ersic&  19.25 $\pm$ 0.05 &  3.6 $\pm$ 0.1 &  0.58 $\pm$ 0.02   & 1.50 $\pm$ 0.03 &\\
&Disc&exponential  & 19.88&  22.2& \\
&Spiral Arm&3-parameter Gaussian ring model& 	21.35&  	24.0 &
                                                                       26.3&\\\\

 N2276  (2D)&{\it HST} Instrument = WFPC2 &Filter = F814W  && \# of comp = 3 && Quality = 1  \\  
&Bulge &S\'ersic& 19.7 & 4.5 &  0.74 & 1.71 &\\
&Disc&exponential  & 20.05 &  20.0 & \\
\hline
\hline
\end{tabular} 
Note that our 2D decomposition for NGC~2859 contains a seventh  component:  an inner Gaussian ring ($\mu_{r} \sim   21.89$ mag arcsec$^{-2}$, $\sigma \sim 5.8\arcsec$, $R_{r} \sim 38.49 \arcsec$). 
 \end {minipage}
\end{table*}
\end{center}

\setcounter{table}{2}
\begin{center}
\begin{table*} 
\setlength{\tabcolsep}{0.000390in}
\begin {minipage}{184mm}
\caption{(\it continued)}
\label{Tab1Ring}
\begin{tabular}{@{}llcccccccccccccccccccccc@{}}
\hline        
\hline   
N3185&{\it HST} Instrument = WFPC2& Filter = F814W &Outer Data = SDSS& \# of comp  = 6 && Quality = 1  \\  
&Bulge &S\'ersic& 21.29 $\pm$ 0.13 & 19.68 $\pm$ 2.4&2.09 $\pm$ 0.25 &3.25 $\pm$ 0.51 &\\
&Disc&exponential  & 21.39& 14.2&   \\
&inner Disc/Ring?&S\'ersic& 17.26& 1.4& 0.4 &\\
&Bar&S\'ersic& 21.42& 	35.9&	0.5&\\
&Spiral Arm&3-parameter Gaussian ring model& 22.78 &  	10.7 & 36.3&\\
&Nucleus & 2-parameter Gaussian function & 15.3& 17.1&  \\\\
\hline
\hline
\end{tabular} 
\end {minipage}
\end{table*}
\end{center}

\setcounter{table}{2}
\begin{center}
\begin{table*} 
\setlength{\tabcolsep}{00.000790in}
\begin {minipage}{184mm}
\caption{(\it continued)}
\label{Tab1Ring}
\begin{tabular}{@{}llcccccccccccccccccccccc@{}}
\hline        
\hline   
 N3631 (2D)&{\it HST} Instrument = WFPC2 & Filter = F814W &&\# of comp  = 4 && Quality = 1  \\  
&Bulge &S\'ersic& 19.29 & 10.7 & 0.99 & 2.1 &\\
&Disc&exponential  & 21.28& 37.9&   \\
&Spiral Arm&3-parameter Gaussian ring model & 21.83 & 11.3 & 57.68 &\\
&Nucleus & 2-parameter Gaussian function & 16.0& 16.9 &  \\
&$\epsilon$/P.A.&Bul: 0.03, 137 &Dis: 0.35, 180 &SAr: 0.25, 67 &Nuc: 0.01, 175&\\\\

N3642 (1D)&{\it HST} Instrument = ACS & Filter = F814W && \# of comp  = 4 && Quality = 1  \\
\hline
\hline
\end{tabular} 
\end {minipage}
\end{table*}
\end{center}

\setcounter{table}{2}
\begin{center}
\begin{table*} 
\setlength{\tabcolsep}{0.000790in}
\begin {minipage}{184mm}
\caption{(\it continued)}
\label{Tab1Ring}
\begin{tabular}{@{}llcccccccccccccccccccccc@{}}
\hline
\hline
N3884 (1D)&{\it HST} Instrument = WFPC2 & Filter = F547M & &\# of comp = 4 && Quality = 1  \\  
&Bulge &S\'ersic&  20.57 $\pm$ 0.15 &  5.0 $\pm$ 0.4 & 2.47 $\pm$ 0.18  & 2.65 $\pm$ 0.32 &\\
&Disc&exponential  & 21.60 & 25.8   \\
&Ring&3-parameter Gaussian ring model &23.27 &  10.5 & 20.3&\\
&Nucleus & 2-parameter Gaussian function & 16.4& 18.9&  \\\\

N3884 (2D)&{\it HST} Instrument = WFPC2 & Filter = F547M & &\# of comp  = 4 && Quality = 1  \\  
&Bulge &S\'ersic& 20.6 &  5.1 & 2.51  & 2.68 &\\
\hline
\hline
\end{tabular} 
\end {minipage}
\end{table*}
\end{center}

\setcounter{table}{2}
\begin{center}
\begin{table*} 
\setlength{\tabcolsep}{0.000790in}
\begin {minipage}{184mm}
\caption{(\it continued)}
\label{Tab1Ring}
\begin{tabular}{@{}llcccccccccccccccccccccc@{}}
\hline
\hline
N4096 (1D)&{\it HST} Instrument = WFPC2 & Filter = F606W  &&\# of comp  = 3 && Quality = 1  \\  
&Bulge &S\'ersic&  19.88 $\pm$ 0.21 &  4.8 $\pm$ 1.2 & 0.25 $\pm$ 0.06 & 1.00 $\pm$ 0.11 &\\
&Disc&exponential  & 19.84 & 44.3   \\
&Spiral Arm&3-parameter Gaussian ring model & 22.44 & 39.6 & 65.8&\\\\

N4096 (2D)&{\it HST} Instrument = WFPC2 & Filter = F606W  &&  \# of comp = 3 && Quality = 1  \\  
&Bulge &S\'ersic& 19.66 &  4.3 & 0.23 & 0.90 &\\
&Disc&exponential  & 19.66 & 40.1&   \\
\hline
\hline
\end{tabular} 
\end {minipage}
\end{table*}
\end{center}

\setcounter{table}{2}
\begin{center}
\begin{table*} 
\setlength{\tabcolsep}{0.000790in}
\begin {minipage}{184mm}
\caption{(\it continued)}
\label{Tab1Ring}
\begin{tabular}{@{}llcccccccccccccccccccccc@{}}
\hline        
\hline   
 N4274 (2D)&{\it HST} Instrument = WFPC2& Filter = F555W  &&\# of comp  = 5 && Quality = 1  \\  
&Bulge &S\'ersic&  18.50 &  3.7 & 0.31 & 1.00 &\\
&Disc&exponential  & 21.10 & 24.2&   \\
&Bar&S\'ersic& 20.35 & 17.0 & 0.55 &\\
&Spiral Arm&3-parameter Gaussian ring model& 21.66 & 23.0 & 52.02 &\\
&Nucleus &S\'ersic& 18.49 & 0.4 & 0.32 & \\
&$\epsilon$/P.A.&Bul: 0.50, 100 &Dis: 0.64, 127 &Bar: 0.31, 72 &SAr: 0.54, 98 &Nuc: 0.35, 178&\\\\
\hline
\hline
\end{tabular} 
\end {minipage}
\end{table*}
\end{center}

\setcounter{table}{2}
\begin{center}
\begin{table*} 
\setlength{\tabcolsep}{0.000790in}
\begin {minipage}{184mm}
\caption{(\it continued)}
\label{Tab1Ring}
\begin{tabular}{@{}llcccccccccccccccccccccc@{}}
\hline        
  \hline
N4800&{\it HST} Instrument = ACS & Filter = F814W  && \# of comp = 5 && Quality = 1  \\  
&Bulge &S\'ersic&  19.06 $\pm$ 0.15&  14.2 $\pm$ 1.3 & 1.06 $\pm$ 0.10 & 1.28 $\pm$ 0.18 &\\
&Disc&exponential  & 22.25 & 32.1    \\
&circumnuclear Ring&S\'ersic& 16.51& 1.3& 0.8&\\
&Spiral Arm&3-parameter Gaussian ring model& 20.68 & 12.3& 22.0&\\
&Nucleus &2-parameter Gaussian function& 13.5& 17.3&  \\

N4826&{\it HST} Instrument = WFPC2& Filter = F814W  &Outer Data = SDSS& \# of comp  = 4 && Quality = 1  \\  
&Bulge &S\'ersic&  17.37 $\pm$ 0.17&  9.8 $\pm$ 0.8 & 0.47 $\pm$ 0.04 & 2.30 $\pm$ 0.24 &\\
\hline
\hline
\end{tabular} 
\end {minipage}
\end{table*}
\end{center}

\setcounter{table}{2}
\begin{center}
\begin{table*} 
\setlength{\tabcolsep}{0.000790in}
\begin {minipage}{184mm}
\caption{(\it continued)}
\label{Tab1Ring}
\begin{tabular}{@{}llcccccccccccccccccccccc@{}}
  \hline
  \hline
 N5457 &  {\it HST} Instrument = WFPC2& Filter = F547M&Outer Data = SDSS&\# of comp  = 5 && Quality = 1  \\ 
&Bulge &S\'ersic& 22.10 $\pm$  0.25    &  11.6 $\pm$ 2.1 &0.29 $\pm$  0.01 &  3.84 $\pm$  0.13 \\
&Disc&exponential &20.57 & 25.8 \\
&inner Spiral Arm/Disc&S\'ersic&  20.91  & 8.5  &  0.6   \\
&Spiral Arm&3-parameter Gaussian ring model& 22.44& 13.3  & 38.0\\
&Nucleus &2-parameter Gaussian function& 16.6  & 18.5 \\\\

N5474  (1D) &  {\it HST} Instrument = WFPC2& Filter = F814W &Outer Data = SDSS&\# of comp = 4 && Quality = 1 \\
\hline
\hline
\end{tabular} 
\end {minipage}
\end{table*}
\end{center}

\setcounter{table}{2}
\begin{center}
\begin{table*} 
\setlength{\tabcolsep}{0.000790in}
\begin {minipage}{184mm}
\caption{(\it continued)}
\label{Tab1Ring}
\begin{tabular}{@{}llcccccccccccccccccccccc@{}}
  \hline
  \hline
 N6412 (1D)& {\it HST} Instrument =WFPC2& Filter = F814W &&\# of comp  = 4 && Quality = 1  \\ 
&Bulge &S\'ersic&20.57 $\pm$  0.14 &   9.3 $\pm$ 0.48 &  0.81 $\pm$  0.04  & 1.55 $\pm$ 0.10  \\
&Disc&exponential & 20.14&    23.6 \\
&Spiral Arm&3-parameter Gaussian ring model&21.99&  12.7 &  33.7  \\
&Nucleus &2-parameter Gaussian function& 15.6  &19.7 \\\\

 N6412 (2D) & {\it HST} Instrument =WFPC2& Filter = F814W &&\# of comp  = 4 && Quality = 1  \\ 
&Bulge &S\'ersic& 20.54 &   9.6 &  0.83  & 1.46&  \\
\hline
\hline
\end{tabular} 
\end {minipage}
\end{table*}
\end{center}

\setcounter{table}{2}
\begin{center}
\begin{table*} 
\setlength{\tabcolsep}{0.000790in}
\begin {minipage}{184mm}
\caption{(\it continued)}
\label{Tab1Ring}
\begin{tabular}{@{}llcccccccccccccccccccccc@{}}
  \hline
  \hline

N7741 (1D) & {\it HST} Instrument =WFPC2& Filter = F814W &Outer Data = SDSS&\# of comp  3 && Quality = 1  \\ 
&Bulge/Bar? &S\'ersic&20.01 $\pm$  0.24  &  10.9 $\pm$ 1.3 &  0.31 $\pm$  0.04  &   0.30 $\pm$  0.15  \\
&Disc&exponential & 19.72 &  10.5    \\
&Spiral Arm&3-parameter Gaussian ring model&21.20 &  19.2&  24.9 \\\\

N7741 (2D)& {\it HST} Instrument =WFPC2& Filter = F814W &&\# of comp  = 3 && Quality = 1  \\ 
&Bulge/Bar? &S\'ersic& 20.94  &  9.2 &  0.26  &  0.28&  \\
&Disc&exponential & 18.81 & 11.2&    \\
&Spiral Arm&3-parameter Gaussian ring model&21.74 &  20.4 &  24.75& \\
&$\epsilon$/P.A.&Bul: 0.82, 101 &Dis: 0.60, 101 &SAr: 0.15, 100 &\\

\hline
\hline
\end{tabular} 
\end {minipage}
\end{table*}
\end{center}

\begin{center}
\begin{table*} 
\setlength{\tabcolsep}{0.000790in}
\begin {minipage}{180mm}
\caption{Structural parameters for core-S\'ersic LeMMINGs galaxies. }
\label{Tab1cS}
\begin{tabular}{@{}llcccccccccccccccccccccc@{}}
\hline        
\hline 
N0315&{\it HST} Instrument = WFPC2& Filter = F814W  &Outer Data = SDSS&\# of comp  = 2 && Quality = 1  \\  
&Bulge &core-S\'ersic&15.91&0.84&0.266& 3.26&  \\
&&&25.8&8.11& 0.03&5.0&\\
&Halo&exponential  & 21.07&43.9    \\\\

N0410&{\it HST} Instrument =  WFC3 IR& Filter = F110W  &Outer Data = SDSS &\# of comp  = 3 && Quality = 1  \\  
&Bulge &core-S\'ersic &14.49&0.52&0.178&7.19&\\
&&&43.0&14.59& -0.01&2.0&\\
\hline
\hline
\end{tabular} 
Notes.--- As Table \ref{Tab1} but here for core-S\'ersic LeMMINGs
galaxies (see Figs.~\ref{FigCSer} and and \ref{2Dfits}).   A `?' shows that we tentatively
classify the lenticular galaxies {\mbox NGC 1167} and {\mbox NGC 3665}
as core-S\'ersic.   The fitted  core-S\'ersic model parameters  (see Section~\ref{Mods}) are $\mu_{\rm b}$, $R_{\rm b}$ [arcsec], $R_{\rm b}$  [\mbox{kpc}], $n$,  $R_{\rm e}$  [arcsec], $R_{\rm e}$  [\mbox{kpc}], $\gamma$,  and $\alpha$. The errors associated with the best-fitting  core-S\'ersic parameters  are  0.15,  5\%,   5\%,  10\%,  10\%,    10\% and 10\%,   respectively.   
\end {minipage}
\end{table*}
\end{center}

\setcounter{table}{3}
\begin{center}
\begin{table*} 
\setlength{\tabcolsep}{0.000790in}
\begin {minipage}{184mm}
\caption{(\it continued)}
\label{Tab1cS}
\begin{tabular}{@{}llcccccccccccccccccccccc@{}}
\hline        
\hline
N3665?   &{\it HST} Instrument = NICMOS NIC2   & Filter = F160W&Outer Data = SDSS & \# of comp = 2 && Quality = 2  \\  
&Bulge &core-S\'ersic & 10.76 &0.31  & 0.049 &   3.85& \\
&&& 38.8 &   6.2& -0.02 &   2.0 &  \\
&Disc&exponential  &  19.25  &179.1  &       \\\\

N4278$^{\mbox{b}}$ &{\it HST} Instrument = WFPC2& Filter = F555W  &&  \# of comp  = 2  &&  Quality = 1  \\  
&Bulge &core-S\'ersic & 15.84&0.83&0.052&3.78\\
&&&20.2&1.25&0.22&5.0   \\
&Nucleus &2-parameter Gaussian function &  13.5&19.4 \\\\

N4291$^{\mbox{b}}$ &{\it HST} Instrument = WFPC2& Filter = F555W  && \# of comp = 1 &&  Quality = 1  \\  
&Bulge &core-S\'ersic &   15.14&0.30&0.036&4.43&\\
&&& 13.6&1.64&0.10&5.0&\\\\

N4589$^{\mbox{b}}$ &{\it HST} Instrument = WFPC2& Filter = F555W  && \# of comp  = 1 &&  Quality = 1  \\  
&Bulge &core-S\'ersic & 15.33&0.20&0.027&5.60&\\
&&&70.8&9.56&0.30&5.0&  \\\\

N4914  &{\it HST} Instrument = WFC3 IR& Filter = F110W &Outer Data = SDSS &\# of comp = 1 &&Quality = 1  \\  
&Bulge &core-S\'ersic & 12.54&0.11&0.035&6.95&\\
&&& 42.7&14.2&-0.01&  2.0&  \\\\

N5322$^{\mbox{c}}$    &{\it HST} Instrument = WFPC2& Filter = F814W&Outer Data = SDSS & \# of comp  = 3 && Quality = 1  \\  
&Bulge &core-S\'ersic &14.36&0.37&0.054&4.03&\\
&&& 30.4&4.4&0.17&2.0&   \\
&Halo&exponential  &   20.79&53.6&   \\
&inner Disc &S\'ersic &    17.21&2.46&1.0  \\\\

N5557$^{\mbox{b}}$ &{\it HST} Instrument = WFPC2& Filter = F555W  &&  \# of comp  = 1 &&  Quality = 1  \\  
&Bulge &core-S\'ersic & 15.46&0.23&0.051&4.46&\\
&&&  30.2&6.80&0.19&5.0&  \\\\

N5631 &{\it HST} Instrument = WFC3 UVIS& Filter = F814W & &\# of comp = 2 &&Quality = 1  \\  
&Bulge &core-S\'ersic &  12.63 &  0.04 & 0.006 &   6.39 &\\
&&& 30.0 &   4.2 & -0.01 &   2.0 & \\
&Disc&exponential  &   19.12 &   7.5       \\\\

N5982$^{\mbox{b}}$ &{\it HST} Instrument = WFPC2& Filter = F555W  &&  \# of comp = 1 &&  Quality = 1  \\  
&Bulge &core-S\'ersic &15.48&0.25&0.051&4.30&\\
&&&26.8&5.45&0.09&5.0&   \\\\

N6482 (1D)&{\it HST} Instrument = ACS& Filter = F850LP && \# of comp  = 2  &&  Quality = 1  \\  
&Bulge &core-S\'ersic &   13.27 &  0.04 & 0.010 &   4.16 &\\
 &&& 19.0 &   5.0 & 0.07 &   2.0 &  \\
                   &Halo&exponential  &   18.04 &   8.3 &\\\\

N6482 (2D)&{\it HST} Instrument = ACS& Filter = F850LP &&  \# of comp  = 2  &&  Quality = 1  \\  
&Bulge &core-S\'ersic &   13.15 &  0.04 & 0.010 &   4.28 &\\
 &&& 18.0 &   4.7 & 0.02&   2.7 &  \\
&Halo&exponential  &   17.75 &  9.5 &\\
&$\epsilon$/P.A./c$_{0}$&Bul: 0.14, 63, --- &Hol: 0.40, 64, $-$0.41 &\\\\
\hline
\hline
\end{tabular} 
\end {minipage}
\end{table*}
\end{center}

\newpage
\begin{center}
\begin{table*} 
\begin{sideways}
\setlength{\tabcolsep}{0.0405983in}
\begin {minipage}{251mm}
\caption{LeMMINGs data }
\label{Tab4}
\begin{tabular}{@{}llcccccccccccccccccccccc@{}}
\hline
\hline
Galaxy&Type& Dist&Class&  $\sigma$ &$M_{\rm V,bul}$&$M_{\rm V,nuc}$&$M_{\rm V,glxy}$&$M/L$&log$M_{\rm *,bul}$&log$M_{\rm *,nuc}$&log$M_{\rm *,glxy}$ &log $M_{\rm BH}$  &log$L_{\rm Radio,core}$ &$<\epsilon>$& Med ($B_{4}$)&$\gamma$&Det\\
&&&&&1D/2D&&&&&&&&&&\\
&&(Mpc)&&(km s$^{-1}$)&(mag)&(mag)&(mag)&&($M_{\sun}$)&($M_{\sun}$)&($M_{\sun}$)&($M_{\sun}$)&(erg s$^{-1}$)&&\\
(1)&(2)&(3)&(4)&(5)&(6)&(7)&(8)&(9)&(10)&(11)&(12&(13)&(14)&(15)&(16)&(17)&(18)\\   
\hline     
I0239    & SAB(rs)c &   9.9 & H {\sc{ii}}    &      92.3 &    -16.82$\pm0.29$/-17.44 &     -9.60 &    -18.80 & 1.8&  9.0$\pm0.14$ &       6.2 &       9.8 &       --- &   <35.0&   0.18&0.003  &0.03&U     \\
I0356    & SA(s)abp &  11.7 & H {\sc{ii}}    &     156.6 &    -16.63$\pm0.71$/-16.77 &    -9.10&    -19.03 & 9.9&   10.6$\pm0.29$ &       7.6 &      11.5 &      --- &        35.5 &0.18&0.006 &0.40&I\\
I0520    & SAB(rs)a &  50.9 & L     &     138.1 &    -20.74$\pm0.52$ &    -15.48 &   -22.62&    3.7& 10.7$\pm0.21$  &       8.6 &      11.5 &       --- &     <36.1& 0.07 & 0.005 &0.62&U   \\
I2574\textsuperscript{$\dagger$}      & SAB(s)m  &   2.1 & H {\sc{ii}}    &      33.9 &     --- &      -4.83  &    -19.87 &0.7&      --- &        3.5  &       9.5 &      --- &        <33.7 &     ---&---& ---&U   \\
N0147&E5 pec& 0.8$^{\rm k}$&ALG &22.0 & -15.76$\pm0.17$  &---& -15.76  &2.0&9.1$\pm0.12$ &---&9.1 &---&<32.4 & 0.30&0.002&0.00&U\\
N0205&E5 pec& 0.8$^{\rm k}$ &ALG &23.3 &-18.22$\pm0.27$/-18.03  &---& -18.22& 2.0&9.5$\pm 0.13$&---& 9.5 &3.83$^{+1.18}_{-1.83}$&<32.4 &0.16&0.001&0.07&U\\
N0221&compact E2& 0.8$^{\rm k}$&ALG &72.1 &-15.52$\pm 0.18$  &---& -16.48 &1.3&8.6$\pm 0.11$ &---&9.0 &6.39$^{+0.19}_{-0.19}$&<32.3 &0.24&0.006&0.24&U\\
N0266&SB(rs)ab&  62.9&L &229.6& -22.93$\pm 0.20$ &-14.36&-23.36 &4.9&11.8$\pm 0.12$ &  8.4  &  12.0  &---&36.9 &0.18&0.003&0.39&I\\
N0278&SAB(rs)b&5.3 &H &47.6&-16.04$\pm 0.21$  &-10.82& -17.65 &1.9&8.8$\pm 0.12$&6.7  &  9.5  &---&34.8 & 0.10&0.001&0.15&I\\
N0315& cD&67.0 &L/RL &303.7&-23.59$\pm 0.27$ &---& -23.79&1.1& 11.7$\pm 0.14$&---&11.8 &8.99$^{+0.32}_{-0.32}$&39.6 &0.29&-0.021& 0.03&I\\
N0404&E&3.1$^{\rm k}$ &L &40.0 &-17.39$\pm 0.23$   &  -12.06&  -17.40&1.9&9.3$\pm  0.13 $ &7.2&9.3&5.65$^{+0.25}_{-0.25}$&<33.3 & 0.14&0.005&0.31&U\\
N0410& cD&72.4&L&299.7 & -20.72$\pm 0.13  $    &---& -21.51&1.5& 11.8$\pm 0.10   $ &---&12.1 &---&37.3 &0.17&-0.020&-0.02&I\\
N0507&SA0$^\wedge$$^{-}$0(r)&63.7 &ALG&292&-22.56$\pm 0.27$ &---&-23.48&5.5&11.7$\pm  0.13$  &---&12.1&---&36.9 &0.25&0.002&0.07&I\\
N0598&SA(s)cd&1.0$^{\rm b}$&H &21.0 &-11.84$\pm 0.21$ &---& -18.59&1.5&6.9$\pm 0.12$   &---&9.6 &---&<32.4 &0.10&0.001&1.21&U\\
N0672\textsuperscript{$\dagger$}  &SB(s)cd&7.2$^{\rm k}$ &H &64.3 &---  &---& -19.51 &1.2 &--- &---&9.8 &---&<34.6 &---&---&---&U\\ 
N0777&E1&68.8 &L/RL &324.1 &-20.68$ \pm 0.22$ &---& -20.68 &1.5&11.8$ \pm 0.12$ &---& 11.8&---&36.8 &0.14&-0.073&0.07&I\\
N0784\textsuperscript{$\dagger$}  &SBdm&5.0$^{\rm k}$ &H &35.5 &--- &---&-16.73 &1.9&--- &---&8.7 &---&<34.2 &---&---&---&U\\
N0841&(R')SAB(s)ab&62.2 &L&159.2&-21.34 $\pm 0.22$ &---& -22.23 &0.6&10.3$\pm 0.12$&---&10.6&---&<36.2 &0.42&0.005&0.01&U\\
N0890&SAB0$^\wedge$$^{-}$(r)?&54.0&ALG &210.9 &-20.12$ \pm 0.19$ &---&-20.46&1.2&11.4$\pm 0.12$&---&  11.6&---&<36.0 &0.31&-0.013&0.50&U\\ 

N0959    & Sdm?     &   5.4 & H {\sc{ii}}     &      43.6 & -15.89 $\pm 0.14$/-15.41  &    -10.43 &    -17.23 & 1.6&     8.6$\pm 0.09$ &       6.4 &       9.1&      ---  &         <34.7 &   0.07&-0.016&0.12&U   \\
N1003    & SA(s)cd  &   6.1 & H {\sc{ii}}   &     --- &   -11.25$\pm 0.02$/-11.72 &       ---  &  -18.12&1.9&      7.0$\pm 0.08$ &        ---  &       9.7 &      ---  &         <34.6 &  0.77&-0.185&0.16&U   \\
N1023    & SB0$^\wedge$$^{-}$(rs) &   6.2 & ALG  &     204.5 &    -19.35$\pm 0.26$   &     -10.69 &   -19.94&4.2&      10.9$\pm 0.13$ &     7.4&      11.1 &       7.38$^{+0.04}_{-0.04}$  &         <34.6 &    0.26&0.007&0.52&U    \\

N1058    & SA(rs)c  &   4.5 & L     &      31.0 &    -13.36$\pm 0.34$/-13.69  &  -12.53&    -16.34 &1.3&       7.36$\pm  0.15$ &      7.0 &    8.5 &       --- &       <34.4 &  0.16&0.006&0.01&U  \\

N1156\textsuperscript{$\dagger$}      & IB(s)m   &   2.5 & H     &      35.9 &      ---  &   -2.8 &   -16.81 & 0.9&      ---&    3.1 &       8.5 &    --- &       <34.2 &      ---& ---&---&U \\
N1161    & S0       &  25.6 & L/RL     &     258.4&    -22.08$\pm 0.07$/-22.25 &  -17.55   &    -22.25 &4.5&     11.4$\pm 0.10$ &   9.6 &      11.5 &     ---  &       36.6 &0.35&0.006 &0.13&I\\
N1167    & SA0$^\wedge$$^{-}$&  68.0 & L/RL     &     216.9 &    -19.34$\pm 0.27$  &    -13.61  &       -20.23    &2.5&      11.4$\pm 0.12$  &       9.1 &      11.8 &    --- &       39.7 &     0.08&-0.001&0.23 &I\\

\hline
error   & ---&   --- &  ---   &    ---   &   ---  & 0.35&   0.33& 25\%&---&  0.17 dex  &0.17 dex    &  ---&     1 dex& 20\%&  35\%&---& ---& \\
\hline
\end{tabular} 
Notes.---Col. (1) galaxy name. Col. (2) morphological classification
from RC3 \citep{1991rc3..book.....D}. Col. (3) while distance
($D$) are primarily from the NASA/IPAC Extragalactic Database NED (3K
CMB), other sources are \citet[][k]{2004AJ....127.2031K} and
\citet[][b]{2006ApJ...652..313B}. Col. (4) optical spectral class from
\citet{2021MNRAS.500.4749B}: \mbox{H {\sc{ii}}}, L = LINER, S =
Seyfert and ALG = Absorption Line Galaxy.  `RL' indicates radio loud
galaxies \citep{2021MNRAS.508.2019B}. Col. (5) the central velocity
dispersion
($\sigma$) from \citet{2009ApJS..183....1H}. Cols. (6)-(8)
$V$-band bulge, nucleus and galaxy magnitudes calculated by
integrating the best-fit S\'ersic, Gaussian or core-S\'ersic functions
and applying our colour transformation equations listed in
Table~\ref{Tab04} (see Section~\ref{Sec3.4}). Cols. (9) the stellar
mass-to-light ratios ($M/L$) measured in the {\it HST} bands listed in 
Tables~\ref{Tab1}$-$\ref{Tab1cS}. Cols. (10-12) we did not convert the $V$-band magnitudes into stellar masses, instead 
the luminosities of the bulge, nucleus
and the galaxy measured in the {\it HST} bands (listed in 
Tables~\ref{Tab1}$-$\ref{Tab1cS}) were converted  into stellar masses ($M_{*}$) using our 
$M/L$ values (col.\ 9). Cols. (13) logarithm of the SMBH mass
($M_{\rm BH}$) for galaxies with measured $M_{\rm
  BH}$ from \citet{2016ApJ...831..134V}, except for \mbox{NGC 205}
which is from \citet{2019ApJ...872..104N}. The BH masses are adjusted
here to our distance.  Col. (14) logarithm of the 1.5 GHz radio core
luminosity ($L_{\rm
  R,core}$).  Col. (15) average ellipticity of the bulge inside
$R_{\rm e}$ omitting PSF-affected region
($<\epsilon>$). Col. (16) median of the isophote shape parameter
inside $R_{\rm e}$ (Med
($B_{4}$)). Col. (17) inner logarithmic slope of the bulge light
profile.  Col. (18) radio detection of the galaxies are based on
\citet{2021MNRAS.500.4749B} and following their nomenclature `I' =
detected and core identified; `unI?' = detected but core unidentified;
`U' = undetected; `I+unI?' = detected and core identified having
additional unknown source(s) in the FOV.
\end {minipage}
\end{sideways}
\end{table*}
\end{center}

\setcounter{table}{4}
\begin{center}
\begin{table*} 
\begin{sideways}
\setlength{\tabcolsep}{.0405983in}
\begin {minipage}{251mm}
\caption{(\it continued)}
\label{Tab4}
\begin{tabular}{@{}llcccccccccccccccccccccc@{}}
\hline
\hline
Galaxy&Type& Dist&Class&  $\sigma$ &$M_{\rm V,bul}$&$M_{\rm V,nuc}$&$M_{\rm V,glxy}$&$M/L$&log$M_{\rm *,bul}$&log$M_{\rm *,nuc}$&log$M_{\rm *,glxy}$ &log $M_{\rm BH}$  &log$L_{\rm Radio,core}$ &$<\epsilon>$&Med ($B_{4}$)& $\gamma$&Det\\
&&&&&1D/2D&&&&&&&&&&\\
&&(Mpc)&&(km s$^{-1}$)&(mag)&(mag)&(mag)&&($M_{\sun}$)&($M_{\sun}$)&($M_{\sun}$)&($M_{\sun}$)&(erg s$^{-1}$)&&\\
(1)&(2)&(3)&(4)&(5)&(6)&(7)&(8)&(9)&(10)&(11)&(12&(13)&(14)&(15)&(16)&(17)&(18)\\          
\hline   
N1275    & cDpec      &  73.9 & L/RL    &     258.9 &    -23.50$\pm 0.43$ &    -17.77 &    -23.77 &1.4&      11.5$\pm 0.18$  &       9.2 &      11.6 &       9.00$^{+0.20}_{-0.20}$  &       41.0 &       0.23&-0.006  &0.24&I \\
N1961    & SAB(rs)c &  56.4 & L/RL     &     241.3 &   -22.96$\pm 0.47$  &     -16.93 &  -23.49 &  1.8&    11.3$\pm 0.20$  &       8.9 &      11.5 &       8.35$^{+0.35}_{-0.35}$  &        37.2 &      0.21&0.014&0.30&I  \\
N2273    & SB(r)a?  &  26.9 & S     &     148.9 &     -20.74$\pm 0.35$  &   -15.37 &  -21.27 & 3.9&     10.7$\pm 0.16$ &       8.6 &      10.9 &       6.89$^{+0.04}_{-0.04}$  &        36.6 &   0.37&0.005 &0.32&I\\ 
N2276    & SAB(rs)c &  34.3 & H {\sc{ii}}     &      83.5 &    -19.13$\pm 0.08$/-19.65  &  ---  &    -21.43 & 0.9&    9.3$\pm 0.09$  &       --- &      10.2 &      --- &        <35.9 &      0.39&-0.017&0.17&U   \\  
N2300&SA0$^\wedge$$^{-}$0&25.7&ALG&266&-21.33$\pm 0.20$  &---&-21.94&5.0&11.2$\pm 0.18$ &---& 11.4&---&36.2&0.18&-0.007 &0.08&I \\
N2342    & Spec     &  78.6 & H {\sc{ii}}     &     147.3 &   -19.78$\pm 0.75$  &   -16.54&   -22.52 &  1.9&   10.0$\pm 0.31$  &       8.7 &      10.7 &    ---  &       36.5 &    0.22&0.012&0.58&I \\   
N2366    & IB(s)m   &   1.6 & H {\sc{ii}}   &    ---&   -9.21$\pm 0.44$   &       ---  &   -16.87  &0.9&    5.4$\pm 0.19$  &       ---  &       8.4 &       --- &         <33.6 &      0.11&0.000 &1.71&U \\   
N2403    & SAB(s)cd &   2.6 & H {\sc{ii}}  &      68.4 &    -19.58$\pm 0.15$ &      ---  &    -20.46 &  0.8&    9.5$\pm 0.11$  &       ---  &       9.8 &     ---  &      <34.0 &   0.32&-0.010&0.00&U  \\  
N2500    & SB(rs)d  &   9.0 & H {\sc{ii}}   &      47.1 &    -14.39$\pm 0.49$  &  	-11.21   &    -18.09 &1.4&       7.8$\pm 0.20$  &     6.6 &       9.3 &     --- &      <34.7 &      0.22&0.008 &0.03&U \\   
N2537\textsuperscript{$\dagger$}        & SB(s)mpe &   8.3 & H {\sc{ii}}    &      63.0 &   --- &   --- &    -19.33& 0.7&      --- &      ---&     10.1 &       ---  &       <34.6 &     ---& --- & ---&U  \\
N2541    & SA(s)cd  &   9.8 & H {\sc{ii}}    &      53.0 &    -12.92$\pm 0.80$  &  -8.21 &    -19.15 &0.8&       6.9$\pm 0.32$  &       5.0 &       9.4 &   --- &        <34.7 &     0.33&0.001 &0.88&U \\
N2549    & SA0$^\wedge$$^{-}$0(r) &  16.6 & ALG  &     142.6 &    -21.07$\pm 0.33$  &    -14.70 &    -21.73 & 3.8&     10.9$\pm  0.14$  &       8.4 &      11.2 &       7.28$^{+0.37}_{-0.37}$  &        <35.3 &      0.30&0.006 &0.35&U  \\
N2634    & E1?      &  33.0 & ALG   &     181.1 &    -19.68$\pm 0.38$/-19.57  &    -14.23 &    -20.53 & 2.4&     10.2$\pm 0.16$  &       8.0 &      10.5 &      --- &        35.8 &   0.07&0.004 &0.37&I \\
N2639    & (R)SA(r) &  50.3 & L/RL     &     179.3 &    -20.74$\pm 0.37$ &    -16.80 &    -21.97 & 3.4&   11.2$\pm 0.15$  &       9.6 &      11.7 &      ---  &        37.6 &  0.28&0.009  &0.37&I\\
N2655    & SAB0a(s) &  20.3 & L/RL      &     159.8 &    -21.08$\pm 0.38$/-21.34  &    -17.28 &    -21.72 &3.7&      10.9$\pm 0.16$  &       9.4 &      11.2 &      ---  &        37.6 &     0.33&-0.011&0.32&I \\  
N2681&SAB(rs)0a&12.1&L&121&-19.00$\pm 0.63$ &-15.20&-20.22&1.6& 9.9$\pm 0.26$ & 8.3&10.3& ---&35.5& 0.20&-0.013 &0.68&I \\
N2683    & SA(rs)b  &   9.1 & L     &     130.2 &    -19.64$\pm 0.73$/-20.83&   	-16.07 &    -21.51 &2.6&     10.5$\pm 0.30$  &     9.1 &      11.3 &       ---  &       34.5 &     0.37&0.003 &0.16&I \\
N2685    & (R)SB0$^\wedge$$^{-}$+ &  14.4 & L     &      93.8 &    -18.48$\pm 0.13$  &    -12.71 &    -19.88 &2.6&      10.1$\pm 0.07$  &       7.8 &      10.6 &       6.65$^{+0.41}_{-0.41}$ &       <34.8 &    0.51&0.024&0.20&U  \\
N2748    & SAbc     &  21.6 & H {\sc{ii}}    &      83.0 &    -18.47$\pm 0.78$/-17.79 &    -10.36      &  -21.18    & 0.8&    9.7$\pm 0.32$  &       6.5 &      10.8 &       7.62$^{+0.24}_{-0.24}$  &       <35.4 &     0.31&0.010&0.24&U   \\
N2768    & E6?      &  21.6 & L/RL    &     181.8 &    -21.09$\pm 0.59$  &    -13.84 &    -21.27 & 2.6&     11.4$\pm 0.24$  &       8.5 &      11.5 &       ---  &       37.1 &    0.36&-0.006&0.41&I\\
N2770    & SA(s)c?  &  31.4 & H {\sc{ii}}     &      81.0 &    -18.62$\pm 0.93$  &    -14.24 &    -22.46 &0.9&    9.2$\pm 0.38$  &       7.4 &      10.7 &      ---  &        <35.5 &     0.48&-0.117 &0.02&U \\
N2782    & SAB(rs)a &  39.7 & H {\sc{ii}}    &     183.1 &    -20.38$\pm 0.49$ &    -15.05 &    -21.51&1.7&      10.4$\pm 0.20$  &       8.3 &      10.9 &      --- &        36.8 &    0.31&0.004 &0.37&I \\
N2787    & SB0$^\wedge$$^{-}$+(r) &  11.0 & L     &     202.0 &    -18.10$\pm 0.31$ &    -12.62&    -19.34& 3.9&     10.3$\pm 0.13$  &       8.1 &      10.8 &      7.78$^{+0.09}_{-0.09}$  &        36.3 & 0.21&0.006&0.06&I \\
N2832    & cD2?     & 105.0 & L    &     334.0 &    -24.58$\pm 0.26$  &    ---  &    -24.69 &4.1&     12.4$\pm 0.11$  &   ---  &      12.4 &      --- &        36.8 &     0.18&-0.001 &0.05&I+unI \\
N2841    & SA(r)b?  &  11.6 & L    &     222.0 &    -21.08$\pm 0.19$/-21.41 &       ---  &    -21.98 &2.9&    11.2$\pm 0.09$  &       --- &      11.6 &     --- &        34.8 &     0.27&-0.001 &0.48&I\\
N2859    & (R)SB0$^\wedge$$^{-}$+ &  27.8 & L     &     188.2 &    -20.15$\pm 0.08$/-20.11  &    -11.89 &    -20.87 &3.9&     11.1$\pm 0.05$  &       7.8 &      11.4 &      ---  &        <35.3 &     0.14&0.000&0.51&U   \\
N2903    & SAB(rs)b &  12.1 & H {\sc{ii}}     &      89.0 &    -15.04$\pm 1.01$   &    -11.98 &    -22.03 &1.5&       8.2$\pm 0.41$  &       7.0 &      11.0 &       7.13$^{+0.28}_{-0.28}$  &       <34.3 &     0.42&0.001 &0.81&U  \\
N2950    & (R)SB0$^\wedge$$^{-}$0 &  20.9 & ALG   &     163.0 &    -19.57$ \pm 0.41$   &    -15.00 &    -20.45 &3.5&      10.8$\pm 0.17$  &       9.0 &      11.1 &      --- &         <35.5 &      0.25&0.003 &0.27&U  \\
N2964    & SAB(r)bc &  22.9 & H {\sc{ii}}     &     109.4 &    -17.31$\pm 0.22$/-17.71 &     -17.04&    -20.85 &1.8&      9.3$\pm 0.10$  &      9.2 &      10.7 &       6.80$^{+0.61}_{-0.61}$  &       36.0 &     0.26&-0.002&0.00&I\\
N2976    & SAcpec   &   1.3 & H {\sc{ii}}     &      36.0 &     -8.94$\pm 0.77$/-8.93  &      --- &    -16.64 &1.4&     5.7$\pm 0.31$  &       --- &       8.8 &      ---  &      <33.2 &     0.27&0.006  &1.24&U  \\
N2985    & (R')SA(r &  19.8 & L   &     140.8 &    -20.70$\pm 0.78$/-20.58 &    -13.20 &    -21.13 &2.2&      10.8$\pm 0.32$  &       7.8 &      11.0 &    --- &      35.8 &    0.14&-0.003&0.40&I&   \\
N3031    & SA(s)ab  &   0.7 & L     &     161.6 &    -16.62$\pm 0.42$  &     -7.91&    -17.63 & 3.2&     9.5$\pm 0.17$  &       6.0 &       9.9 &       7.09$^{+0.13}_{-0.13}$ &        35.5 &    0.19&0.002 &0.36&I  \\
N3034    & I0edge-on &   4.0 & H {\sc{ii}}     &     129.5 &    -15.67$\pm 0.45$  &      ---  &    -21.10 &2.1&       8.7$\pm 0.19$  &       ---  &      10.9 &      --- &       34.3 &    0.19&0.033 &0.37&unI   \\
N3073    & SAB0$^\wedge$$^{-}$   &  19.1 & H {\sc{ii}}   &      35.6 &    -18.14$\pm 0.16$/-18.13  &    -9.73 &    -18.70 &1.9&      9.7$\pm 0.08$  &       6.3 &       9.9 &      ---  &        <35.3 &     0.34&0.002&0.33&U \\
N3077\textsuperscript{$\dagger$}      & I0pec    &   1.4 & H     &       32.4 &  --- &       -8.66&-16.24&2.1&     --- &     5.4&      9.0&     ---&        33.3 & ---&---&---& I+unI    \\
N3079    & SB(s)c edge-on &  18.3 & L    &     182.3 &    -20.34$\pm 0.32$  &      -14.38 &    -22.33 &1.3&      10.2$\pm 0.15$  &       7.8 &      11.0 &       6.46$^{+0.05}_{-0.05}$  &        37.3 &     0.24&-0.03&0.17&I    \\
N3184    & SAB(rs)c &  11.4 & H {\sc{ii}}     &      43.3 &  -16.62$\pm 0.35$  &       -13.55  &    -20.24 &1.8&        8.7$\pm 0.15$  &     7.5&      10.2 &      --- &        <34.6 &      0.30&-0.006&0.24&U   \\

\hline
error   & ---&   --- &  ---   &    ---   &   ---  & 0.35&   0.33& 25\%&   ---&     0.17 dex    &0.17 dex &  ---&    1 dex & 20\%&  35\%&---& ---& \\
\hline

\end{tabular} 
\end {minipage}
\end{sideways}
\end{table*}
\end{center}

\setcounter{table}{4}
\begin{center}
\begin{table*} 
\begin{sideways}
\setlength{\tabcolsep}{0.0405983in}
\begin {minipage}{251mm} 
\caption{(\it continued)}
\begin{tabular}{@{}llcccccccccccccccccccccc@{}}
\hline
\hline
Galaxy&Type& Dist&Class&  $\sigma$ &$M_{\rm V,bul}$&$M_{\rm V,nuc}$&$M_{\rm V,glxy}$&$M/L$&log$M_{\rm *,bul}$&log$M_{\rm *,nuc}$&log$M_{\rm *,glxy}$ &log $M_{\rm BH}$  &log$L_{\rm Radio,core}$ &$<\epsilon>$& Med ($B_{4}$)&$\gamma$&Det\\
&&&&&1D/2D&&&&&&&&&&\\
&&(Mpc)&&(km s$^{-1}$)&(mag)&(mag)&(mag)&&($M_{\sun}$)&($M_{\sun}$)&($M_{\sun}$)&($M_{\sun}$)&(erg s$^{-1}$)&&\\
(1)&(2)&(3)&(4)&(5)&(6)&(7)&(8)&(9)&(10)&(11)&(12&(13)&(14)&(15)&(16)&(17)&(18)\\    
\hline  
N3185    & (R)SB(r) &  22.0 & S     &      79.3 &    -19.16$\pm 0.39$  &    -13.98 &    -19.94& 2.5&   10.3$\pm 0.17$ &       8.2 &      10.6 &      ---  &       <35.2 &      0.20&0.001 &0.37&U   \\
N3190    & SA(s)ape &  21.8 & L     &     188.1 &    -19.42$\pm 0.02$ &       ---  &     -21.79 & 4.1&  10.8$\pm 0.04$  &       ---  &      11.7 &      ---  &       <35.3 &   0.20&0.010  &0.16&U   \\
N3193    & E2       &  24.3 & L     &     194.3 &    -21.79$\pm 0.20$&    -14.40 &    -21.93 &4.8&      11.4$\pm 0.09$  &       8.4 &      11.4 &      --- &        <35.2 &      0.19&0.004 &-0.02&U \\
N3198    & SB(rs)c  &  12.6 & H {\sc{ii}}    &      46.1 &     -20.74$\pm 0.47$&    -12.98 &    -21.51 &0.5&      10.5$\pm 0.19$ &      7.4&     	 10.8&       ---  &       35.0    & 0.32&0.005&0.33&I\\
N3245    & SA0$^\wedge$$^{-}$0(r) &  22.3 & H {\sc{ii}}    &     209.9 &    -20.74$\pm 0.18$    &    -13.50  &      -21.39 & 2.2&       10.6$\pm 0.08$   &     7.7 &   10.8  &       8.40$^{+0.11}_{-0.11}$  &       35.7     & 0.27&0.003 &0.44&I \\
N3319    & SB(rs)cd &  14.1 & L     &      87.4 &    -19.43$\pm 0.74$/-19.06 &    -12.85 &    -20.56 &0.7&       9.5$\pm 0.30$  &       6.9 &      9.9 &     ---  &      <34.8 &      0.47&0.007 &0.05&U \\
N3344    & (R)SAB(r &  12.3 & H {\sc{ii}}     &      73.6 &    -17.83$\pm 0.79$  &    -13.19 &    -20.82 &1.4&      9.3$\pm 0.32$  &       7.4 &      10.5 &       ---  &        <34.1      & 0.07&0.004 &0.56&U   \\
N3348    & E0       &  41.8 & ALG   &     236.4 &    -21.31$\pm 0.24$/-21.28  &       ---  &    -21.31 &2.3&     11.4$\pm 0.11$  &      ---  &      11.4 &      --- &         36.6 &    0.07&0.001 &0.04&I\\
N3414    & S0pec    &  24.4 & L   &     236.8 &    -20.59$\pm 0.15$/-20.83 &    -13.28 &    -20.77 &4.0&      11.3$\pm 0.07$  &       8.4 &      11.4 &       8.39$^{+0.07}_{-0.07}$  &        36.2      & 0.20&0.002 &0.64&I \\
N3448    & I0       &  21.0 & H {\sc{ii}}     &      50.7 &    -20.54$\pm 0.26$/-20.73   &     --- &    -21.39 &0.6&     10.3$\pm 0.11$  &       ---  &      10.6 &       --- &        35.5  & 0.37&-0.016&0.04&I\\
N3486    & SAB(r)c  &  14.1 & S     &      65.0 &    -18.58$\pm 0.47$/-18.24 &    -13.79 &    -20.27 & 1.1&      9.3$\pm 0.19$  &       7.4 &      10.0 &        ---  &       <34.3 &      0.14&-0.002 &0.37&U    \\
N3504    & (R)SAB(s) &  26.2 & H {\sc{ii}}     &     119.3 &    -20.43$\pm 0.38$    &     -15.36  &        -20.90 & 0.9&     10.6$\pm 0.16$  &       8.6 &      10.8 &     ---  &        37.3    & 0.17&0.005  &0.44&I  \\
N3516    & (R)SB0$^\wedge$$^{-}$0 &  37.5 & S     &     181.0 &    -21.01$\pm 0.39$ &    -16.80 &    -22.36 & 2.5&  11.1$\pm 0.16$  &       9.4 &      11.6 &       7.37$^{+0.16}_{-0.16}$  &        36.8 &      0.17&-0.003 &0.47&I \\
N3600    & Sa?      &  13.2 & H {\sc{ii}}      &      49.8 &    -16.23$\pm 0.56$/-16.20&      --- &    -17.57 &2.7&      9.2$\pm 0.23$  &         --- &       9.8 &       ---  &        <34.7 &       0.56&-0.010  &0.05&U   \\
N3610    & E5?      &  25.6 & ALG   &     161.2 &    -20.79$\pm 0.64$ &   -15.68&     -20.88   & 0.5&    10.6$\pm 0.26$  &      8.5&      10.6 &      --- &       <35.6 &     0.33&0.003 &0.38&U  \\
N3613    & E6       &  31.7 & ALG   &     220.1 &    -21.39$\pm 0.34$ &       --- &    -21.99 &2.5&      10.9$\pm 0.14$  &     --- &      11.2 &       ---   &     <35.7 &     0.27&-0.002  &0.09&U  \\
N3631    & SA(s)c   &  19.2 & H {\sc{ii}}     &      43.9 &    -20.13$\pm 0.22$/-20.08 &    -13.24 &    -20.32 &1.4&     10.1$\pm 0.10$  &       7.4 &      10.2 &        ---  &     <35.2 &   0.05&0.002 &0.27&U    \\
N3642    & SA(r)bc? &  24.9 & L    &      85.0 &    -18.85$\pm 0.03$/-18.98 &    -13.20 &    -20.63&1.1&       9.4$\pm 0.05$  &       7.2 &      10.1 &       7.48$^{+0.04}_{-0.04}$  &        <35.5 &    0.14&0.000   &0.61&U \\
N3665    & SA0$^\wedge$$^{-}$0(s) &  32.1 & H {\sc{ii}}/RL      &     236.8 &     -24.81$\pm 0.69$ &       ---  &   -25.12  &1.1&      12.5$\pm 0.28$  &      ---&      12.6 &       8.73$^{+0.09}_{-0.09}$  &       36.8 &   0.23&0.001 &-0.02&I \\
N3675    & SA(s)b   &  13.8 & L    &     108.0 &    -15.54$\pm 0.71$&        ---  &    -20.93 &4.5&       9.4$\pm 0.29$  &      --- &      11.6 &       7.31$^{+0.29}_{-0.29}$  &       35.0 &    0.30&-0.022  &0.61&I  \\
N3718    & SB(s)a &  16.9 & L/RL    &     158.1 &    -19.62$\pm 0.31$/-19.59 &    -13.01 &    -20.15 &2.4&      10.5$\pm 0.13$  &       7.8  &      10.7 &       ---  &        36.8 &   0.28&-0.020   &0.18&I  \\
N3729    & SB(r)ape &  17.8 & H {\sc{ii}}      &      76.2 &    -16.22$\pm 0.37$ &    -12.89 &    -18.86 &1.9&       8.9$\pm 0.15$  &       7.6 &       9.9 &   ---  &        35.8 &   0.24&0.006  &0.21&I \\
N3756    & SAB(rs)b &  21.4 & H {\sc{ii}}     &      47.6 &    -17.81$\pm 0.78$/-17.80 &    -12.79 &    -20.57 &1.4&       9.2$\pm 0.32$  &       7.2 &      10.4 &    --- &       <35.3 &    0.19&0.015   &0.33&U    \\
N3838    & SA0a?    &  21.0 & ALG  &     141.4 &  -18.69$\pm 0.39$&     -13.01   &  -22.11&4.7&   10.0$\pm 0.16$ &    7.7 &       11.4 &       --- &         35.4 &   0.36&0.019 &0.74& unI    \\
N3884    &SA(r)0/a&	107.0&L&208.3&  -21.92$\pm 0.29$/-22.15&     -16.74&      -23.11    &7.1&  11.6$\pm 0.13$     &  9.5&      12.0&&37.8&0.15&0.005 & 0.43&I\\
N3898    & SA(s)ab  &  19.2 & L   &     206.5 &    -20.25$\pm 0.59$ &    -14.71 &    -20.59 & 3.3&     11.0$\pm 0.24$  &       8.8 &      11.1 &        ---  &       35.8 &     0.28&0.004   &0.44&I \\
N3900    & SA0$^\wedge$$^{-}$+(r) &  30.2 & ALG   &     139.2 &    -20.83$\pm 0.04$ &    -14.00 &    -21.61 &3.6&      10.7$\pm 0.07$  &       8.0 &      11.1 &      ---  &      <35.6 &0.33&0.006    &0.72&U \\
N3945    &(R)SB0$^\wedge$$^{-}$+(rs)   &20.3&L &191.5 &-17.68$\pm 0.38$ & -17.83    &    -20.20  &2.5&  10.0$\pm 0.16$&   10.1 &  11.0&6.96$^{+0.47}_{-0.47}$ &35.8& 0.20&0.018& 0.53&I\\
N3949    & SA(s)bc? &  14.5 & H {\sc{ii}}      &      82.0 &    -19.92$\pm 0.29$/-19.72 &    -13.49 &    -20.43 &0.9&       9.6$\pm 0.13$  &       7.0 &       9.8 &        --- &       <35.0 &  0.35&-0.004&0.06&U   \\
N3982    & SAB(r)b? &  18.3 & S     &      73.0 &    -19.40$\pm 0.34$ &    -14.55 &    -20.77 &0.4&       8.7$\pm 0.15$  &       6.7 &       9.2  &       7.01$^{+0.26}_{-0.26}$ &        36.2 &     0.15&-0.005&0.27&I   \\
N3992    & SB(rs)bc &  17.6 & L   &     148.4 &    -18.50$\pm 0.39$/-19.00 &    -10.87 &    -21.11 &3.0&      9.8$\pm 0.16$  &       6.8 &      10.9 &       7.57$^{+0.28}_{-0.28}$  &       <35.1 &      0.19&0.004   &0.54&U  \\
N3998    & SA0$^\wedge$$^{-}$0(r) &  17.4 & L/RL       &     304.6 &    -19.27$\pm0.47$/-19.16 &     	-14.16 &    -20.04 &4.3&      10.9$\pm 0.19$ &    8.8 &      11.2 &       9.02$^{+0.05}_{-0.05}$ &        38.0 &   0.12&0.003   &0.36&I  \\
N4026    & S0edge-on &  16.9 & L  &     177.2 &    -22.37$\pm 0.41$ &    -15.80 &    -23.11 &3.3&     11.3$\pm 0.17$  &       8.7 &      11.6 &       8.36$^{+0.12}_{-0.12}$ &         <35.1 &0.28&0.017   &0.43&U   \\

N4036    & S0$^\wedge$$^{-}$     &  21.7 & L     &     215.1 &    -19.23$\pm 0.08$ &     -15.22	  &    -21.45 &3.5&     10.5$\pm 0.09$  &       8.9&      11.4 &       7.95$^{+0.36}_{-0.36}$ &       36.0 & 0.28&0.007  &0.15&I  \\

N4041    & SA(rs)bc &  19.5 & H {\sc{ii}}      &      95.0 &    -19.32$\pm 0.11$/-19.76 &     	-13.54 &    -20.14 &1.4&       9.8$\pm 0.09$  &     7.5&      10.2 &       6.00$^{+0.20}_{-0.20}$ &        35.5 &     0.34&-0.006   &0.01&I \\
N4062    & SA(s)c   &  14.9 & H {\sc{ii}}      &      93.2 &    -15.35$\pm 0.22$/-15.83 &    -13.48 &    -20.31&2.5&       8.4$\pm 0.10$  &       7.6 &      10.4 &       ---  &        <34.6 &    0.32&0.004 &0.00&U     \\
N4096    & SAB(rs)c &  11.1 & H {\sc{ii}}      &      79.5 &    -17.49$\pm 0.49$/-17.88&     --- &    -20.94 & 1.4&      9.0$\pm 0.20$  &       --- &      10.4 &        ---  &        <34.5 &   0.36&-0.006  &0.04&U   \\
\hline 
error   & ---&   --- &  ---   &    ---   &   ---  & 0.35&   0.33& 25\%&  ---&      0.17  dex  &0.17 dex &  ---&     1 dex & 20\%&  35\%&---& ---& \\
\hline
\end{tabular} 
\end {minipage}
\end{sideways}
\end{table*}
\end{center}

\setcounter{table}{4}
\begin{center}
\begin{table*} 
\begin{sideways}
\setlength{\tabcolsep}{0.0405983in}
\begin {minipage}{251mm}
\caption{(\it continued)}
\begin{tabular}{@{}llcccccccccccccccccccccc@{}}
\hline
\hline 
Galaxy&Type& Dist&Class&  $\sigma$ &$M_{\rm V,bul}$&$M_{\rm V,nuc}$&$M_{\rm V,glxy}$&$M/L$&log$M_{\rm *,bul}$&log$M_{\rm *,nuc}$&log$M_{\rm *,glxy}$ &log $M_{\rm BH}$  &log$L_{\rm Radio,core}$ &$<\epsilon>$& Med ($B_{4}$)&$\gamma$&Det&\\
&&&&&1D/2D&&&&&&&&&&\\
&&(Mpc)&&(km s$^{-1}$)&(mag)&(mag)&(mag)&&($M_{\sun}$)&($M_{\sun}$)&($M_{\sun}$)&($M_{\sun}$)&(erg s$^{-1}$)&&\\
(1)&(2)&(3)&(4)&(5)&(6)&(7)&(8)&(9)&(10)&(11)&(12&(13)&(14)&(15)&(16)&(17)&(18)\\    
\hline  
N4102    & SAB(s)b? &  14.7 & H {\sc{ii}}      &     174.3 &   -18.30$\pm 0.44$&     -16.32 &    -18.77 &5.0&     10.6$\pm 0.21$  &     9.8 &      10.8 &       ---  &        35.9 &    0.13&0.008 &0.36&I \\
N4125    & E6pec    &  21.0 & L    &     226.7 &    -21.14$\pm 0.23$ &      -14.39&     -21.14 &2.1&       11.26$\pm 0.10$  &    8.6 &      11.3 &        --- &        <35.4 &    0.24&0.006     &0.28&U  \\
N4138    & SA0$^\wedge$$^{+}$(r) &  16.0 & L     &     120.9 &    -20.01$\pm 0.39$ &    -14.93&    -20.58 &4.2&      10.6$\pm 0.16$  & 8.5 &      10.8 &    --- &        <35.3 &     0.37&0.003 &0.41&U   \\
N4143    & SAB0$^\wedge$0(s &  16.9 & L     &     204.9 &     -20.45$\pm 0.34$/-20.70 &     -13.82 &    -21.22 &7.9&   10.9$\pm 0.14$  &       8.3 &      11.2 &       7.98$^{+0.37}_{-0.37}$  &        36.1 &  0.16&0.005  &0.52&I\\
N4144    & SAB(s)cd &   6.8 & H {\sc{ii}}      &      64.3 &    -10.08$\pm 1.10$ &       ---  &    -19.38 &0.8&       5.5$\pm 0.44$  &     --- &       9.3 &       ---  &        <34.0 &    0.46&0.017 &0.57&U   \\
N4150    & SA0$^\wedge$0(r) &   7.1 & L     &      87.0 &    -15.80$\pm 0.14$/-16.18&    -13.31 &    -17.77 &2.7&       9.0$\pm 0.08$  &       8.0 &       9.8 &       5.68$^{+0.44}_{-0.44}$&       <34.6 &0.33&-0.008  &0.07&U    \\
N4151    & (R')SAB( &  17.8 & S     &      97.0 &    -19.22$\pm 0.84$ &    -17.62 &    -20.85 &2.0&      10.1$\pm 0.34$  &       9.5 &      10.8 &       7.76$^{+0.08}_{-0.08}$ &        37.8 &     0.11&0.000   &0.81&I \\
N4183\textsuperscript{$\dagger$}      & SA(s)cd? &  16.6 & H {\sc{ii}}      &      34.4 &   --- &      ---  &    -20.54 &0.8&     ---  &      --- &       9.8 &      ---  &        <35.2 &    ---&---&---&U    \\
N4203   &SAB0$^\wedge$$^{-}$     &   19.5  &L&    167.0& -19.59$\pm 0.27$ &     -12.75  &  -20.44&3.9&10.9$\pm 0.11$&     8.2     & 11.2&---&36.1&0.07 &0.005& 0.36&I \\
N4217    & Sb-edge on&  17.7 & H {\sc{ii}}      &      91.3 &    -19.88$\pm 0.18$/-19.76 &   ---  &    -21.16 &2.5&    10.6$\pm 0.08$  &      --- &      11.1 &        ---  &        35.2 &     0.58&0.009&0.01&I \\
N4220    & SA0$^\wedge$$^{+}$(r) &  16.0 & L     &     105.5 &    -18.25$\pm 0.14$ &   -13.08&    -19.96 &2.8&      10.1$\pm 0.08$  &     8.0&      10.7 &        ---  &       35.2 &    0.44&-0.008&0.38&I   \\
N4242    & SAB(s)dm &  10.3 & H {\sc{ii}}    &     --- &    -11.74$\pm 0.19$-11.69 &     --- &    -18.80 &1.2&       6.7$\pm 0.09$  &         ---  &       9.5 &       ---  &         34.4 &    0.71&-0.012 &1.68& unI \\
N4244    & SA(s)cd? &   7.1 & H {\sc{ii}}      &      36.8 &    -14.79$\pm 0.09$/-14.10 &      ---  &    -21.99 &1.0&      7.7$\pm 0.06$  &      ---&      10.5 &       ---  &        33.8 &     0.40&0.018 &0.78&I  \\
N4245    & SB0a?(r) &  16.7 & H {\sc{ii}}      &      82.7 &    -19.30$\pm 0.33$ &   -12.80&    -20.18 &5.2&      10.3$\pm 0.14$  &     7.7&      10.6 &       7.25$^{+0.48}_{-0.48}$ &    <34.6 &  0.32&-0.014&0.16&U      \\
N4258    & SAB(s)bc &   9.4 & S     &     148.0 &    -19.94$\pm 0.31$ &     -14.00 &    -21.69 &3.1&      10.8$\pm 0.13$  &    8.4 &      11.5 &    7.69$^{+0.03}_{-0.03}$ &        34.9 &    0.38&0.008 &0.39&I  \\
N4274    & (R)SB(r) &  17.4 & L   &      96.6 &    -17.93$\pm 0.16$/-18.29 &     -12.06 &    -20.10 &2.9&       9.6$\pm 0.10$  &       7.2&      10.4 &       ---  &        <34.6 &      0.45&0.004 &0.05&U\\
N4278 &E1-2&15.6&L/RL  &237 &-20.91$\pm 0.27$&-11.57&-20.91&4.5& 11.0$\pm 0.12$ &  7.2& 11.0&7.98$^{+0.27}_{-0.27}$&37.6&0.16&0.000&0.22&I \\
N4291 &E&25.5&ALG&293 &-20.71$\pm 0.19$&---&-20.71&4.5& 10.9$\pm 0.09$ &---& 10.9&8.97$^{+0.16}_{-0.16}$&<35.5&0.24&-0.006&0.10&U    \\
N4314    & SB(rs)a  &  17.8 & L   &     117.0 &    -20.11$\pm 0.22$ &    -13.20 &    -20.91 &3.1&      10.9$\pm 0.10$  &       8.1&      11.2&       6.97$^{+0.30}_{-0.30}$&        <34.7 &    0.19&0.009&0.16&U    \\
N4414    & SA(rs)c? &  14.2 & L   &     117.0 &    -19.00$\pm 0.79$&     -11.00  &   -20.83&2.7&       10.3$\pm 0.32$  &     7.1&      11.1 &       ---  &        <34.7 &     0.30&0.004&0.42&U   \\
N4448    & SB(r)ab  &  13.5 & H {\sc{ii}}      &     119.8 &    -18.89$\pm 0.02$/-19.30 &     -9.23 &    -19.38 &3.7&      10.6$\pm 0.05$  &       6.7 &      10.76 &      ---  &        <34.5 &     0.18&0.012 &0.19&U   \\
N4449    & IBm      &   6.1 &H {\sc{ii}}    &      17.8 &    -16.61$\pm 0.54$/-16.58 &    -13.21 &   -20.10 &0.8&       8.4$\pm 0.22$  &       7.0 &       9.8 &       --- &        <33.6 &    0.26&-0.010 &0.46&U      \\
N4485    & IB(s)mpe &  10.3 & H {\sc{ii}}     &      52.2 &    -17.58$\pm 0.32$ &       ---  &    -18.09 &0.7&       8.7$\pm 0.14$  &     ---&       8.9 &       ---  &        <34.3 &    0.49&-0.018 &0.00&U     \\
N4490    & SB(s)dpe &  11.4 & H {\sc{ii}}     &      45.1 &    -20.05$\pm 0.14$ &     ---  &    -21.94 & 0.7&      9.4 $\pm 0.07$ &   ---&      10.2 &     ---  &       <34.3 &   0.38&0.003  &0.00&U    \\
N4559    & SAB(rs)c &  15.6 & H {\sc{ii}}     &      49.2 &   -16.16$\pm 0.40$/-16.66&    -14.25 &    -19.64 & 0.8&      8.0$\pm 0.15$  &       7.3 &     9.45 &     --- &      <34.5 & 0.13&0.035 &0.61&U    \\
N4565    & SA(s)b?e &  21.8 & S     &     136.0 &     -21.71$\pm 0.25$ &     	-9.86 &     -23.62&2.1&      11.2$\pm 0.11$  &   6.4&      11.9 &       --- &        35.2 &   0.36&-0.002 &0.69&I \\
N4589    &E2	&	 29.2 &L/RL  &     224.3&-21.04$\pm 0.25$&---&-21.04&4.7&      11.0$\pm 0.11$  &    --- &     11.0&---&37.5&0.34&0.003&0.30&I\\

N4605    & SB(s)cpe &   3.7 & H {\sc{ii}}      &      26.1 &    -17.72$\pm 0.29$/-18.64&       ---  &    -18.48&1.1&       9.0$\pm 0.12$ &      ---&       9.3 &    ---  &        <33.8 &    0.49&0.050&0.00&U     \\

N4648    & E3       &  21.0 & ALG   &     224.5 &    -18.66$\pm 0.39$/-18.65 &    -11.04 &    -19.58 &2.0&      10.1$\pm 0.16$  &       7.0 &      10.4 &       ---  &         <35.5 &    0.16&-0.001 &0.53&U \\
N4656\textsuperscript{$\dagger$}      & SB(s)mpe &  13.0 & H {\sc{ii}}      &      70.4 &    --- &    -10.88&    -21.47 &0.5&   ---  &      5.5&       9.8 &       --- &        <34.3 &---&---&---&U   \\
N4736    & (R)SA(r) &   7.6 & L  &     112.0 &    -19.78$\pm 0.77$ &   -15.33 &  -21.34&2.4&      10.5$\pm 0.12$  &      8.7 &      11.1 &       7.00$^{+0.12}_{-0.12}$&        34.8 &  0.11&-0.007 &0.01&I \\
N4750    & (R)SA(rs &  24.1 & L     &     136.0 &    -19.57$\pm 0.56$ &    -14.03 &    -20.89 &1.9&     10.2$\pm 0.23$  &       8.0 &      10.8 &      --- &        35.8 &     0.23&0.010&0.37&I  \\
N4800    & SA(rs)b  &  15.6 & H {\sc{ii}}      &     111.0 &    -19.24$\pm 0.26$ &    -12.37 &    -19.41 &2.5&      10.3$\pm 0.11$  &       7.6 &      10.4 &       7.12$^{+0.53}_{-0.53}$ &        <34.9 &    0.18&-0.003  &0.04&U  \\
N4826    & (R)SA(rs &  10.0 & L   &      96.0 &    -19.63$\pm 0.33$ &    -14.96 &    -21.45 &2.5&      10.5$\pm 0.14$  &       8.6 &      11.2 &       6.33$^{+0.13}_{-0.13}$ &        33.9  & 0.28&-0.007 &0.25&I \\
N4914    & cD       &  70.8 & ALG   &     224.7 &     -20.94$\pm 0.25$ &    ---&        -20.94 & 0.9&     11.6$\pm 0.11$  &       --- &      11.6&        --- &        <36.2 &   0.31&0.006 &-0.01&U    \\
N5005    & SAB(rs)b &  16.8 & L     &     172.0 &   -21.30$\pm 0.64$ &   -14.22 &    -21.47 &2.2&     11.0$\pm 0.26$  &       8.2 &      11.1 &       8.33$^{+0.23}_{-0.23}$ &       36.3 &       0.45&-0.005&0.21&I \\
N5033    & SA(s)c   &  15.8 & L     &     151.0 &    -20.00$\pm 0.44$ &    -14.31 &    -22.31 & 1.1&      9.90$\pm 0.18$  &       7.6 &  10.8&        ---  &        36.0 &   0.31&-0.002&0.40&I  \\
N5055    & SA(rs)bc &   9.9 & L    &     117.0 &    -17.70$\pm 0.33$/-18.10&    -15.80 &    -21.56 &2.0&      9.5$\pm 0.14$  &       8.8 &    11.1 &       8.97$^{+0.11}_{-0.11}$ &       <34.4 &      0.24&0.001 &0.06&U    \\
\hline  
error   & ---&   --- &  ---   &    ---   &   ---  & 0.35&   0.33&25\%&   ---&     0.17  dex  &0.17 dex &  ---&     1 dex & 20\%&  35\%&---& ---& \\
\hline
\end{tabular} 
\end {minipage}
\end{sideways}
\end{table*}
\end{center}

\setcounter{table}{4}
\begin{center}
\begin{table*} 
\begin{sideways}
\setlength{\tabcolsep}{0.0405983in}
\begin {minipage}{251mm}
\caption{(\it continued)}
\begin{tabular}{@{}llcccccccccccccccccccccc@{}}
\hline
\hline
Galaxy&Type& Dist&Class&  $\sigma$ &$M_{\rm V,bul}$&$M_{\rm V,nuc}$&$M_{\rm V,glxy}$&  $M/L$&log$M_{\rm *,bul}$&log$M_{\rm *,nuc}$&log$M_{\rm *,glxy}$ &log $M_{\rm BH}$  &log$L_{\rm Radio,core}$ &$<\epsilon>$& Med ($B_{4}$)&$\gamma$&Det\\
&&&&&1D/2D&&&&&&&&&&\\
&&(Mpc)&&(km s$^{-1}$)&(mag)&(mag)&(mag)&&($M_{\sun}$)&($M_{\sun}$)&($M_{\sun}$)&($M_{\sun}$)&(erg s$^{-1}$)&&\\
(1)&(2)&(3)&(4)&(5)&(6)&(7)&(8)&(9)&(10)&(11)&(12&(13)&(14)&(15)&(16)&(17)&(18)\\     
\hline   
N5112\textsuperscript{$\dagger$}      & SB(rs)cd &  17.0 & H {\sc{ii}}      &      60.8 & ---&    -11.16 &   -19.51 &0.9&   ---  &       6.2 &     9.50&        ---  &        <35.4 &  ---&---&---&U    \\

N5204    & SA(s)m   &   4.6 & H {\sc{ii}}      &      39.9 &    -17.75$\pm 0.28$/-17.64  &     -6.93 &    -18.31 &0.8&       8.7$\pm 0.12$  &       4.3 &       8.9 &       --- &        <34.1 &      0.45&-0.011&0.01&U    \\

N5273    & SA0$^\wedge$0(s) &  18.6 & S     &      71.0 &    -18.86$\pm 0.09$  &    -13.73 &    -20.24 &4.5&      10.1$\pm 0.06$  &       8.0 &      10.6&       6.69$^{+0.27}_{-0.27}$ &        35.3 &  0.11&0.000 &0.55&unI    \\
N5308    & S0$^\wedge$$^{-}$edge on &  30.1 & ALG   &     249.0 &    19.99$\pm 0.49$  &   -17.52 &    -21.08 &3.1  &    10.8$\pm 0.20$  &      9.8&      11.3 &       ---  &         <35.8 &  0.46&0.009  &0.12&U  \\
N5322 &E3-4 & 27.0&L/RL   & 230&-21.77$\pm 0.24$ &-18.02 &-22.10&2.8& 11.3$\pm 0.11$ &  9.8&11.5&---&37.0&  0.32&-0.004 &0.17&I\\
N5353    & S0edge-on &  36.1 & L/RL       &     286.4 &    -18.04$\pm 0.23$     &      --- &   -19.72   &0.9&      10.4$\pm 0.10$  &      ---&      11.1 &       --- &        37.6 &      0.37&-0.011 &0.11&I  \\
N5354    & S0edge-on &  39.8 & L/RL       &     217.4 &     -20.51$\pm 0.18$/-20.78  &   -15.82   &     -21.34 &1.4&      10.9$\pm 0.08$  &       9.0 &      11.2 &    ---  &        36.9 &   0.17&-0.005 &0.26&I \\
N5377    & (R)SB(s) &  28.0 & L     &     169.7 &   -20.987$\pm 0.33$  &   	-16.48&    -21.77 &2.9&      11.2$\pm 0.15$  & 9.4&      11.5 &     ---  &        35.7 &       0.23&	-0.008  &0.22&I\\
N5448    & (R)SAB(r &  31.1 & L     &     124.5 &    -19.77$\pm 0.22$  &   -16.53 &    -21.84 &0.5&      9.5$\pm 0.10$  &     8.2&      10.4 &   --- &      35.7 &       0.32&-0.008  &0.18&I \\
N5457    & SAB(rs)c &   5.2 & H {\sc{ii}}      &      23.6 &    -15.72$\pm 0.81$  &     -9.96 &    -17.42 &1.1&       8.3$\pm 0.33$  &       6.4 &       9.0 &       6.41$^{+0.08}_{-0.08}$ &        <34.3 &       0.31&0.002 &0.57&U   \\
N5474    &SA(s)cd pec	 &5.7&H&    29.0& -16.22$\pm 0.39$/-16.62&    -8.95 &   -17.77& 1.1&     8.4$\pm 0.16$&    5.5    &  9.0&&<34.4& 0.29&-0.054&  0.00&U\\
N5548    & (R')SA0a &  77.6 & S     &     291.0 &    -21.45$\pm 0.42$  &    -19.06 &    -22.31 &2.1&     10.8$\pm 0.17$  &       9.8 &      11.1 &       7.72$^{+0.13}_{-0.13}$ &        37.0 &     0.07&-0.005&0.33&I \\
N5557   &E1&46.4&ALG&259 &-22.39$\pm 0.32$ &---&-22.39&3.2&11.4$\pm 0.14$ &---& 11.4&---&<35.9&  0.13&0.000&0.19&U\\
N5585    & SAB(s)d  &   5.6 & H {\sc{ii}}      &      42.0 &    -14.11$\pm 0.33 $/-14.00& -10.96&    -17.19 &0.9&       7.3$\pm 0.13$  &      6.1&       8.6 &       --- &        <34.5 &        0.24&0.007&0.01&U \\
N5631    & SA0$^\wedge$0(s) &  29.3 & L    &     168.1 &    -20.84$\pm 0.19 $ &       ---  &    -20.90 &3.8&     11.4$\pm 0.09$  &    --- &      11.4 &       ---    &     <35.9 &      0.12&0.000 &-0.01&U  \\
N5866    & SA0$^\wedge$$^{+}$edge &  11.8 & L/RL   &     169.1 & -19.58$\pm 0.32 $/-19.76  &       --- & -20.57& 1.1&    10.4$\pm 0.14$  &     ---  &      10.8 &       ---  &       36.5 &       0.33&-0.001 &0.08&I   \\
N5879    & SA(rs)bc &  12.0 & L     &      73.9 &    -18.20$\pm 0.95 $/-18.43 &    -11.54 &    -20.29 &1.4&     9.4$\pm 0.38$  &       6.7 &      10.2 &       6.67$^{+0.28}_{-0.28}$ &       35.3 &       0.39&0.008 &0.40&I \\
N5907\textsuperscript{$\dagger$}      & SA(s)c?e &  10.4 & H {\sc{ii}}      &     120.2 &   --- &       ---  &    -22.12 &1.8&     --- &        ---  &      11.2 &        ---  &      35.0 &        ---&---&---& unI   \\
N5982    &E3     &44.0&ALG&239.4 &-22.08$\pm 0.27$&---&-22.08&4.0&11.4$\pm 0.11$   &---&    11.4 &---& <35.8&0.28&0.003&0.09&U\\
N5985    & SAB(r)b  &  36.7 & L     &     157.6 &    -21.22$\pm 0.34$ &    -14.49 &    -22.70 &2.8&     10.8$\pm 0.14$  &       8.1 &      11.4 &      --- &       35.8 &       0.27&0.013   &0.32&I  \\
N6140    & SB(s)cd &  12.9 & H {\sc{ii}}     &      49.4 &    -17.18$\pm 0.19 $/-18.20 &    ---  &    -18.36 &1.2&       8.8$\pm 0.09$  &       --- &       9.3 &  ---  &       <35.2 &        0.58&0.000&0.00&U      \\
N6207    & SA(s)c   &  12.3 & H {\sc{ii}}      &      92.1 &    -17.36$\pm 0.51$/-17.11  &     -9.87 &    -20.20 &1.1&       8.8$\pm 0.21$  &       5.8 &       9.9 &       ---  &       <35.2 &        0.28&0.009&0.10&U     \\
N6217    & (R)SB(rs &  19.1 & H {\sc{ii}}      &      70.3 &    -18.40$\pm 0.71$  &    -15.37 &    -19.85&1.4&       9.5$\pm 0.29$  &       8.3 &      10.1&       --- &       35.7 &       0.18&-0.005 &0.02&I  \\
N6340    & SA0a(s)  &  16.6 & L     &     143.9 &    -19.78$\pm 0.39$/-19.71  &    -14.02 &   -20.20 &3.0&    10.73$\pm 0.16$  &       8.4 &      10.9 &        --- &     35.5 &       0.05&0.001 &0.60&I  \\
N6412    & SA(s)c   &  18.1 & H {\sc{ii}}   &      49.9 &    -17.81$\pm 0.19$/-18.59 &    -11.25 &    -19.35 & 1.1&      9.0$\pm 0.09$  &       6.4 &       9.6 &      --- &       <35.4 &        0.34&0.003 &0.10&U   \\
N6482    & E?       &  55.7 & L   &     310.4 &    -22.30$\pm 0.19$/-22.23 &   ---  &    -22.53 &1.3&     11.2$\pm 0.09$  &     --- &      11.3 &        --- &        36.4 &       0.17&0.006 &0.07&I  \\
N6503    & SA(s)cd  &   4.9 & L   &      46.0 &    -13.75$\pm 0.37$ &    -10.57 &    -19.20 &1.6&       7.7$\pm 0.15$  &       6.4 &       9.9 &       6.27$^{+0.11}_{-0.11}$ &        <34.3 &        0.31&-0.001 &0.45&U     \\
N6654    & (R')SB0a &  25.0 & ALG   &     172.2 &    -18.61$\pm 0.04$  &    -10.83 &    -20.19 &  3.7&  10.5$\pm 0.05$  &       7.3 &      11.1 &     --- &       <35.6 &        0.12&0.001   &0.38&U \\
N6946    & SAB(rs)c &   5.0 & H {\sc{ii}}    &      55.8 &    -16.38$\pm 0.31$/-16.12  &      -14.64&    -18.79& 0.9&      8.3$\pm 0.13$  &      7.6 &       9.3 & --- &       34.4 &       0.32&-0.019  &0.12&I   \\
N6951    & SAB(rs)b &  18.2 & L    &     127.8 &    -20.25$\pm 0.55$  &    -14.34 &    -21.51 &1.6&      10.3$\pm 0.22$  &    8.0&      10.8 &       6.99$^{+0.20}_{-0.20}$ &       35.4 &       0.26&0.001 &0.27&I \\
N7217    & (R)SA(r)ab &   9.0 & L     &     141.4 &    -19.55$\pm 0.20$/-19.61 &  -14.15 &    -19.86 & 2.9&     10.6$\pm 0.09$  &   8.5&      10.7 &  --- &      35.1 &       0.08&0.001  &0.22&I   \\
N7331    & SA(s)b   &   7.0 & L    &     137.2 &    -17.62$\pm 0.47$/-18.65  &     -12.30 &    -19.29 &4.5&      10.2$\pm 0.19$  &     8.1 &      10.9 &       7.78$^{+0.18}_{-0.18}$ &      <35.0 &        0.34&0.020  &0.44&U  \\
N7457    & SA0$^\wedge$$^{-}$(rs) &   7.2 & ALG   &      69.4 &    -16.84$\pm 0.72$/-17.15  &    -11.72&    -18.11 & 3.1&     9.6$\pm 0.30$  &       7.5 &      10.1 &       6.71$^{+0.30}_{-0.30}$ &        <34.9 &        0.24&0.003 &0.48&U    \\
N7640    & SB(s)c   &   0.9 &H {\sc{ii}}     &      48.1 &    -14.45$\pm 0.33$/-14.38 &    ---  &    -16.09 & 0.7&    7.2$\pm 0.14$  &        ---  &       7.9 &       --- &       <34.5 &        0.52&0.011 &0.01&U   \\
N7741    & SB(s)cd  &   5.8 & H {\sc{ii}}     &      29.4 &    -15.33$\pm 0.37$/-15.25  &       --- &    -16.00 & 0.9&      7.9$\pm 0.15$  &      ---  &       8.1 &        ---  &       <34.7 &        0.77&-0.005 &0.00&U  \\
\hline
error   & ---&   --- &  ---   &    ---   &   ---  & 0.35&   0.33& 25\%&   ---&   0.17 dex    &0.17 dex &  ---&    1 dex & 20\%&  35\%&---& ---& \\
\hline
\end{tabular} 
\end {minipage}
\end{sideways}
\end{table*}
\end{center}

\clearpage
\newpage
\section{{\it HST} images}\label{AppendB}

For each galaxy newly studied in this paper, we show a  $15\arcsec \times 15\arcsec$ {\it HST} image (Fig.~\ref{Figcutout}). 

\setcounter{figure}{0}
\renewcommand{\thefigure}{A\arabic{figure}}

\begin{figure*}
\vspace{-.275030cm}
 \hspace*{4.0cm}
\includegraphics[trim= {11cm 0cm 0cm -.29cm},angle=0,scale=0.26804]{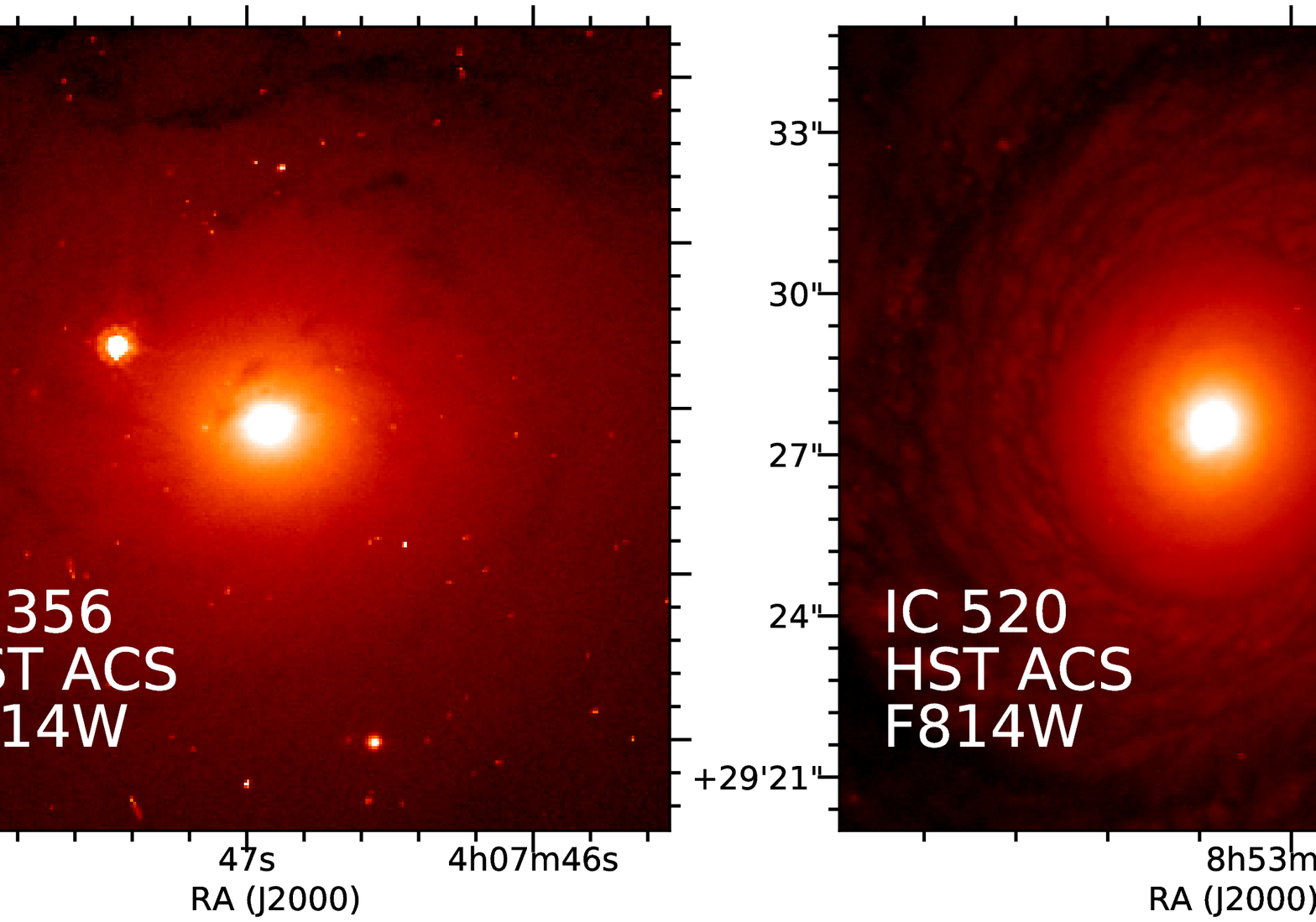}\\
\hspace*{4.0cm}
\includegraphics[trim= {11cm 0cm 0cm 0cm},angle=0,scale=0.26804]{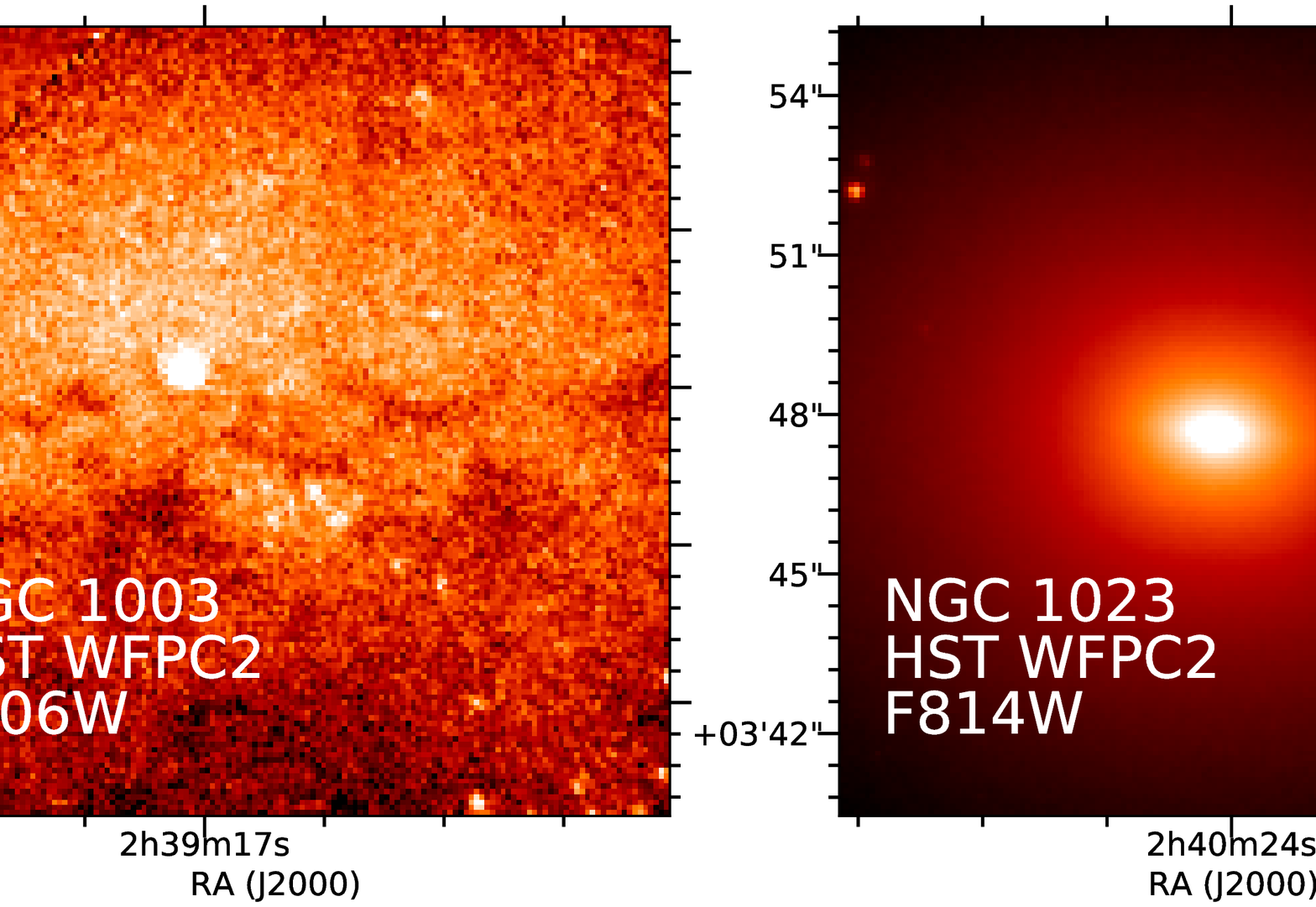}\\
\hspace*{4.0cm}
\includegraphics[trim= {11cm 0cm 0cm 0cm},angle=0,scale=0.26804]{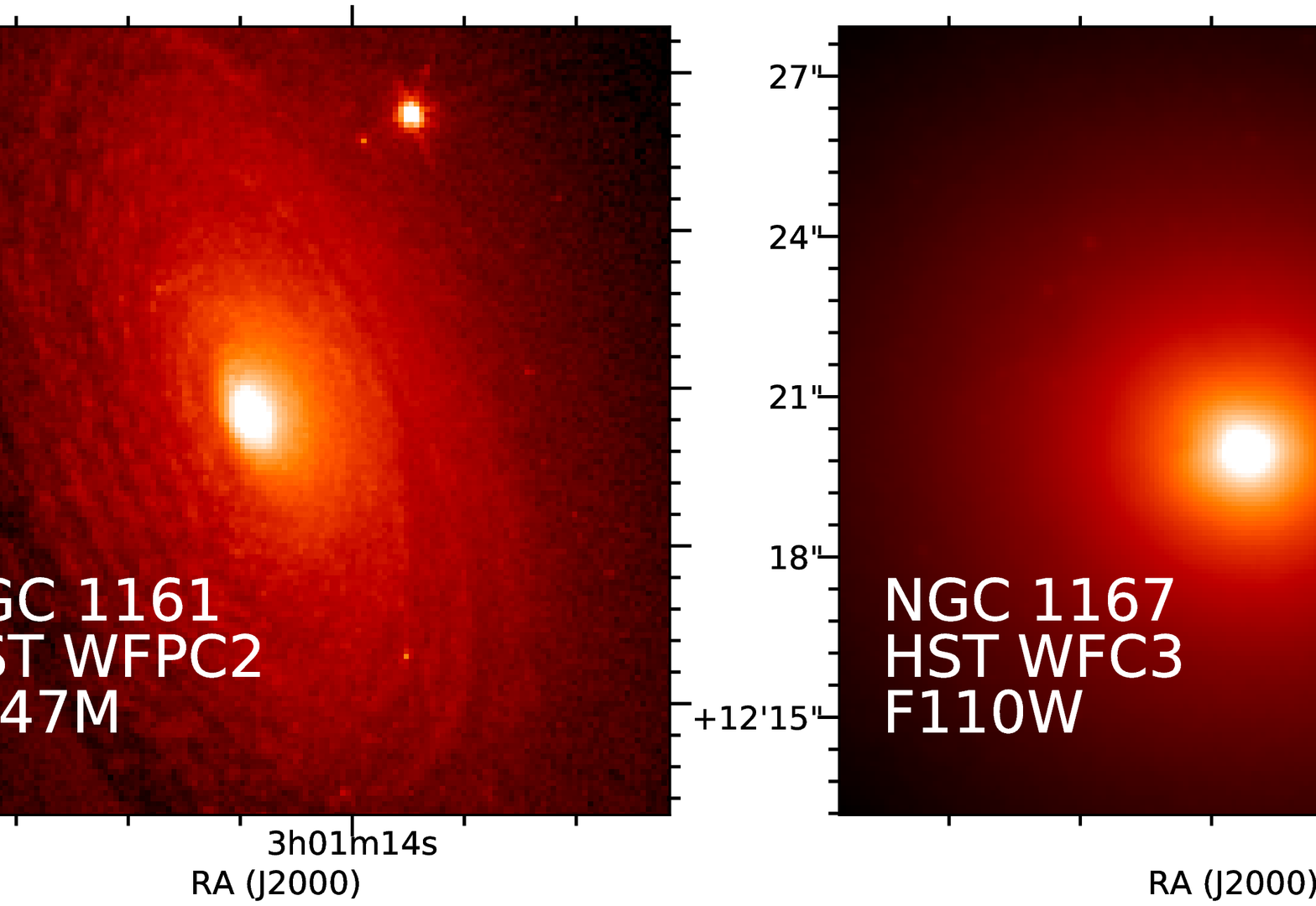}\\
\hspace*{4.0cm}
\includegraphics[trim= {11cm 0cm 0cm  0cm},angle=0,scale=0.26804]{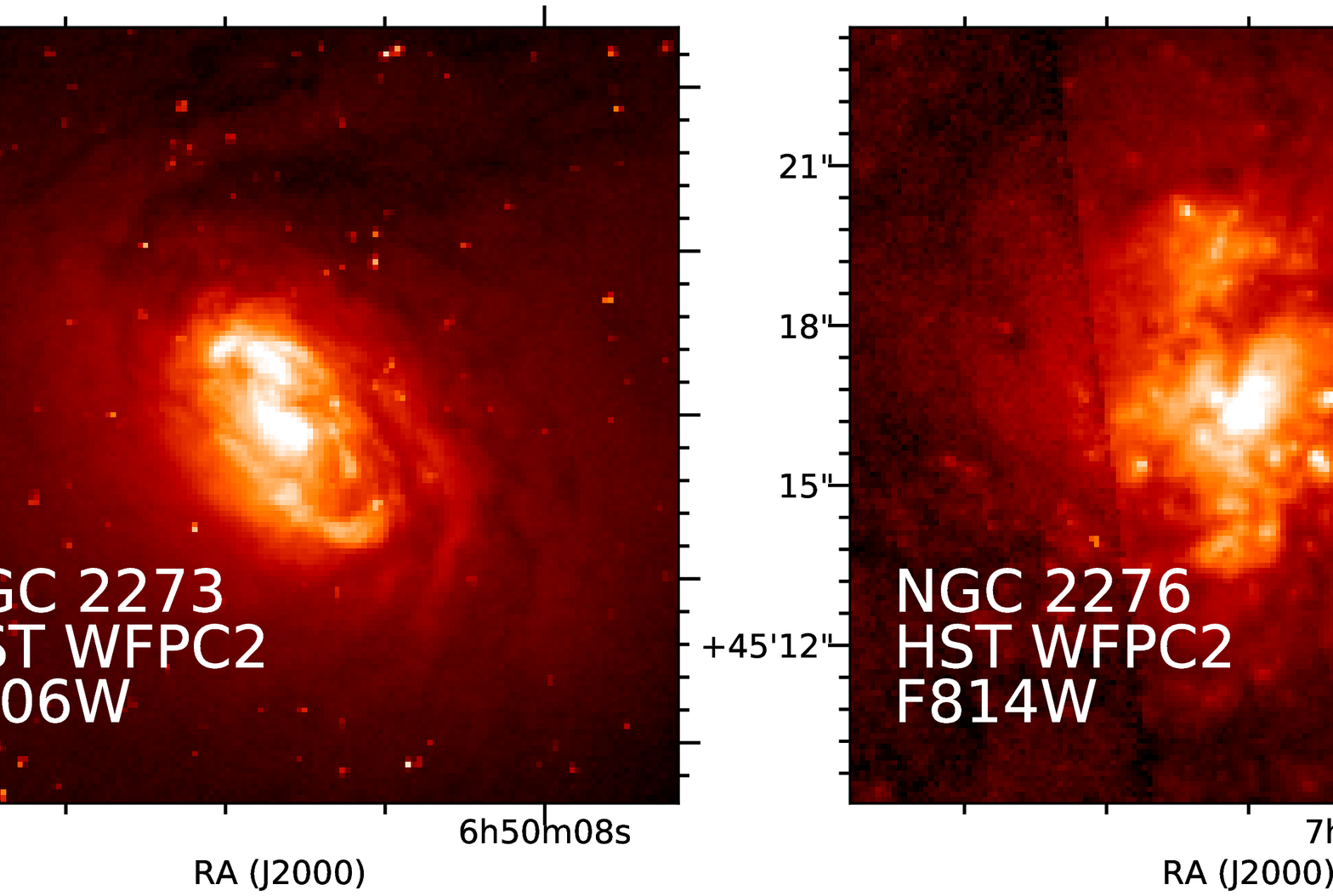}\\
\hspace*{4.0cm}
\includegraphics[trim= {11cm 0cm 0cm  0cm},angle=0,scale=0.26804]{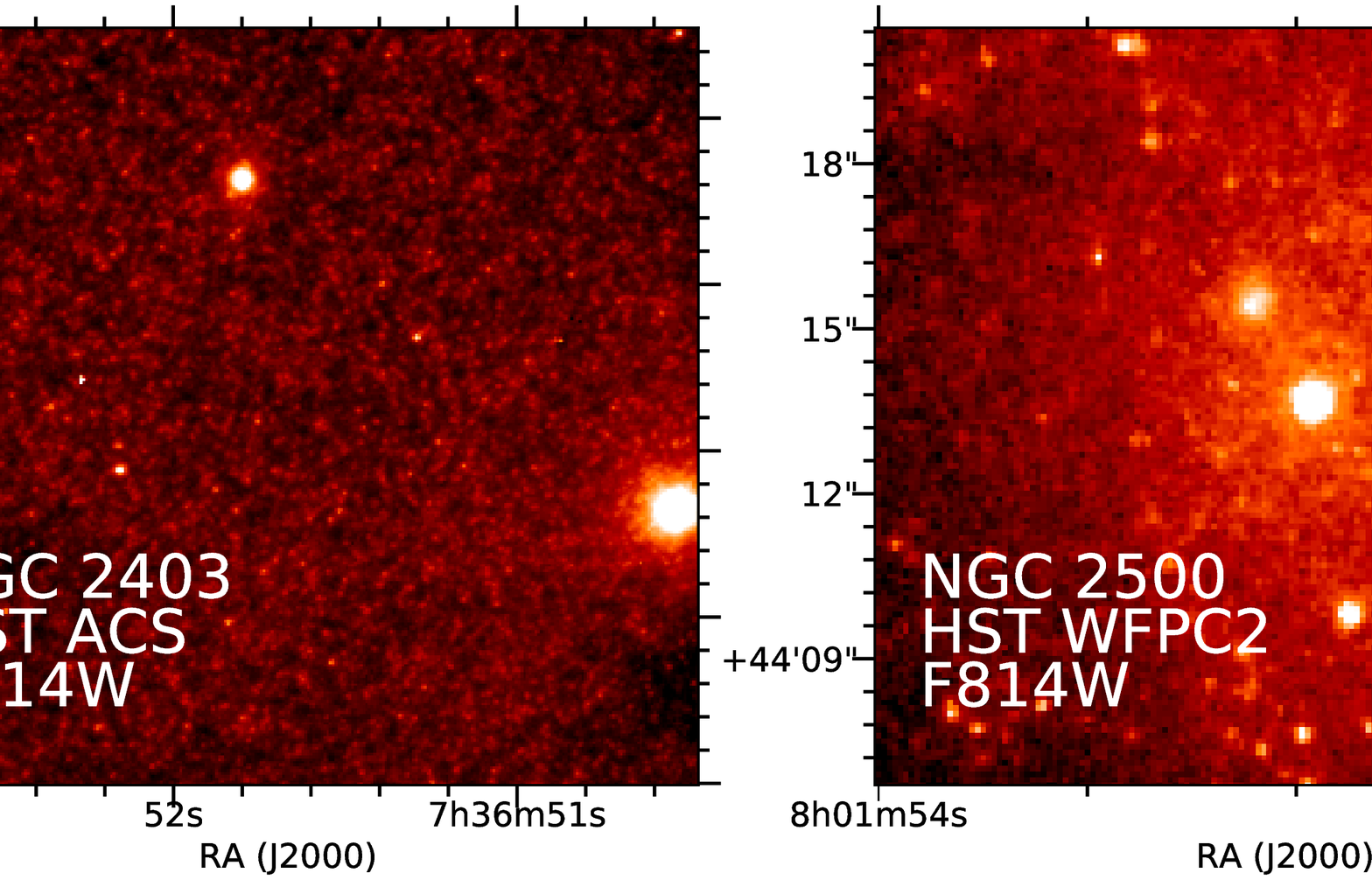}\\
\hspace*{4.0cm}
\includegraphics[trim= {11cm 0cm 0cm 0cm},angle=0,scale=0.26804]{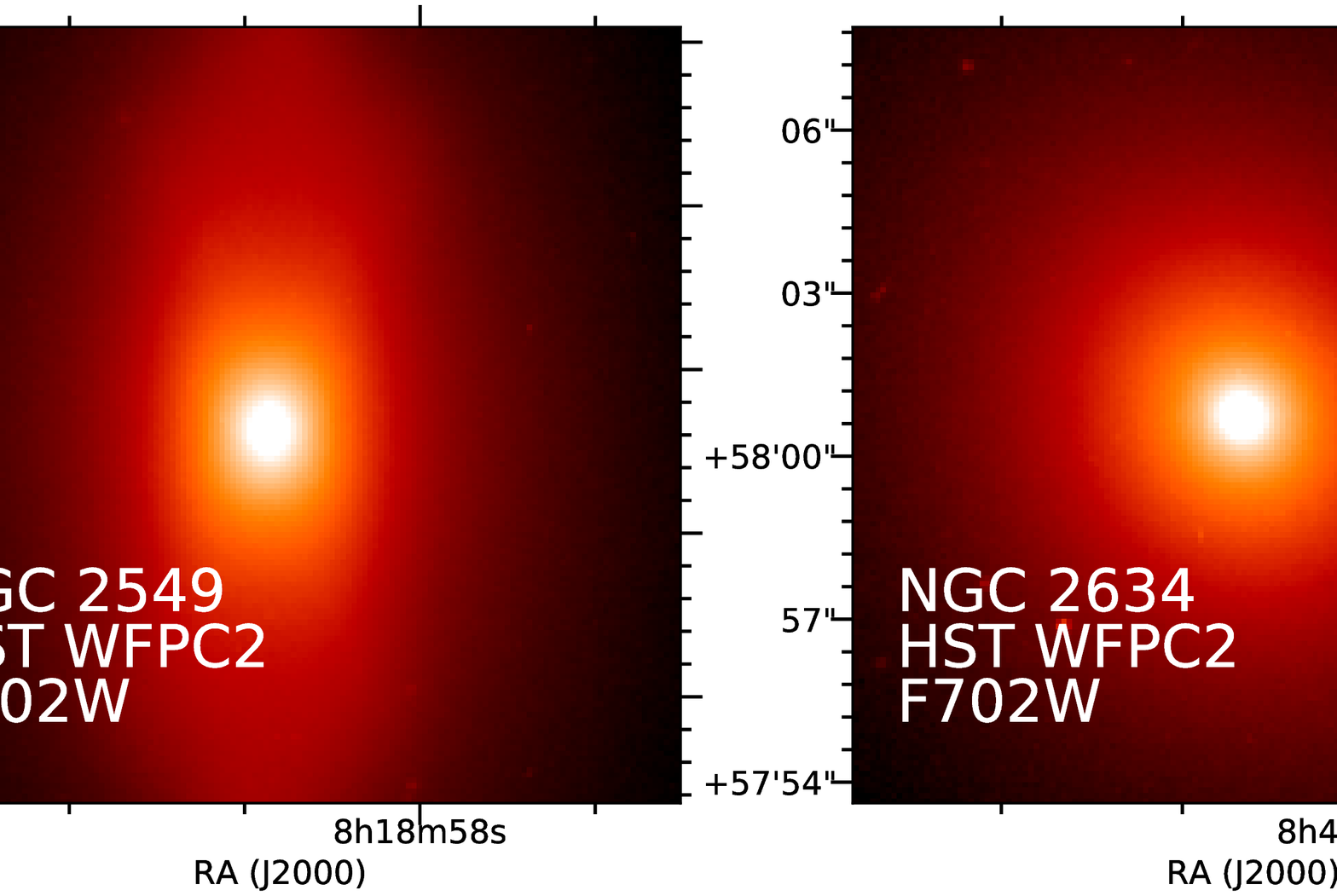}
\vspace*{-.2630cm}
\caption{ {\it HST} images for our sample of 163 newly  analysed LeMMINGs galaxies
  (Table~\ref{NewTabA1}) showing their central
  $15\arcsec \times 15\arcsec$ regions. North is up and east is to the 
  left.}
\label{Figcutout}
\end{figure*}

\begin{figure*}
\setcounter{figure}{0} 
\vspace{-.275030cm}
\hspace*{4.0cm}
\includegraphics[trim= {11cm 0cm 0cm -0.29cm},angle=0,scale=0.2604]{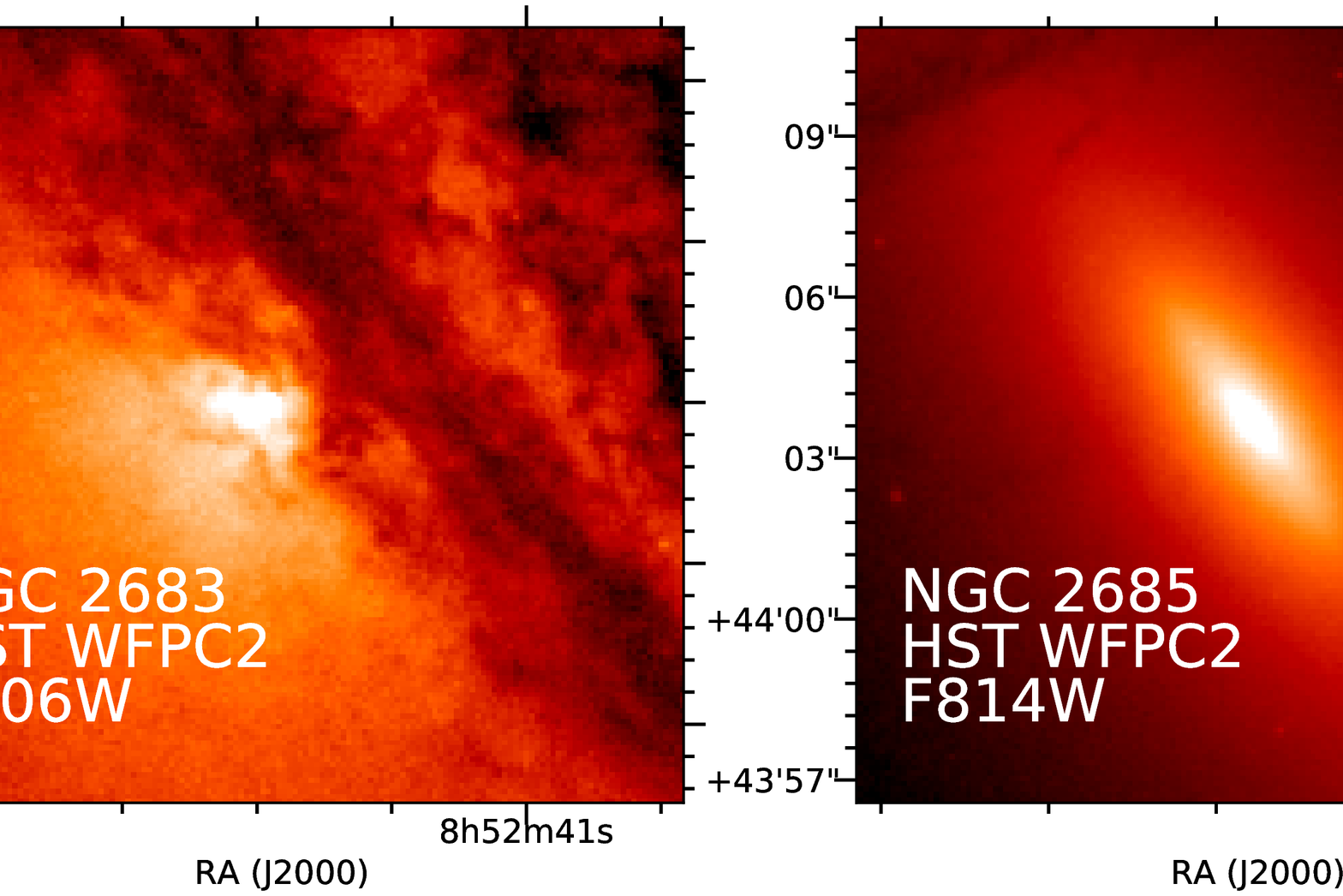}\\
\hspace*{4.0cm}
\includegraphics[trim= {11cm 0cm 0cm 0cm},angle=0,scale=0.2604]{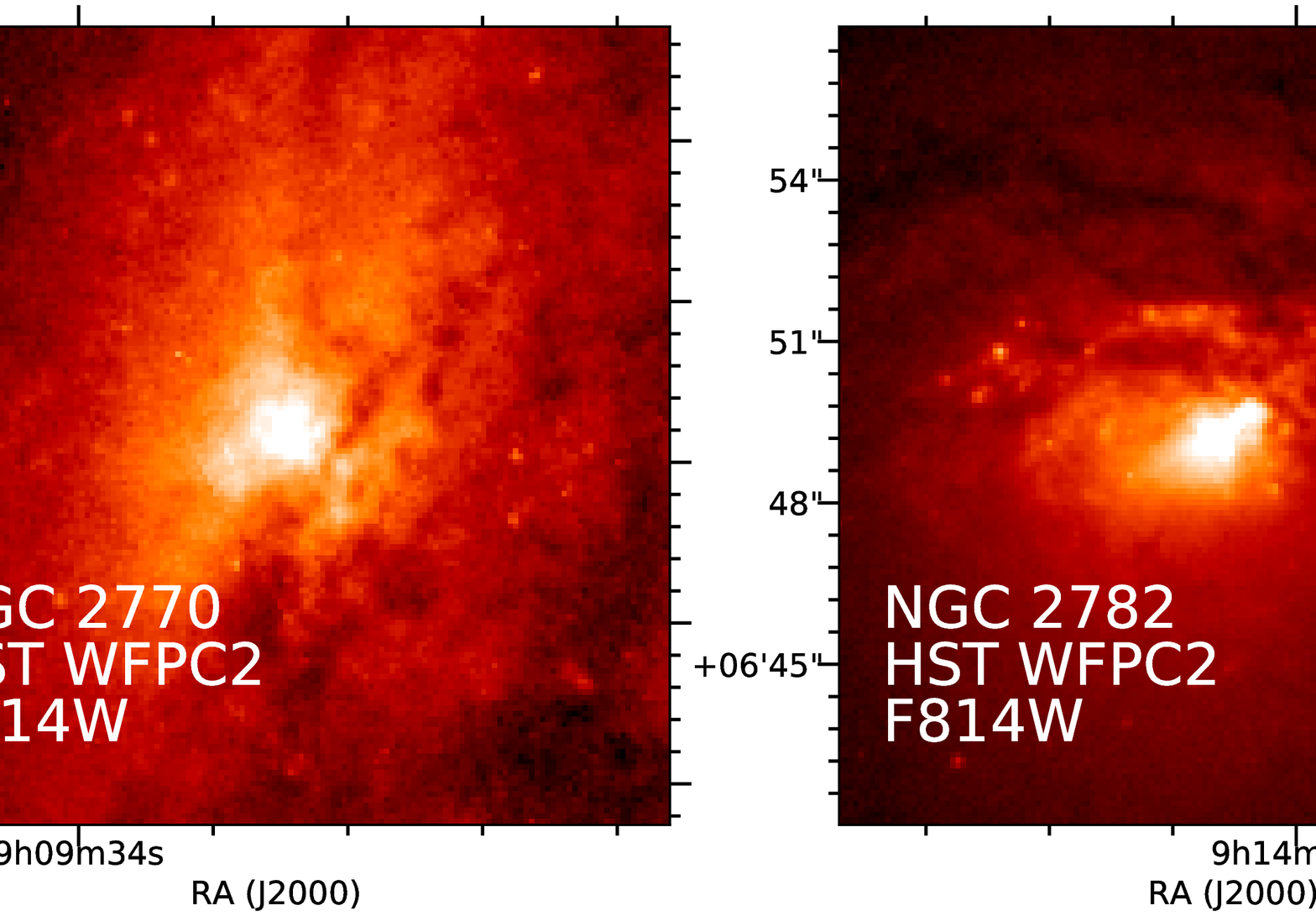}\\
\hspace*{4.0cm}
\includegraphics[trim= {11cm 0cm 0cm 0cm},angle=0,scale=0.2604]{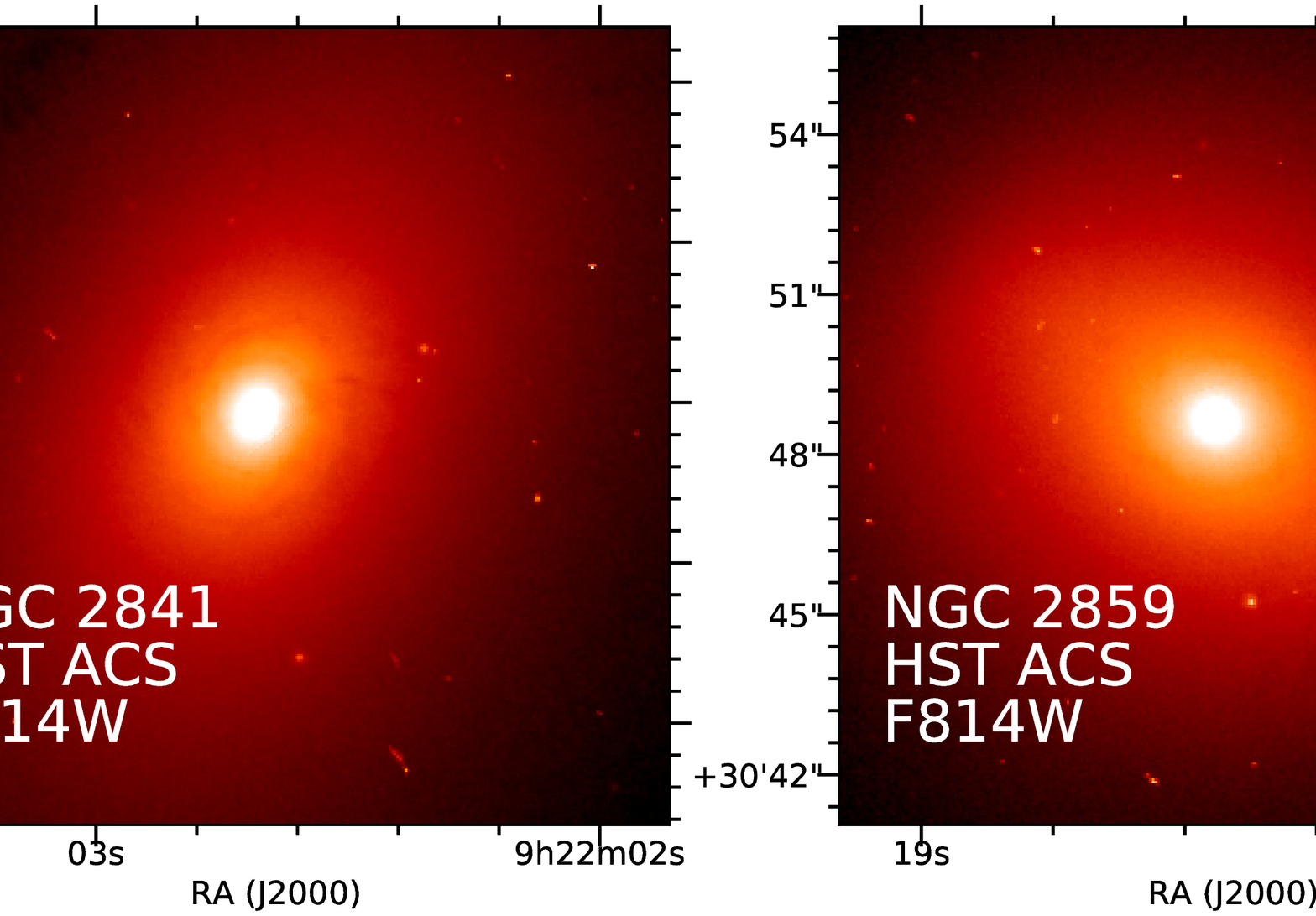}\\
\hspace*{4.0cm}
\includegraphics[trim= {11cm 0cm 0cm 0cm},angle=0,scale=0.2604]{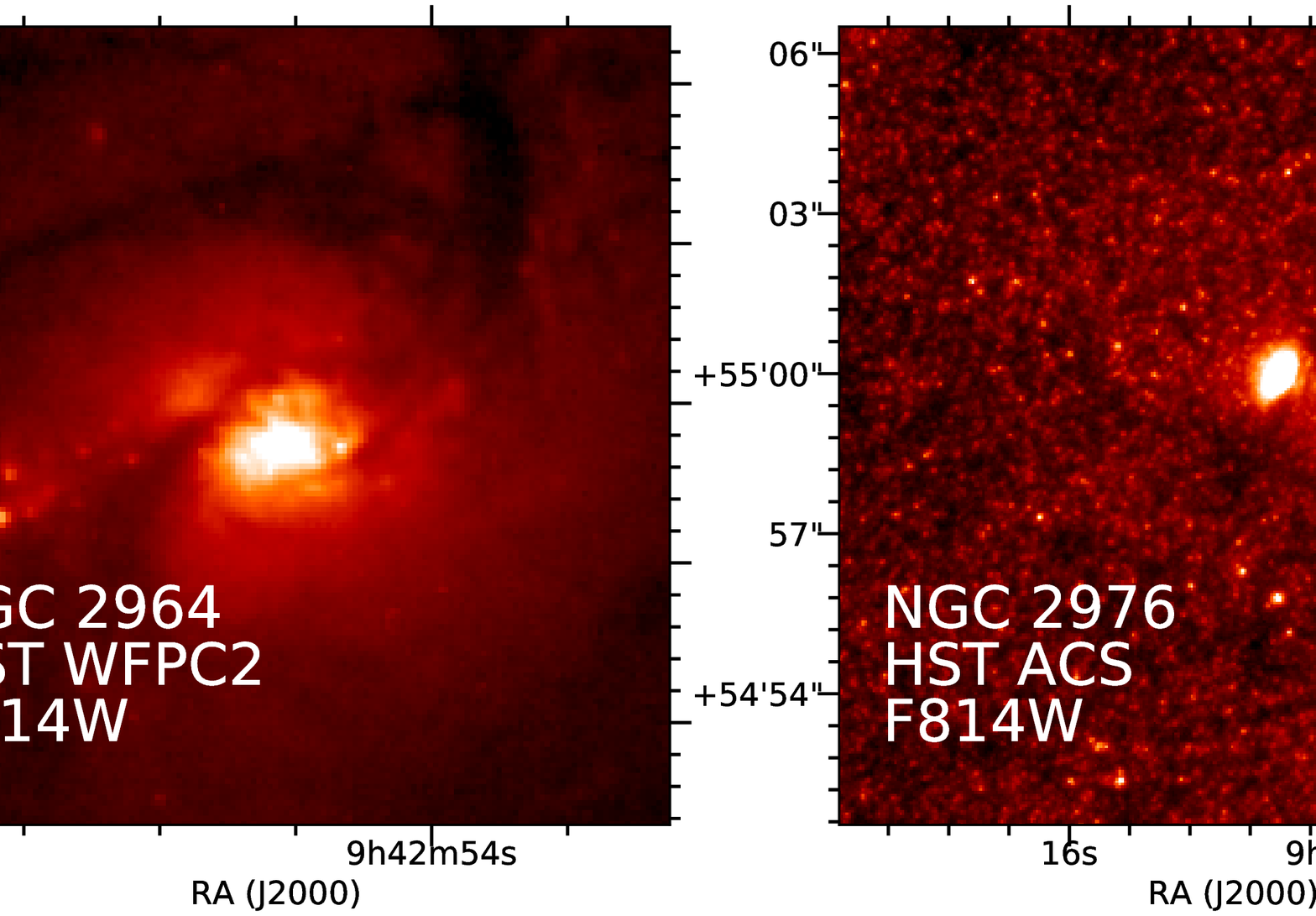}\\
\hspace*{4.0cm}
\includegraphics[trim= {11cm 0cm 0cm 0cm},angle=0,scale=0.2604]{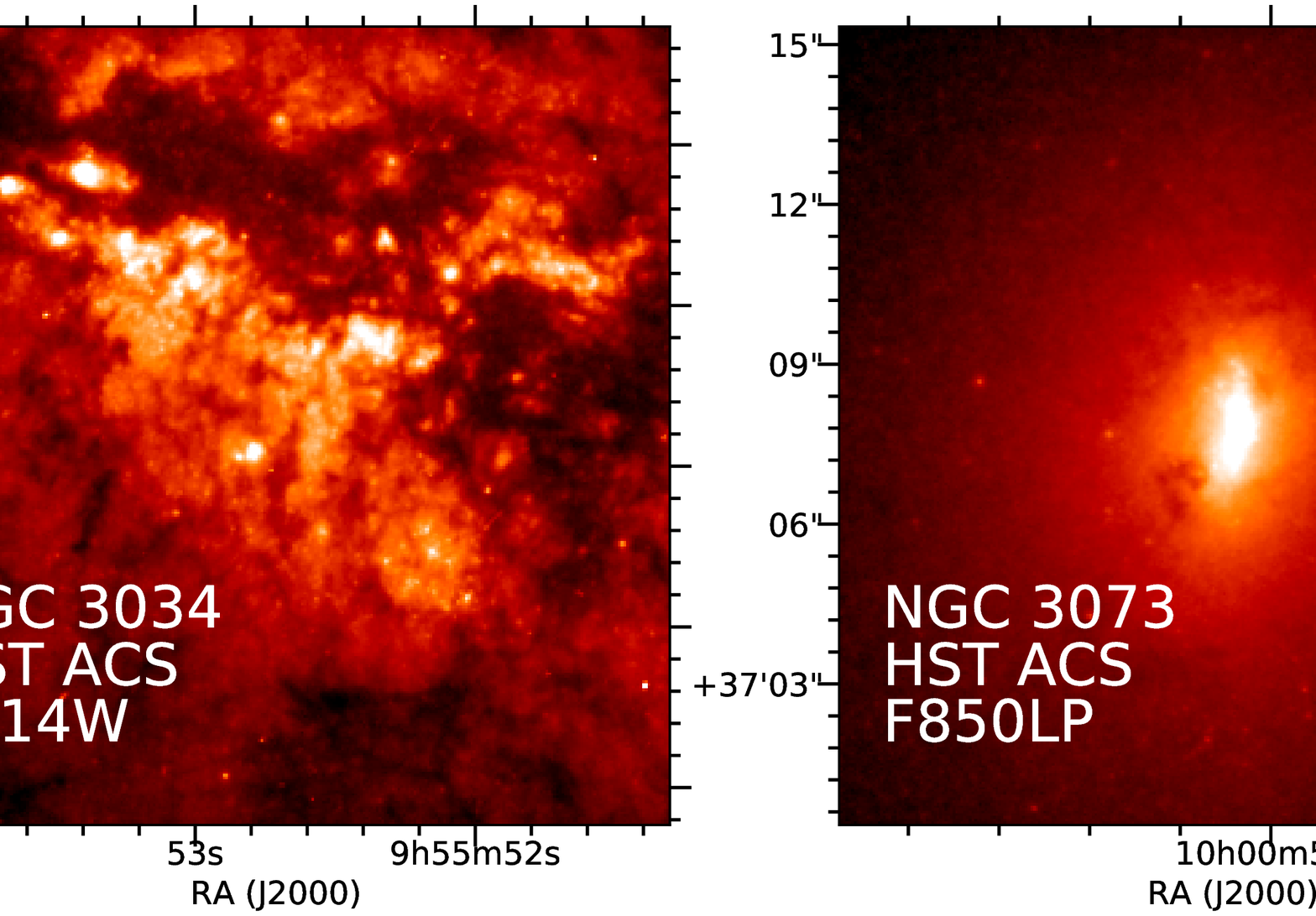}\\
\hspace*{4.0cm}
\includegraphics[trim= {11cm 0cm 0cm 0cm},angle=0,scale=0.2604]{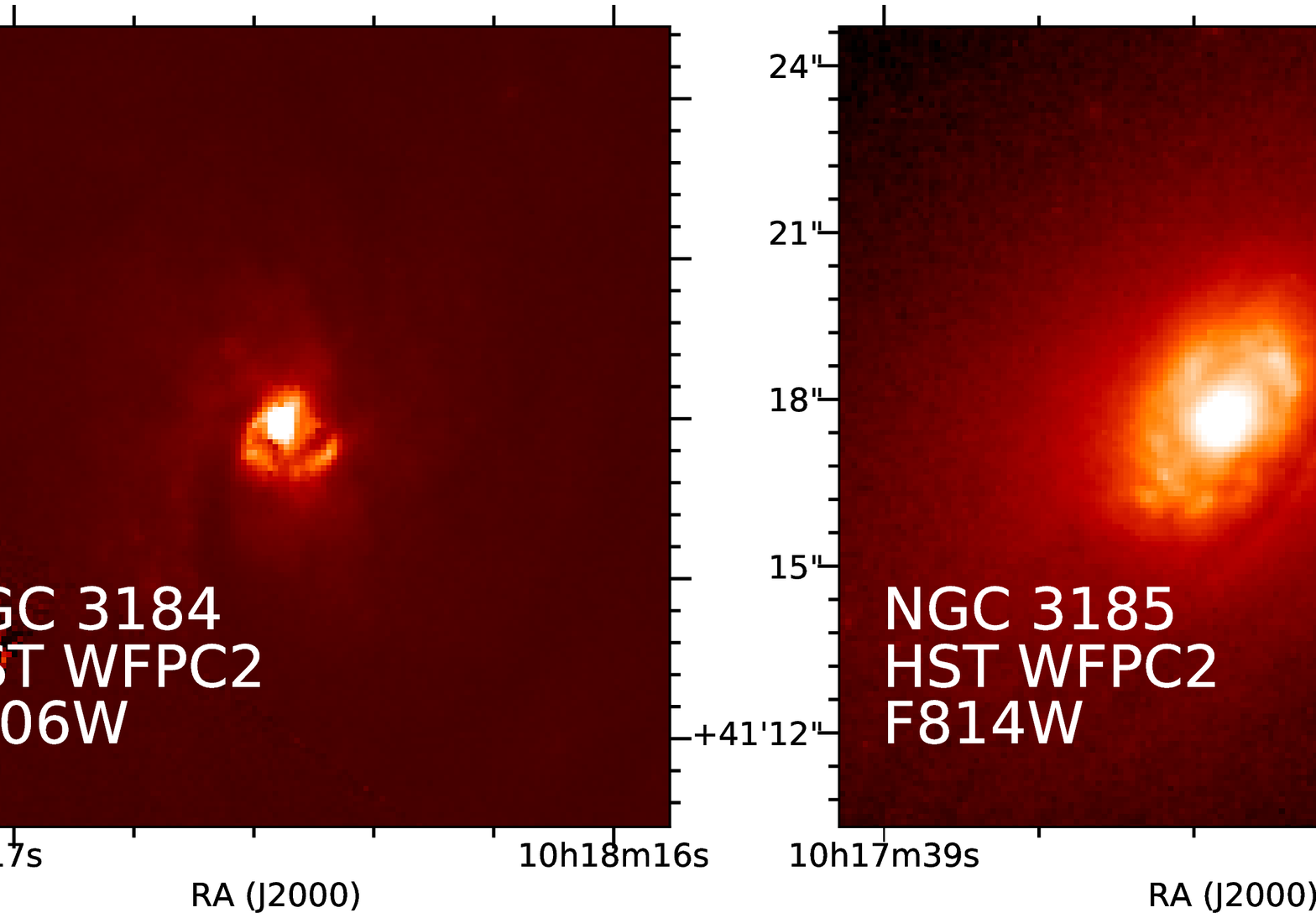}
\vspace{-.2630cm}
\caption{\it continued.}
\end{figure*}

\begin{figure*}
\setcounter{figure}{0} 
\vspace{-.275030cm}
\hspace*{4.0cm}
\includegraphics[trim= {11cm 0cm 0cm -0.29cm},angle=0,scale=0.24804]{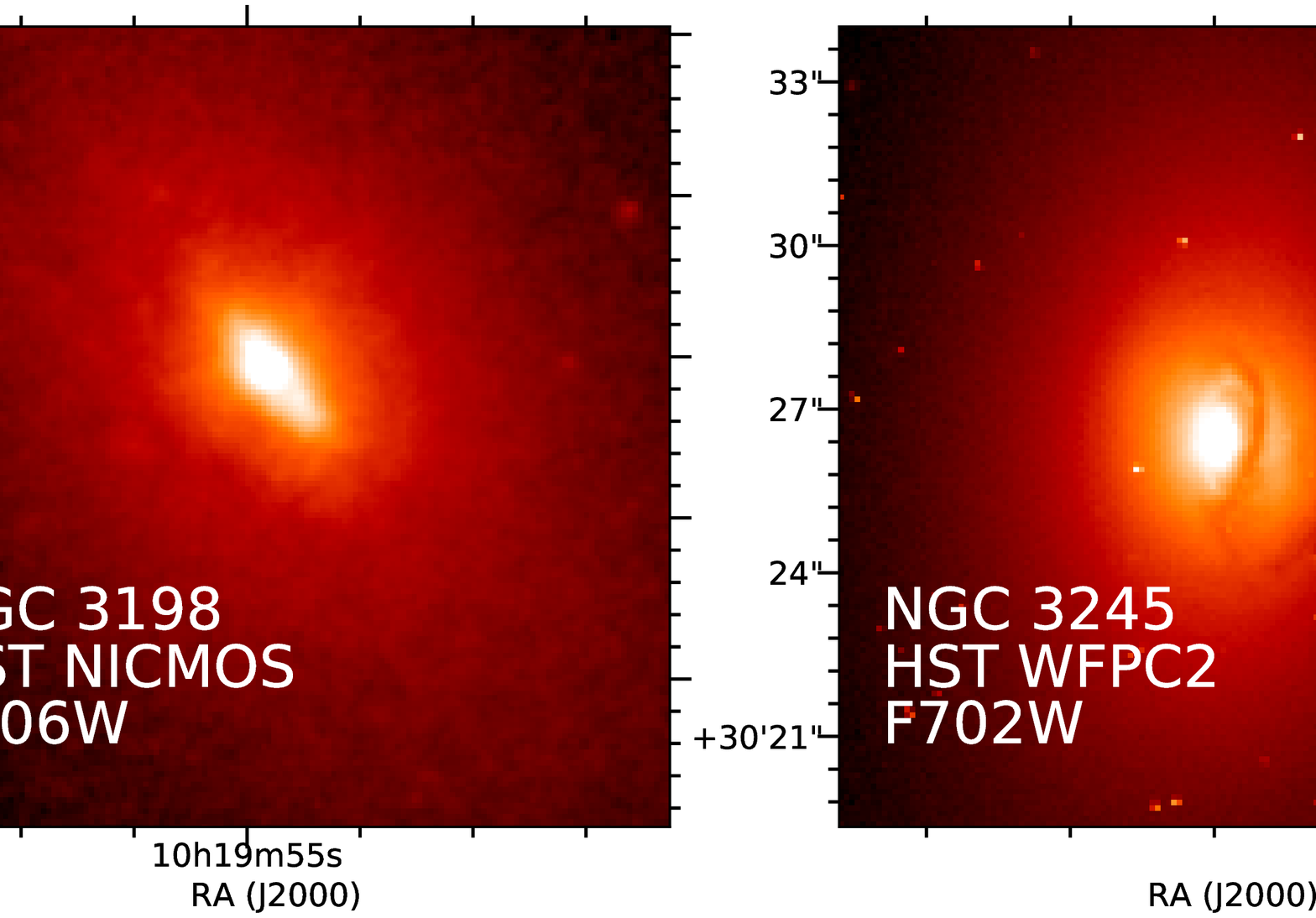}\\
\hspace*{4.0cm}
\includegraphics[trim= {11cm 0cm 0cm 0cm},angle=0,scale=0.24804]{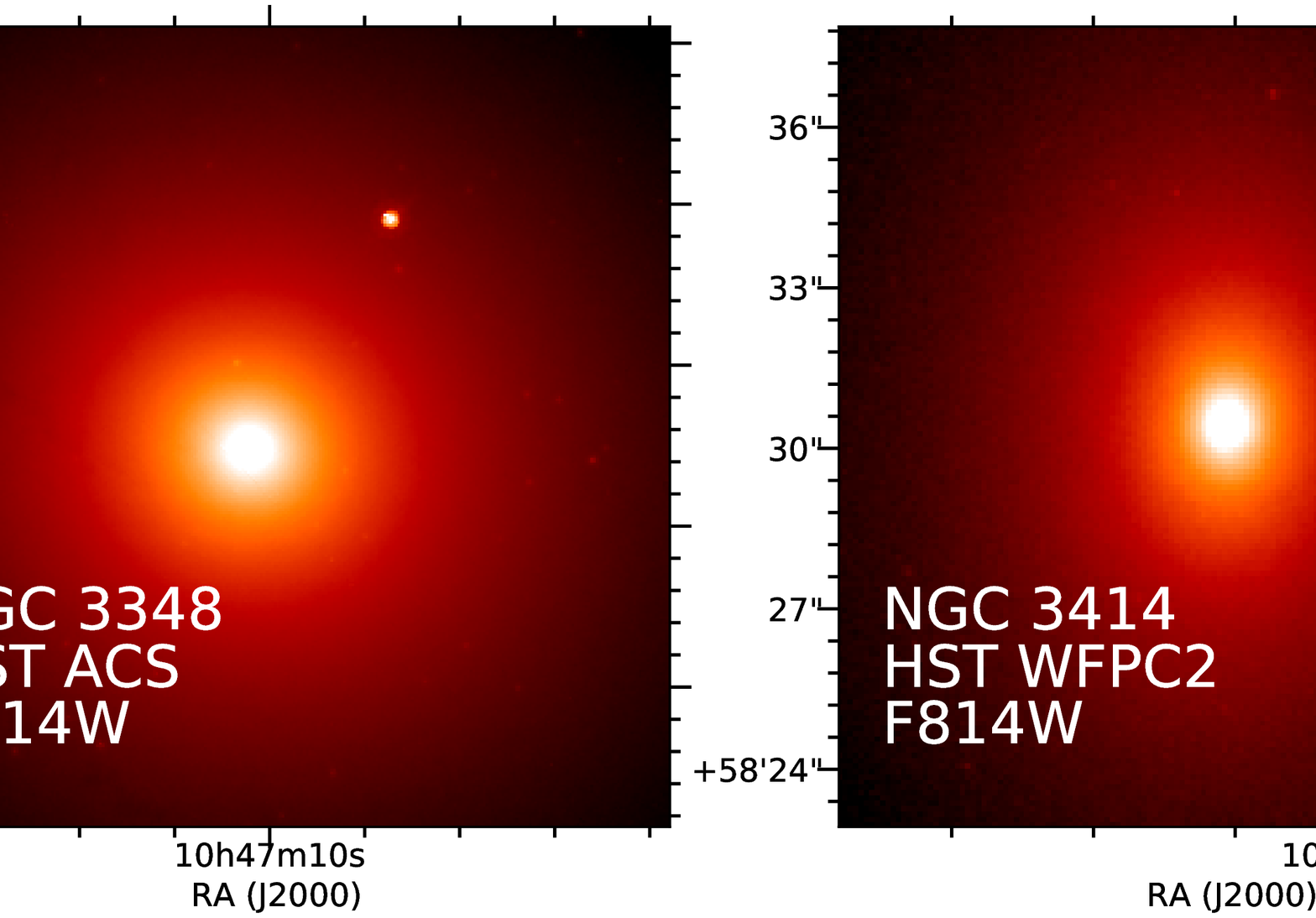}\\
\hspace*{4.0cm}
\includegraphics[trim= {11cm 0cm 0cm 0cm},angle=0,scale=0.24804]{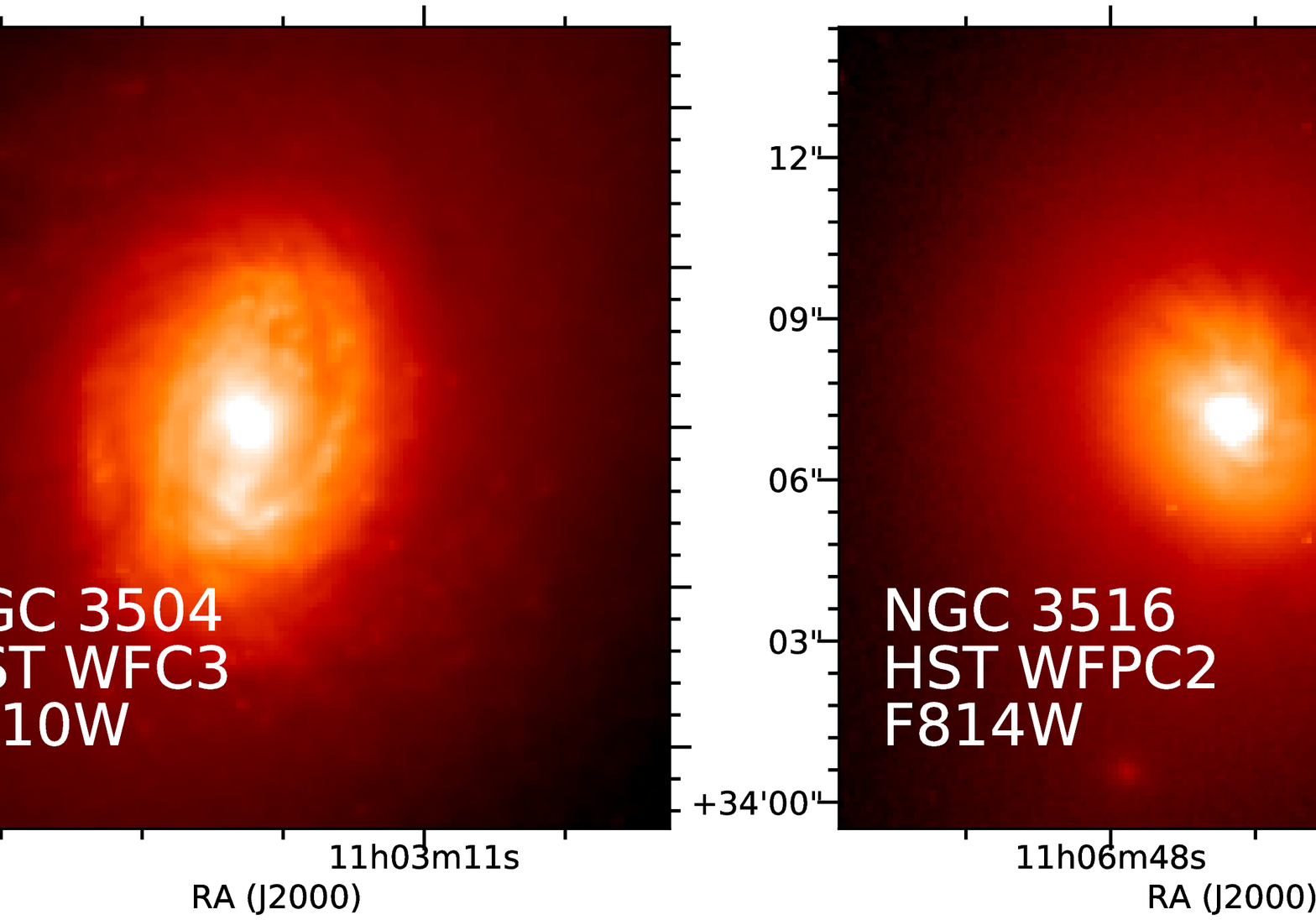}\\
\hspace*{4.0cm}
\includegraphics[trim= {11cm 0cm 0cm 0cm},angle=0,scale=0.24804]{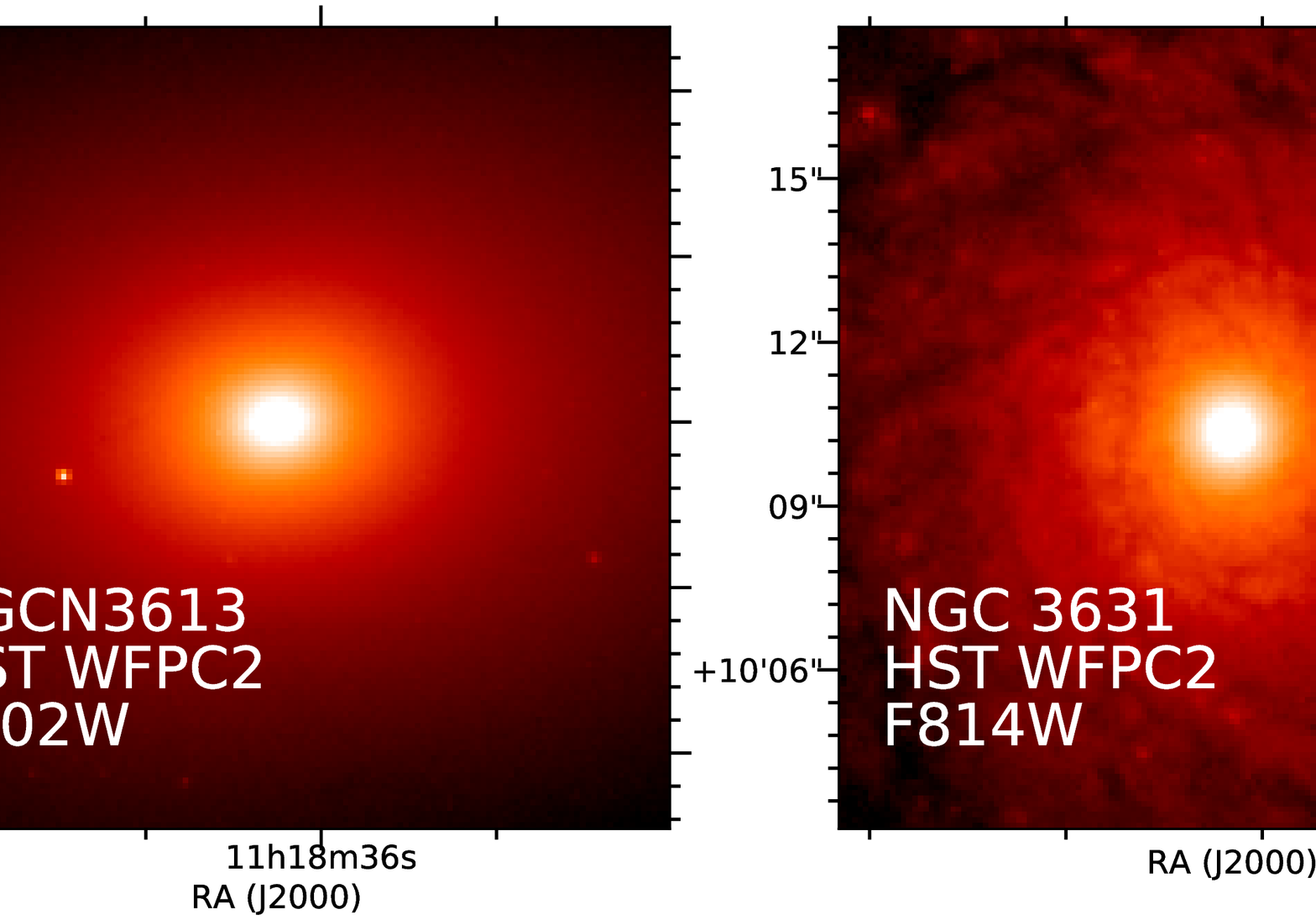}\\
\hspace*{4.0cm}
\includegraphics[trim= {11cm 0cm 0cm 0cm},angle=0,scale=0.24804]{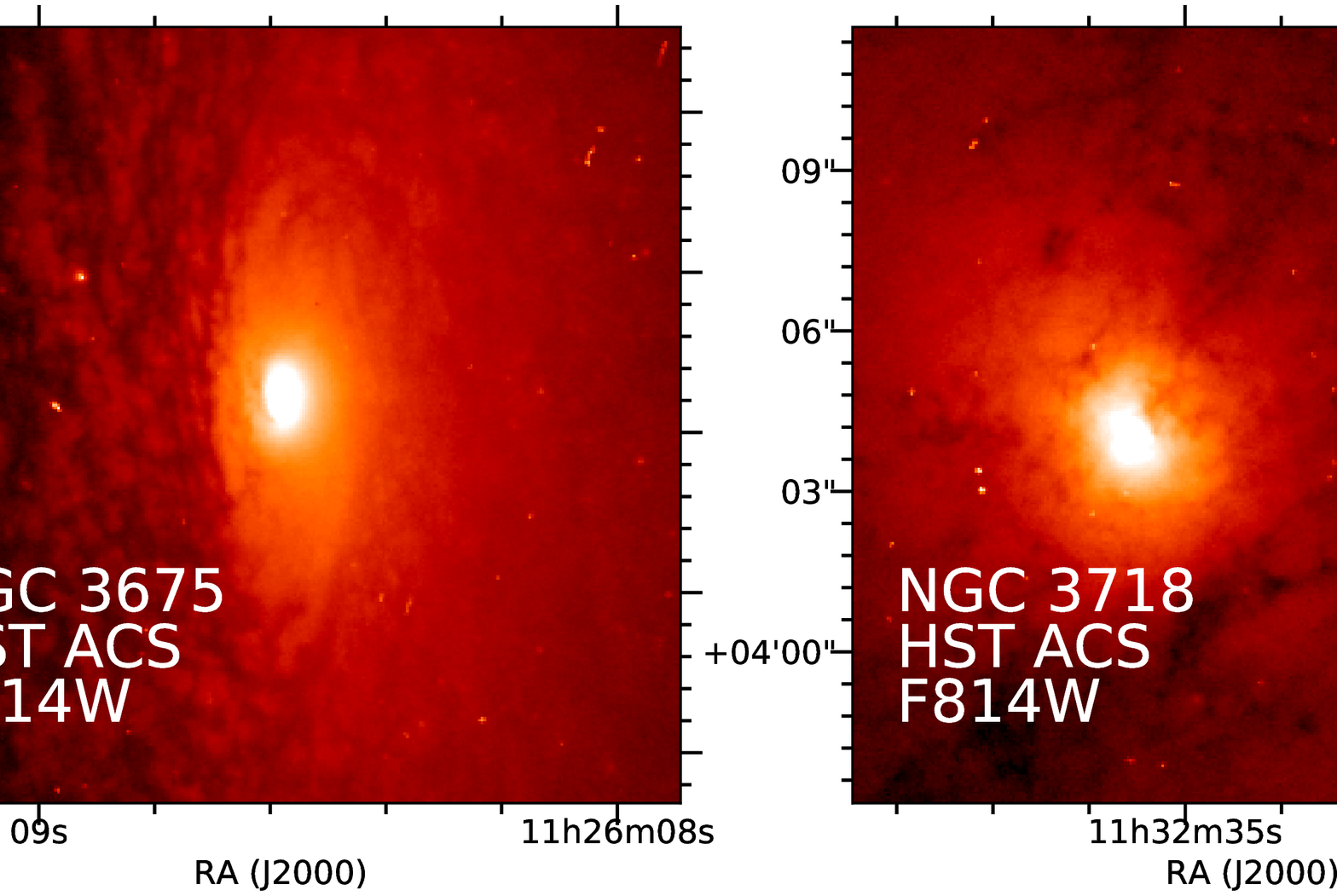}\\
\hspace*{4.0cm}
\includegraphics[trim= {11cm 0cm 0cm 0cm},angle=0,scale=0.24804]{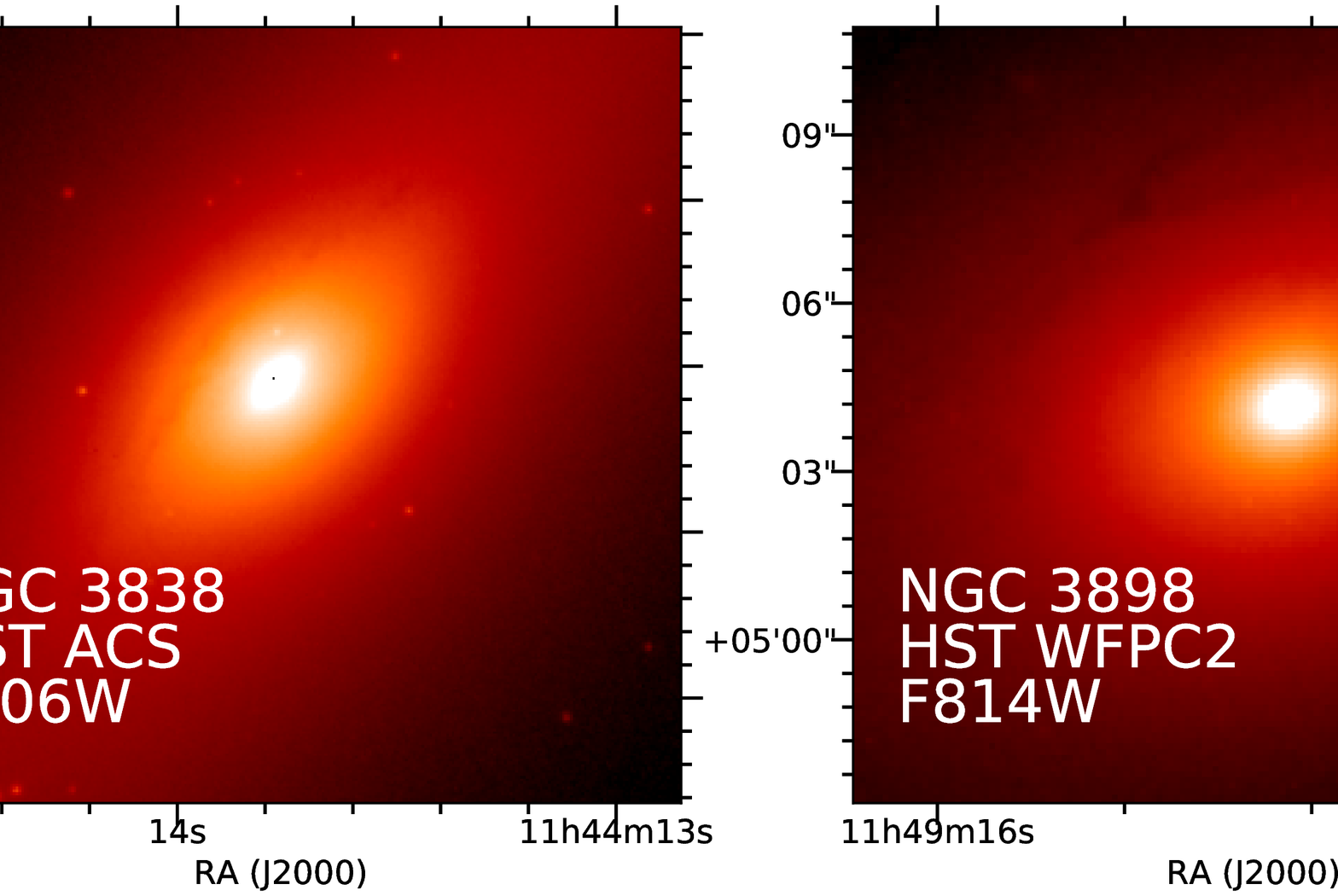}
\vspace{-.2630cm}
\caption{\it continued.}
\end{figure*}
\begin{figure*}
\setcounter{figure}{0} 
\vspace{-.275030cm}
\hspace*{4.0cm}
\includegraphics[trim= {11cm 0cm 0cm -0.29cm},angle=0,scale=0.24804]{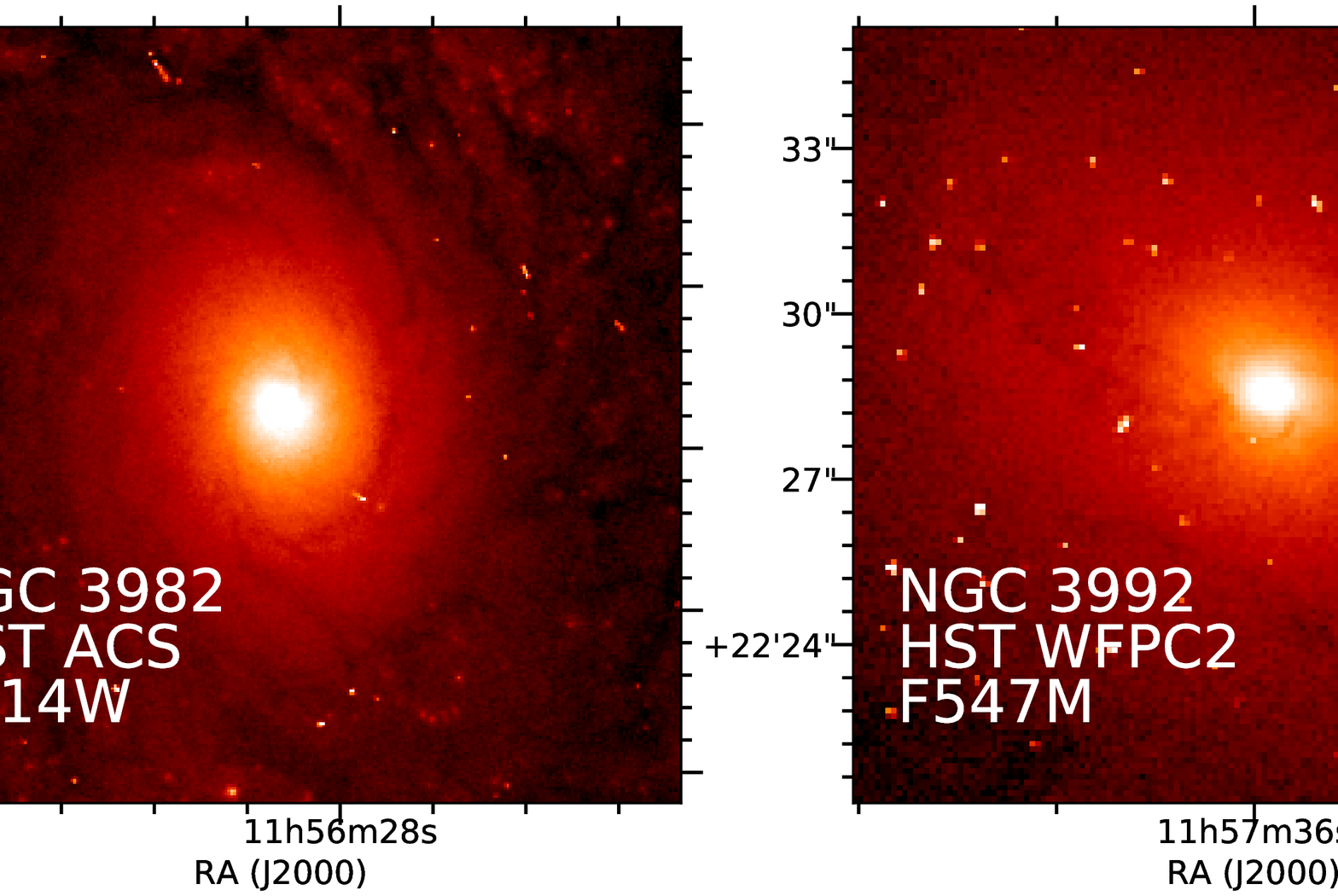}\\
\hspace*{4.0cm}
\includegraphics[trim= {11cm 0cm 0cm 0cm},angle=0,scale=0.24804]{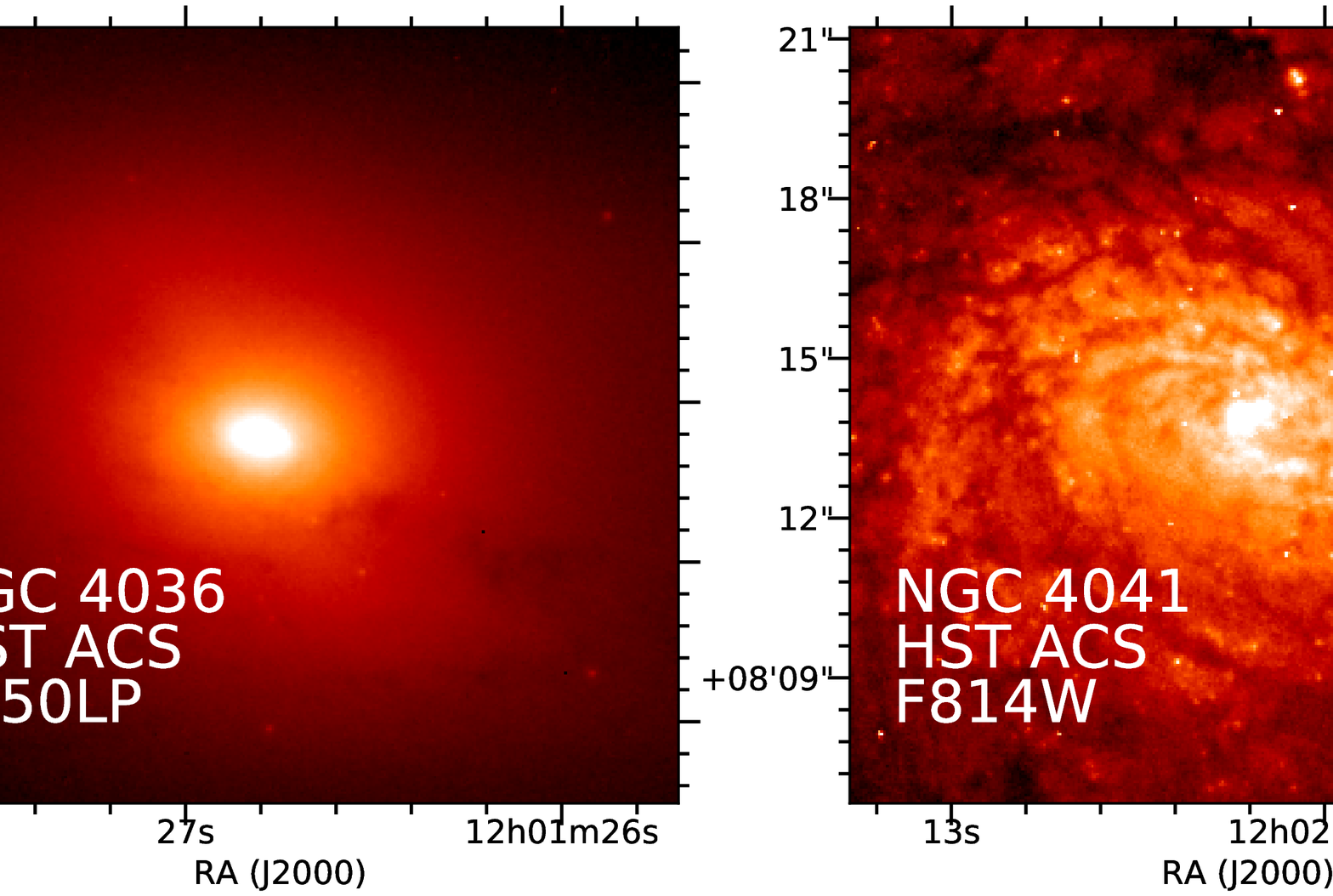}\\
\hspace*{4.0cm}
\includegraphics[trim= {11cm 0cm 0cm 0cm},angle=0,scale=0.24804]{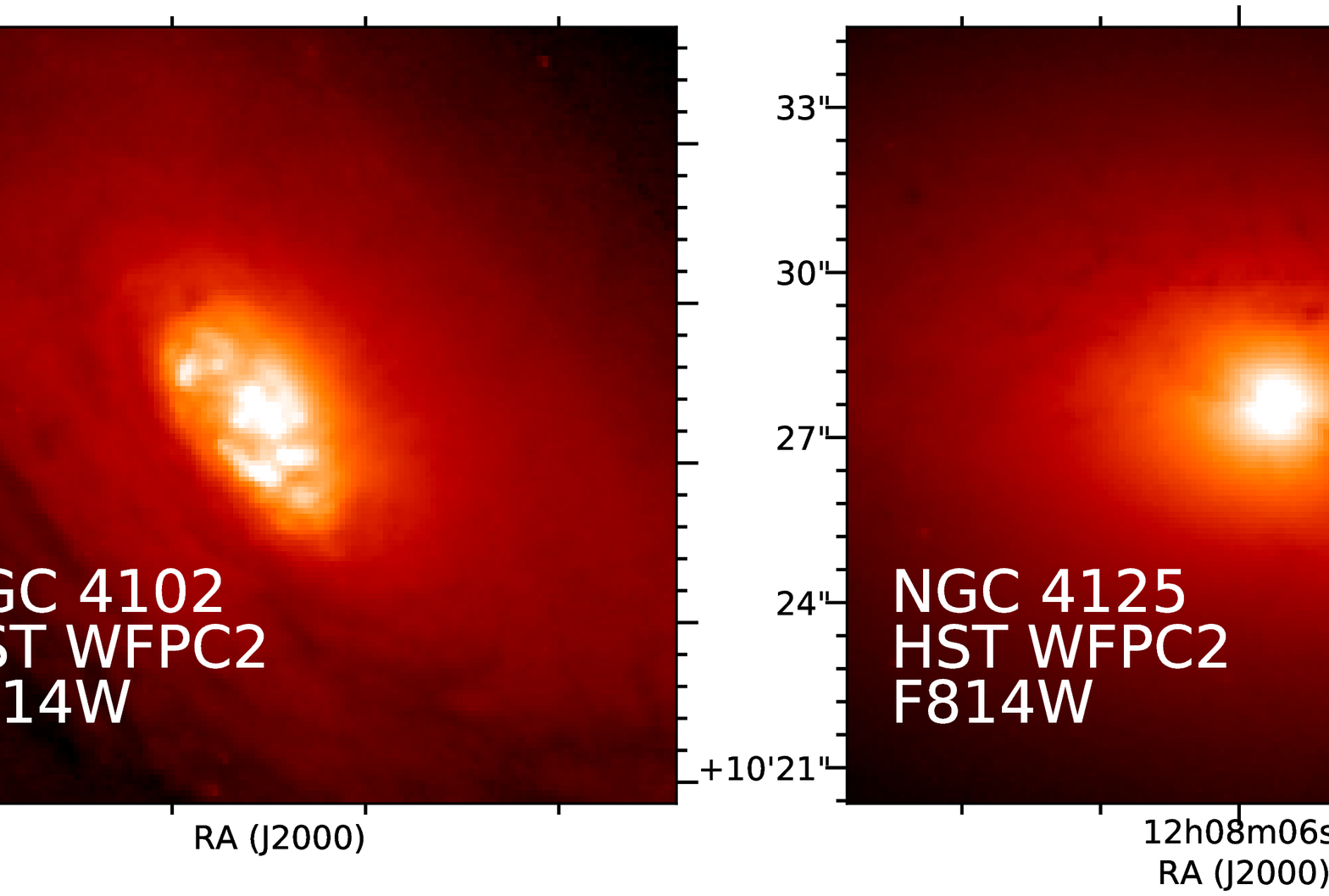}\\
\hspace*{4.0cm}
\includegraphics[trim= {11cm 0cm 0cm 0cm},angle=0,scale=0.24804]{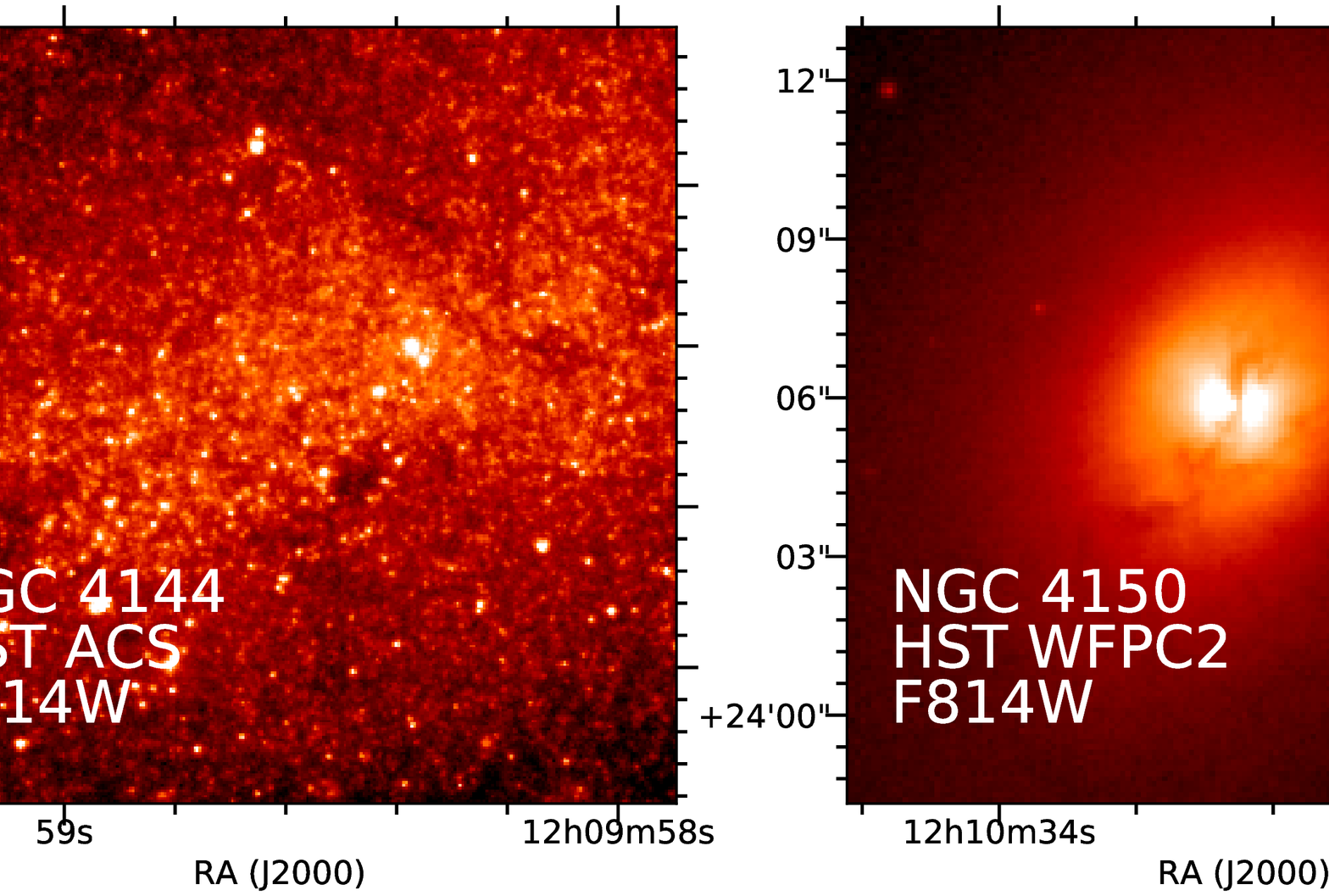}\\
\hspace*{4.0cm}
\includegraphics[trim= {11cm 0cm 0cm 0cm},angle=0,scale=0.24804]{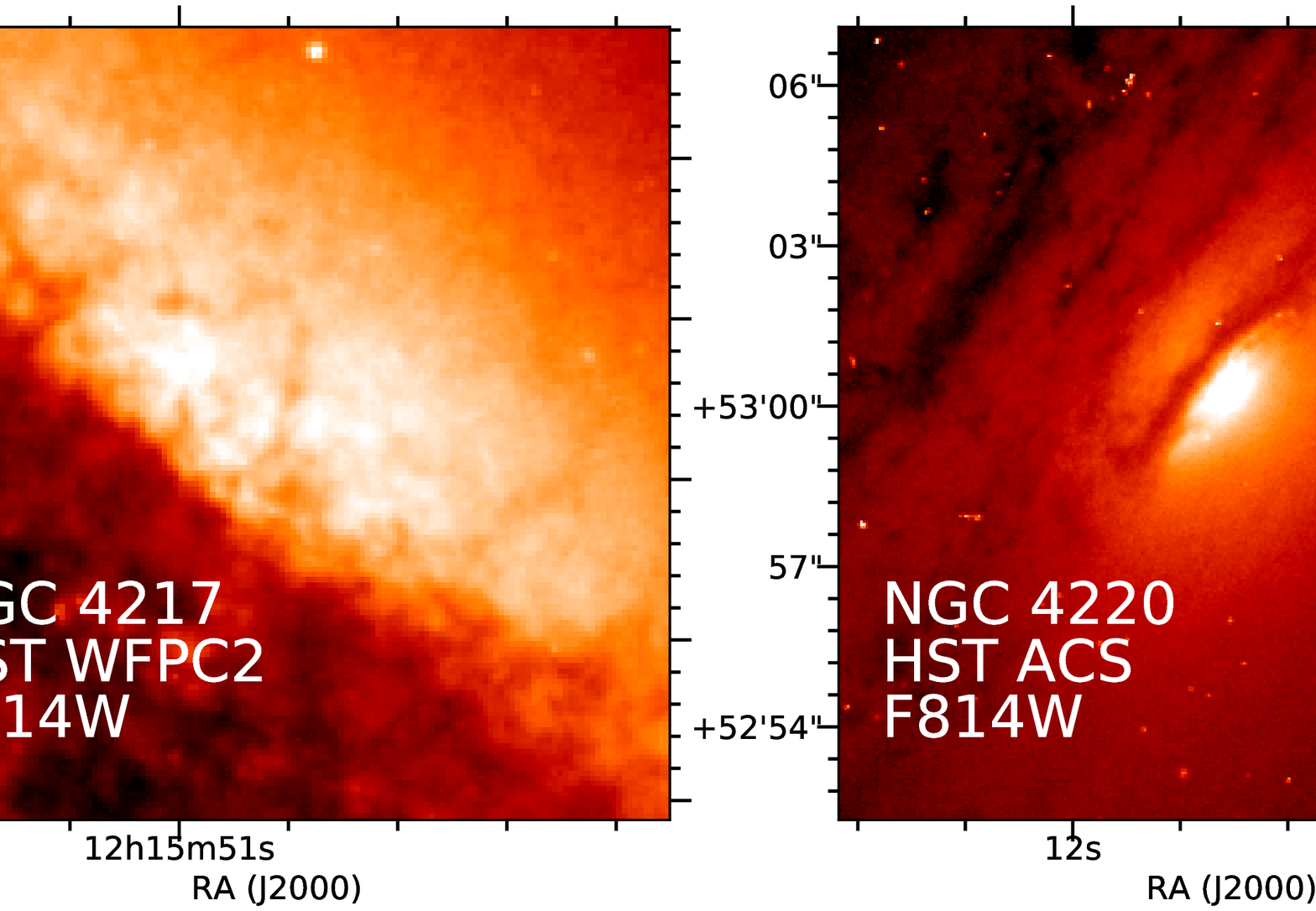}\\
\hspace*{4.0cm}
\includegraphics[trim= {11cm 0cm 0cm 0cm},angle=0,scale=0.24804]{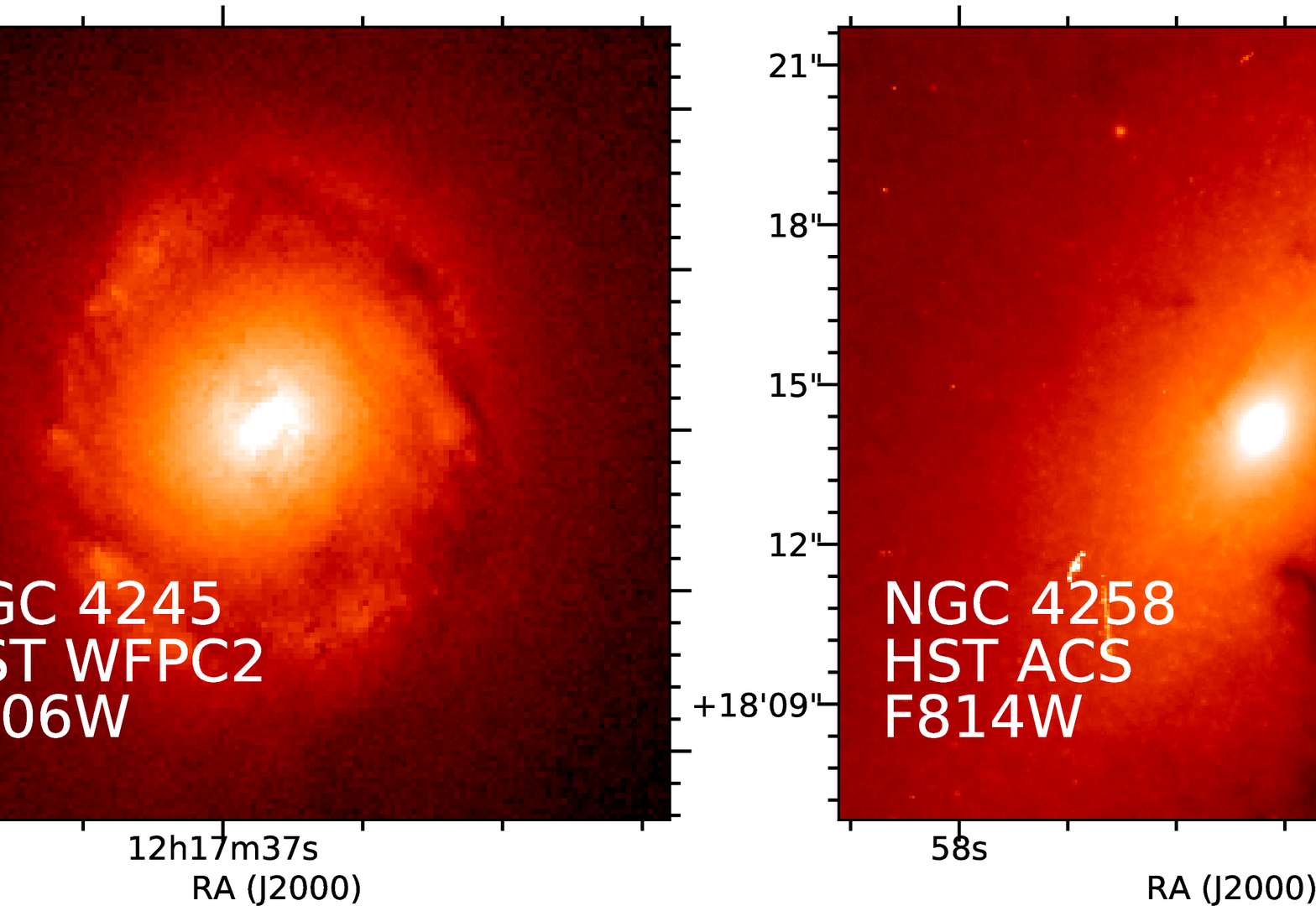}
\vspace{-.2630cm}
\caption{\it continued.}
\end{figure*}
\begin{figure*}
\setcounter{figure}{0} 
\vspace{-.275030cm}
\hspace*{4.0cm}
\includegraphics[trim= {11cm 0cm 0cm -0.29cm},angle=0,scale=0.24804]{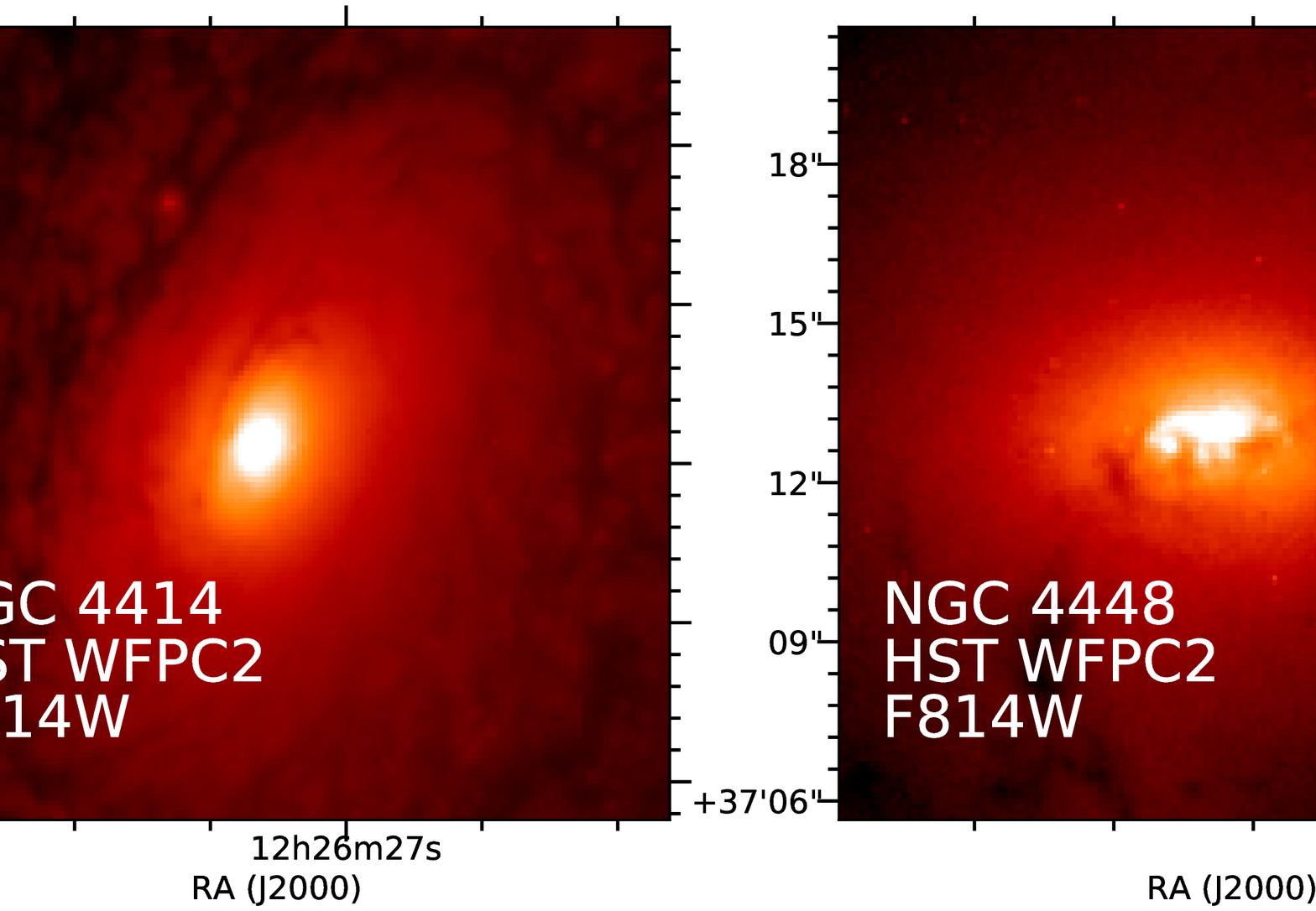}\\
\hspace*{4.0cm}
\includegraphics[trim= {11cm 0cm 0cm 0cm},angle=0,scale=0.24804]{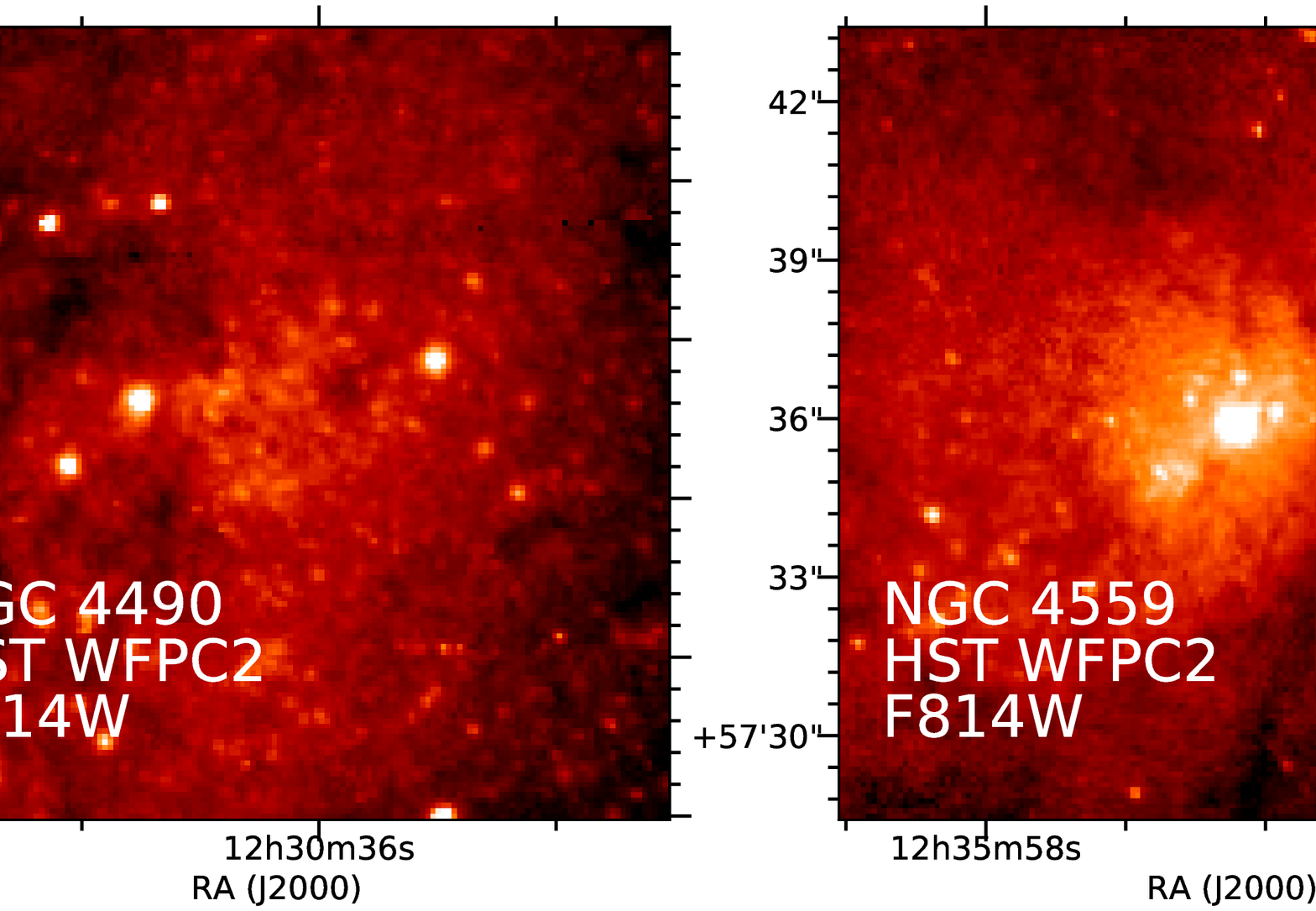}\\
\hspace*{4.0cm}
\includegraphics[trim= {11cm 0cm 0cm 0cm},angle=0,scale=0.24804]{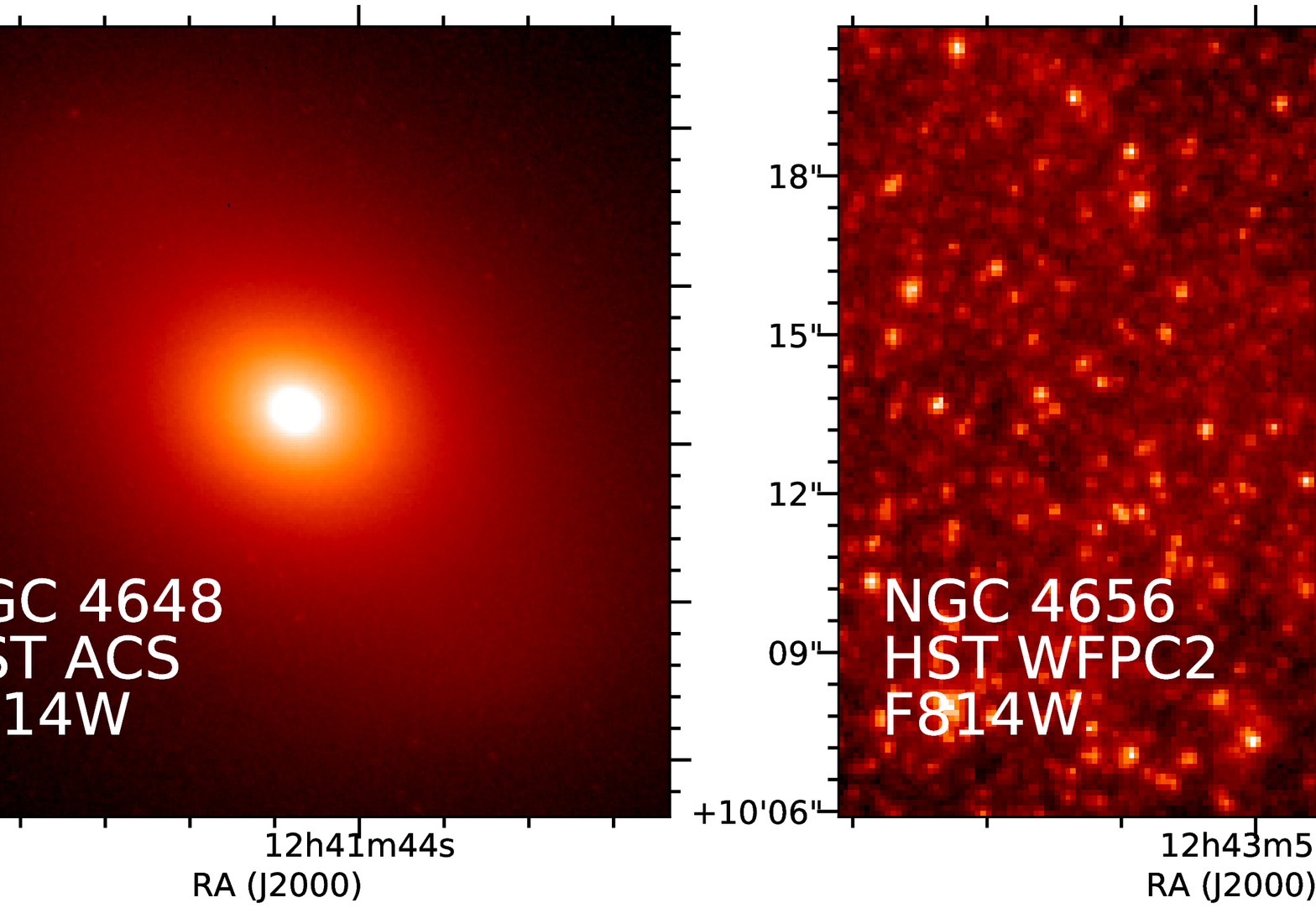}\\
\hspace*{4.0cm}
\includegraphics[trim= {11cm 0cm 0cm 0cm},angle=0,scale=0.24804]{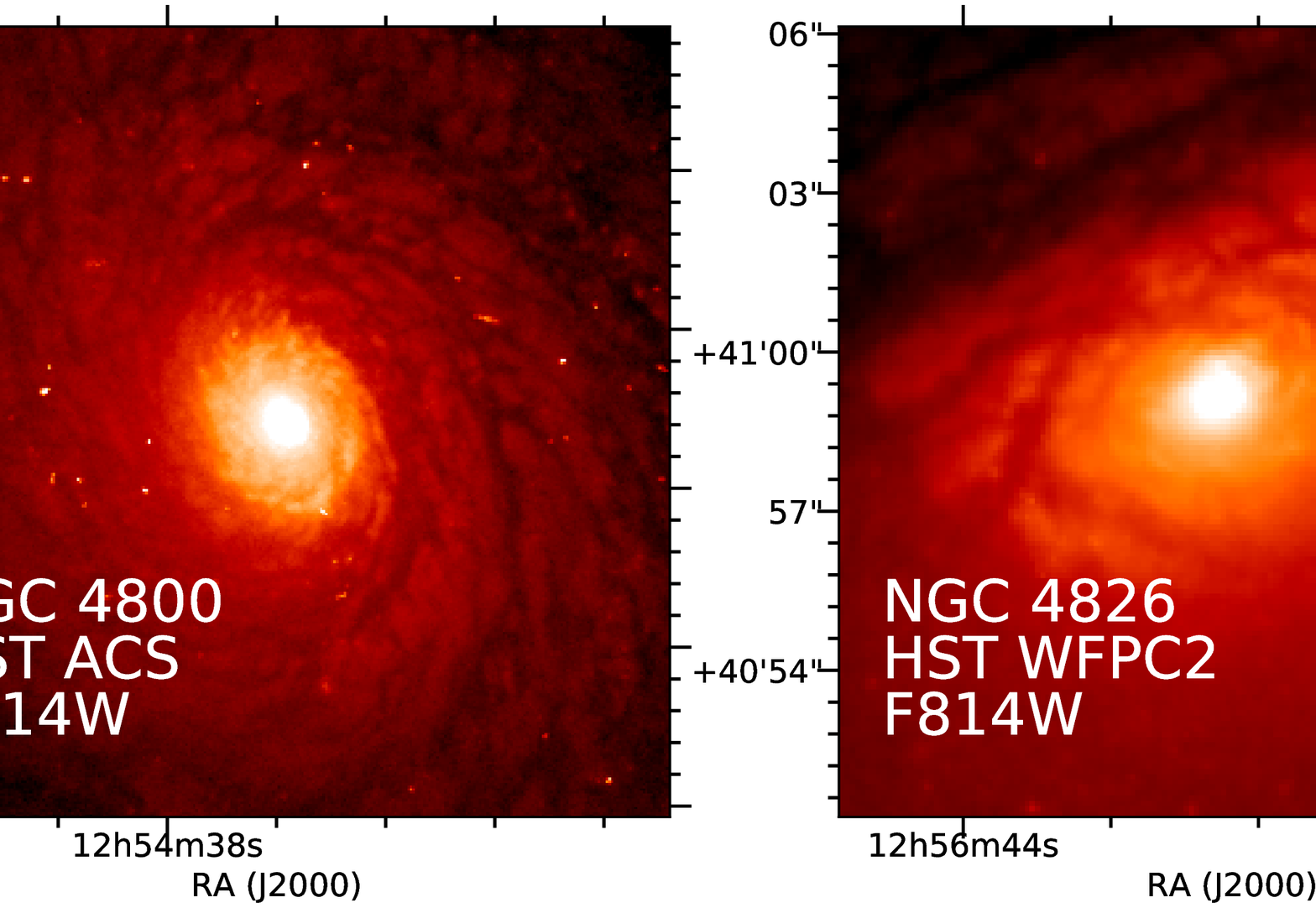}\\
\hspace*{4.0cm}
\includegraphics[trim= {11cm 0cm 0cm 0cm},angle=0,scale=0.24804]{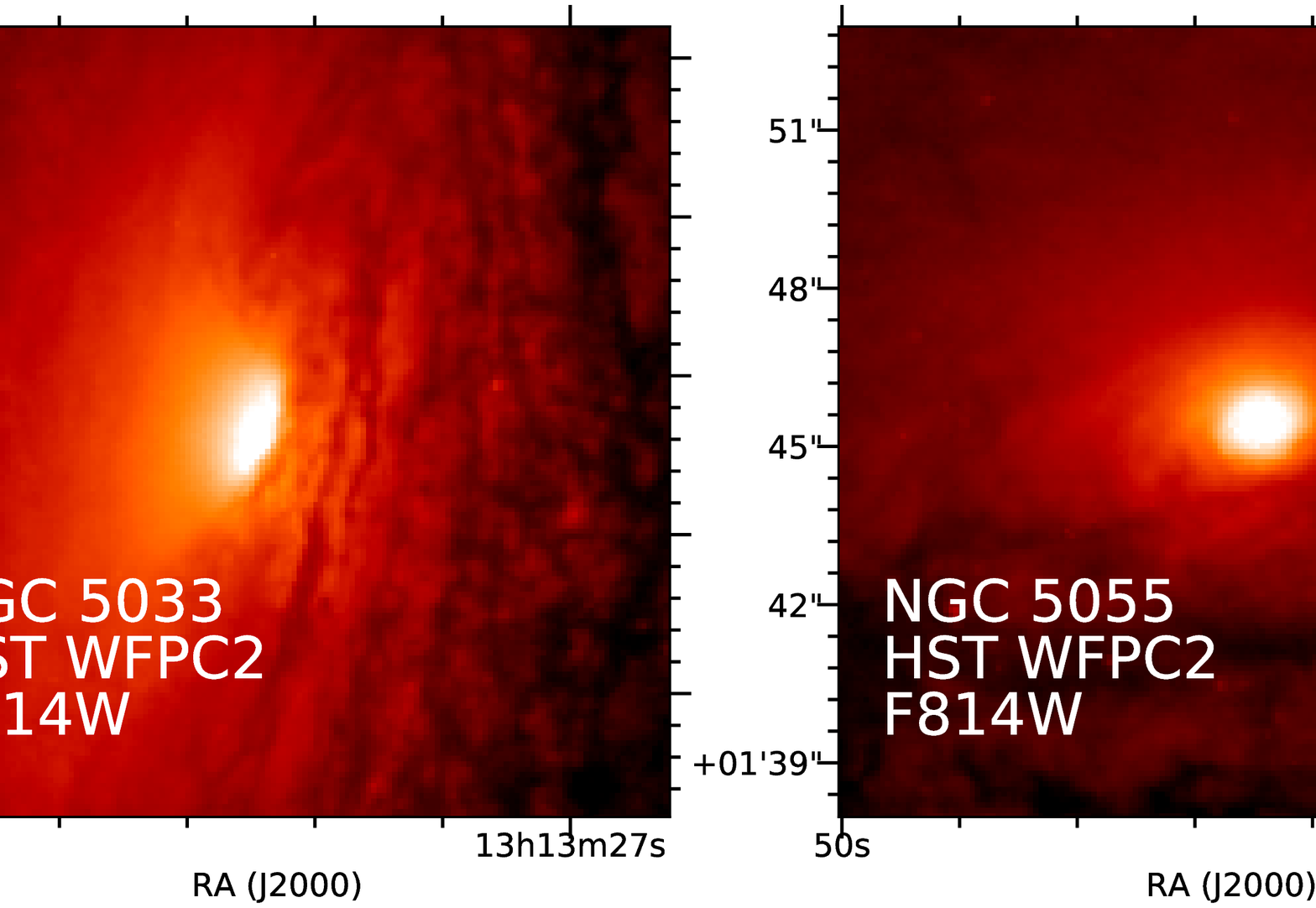}\\
\hspace*{4.0cm}
\includegraphics[trim= {11cm 0cm 0cm 0cm},angle=0,scale=0.24804]{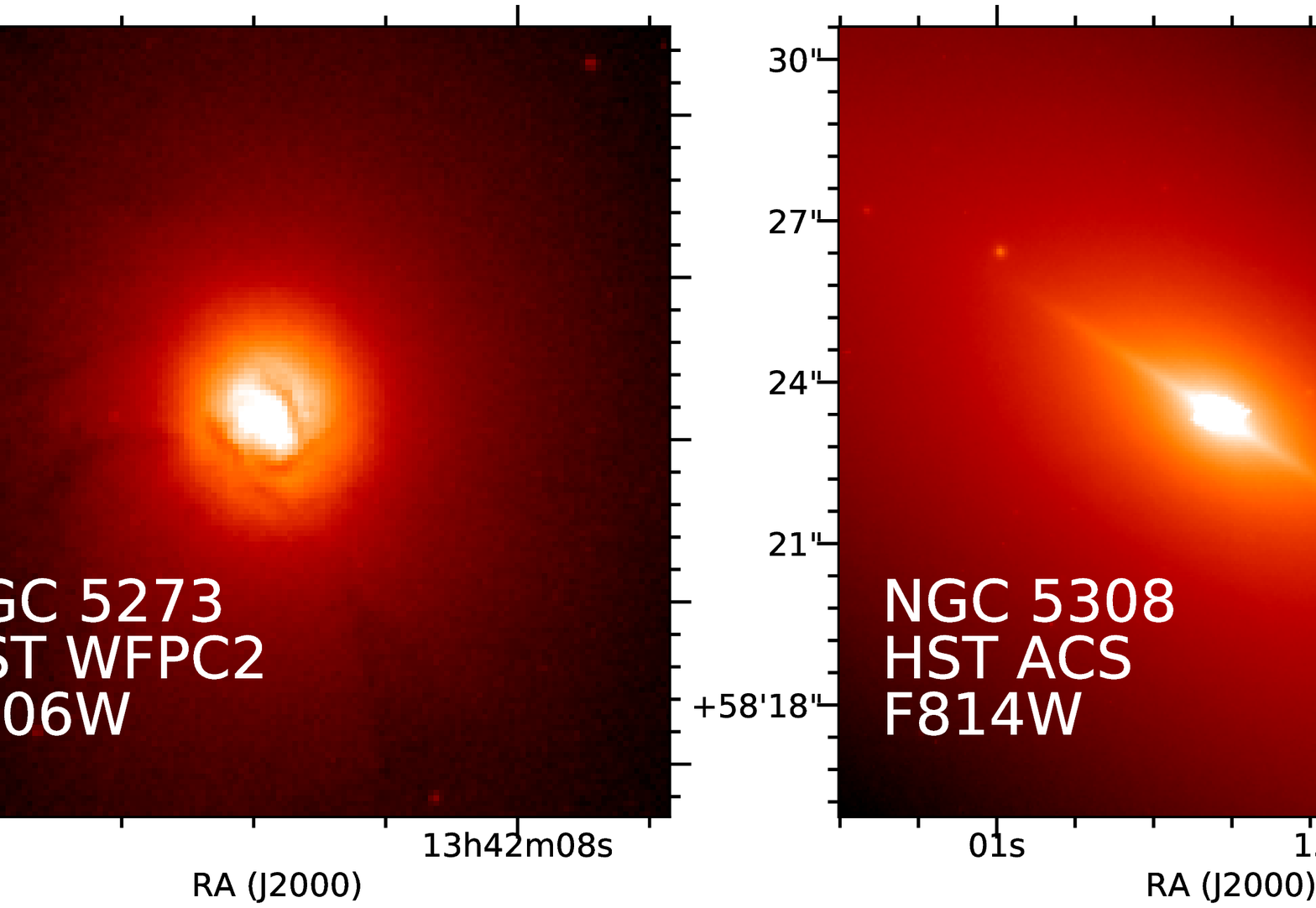}
\vspace{-.2630cm}
\caption{\it continued.}
\end{figure*}

\begin{figure*}
\setcounter{figure}{0} 
\vspace{-.275030cm}
\hspace*{4.0cm}
\includegraphics[trim= {11cm 0cm 0cm -0.29cm},angle=0,scale=0.24804]{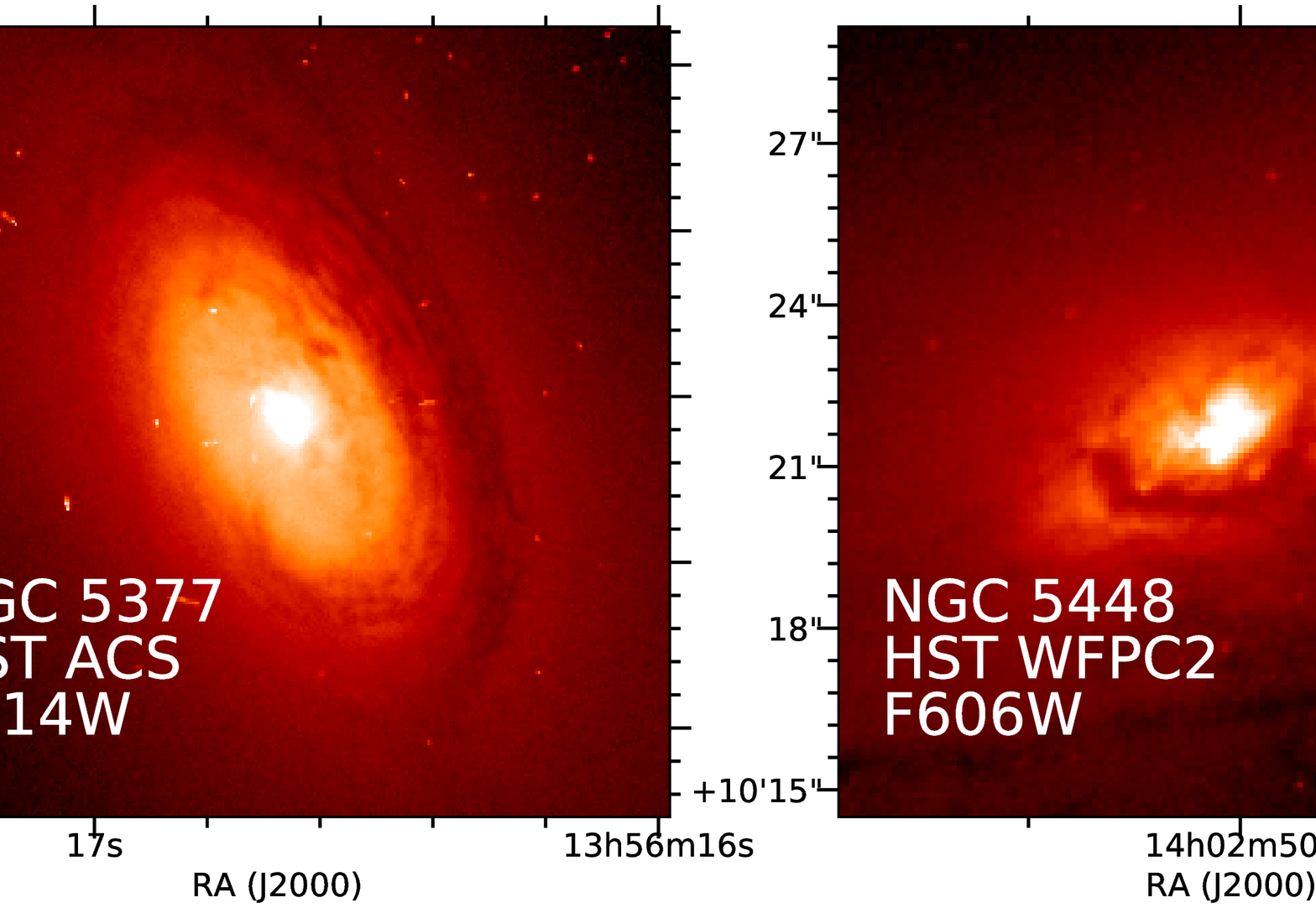}\\
\hspace*{4.0cm}
\includegraphics[trim= {11cm 0cm 0cm 0cm},angle=0,scale=0.24804]{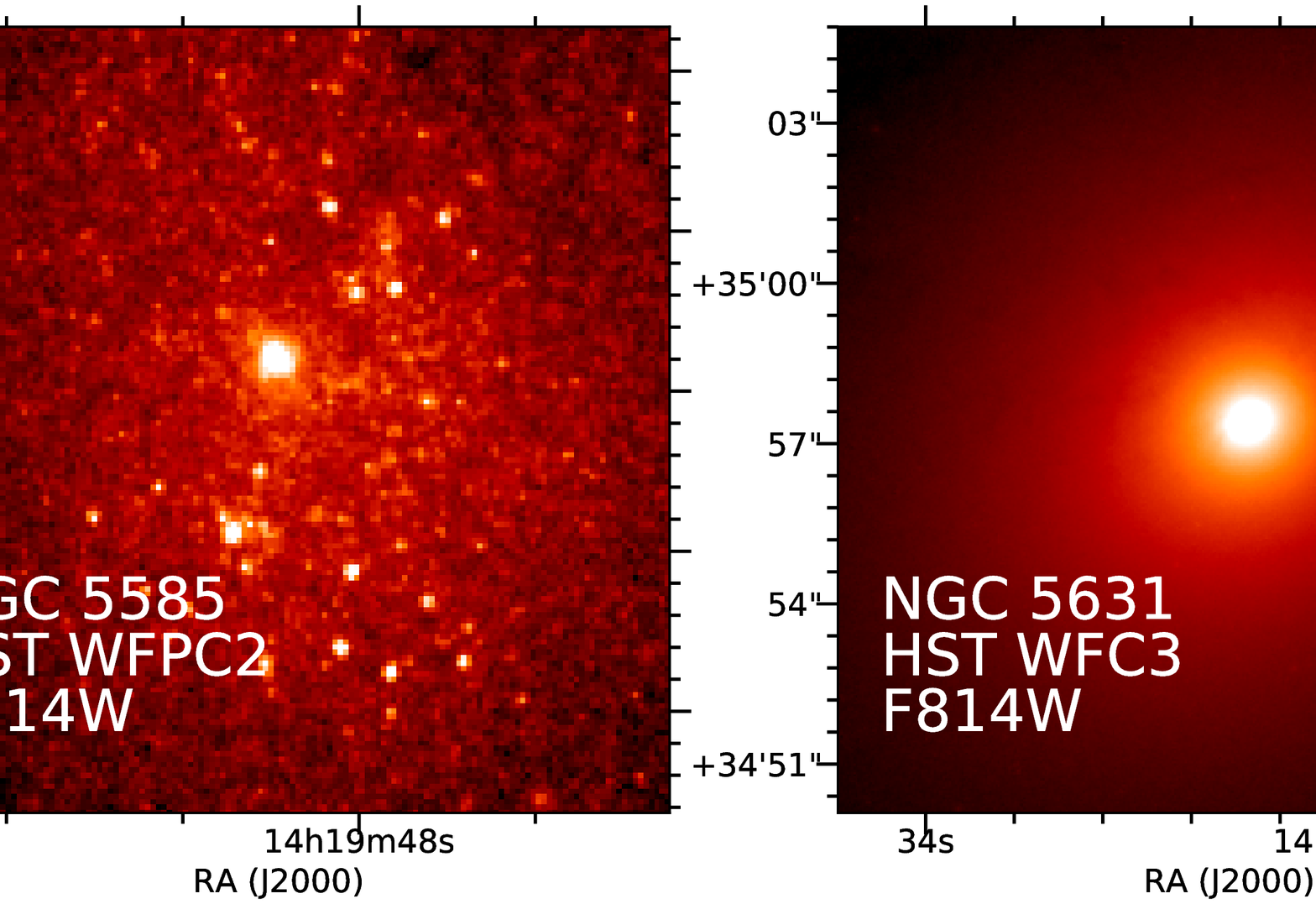}\\
\hspace*{4.0cm}
\includegraphics[trim= {11cm 0cm 0cm 0cm},angle=0,scale=0.24804]{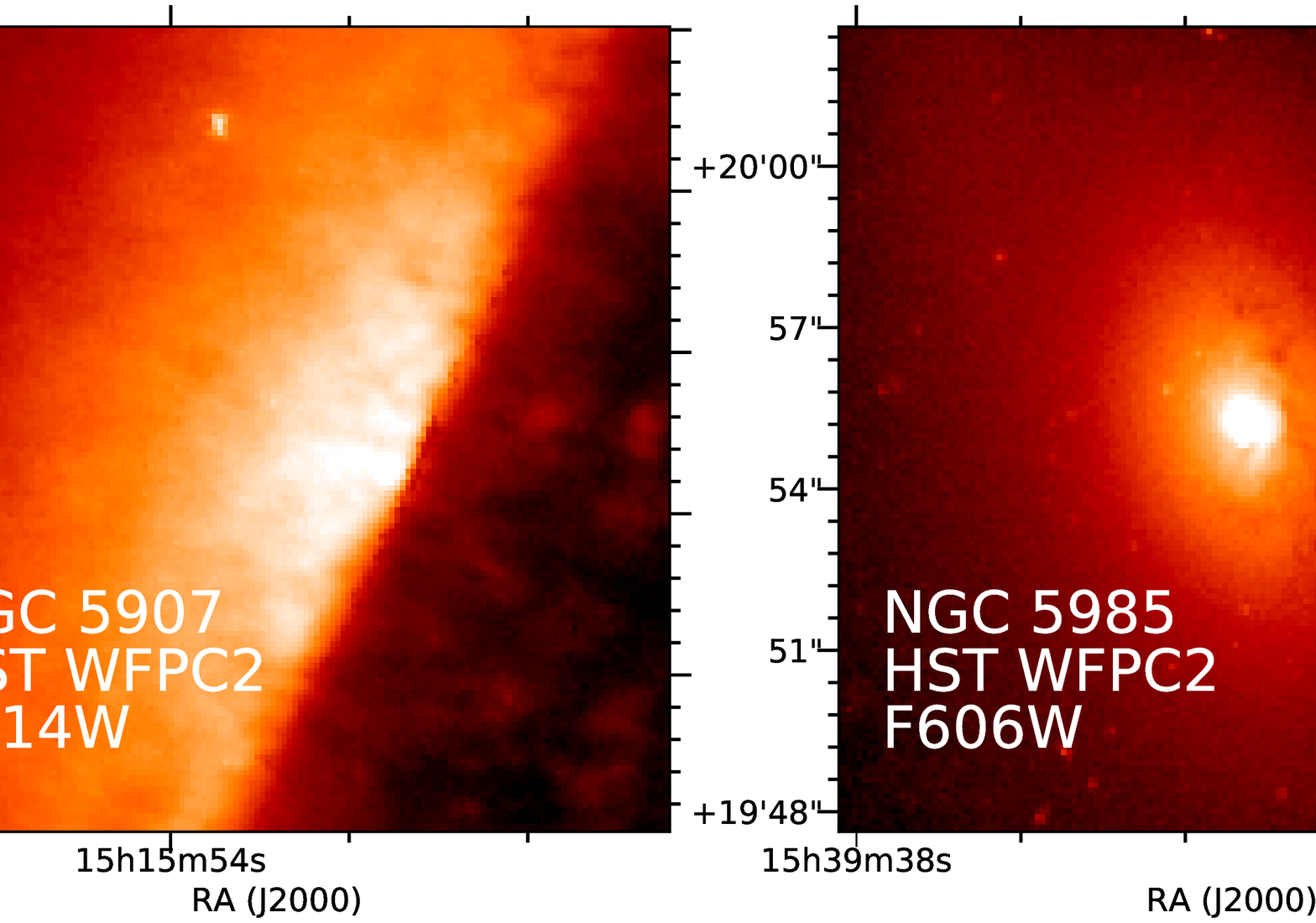}\\
\hspace*{4.0cm}
\includegraphics[trim= {11cm 0cm 0cm 0cm},angle=0,scale=0.24804]{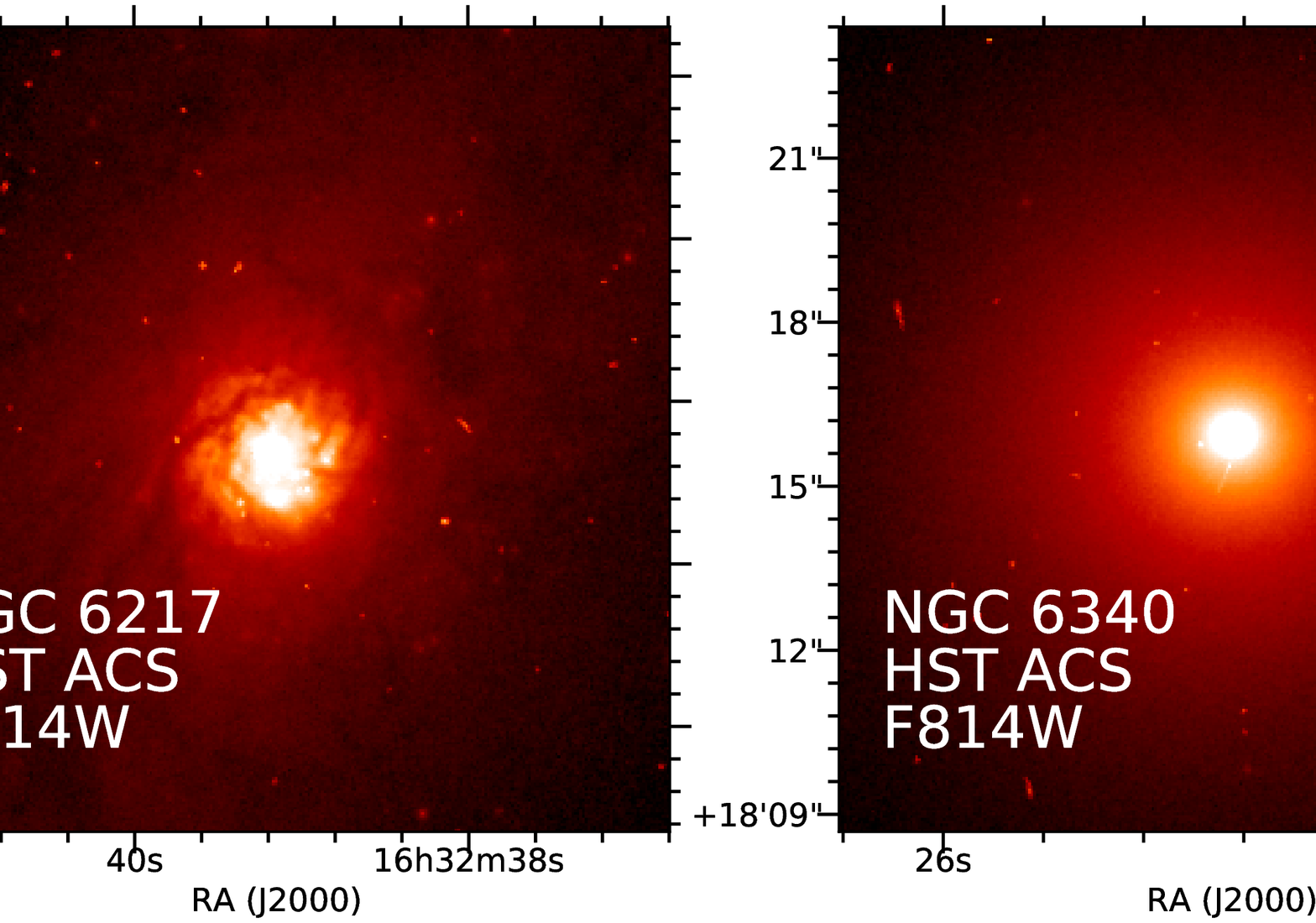}\\
\hspace*{4.0cm}
\includegraphics[trim= {11cm 0cm 0cm 0cm},angle=0,scale=0.24804]{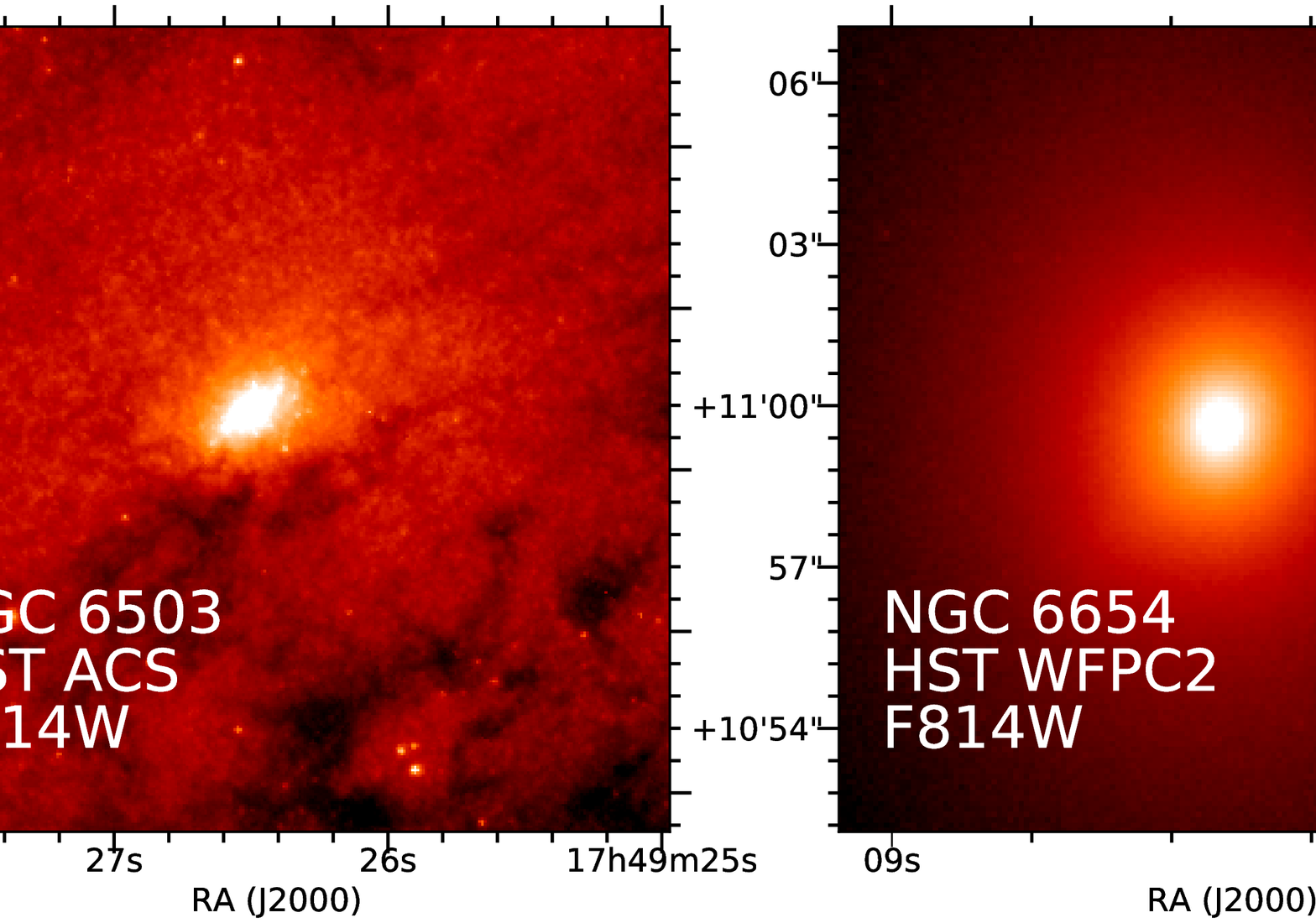}\\
\hspace*{4.0cm}
\includegraphics[trim= {11cm 0cm 0cm 0cm},angle=0,scale=0.24804]{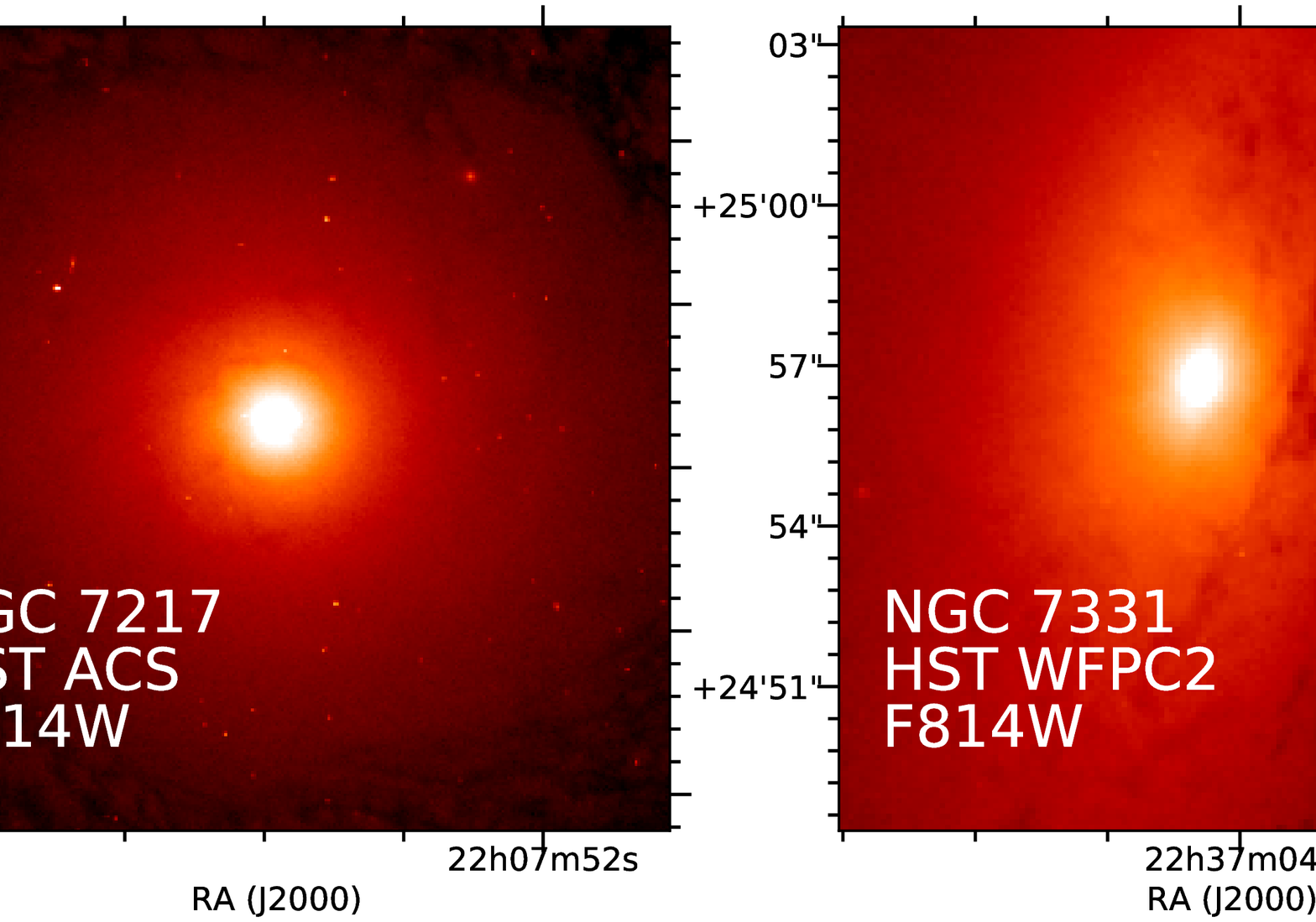}
\vspace{-.2630cm}
\caption{\it continued.}
\end{figure*}
\begin{figure*}
\setcounter{figure}{0} 
\hspace*{4.0cm}
\includegraphics[trim= {11cm 0cm 0cm 0cm},angle=0,scale=0.24804]{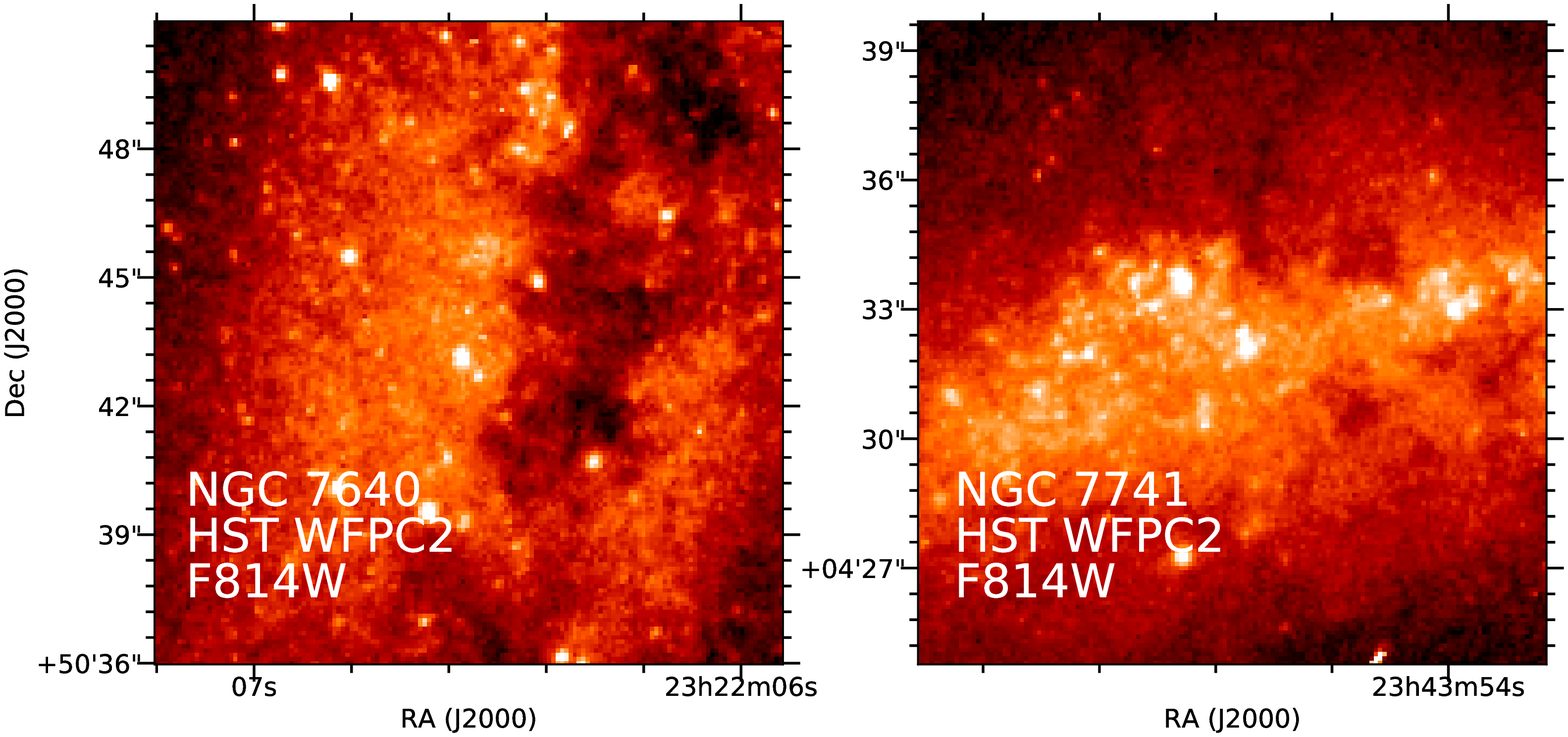}
\vspace{-.2630cm}
\caption{\it continued.}
\end{figure*}

\setcounter{section}{3}
\renewcommand{\thesection}{A\arabic{section}}

\section{1D multi-component decomposition}\label{AppendC}

1D multi-component decompositions of the major-axis surface brightness
profiles (black dots) of LeMMINGs galaxies listed in
Tables~\ref{NewTabA1}-\ref{Tab1cS}. Figs.~\ref{FigSer1} and
\ref{FigSerR} show S\'ersic LeMMINGs galaxies without and with a
large-scale Gaussian ring component, respectively. Fig.~\ref{FigCSer}
shows core-S\'ersic LeMMINGs galaxies.  Furthermore, in
Fig.~\ref{FigMoff} we show light profile decompositions of a dozen
LeMMINGs galaxies using multiple S\'ersic models that are
PSF-convolved with a Moffat function.

\section{2D  decomposition}\label{Append2D}

Fig.~\ref{2Dfits} shows the residual images from our 2D fitting which
were generated after subtracting the 2D {\sc imfit} model images from
the original {\it HST} images for 28 sample galaxies, which are
representative of the 65 sample galaxies with 2D decompositions,
listed in Tables~\ref{NewTabA1}-\ref{Tab1cS}. We note that the 2D
multi-component modelling of the full {\it HST} images for the 65
galaxies are performed using the same type and number of galaxy
structural components as the corresponding 1D modelling. 

\begin{figure*}
\vspace{-.295030cm}
\hspace*{1.24811cm}
\includegraphics[angle=270,scale=0.54]{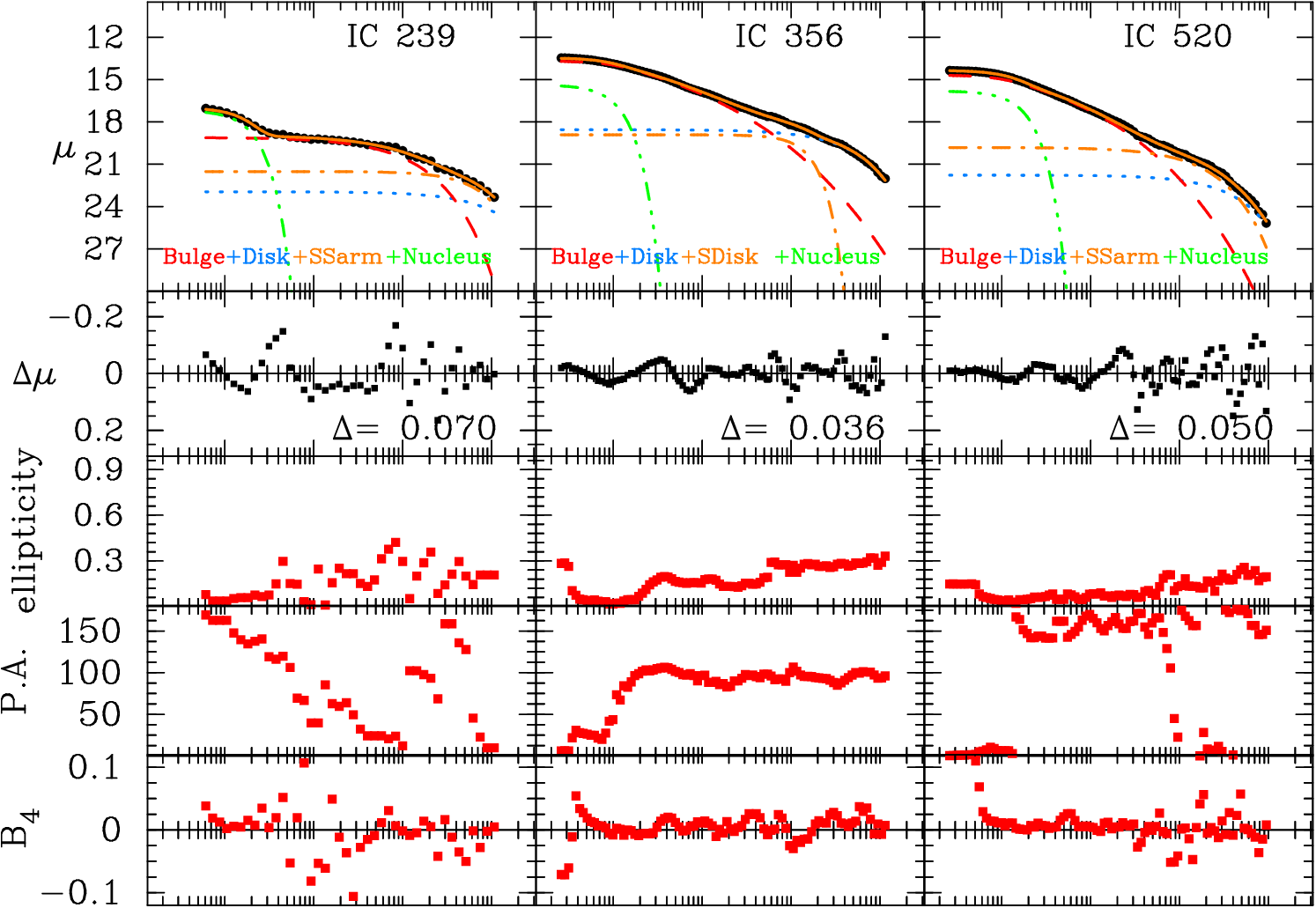}\\
\hspace*{1.24811cm}
\includegraphics[angle=270,scale=0.54]{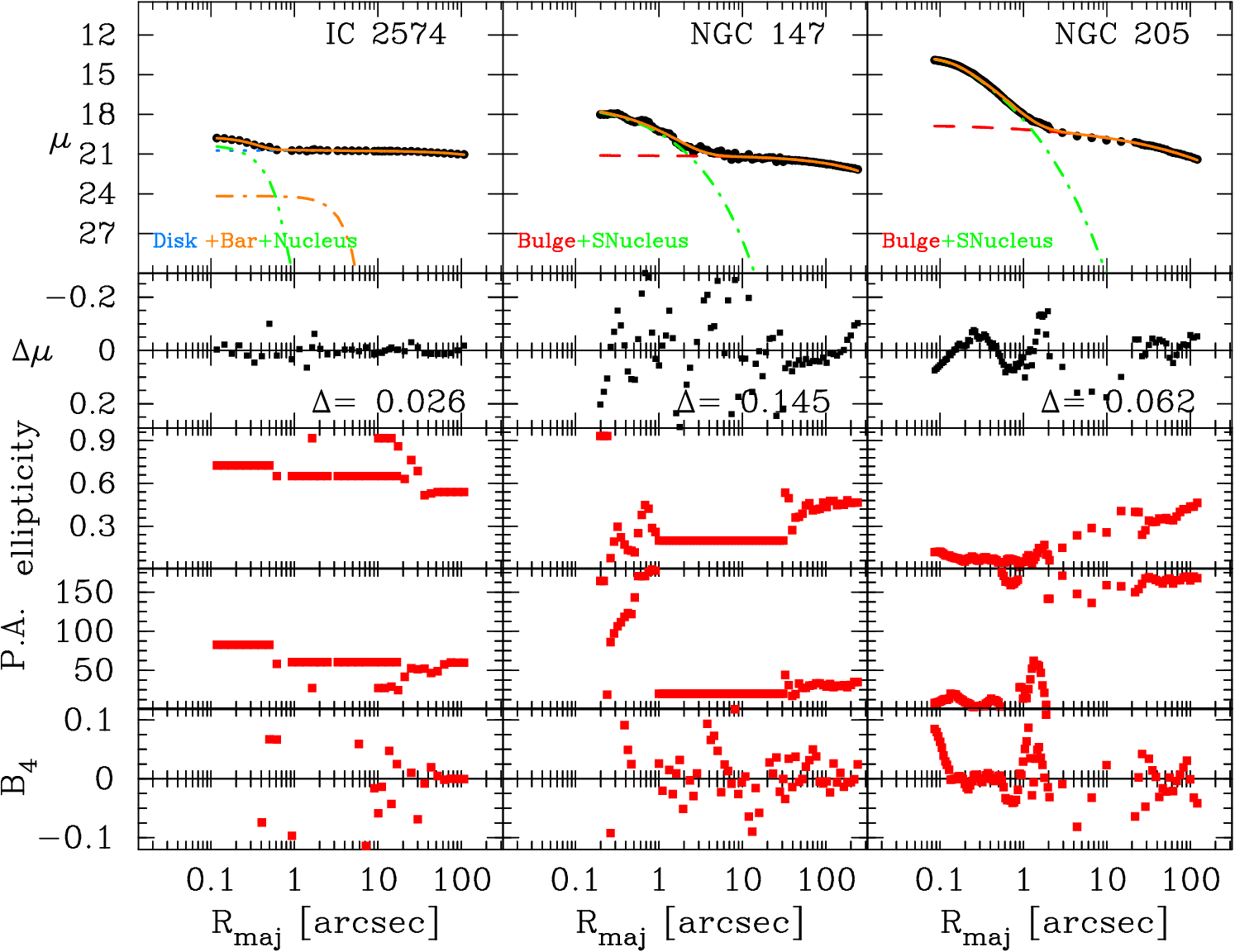}\\
\vspace{.2030cm}
\caption{Multi-component decompositions of the major-axis surface
  brightness profiles (black dots) of the S\'ersic LeMMINGs galaxies
  in our sample without a large-scale Gaussian ring component (see
  Table ~\ref{Tab1}; galaxies with a large-scale Gaussian ring
  component are shown in Fig.~\ref{FigSerR}, whereas core-S\'ersic
  galaxies are in Fig.~\ref{FigCSer}).  The residual profiles along
  with the rms residual $ \Delta $ for each fit are shown.  The
  ellipticity, position angle (P.A., measured in degrees from north to
  east), and isophote shape parameter ($B_{4}$) profiles (red boxes)
  are given in the lower panels. The {\it HST} filter associated with
  each galaxy profile is listed in Table~\ref{Tab1}. The surface
  brightnesses $(\mu)$ are in units of mag arcsec$^{-2}$. The dashed
  red curves represent the bulges, while the dotted blue curve shows
  the large-scale discs or stellar halo light which we model with an
  exponential function.  Nuclear components (i.e., AGN, NSCs and inner
  discs and bars) were typically fit with the two-parameter Gaussian
  model (dash-dot-dot-dot green curve). `SNucleus' (`SNuc')
  (dot-dashed green curve) and `SDisc' (dot-dashed blue curve) denote
  nuclei and large-scale disc components that were fitted with
  S\'ersic models, respectively.  Galaxy components such as bars and
  small-scale discs, rings, spirals and lenses are described by
  S\'ersic models (i.e., orange and magenta dot-dashed curves). For
  two galaxies in our sample (IC~239 and IC~520) their spiral arm
  components were modelled with a S\'ersic model (i.e., `SSarm'). The
  complete fits (solid orange curves) fit the observed galaxy profiles
  with a median rms residual $ \Delta \sim$ 0.065 mag arcsec$^{-2}$.
  We fit up to six model components which are summed up to a full
  model with (up to) 16 free parameters.  }
\label{FigSer1}
\end{figure*}

\begin{figure*}
\setcounter{figure}{1} 
\vspace{-.295030cm}
\hspace*{.4811cm}
\includegraphics[angle=270,scale=0.60504]{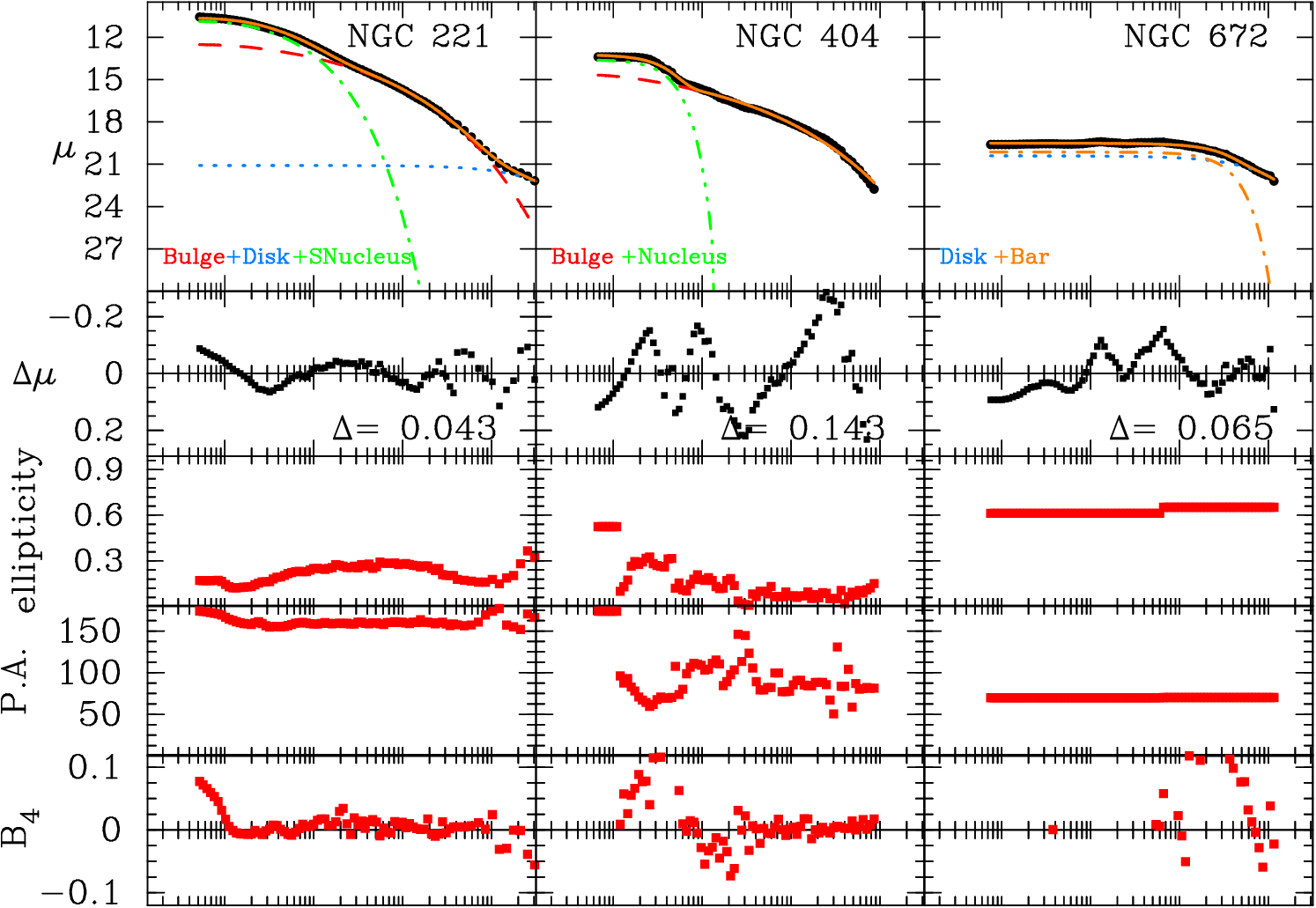}\\
\hspace*{.4811cm}
\includegraphics[angle=270,scale=0.60504]{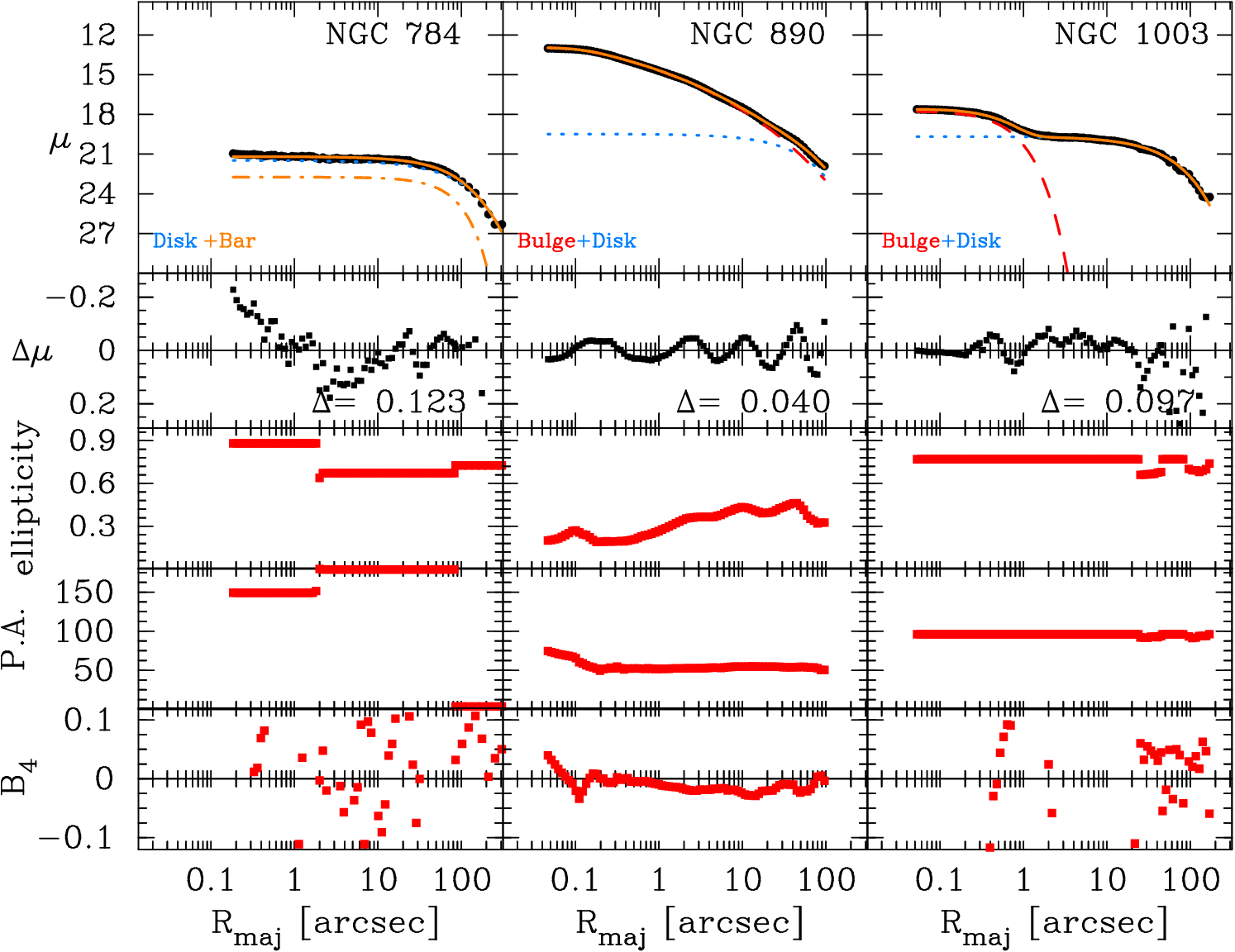}\\
\vspace{.2030cm}
\caption{\it continued.}
\end{figure*}

\begin{figure*}
\setcounter{figure}{1}
\vspace{-.295030cm}
\hspace*{.4811cm}
\includegraphics[angle=270,scale=0.60504]{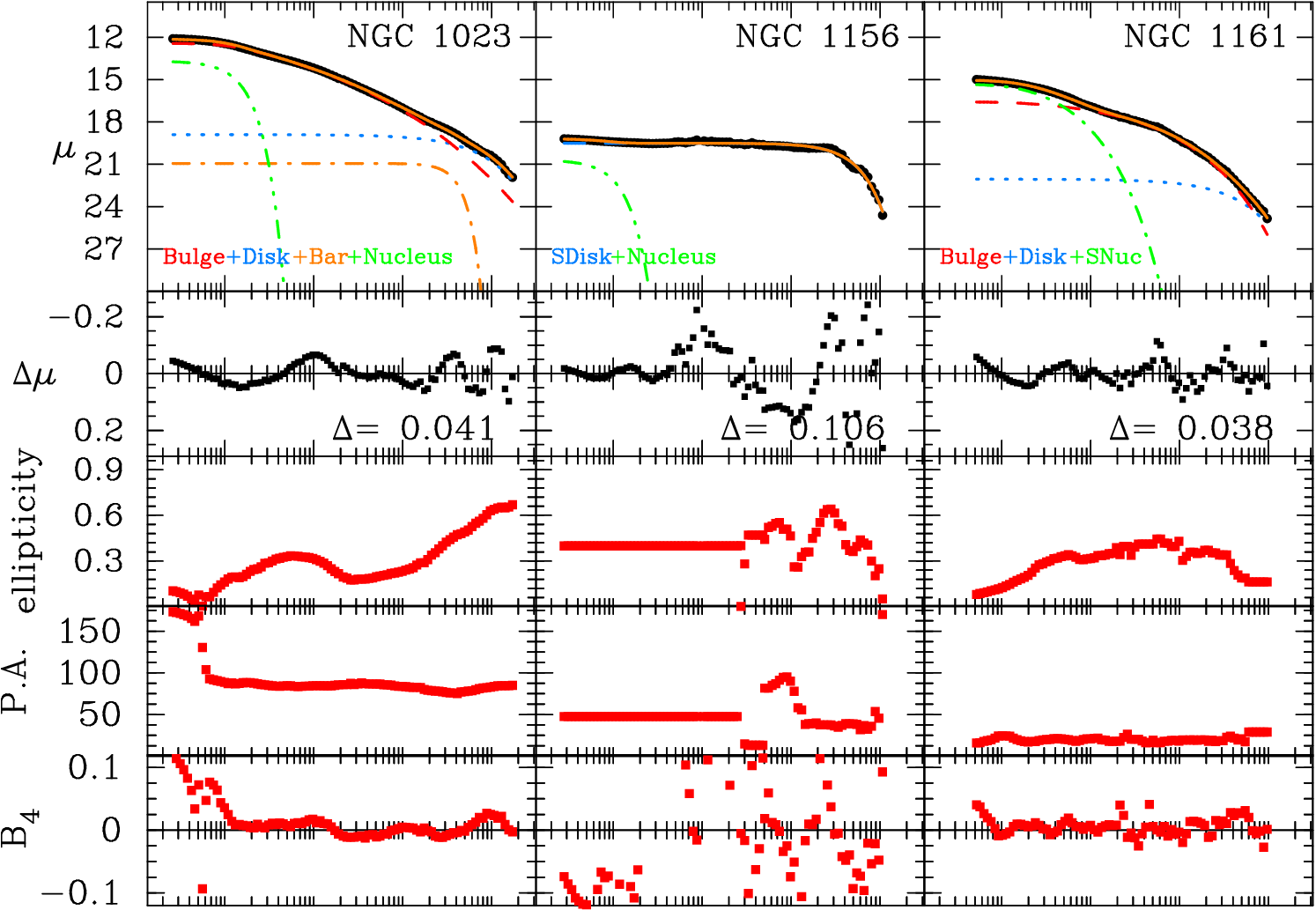}\\
\hspace*{.4811cm}
\includegraphics[angle=270,scale=0.60504]{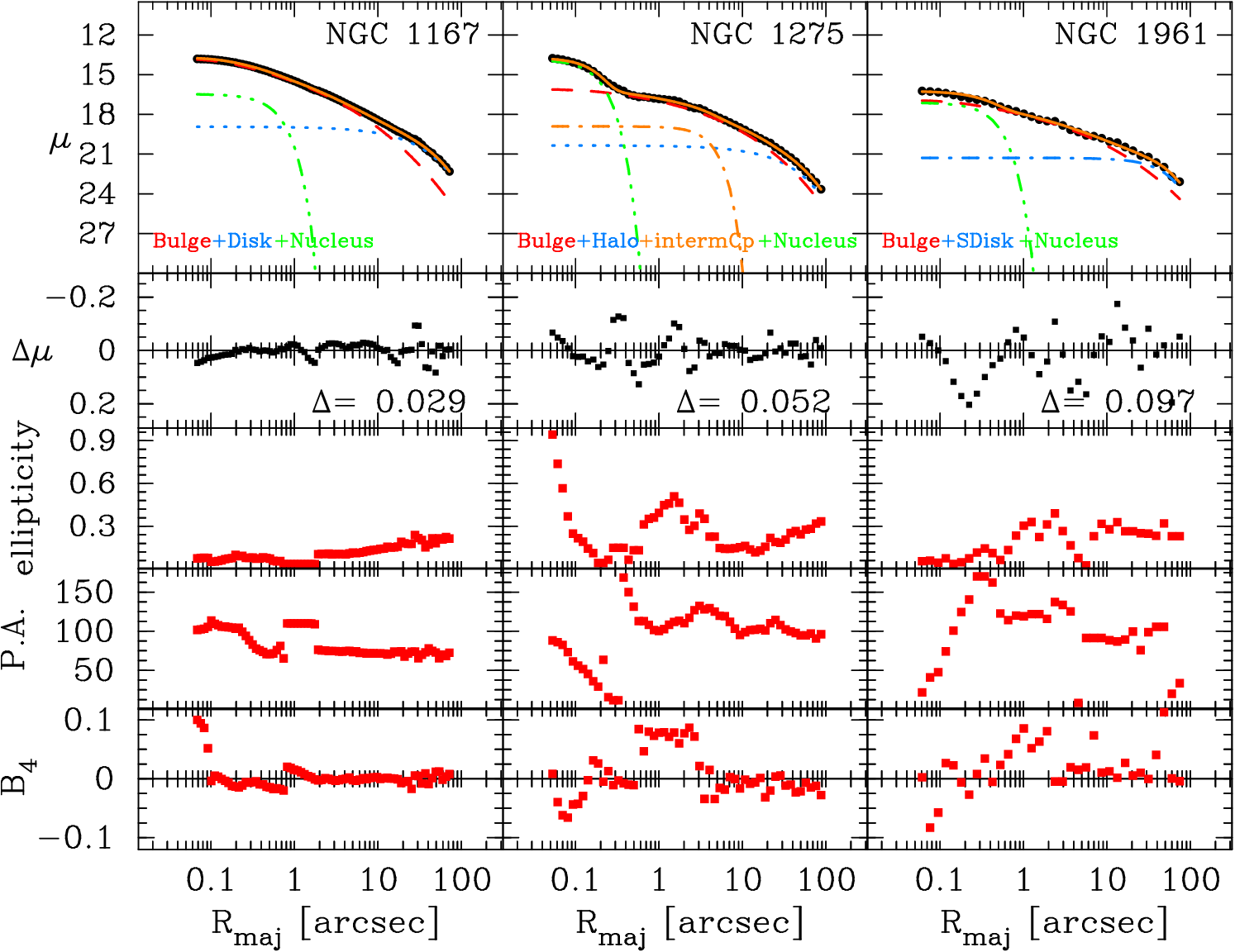}\\
\vspace{.2030cm}
\caption{\it continued.}
\end{figure*}

\begin{figure*}
\setcounter{figure}{1}
\vspace{-.295030cm}
\hspace*{.4811cm}
\includegraphics[angle=270,scale=0.60504]{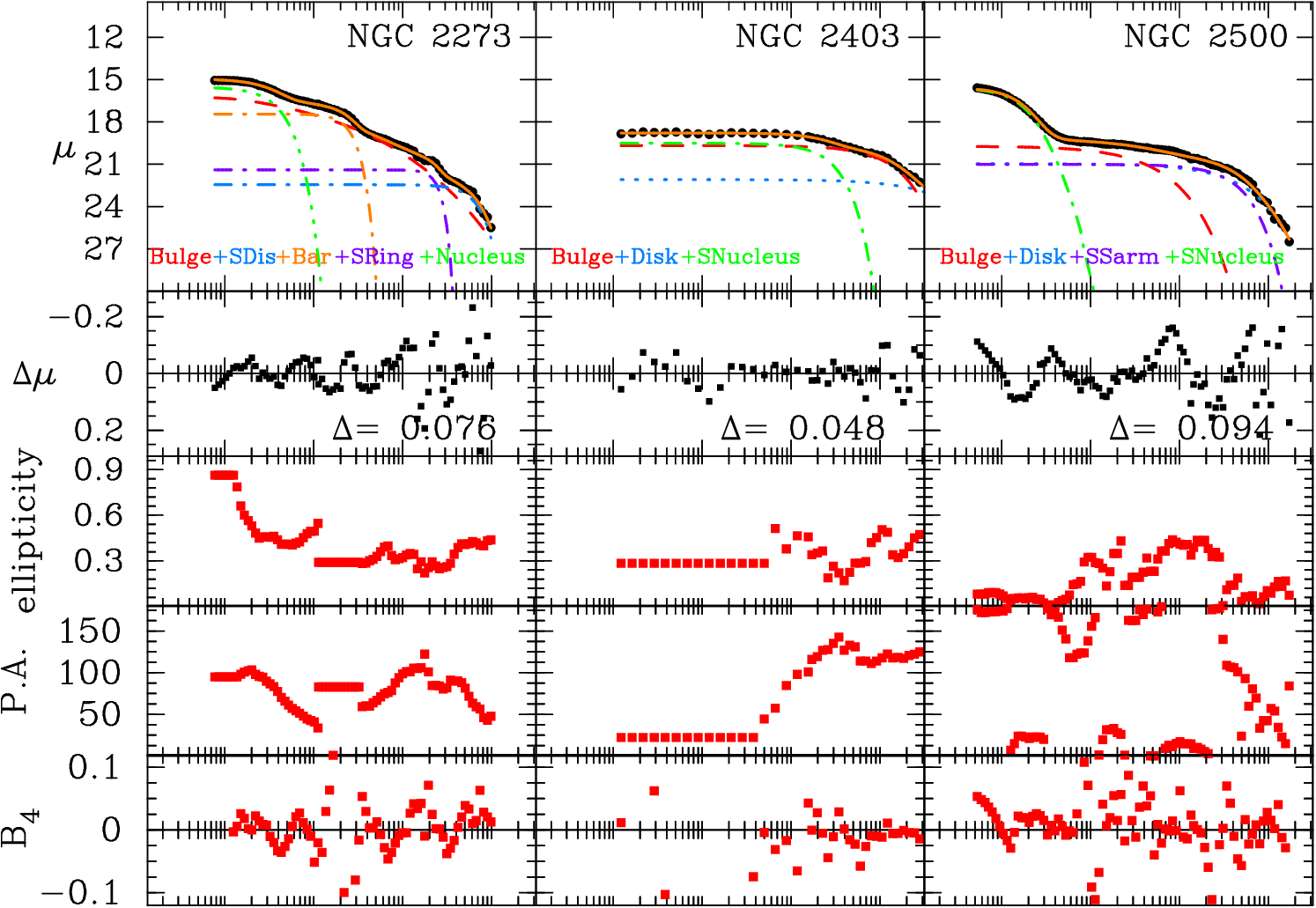}\\
\hspace*{.4811cm}
\includegraphics[angle=270,scale=0.60504]{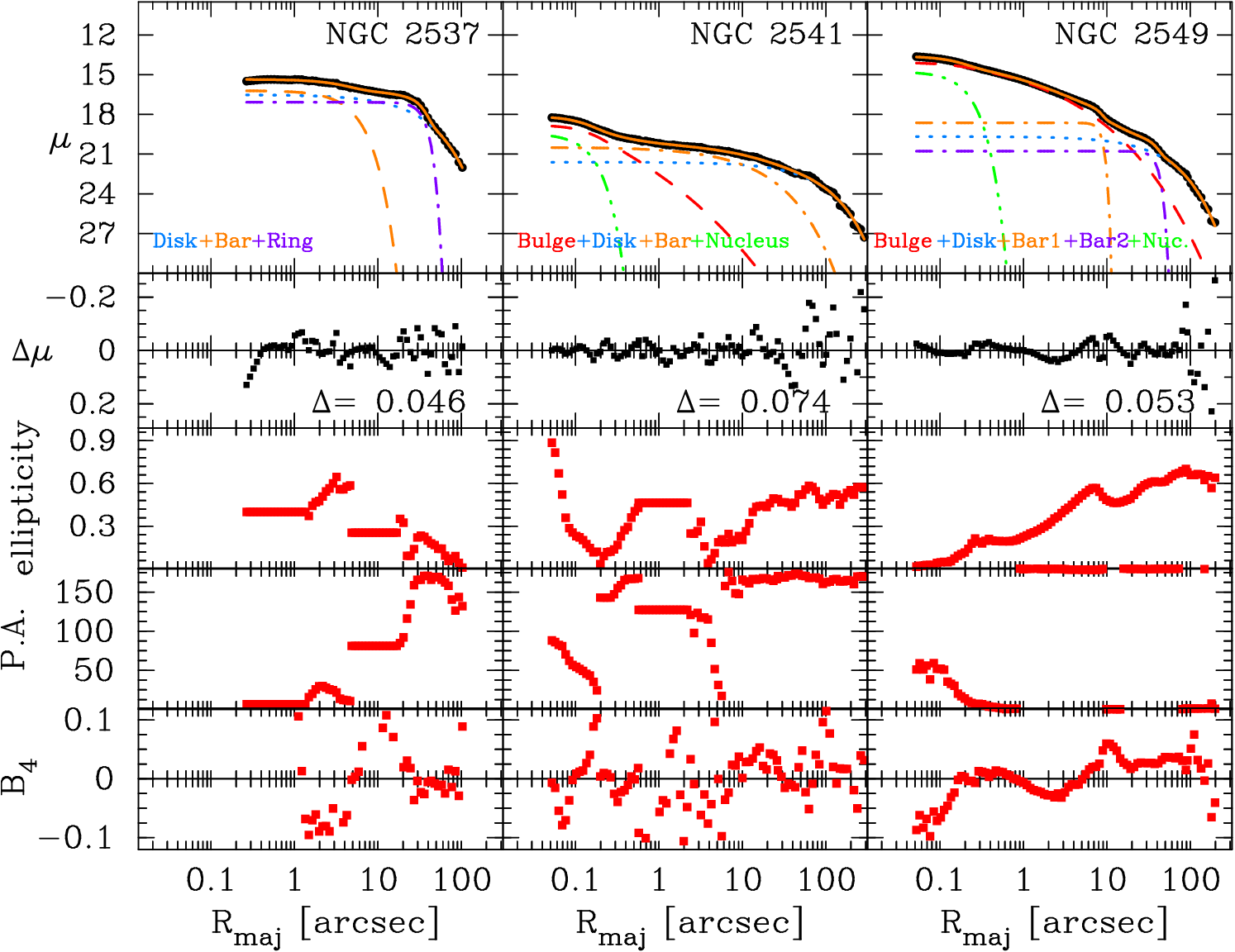}\\
\vspace{.2030cm}
\caption{\it continued.}
\end{figure*}

\begin{figure*}
\setcounter{figure}{1}
\vspace{-.295030cm}
\hspace*{.4811cm}
\includegraphics[angle=270,scale=0.6504]{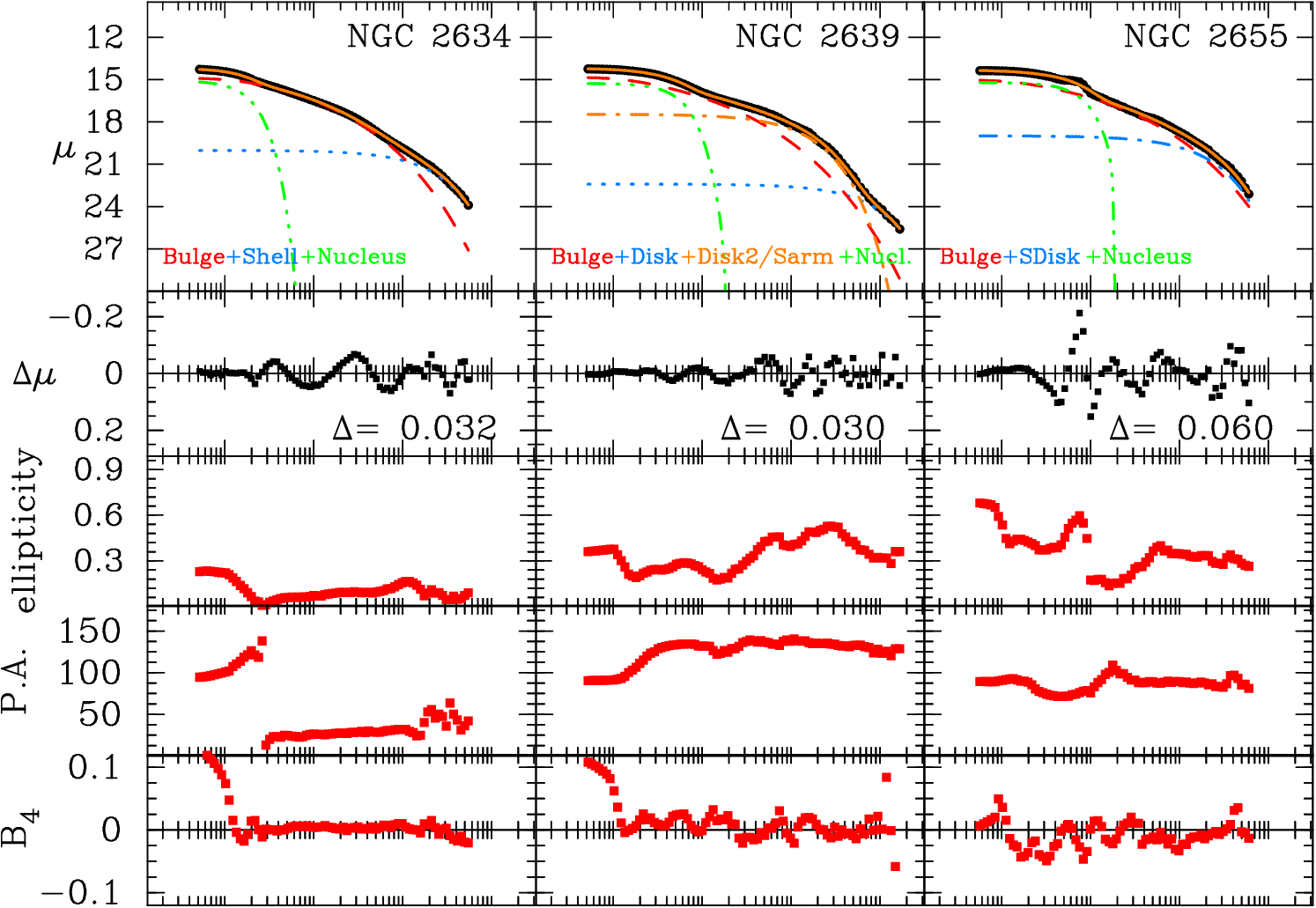}\\
\hspace*{.4811cm}
\includegraphics[angle=270,scale=0.6504]{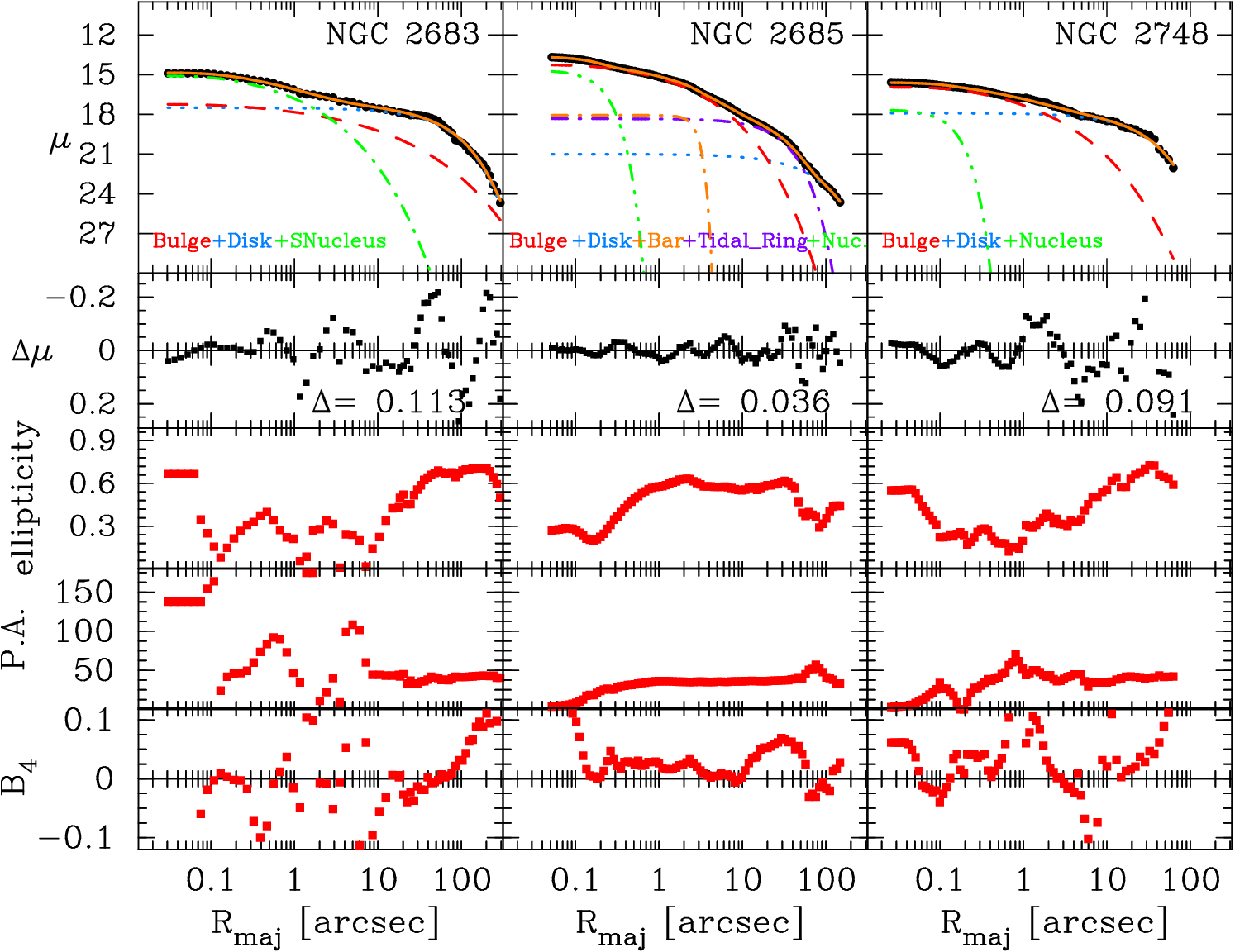}\\
\vspace{.2030cm}
\caption{\it continued.}
\end{figure*}

\begin{figure*}
\setcounter{figure}{1}
\vspace{-.295030cm}
\hspace*{.4811cm}
\includegraphics[angle=270,scale=0.60504]{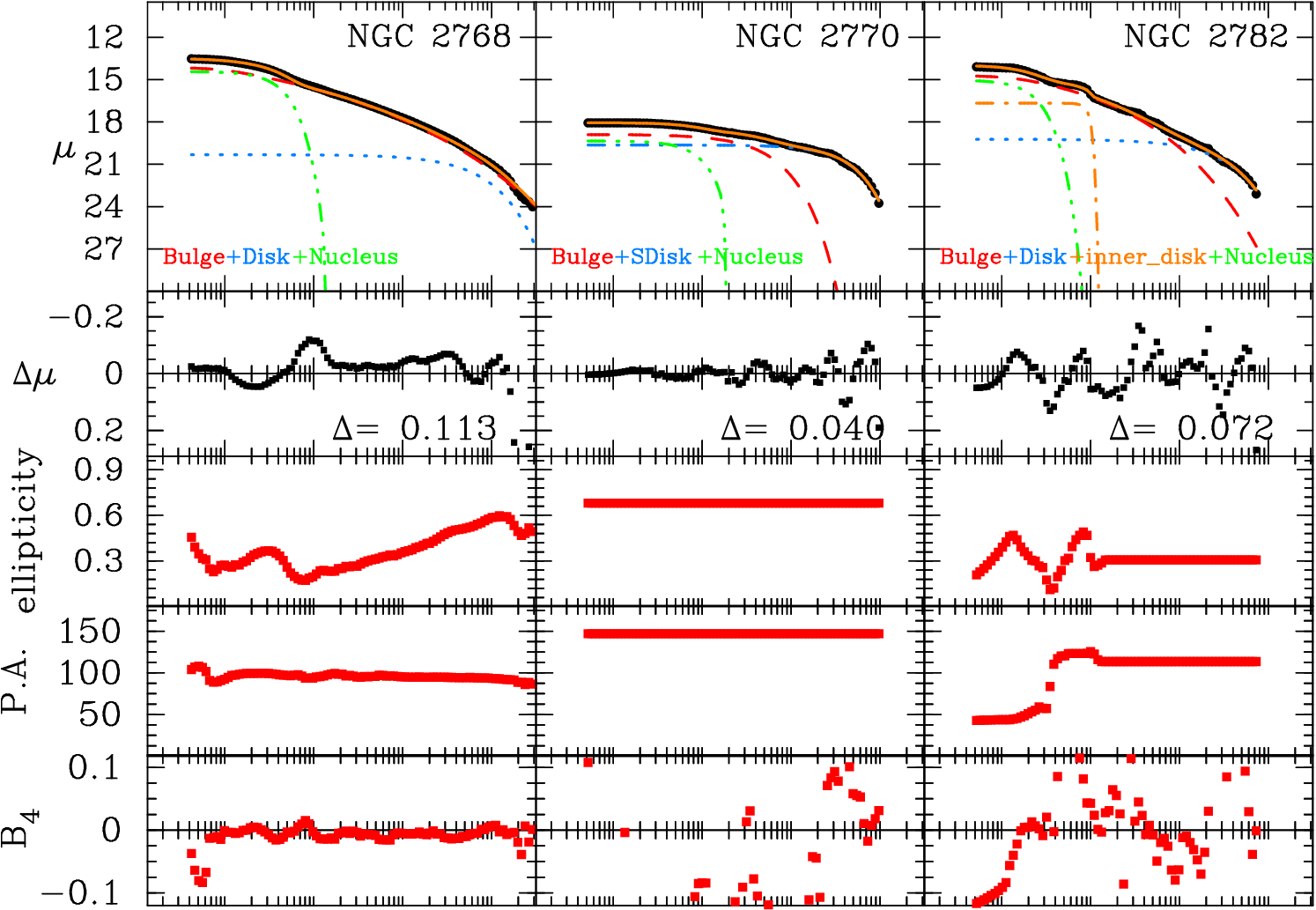}\\
\hspace*{.4811cm}
\includegraphics[angle=270,scale=0.60504]{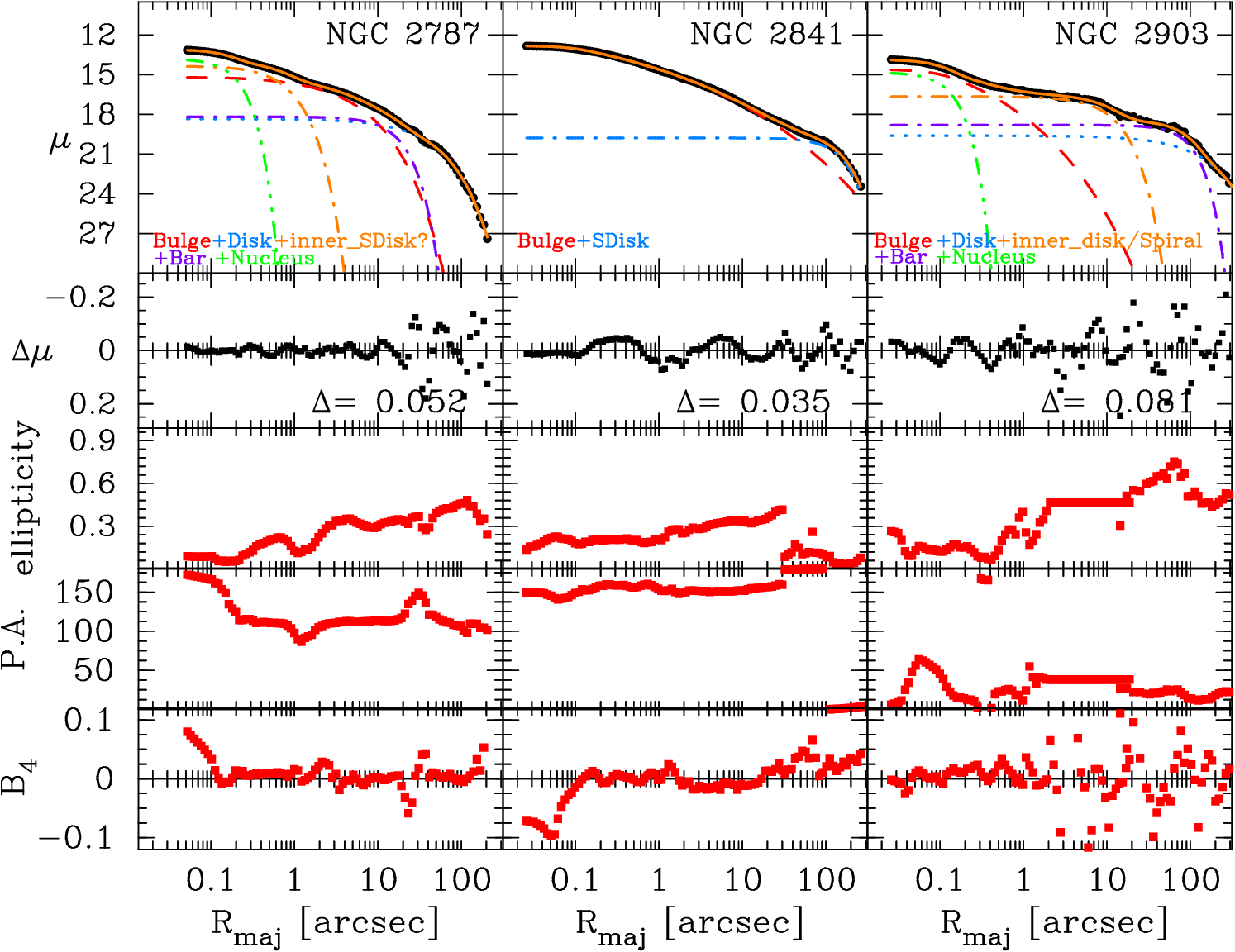}\\
\vspace{.2030cm}
\caption{\it continued.}
\end{figure*}

\begin{figure*}
\setcounter{figure}{1}
\vspace{-.295030cm}
\hspace*{.4811cm}
\includegraphics[angle=270,scale=0.60504]{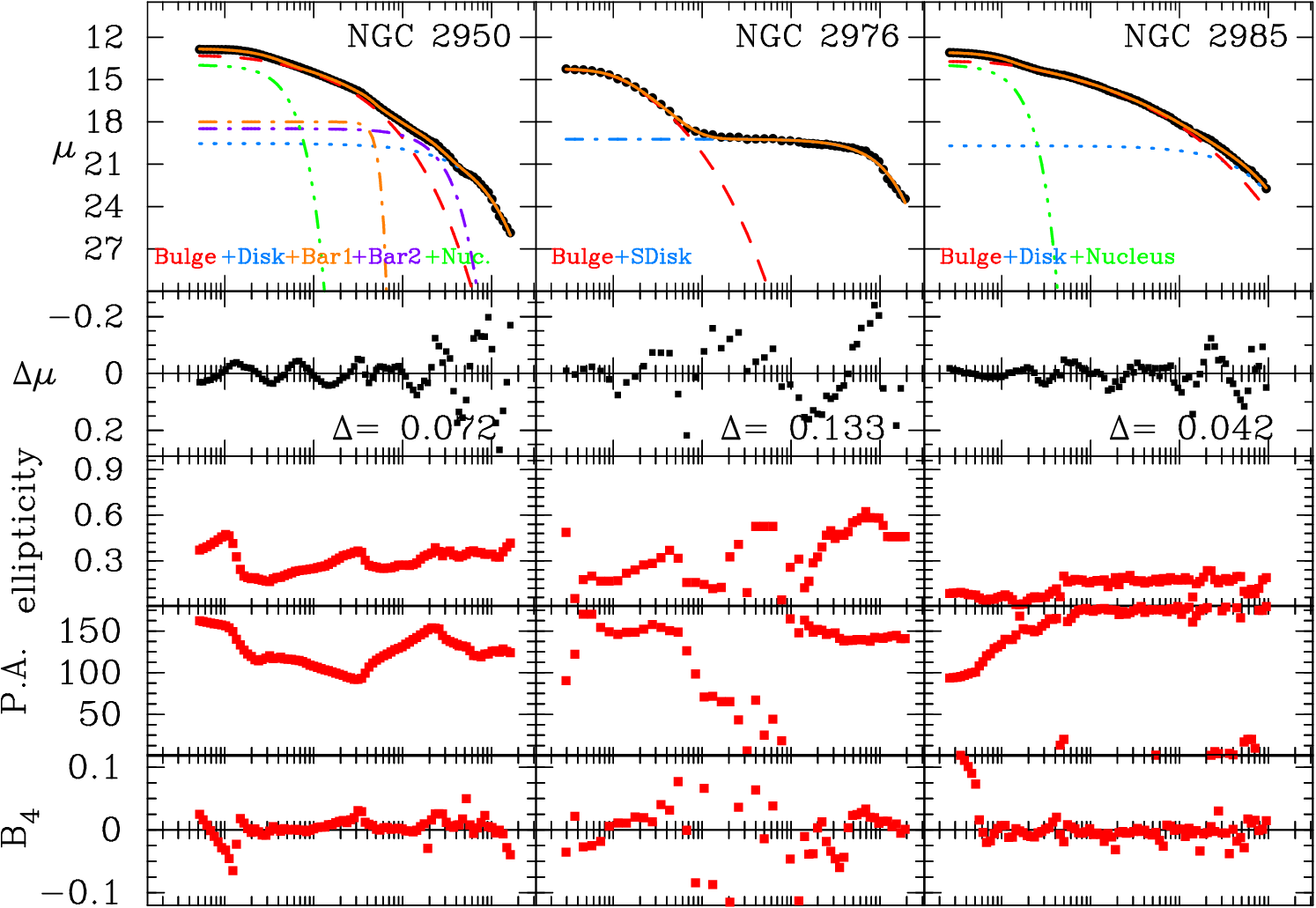}\\
\hspace*{.4811cm}
\includegraphics[angle=270,scale=0.60504]{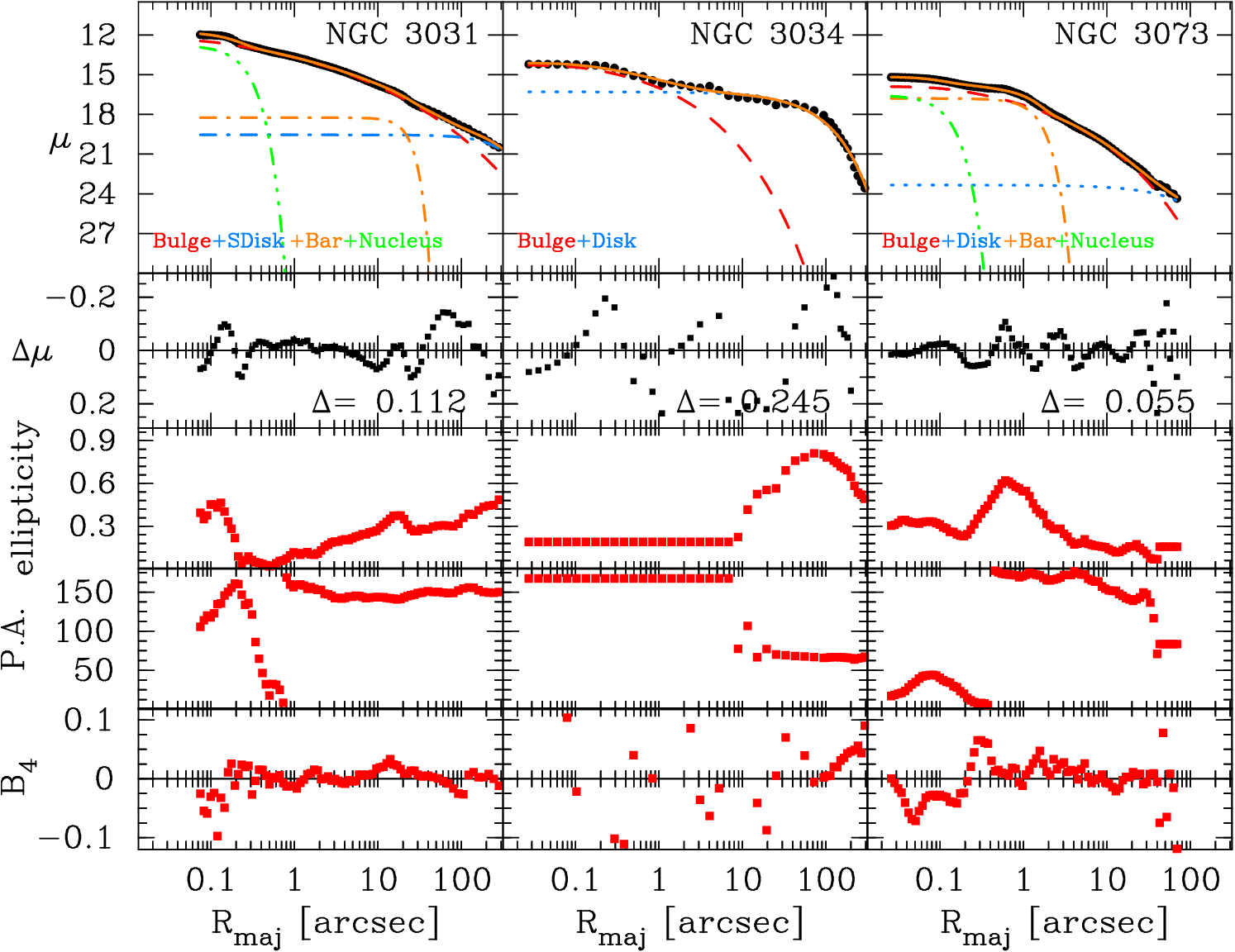}\\
\vspace{.2030cm}
\caption{\it continued.}
\end{figure*}

\begin{figure*}
\setcounter{figure}{1}
\vspace{-.295030cm}
\hspace*{.4811cm}
\includegraphics[angle=270,scale=0.60504]{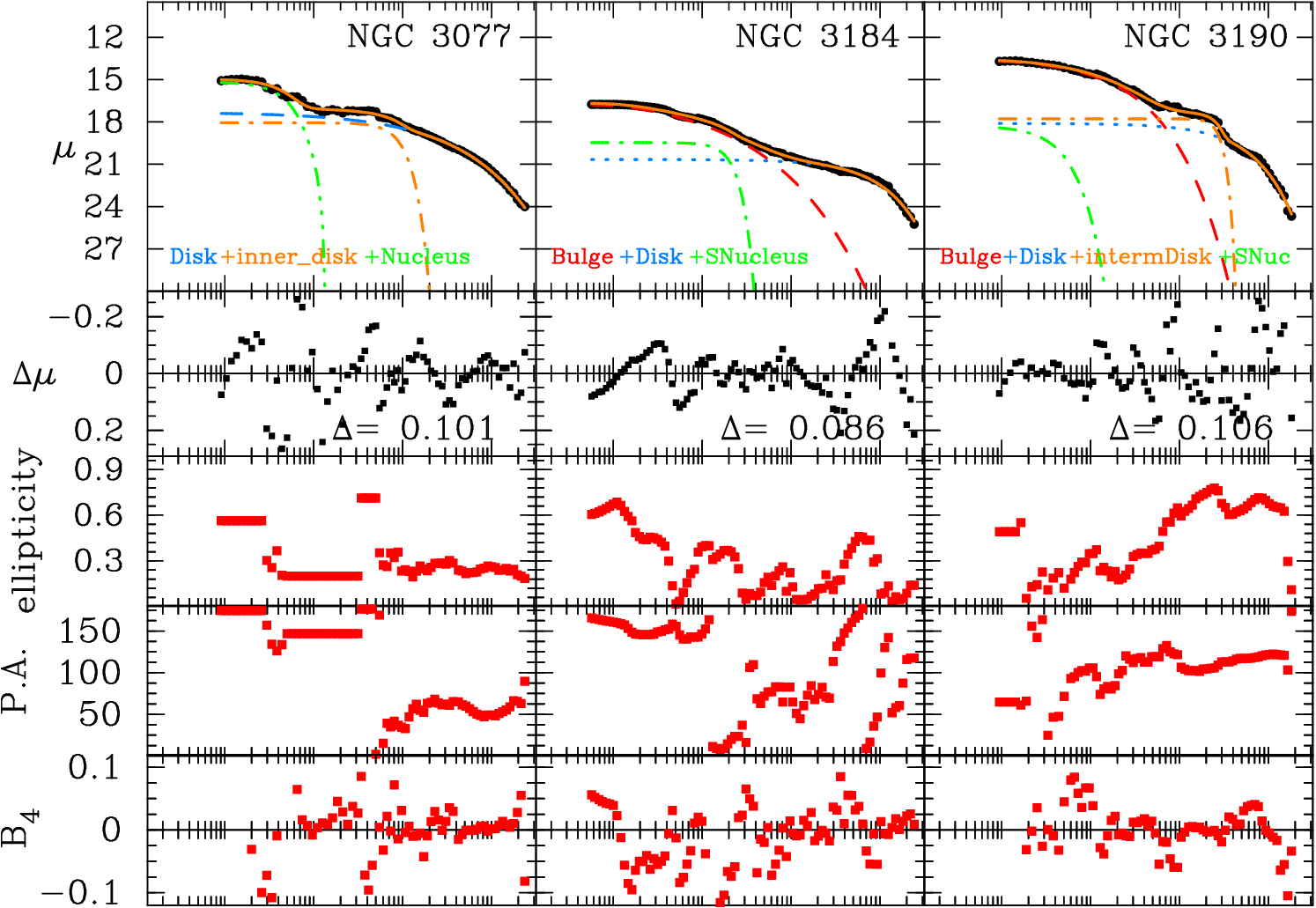}\\
\hspace*{.4811cm}
\includegraphics[angle=270,scale=0.60504]{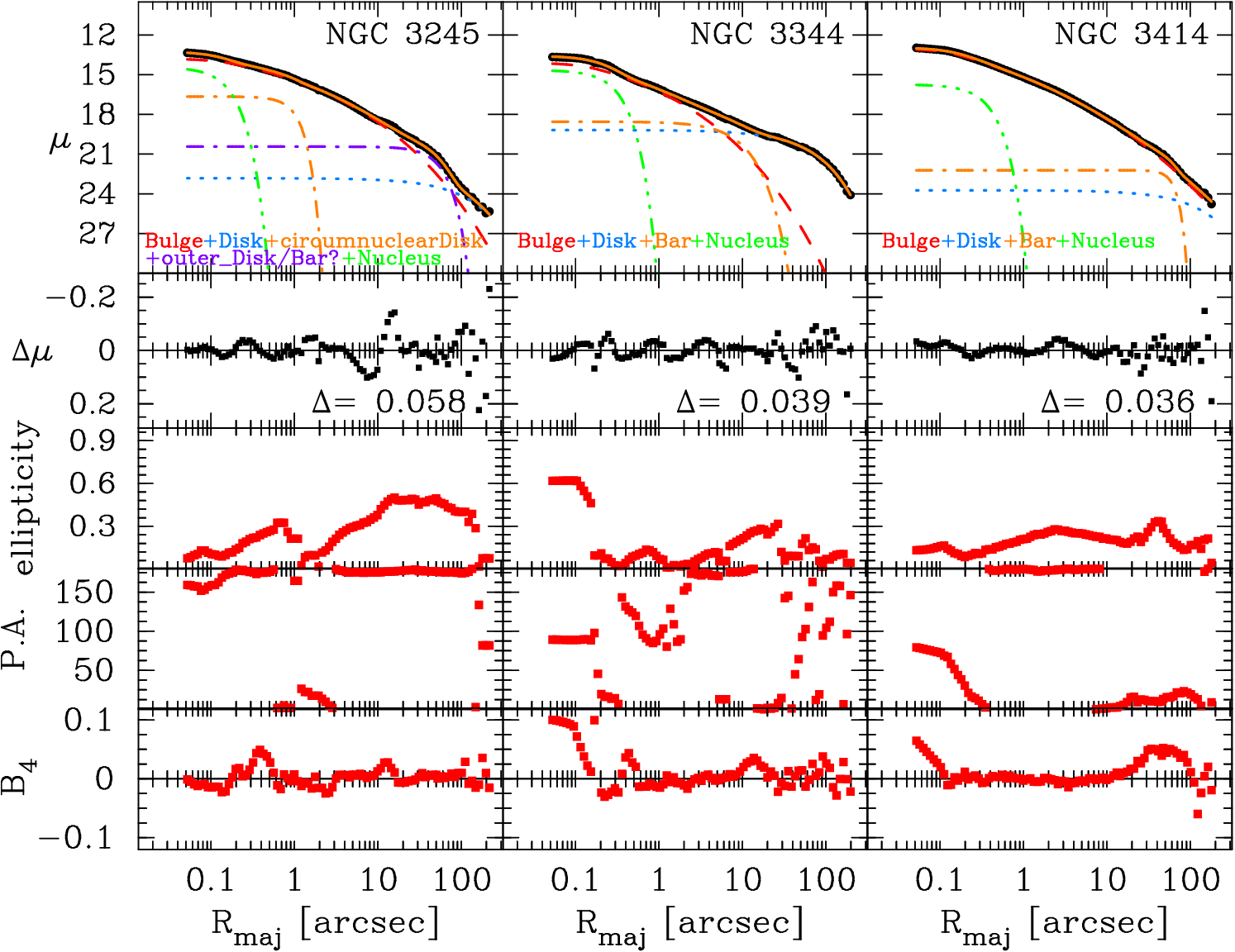}\\
\vspace{.2030cm}
\caption{\it continued.}
\end{figure*}

\begin{figure*}
\setcounter{figure}{1}
\vspace{-.295030cm}
\hspace*{.4811cm}
\includegraphics[angle=270,scale=0.60504]{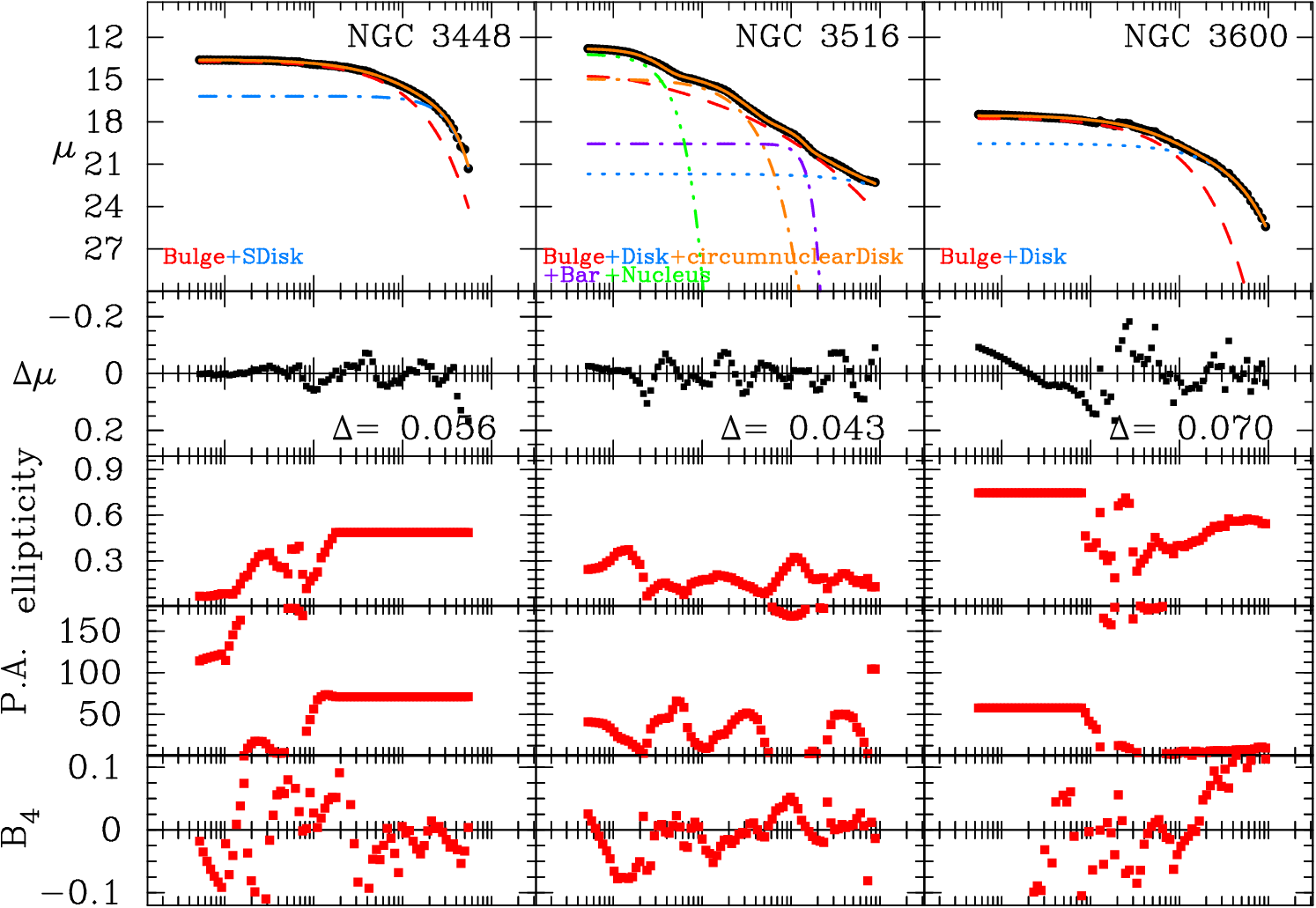}\\
\hspace*{.4811cm}
\includegraphics[angle=270,scale=0.60504]{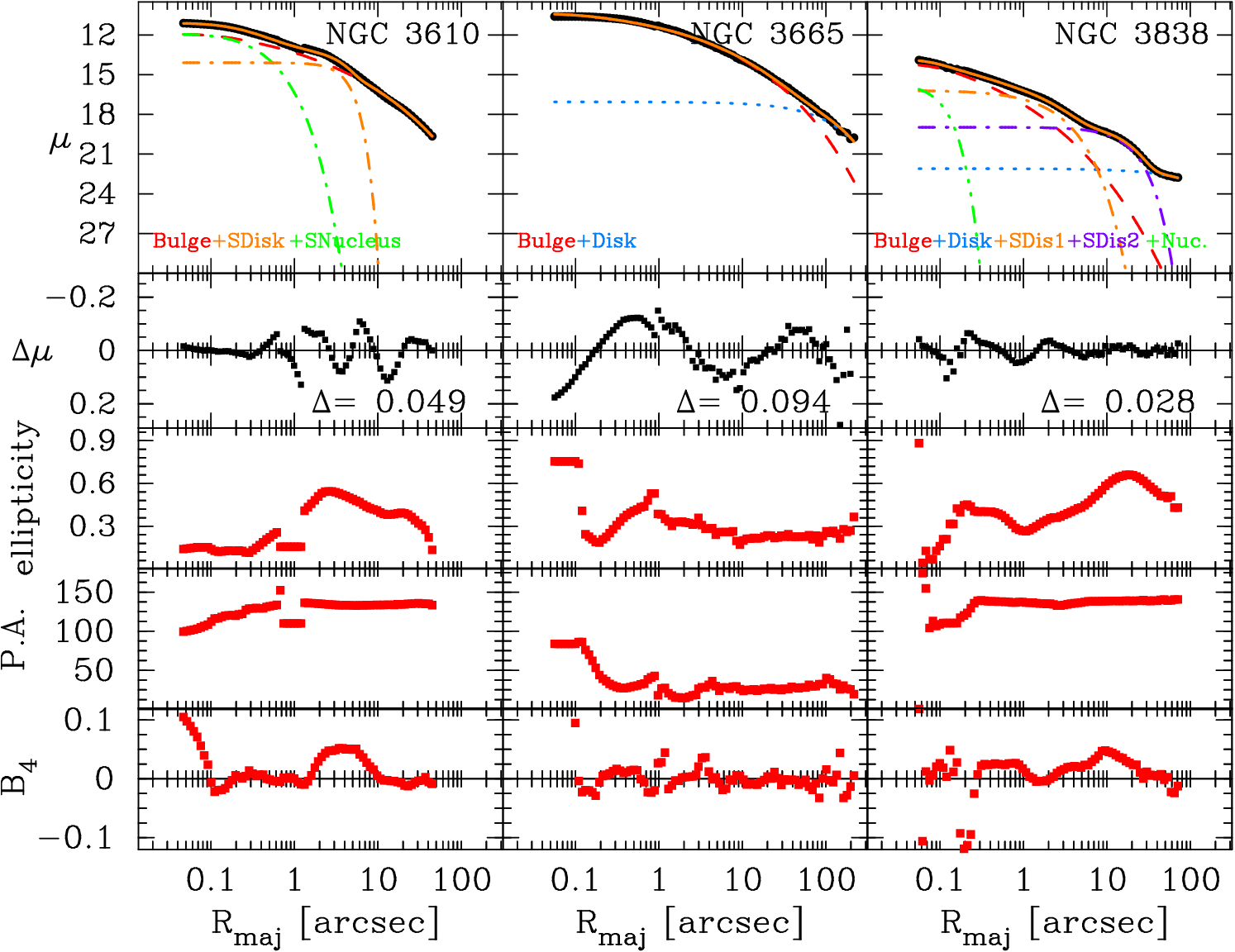}\\
\vspace{.2030cm}
\caption{\it continued.}
\end{figure*}

\begin{figure*}
\setcounter{figure}{1}
\vspace{-.295030cm}
\hspace*{.4811cm}
\includegraphics[angle=270,scale=0.60504]{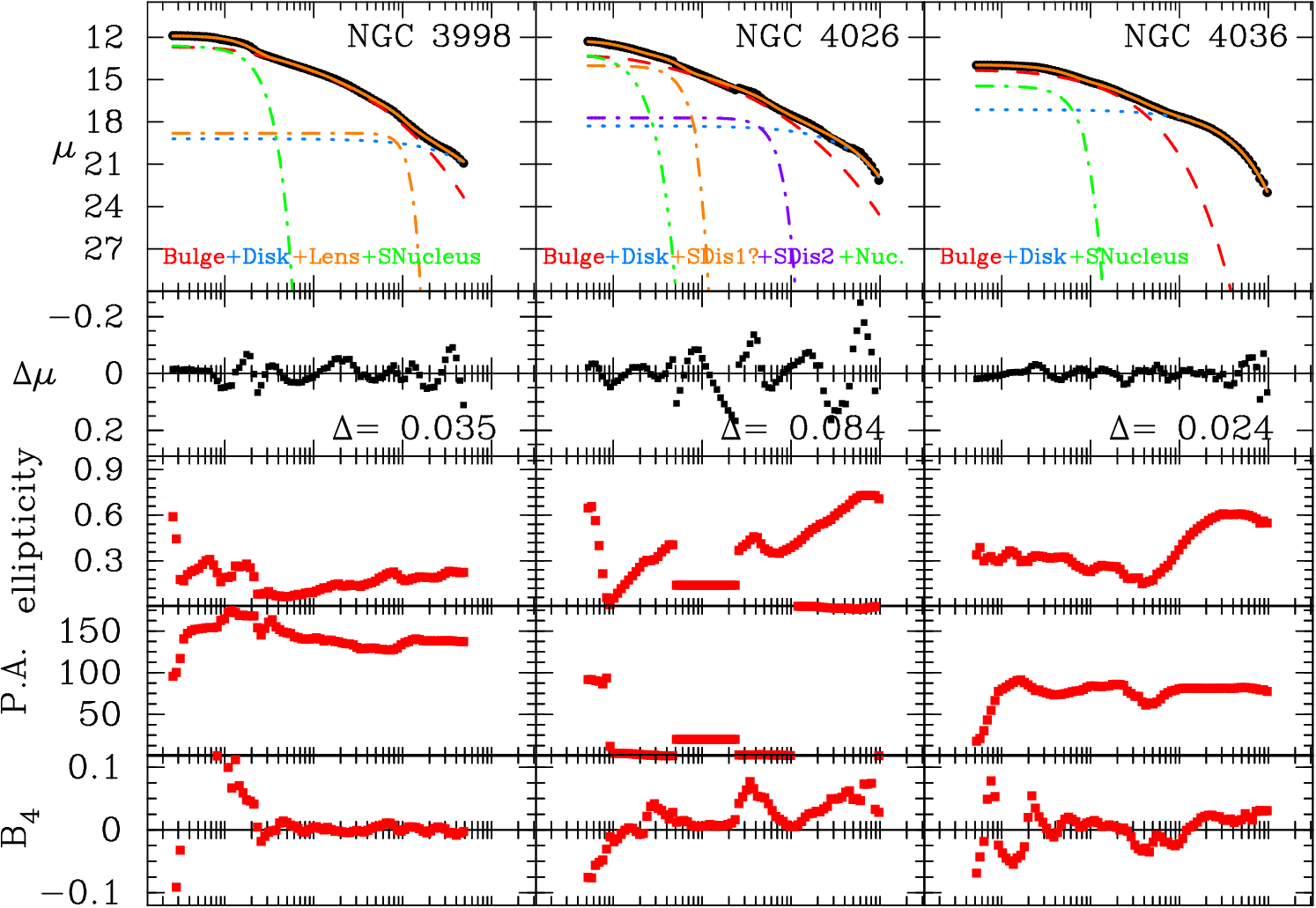}\\
\hspace*{.4811cm}
\includegraphics[angle=270,scale=0.60504]{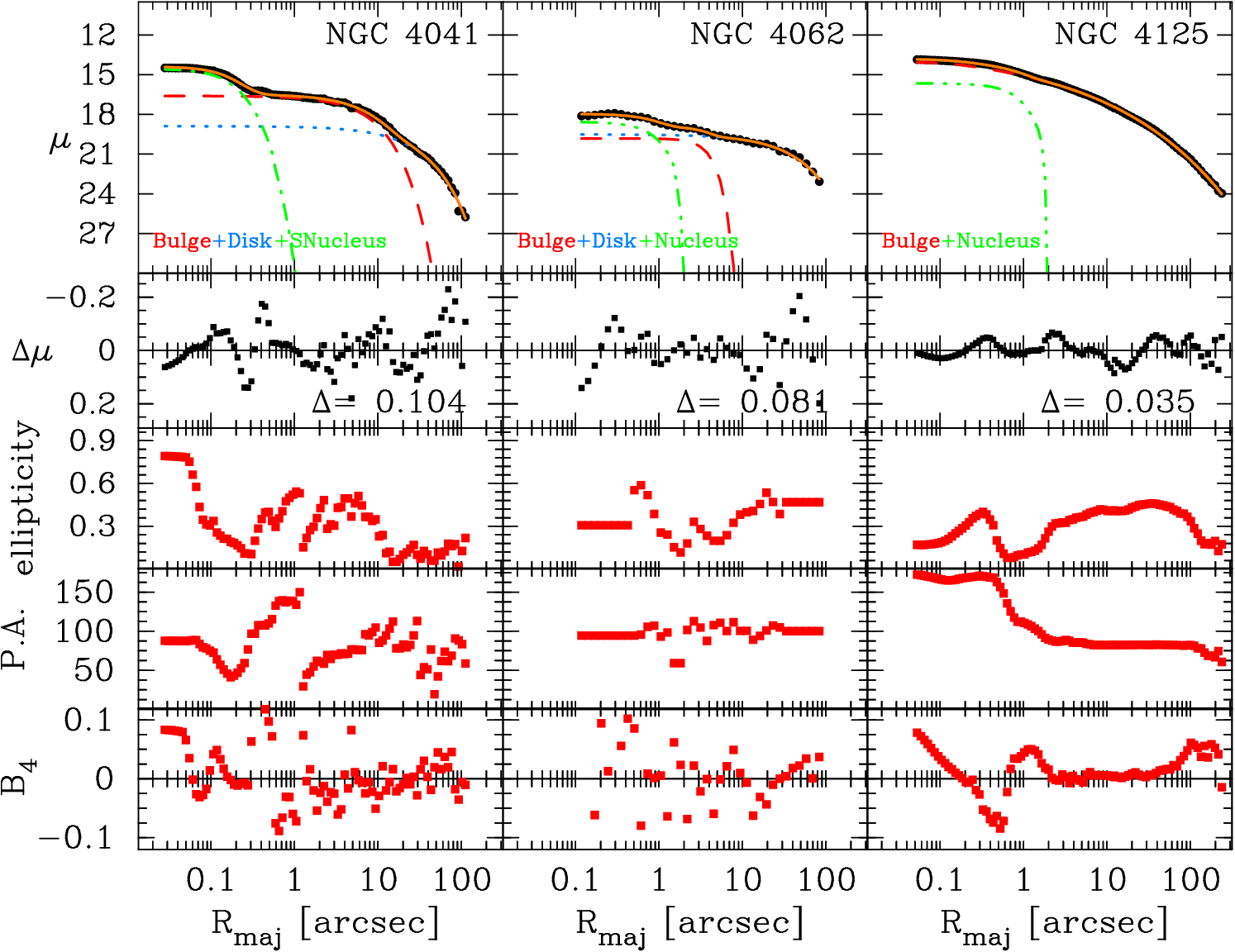}\\
\vspace{.2030cm}
\caption{\it continued.}
\end{figure*}

\begin{figure*}
\setcounter{figure}{1}
\vspace{-.295030cm}
\hspace*{.4811cm}
\includegraphics[angle=270,scale=0.60504]{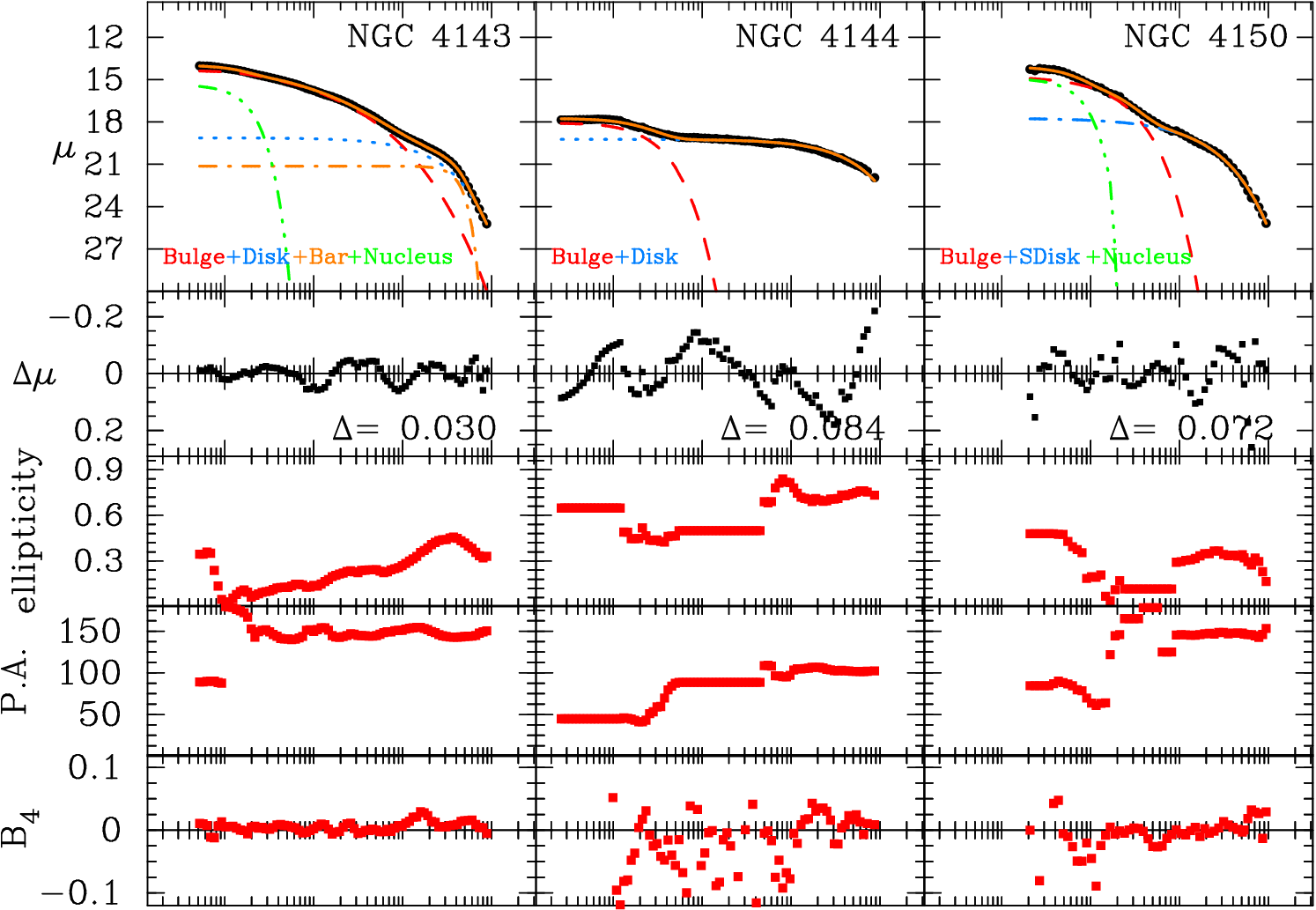}\\
\hspace*{.4811cm}
\includegraphics[angle=270,scale=0.60504]{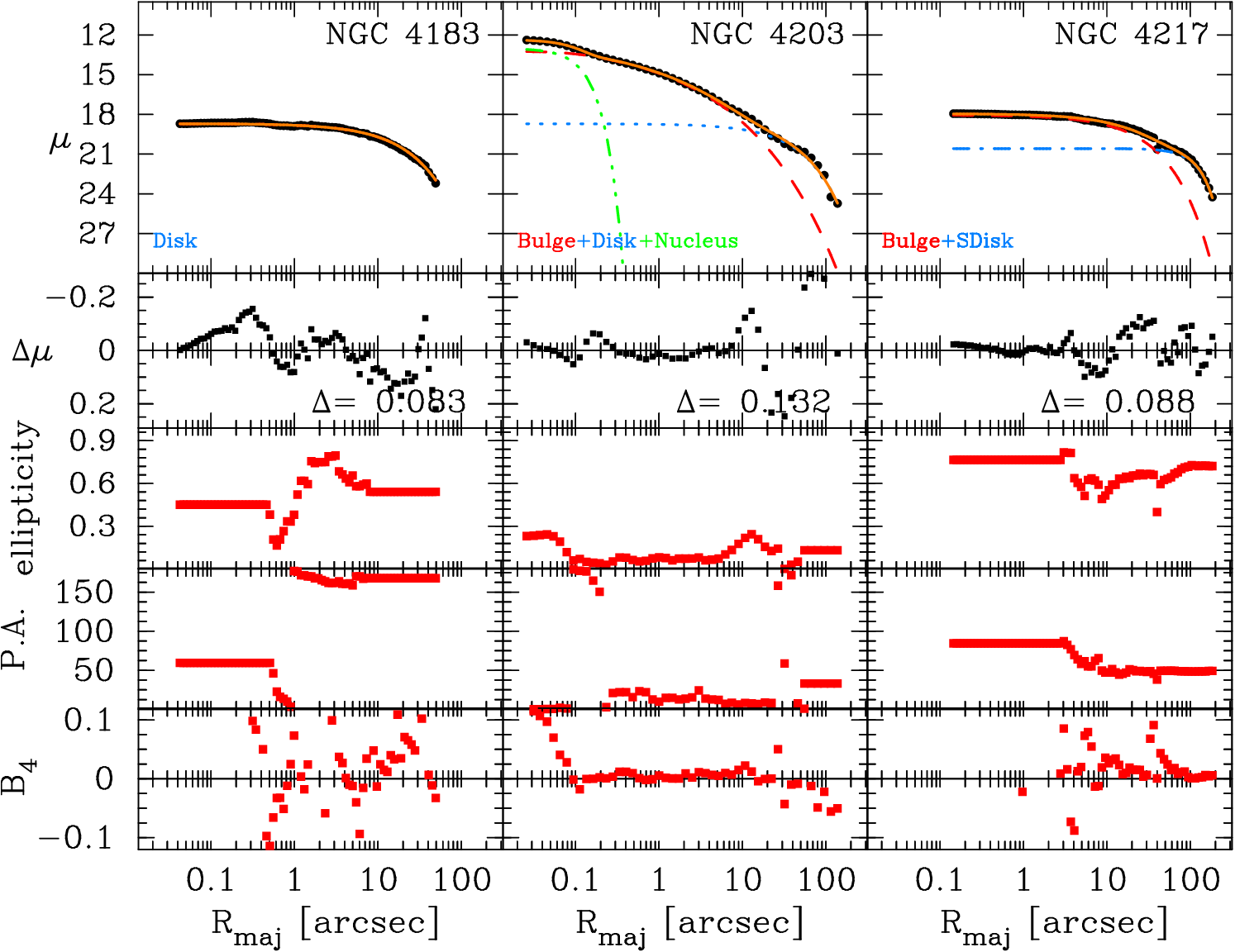}\\
\vspace{.2030cm}
\caption{\it continued.}
\end{figure*}

\begin{figure*}
\setcounter{figure}{1}
\vspace{-.295030cm}
\hspace*{.4811cm}
\includegraphics[angle=270,scale=0.60504]{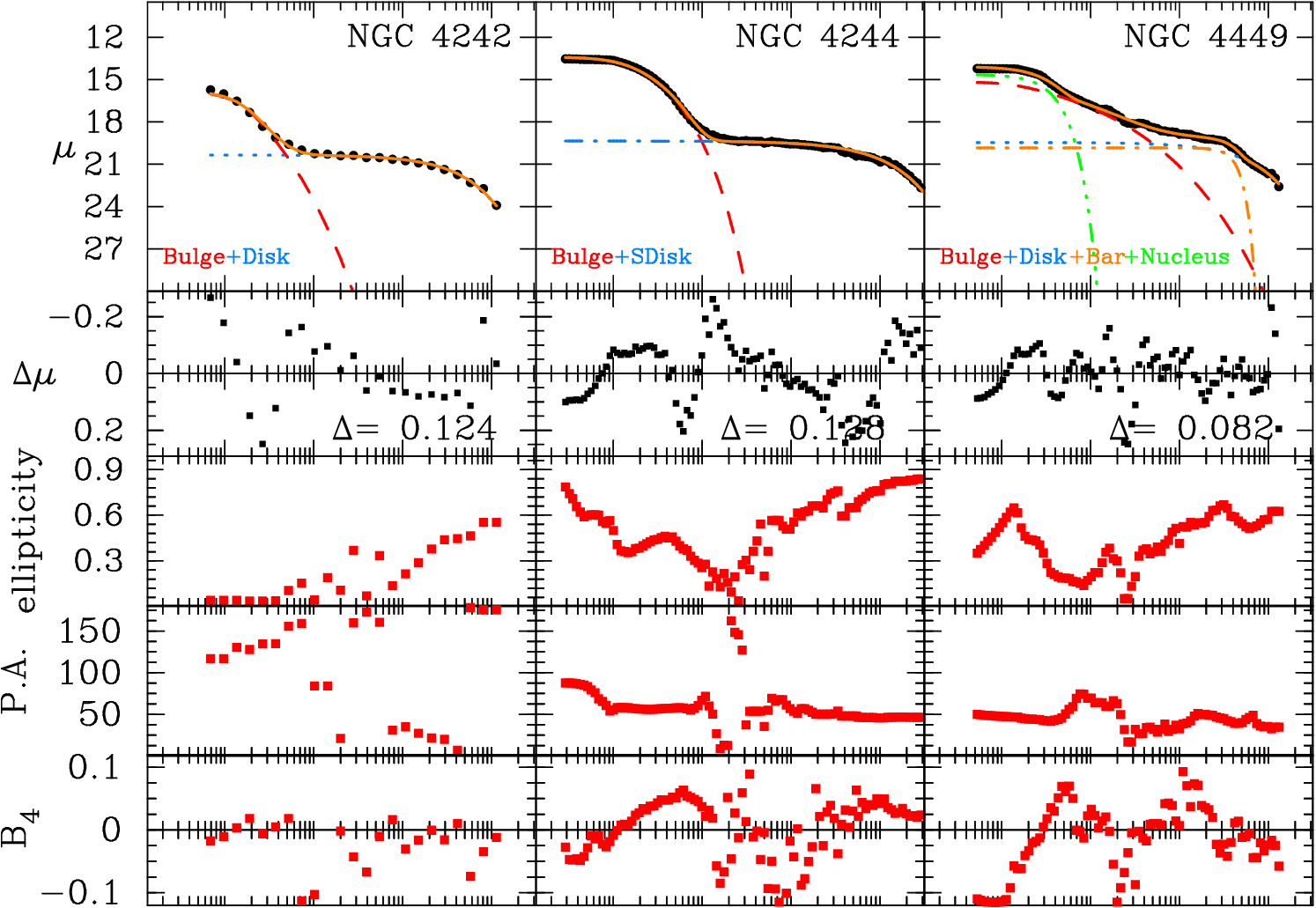}\\
\hspace*{.4811cm}
\includegraphics[angle=270,scale=0.60504]{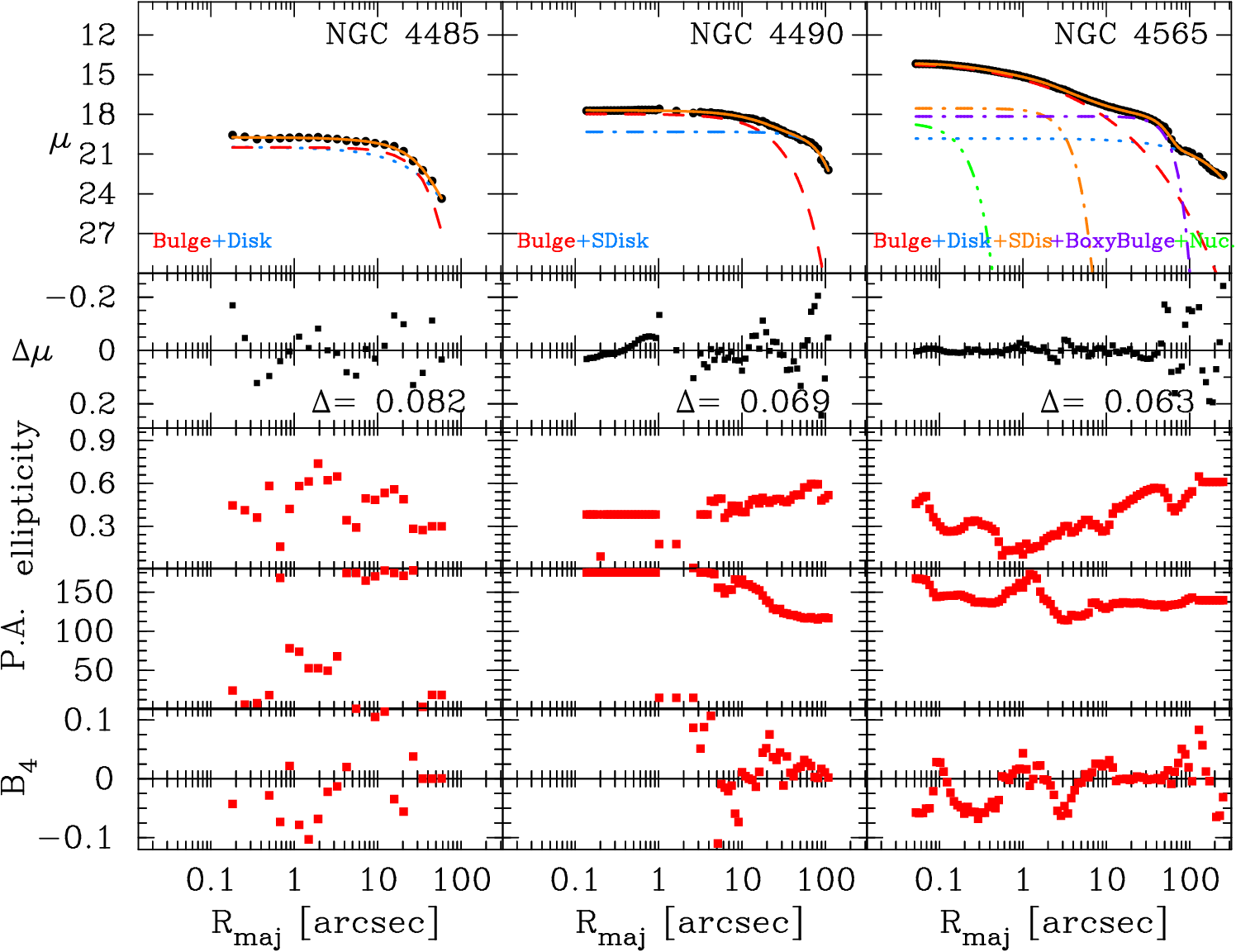}\\
\vspace{.2030cm}
\caption{\it continued.}
\end{figure*}

\begin{figure*}
\setcounter{figure}{1}
\vspace{-.295030cm}
\hspace*{.4811cm}
\includegraphics[angle=270,scale=0.60504]{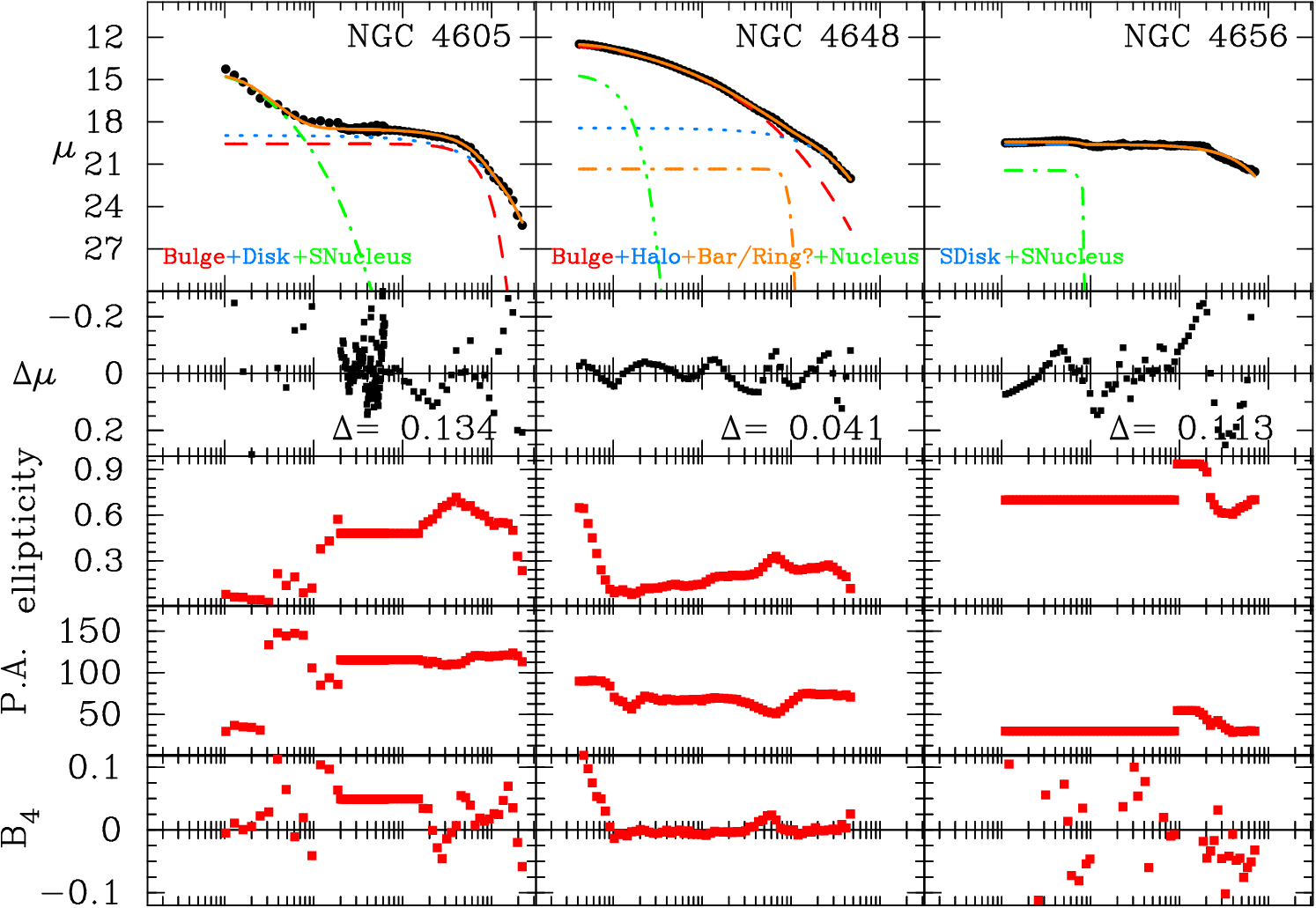}\\
\hspace*{.4811cm}
\includegraphics[angle=270,scale=0.60504]{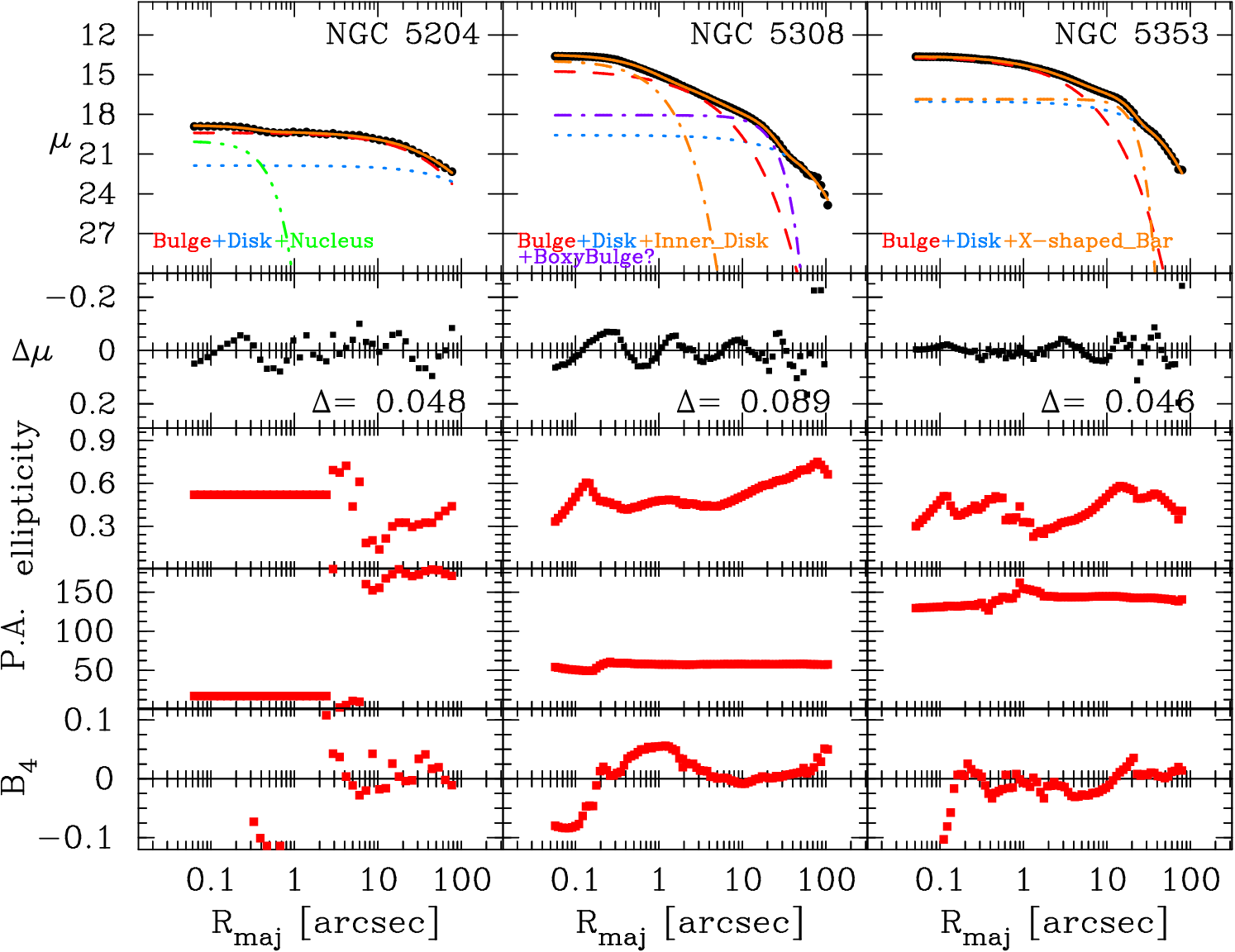}\\
\vspace{.2030cm}
\caption{\it continued.}
\end{figure*}

\begin{figure*}
\setcounter{figure}{1}
\vspace{-.295030cm}
\hspace*{.4811cm}
\includegraphics[angle=270,scale=0.60504]{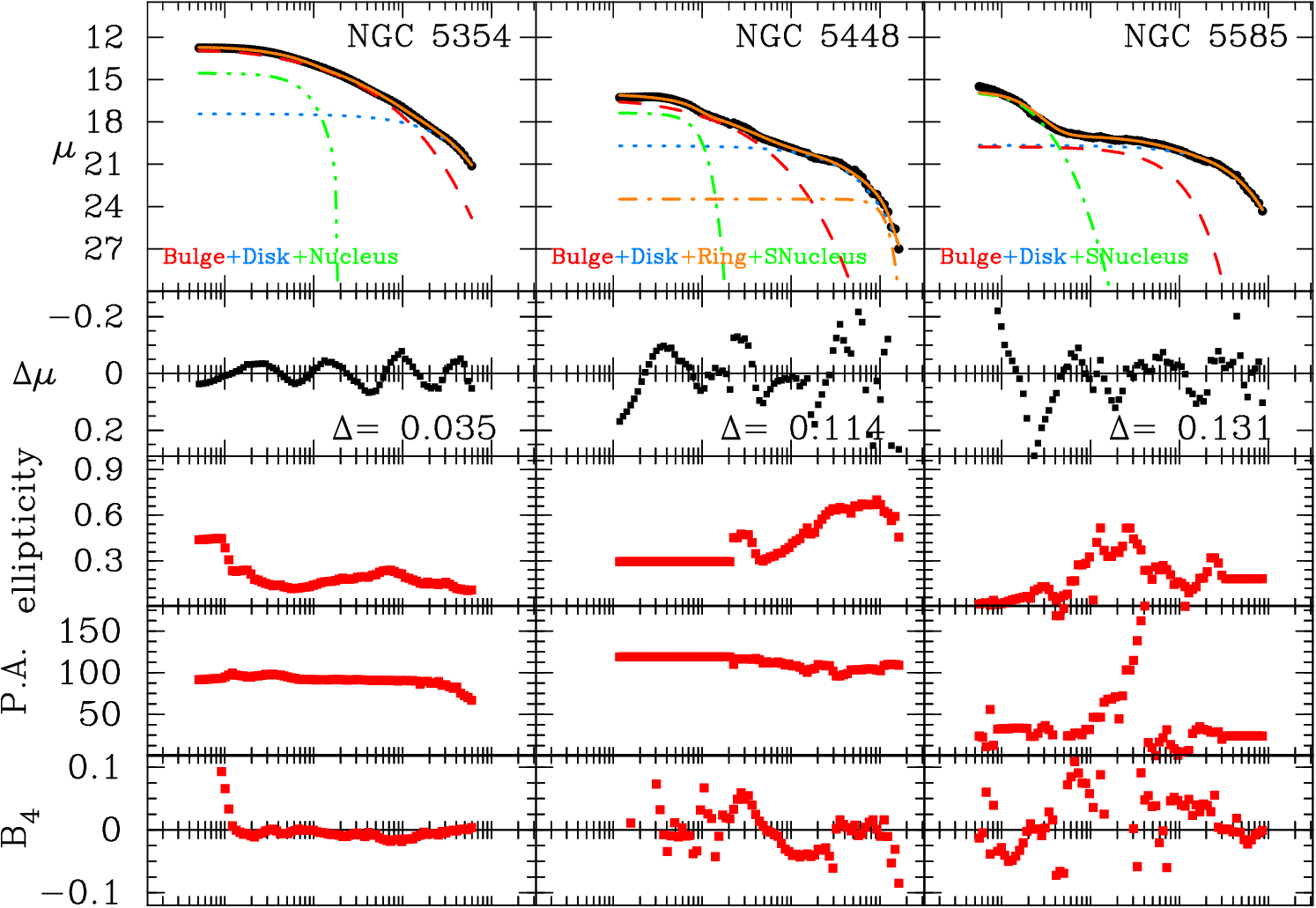}\\
\hspace*{.4811cm}
\includegraphics[angle=270,scale=0.60504]{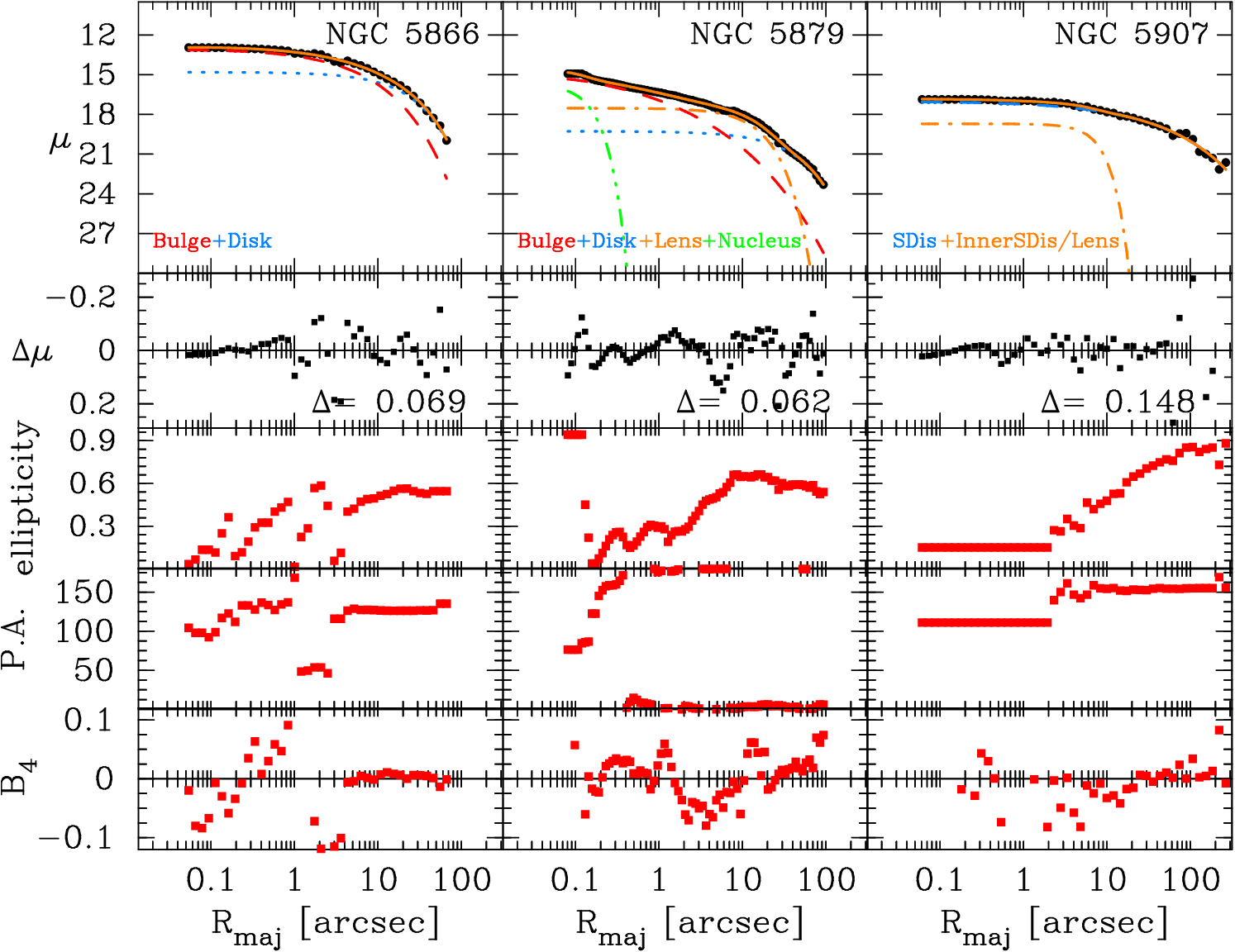}\\
\vspace{.2030cm}
\caption{\it continued.}
\end{figure*}

\begin{figure*}
\setcounter{figure}{1}
\vspace{-.295030cm}
\hspace*{.4811cm}
\includegraphics[angle=270,scale=0.6504]{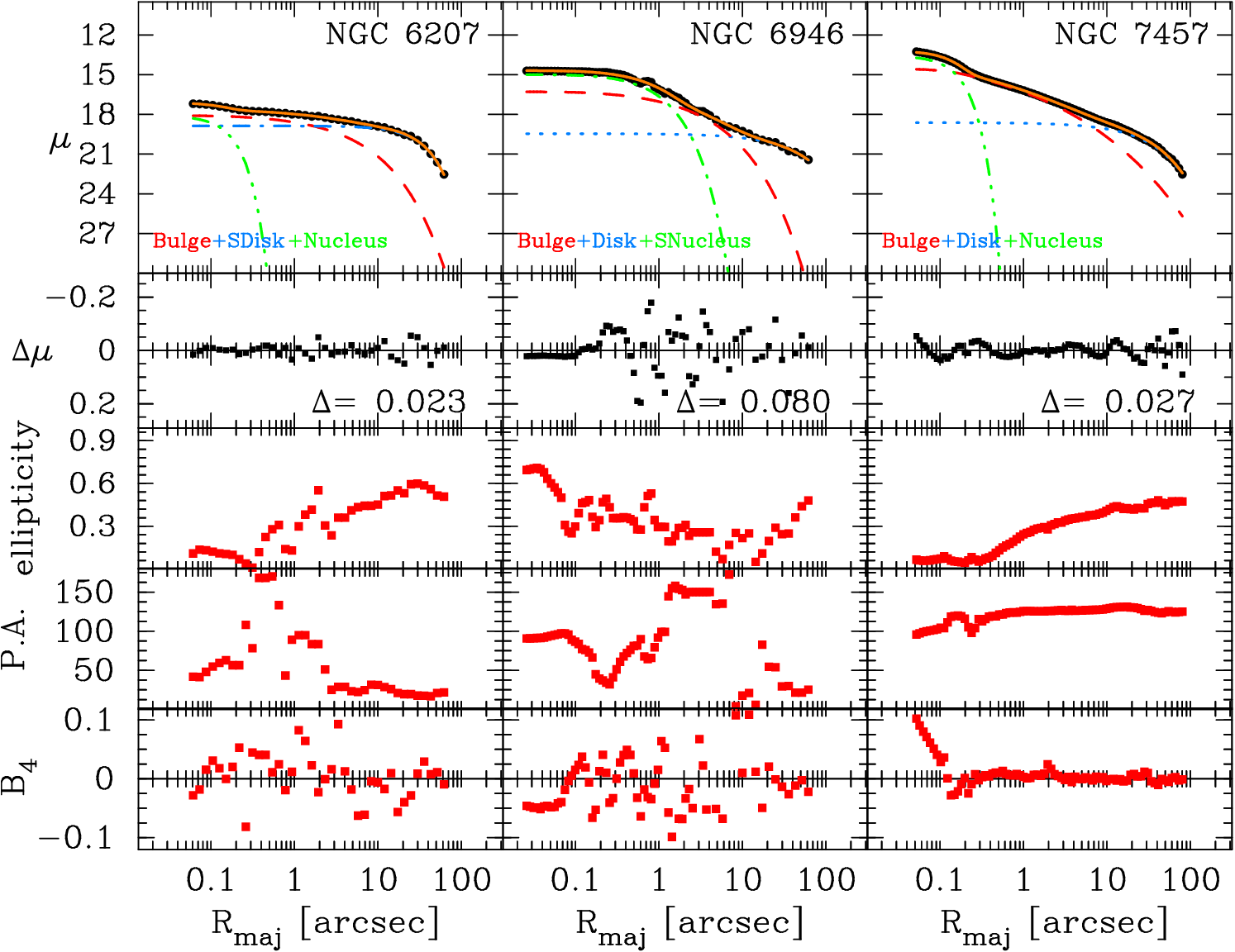}\\
\hspace*{.4811cm}
\includegraphics[angle=270,scale=0.6504]{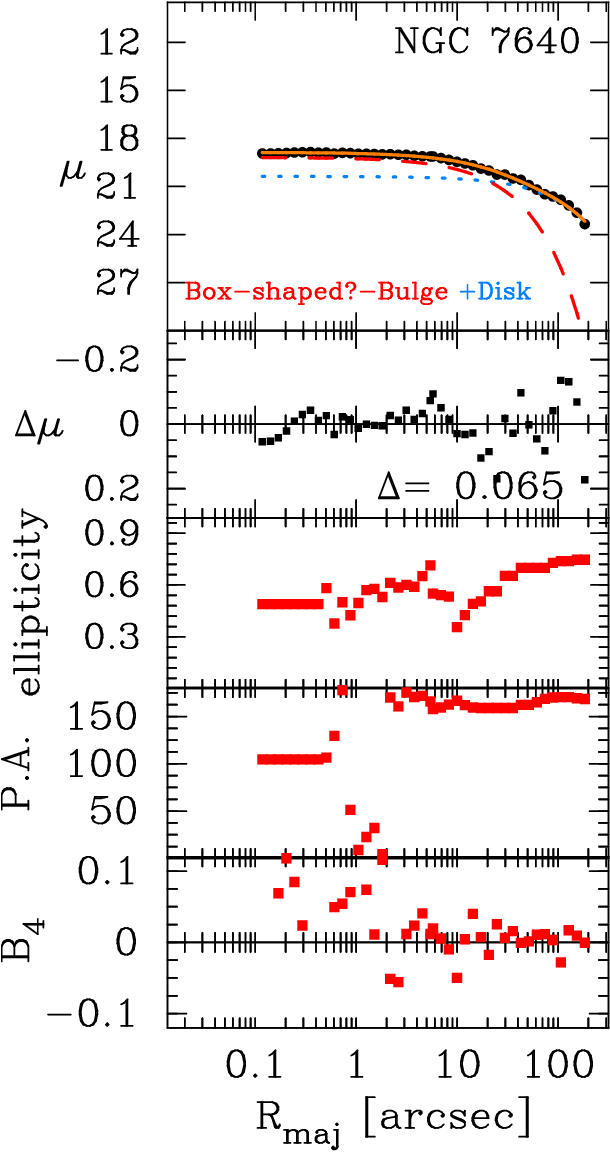}\\
\vspace{.2030cm}
\caption{\it continued.}
\end{figure*}

\begin{figure*}
\vspace{-.295030cm}
\hspace*{.204811cm}
\includegraphics[angle=270,scale=0.634]{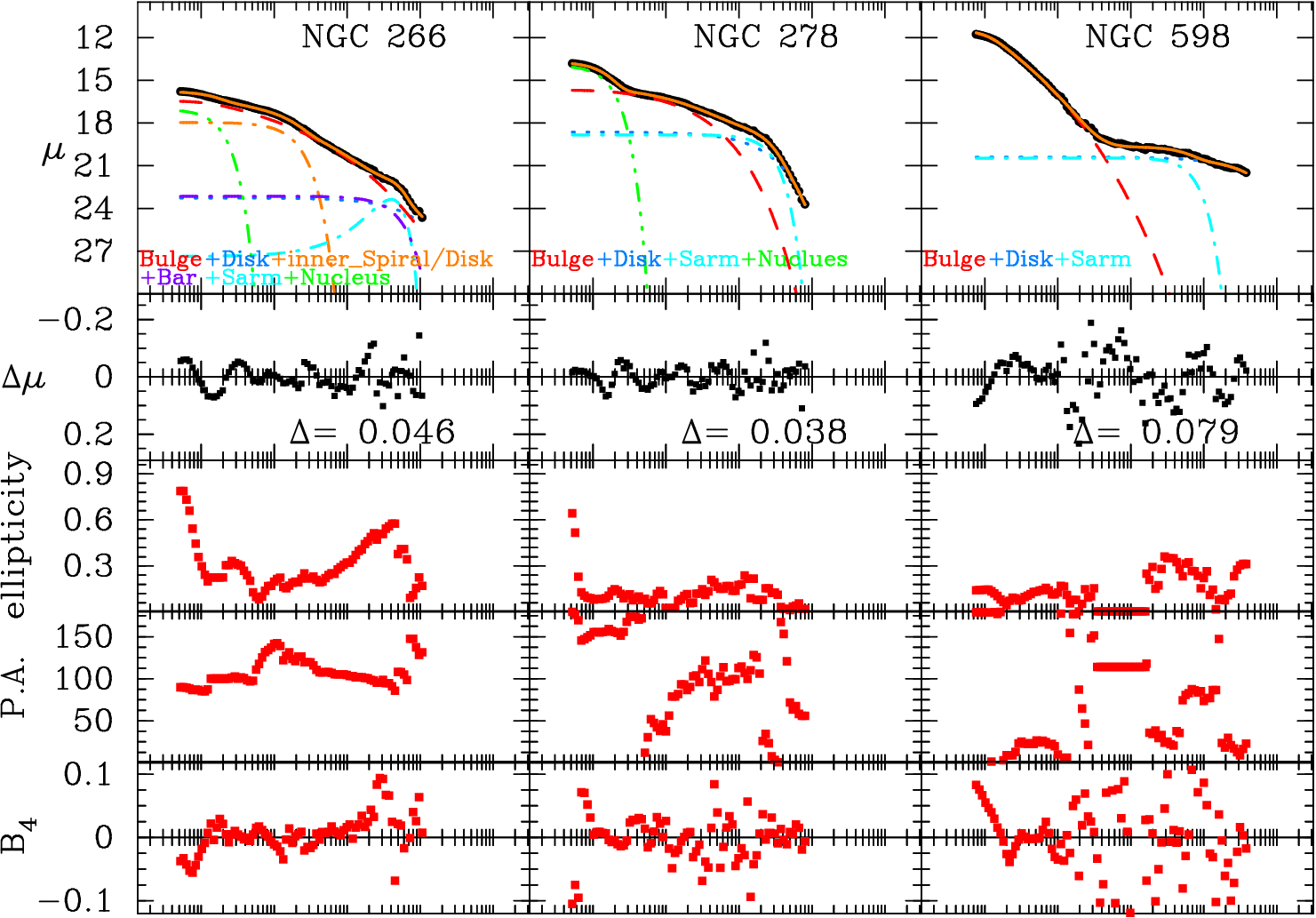}\\
\hspace*{.204811cm}
\includegraphics[angle=270,scale=0.634]{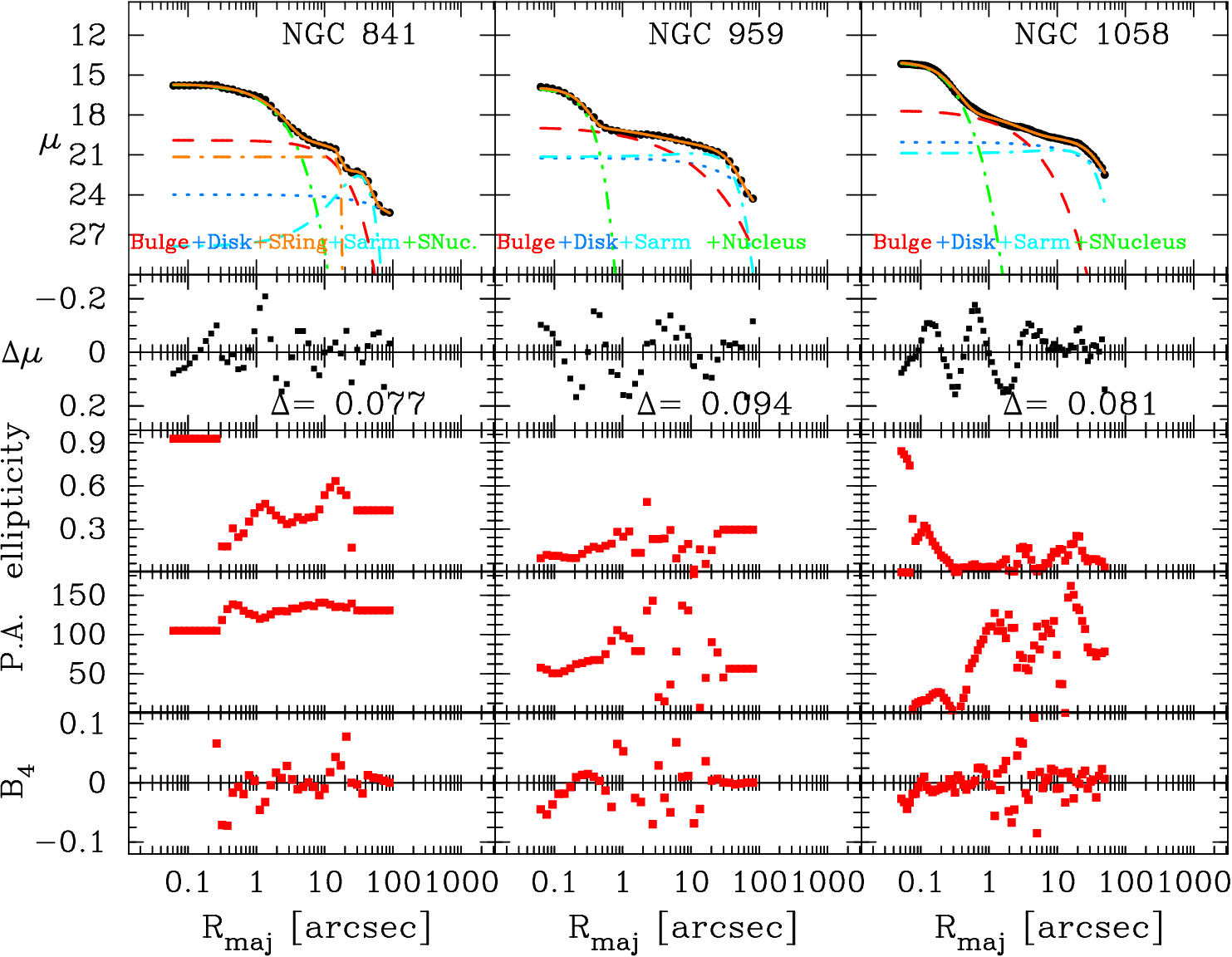}\\
\vspace{.2030cm}
\caption{Similar to Fig.~\ref{FigSer1} but showing here multi-component
  decompositions of S\'ersic
 galaxies that contain a large-scale Gaussian ring component (see  Table~\ref{Tab1Ring}). The three-parameter Gaussian 
 ring model  (dot-dashed cyan curve) gives a good fit to the spiral arm or  ring profiles.}
\label{FigSerR}
\end{figure*}

\begin{figure*}
\setcounter{figure}{2}
\vspace{-.295030cm}
\hspace*{.4811cm}
\includegraphics[angle=270,scale=0.60504]{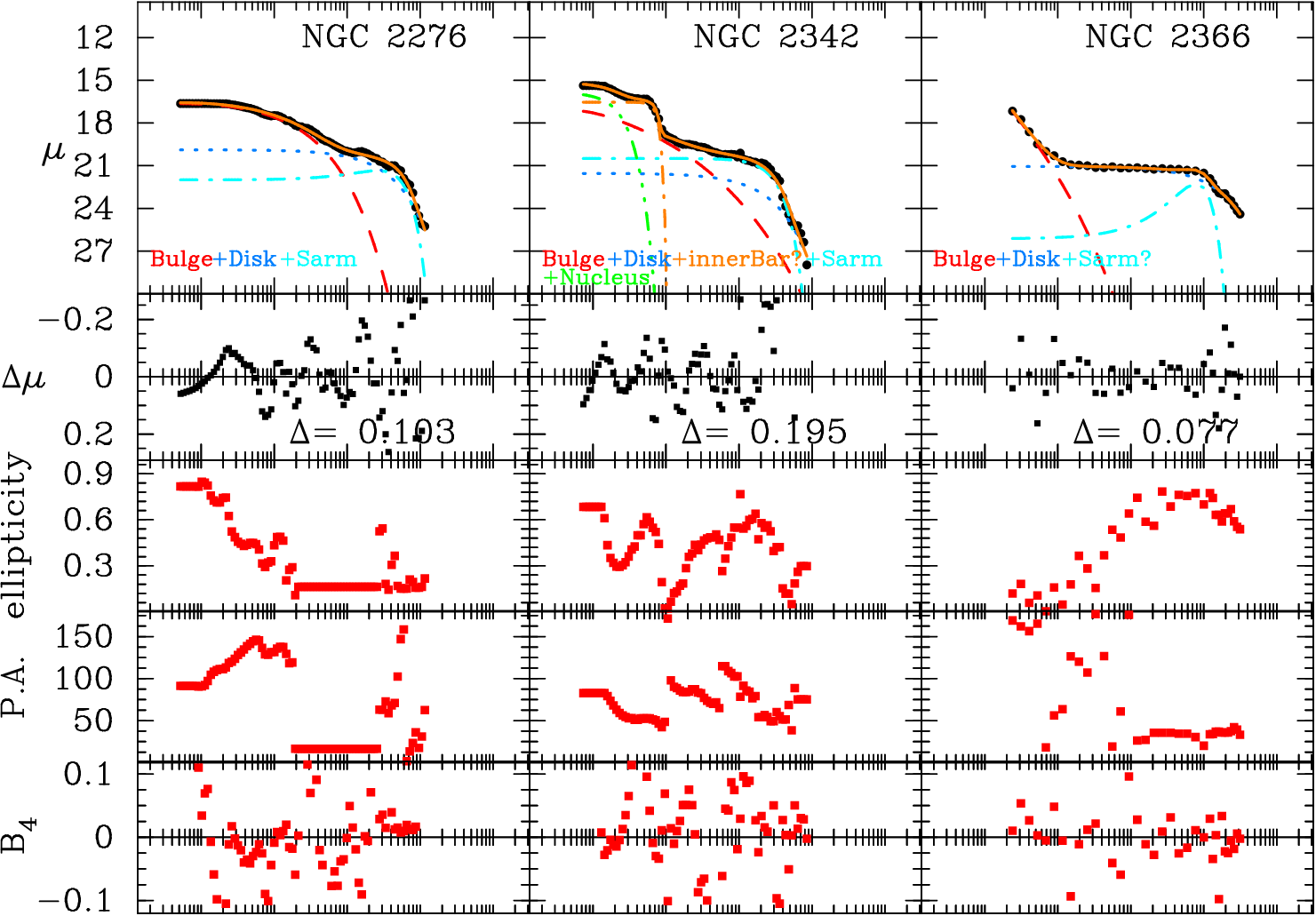}\\
\hspace*{.4811cm}
\includegraphics[angle=270,scale=0.60504]{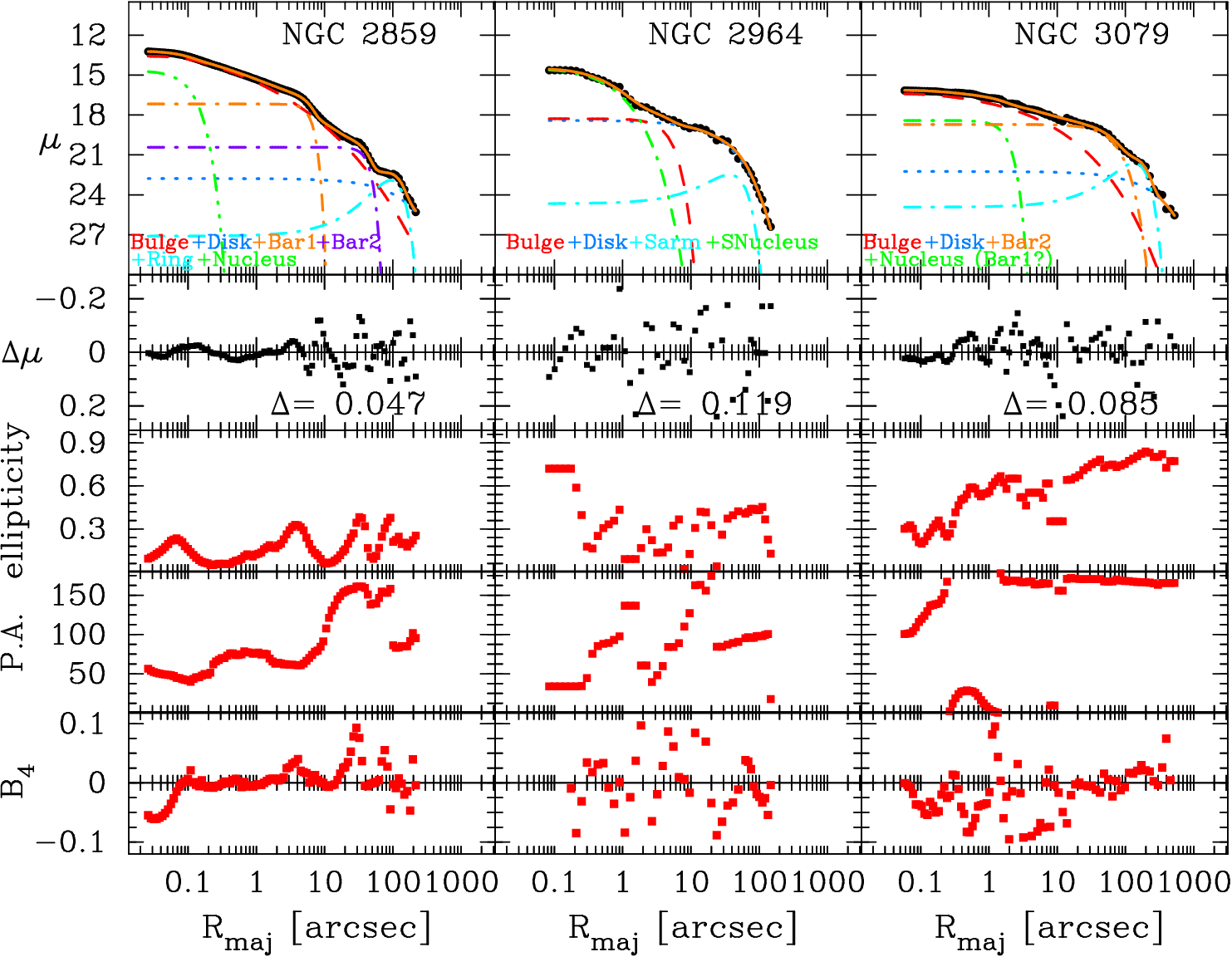}\\
\vspace{.2030cm}
\caption{\it continued.}
\end{figure*}

\begin{figure*}
\setcounter{figure}{2}
\vspace{-.295030cm}
\hspace*{.4811cm}
\includegraphics[angle=270,scale=0.60504]{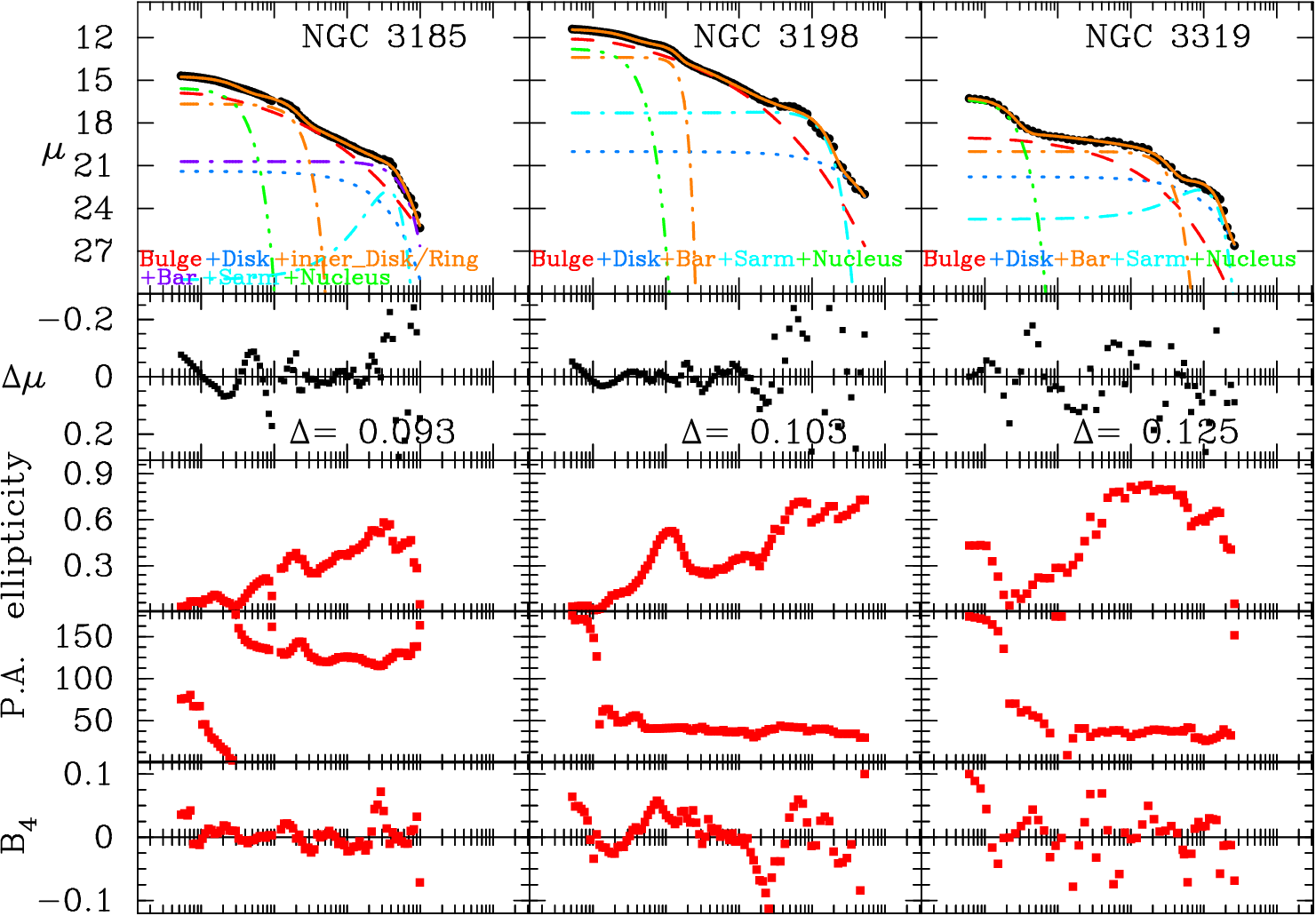}\\
\hspace*{.4811cm}
\includegraphics[angle=270,scale=0.60504]{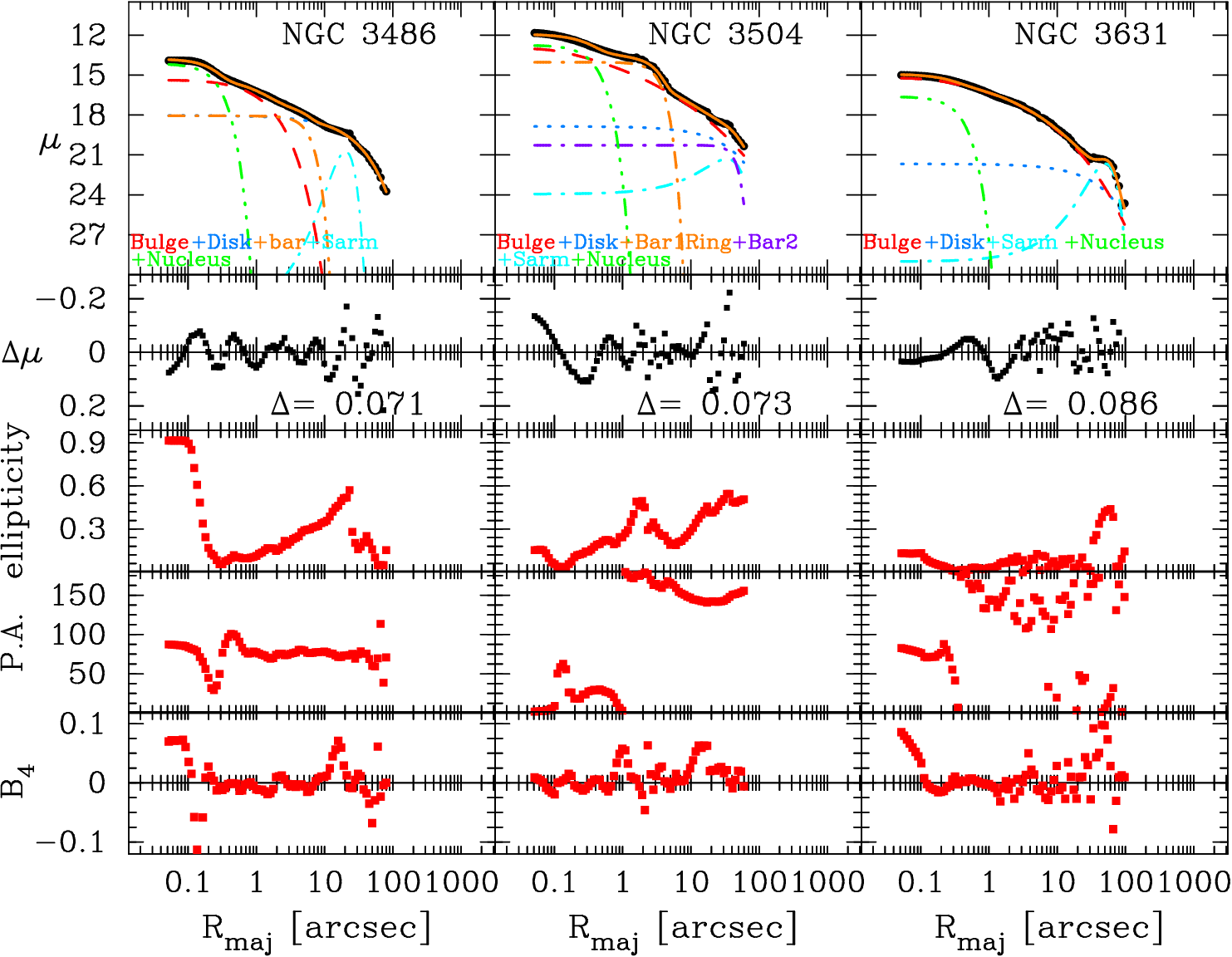}\\
\vspace{.2030cm}
\caption{\it continued.}
\end{figure*}

\begin{figure*}
\setcounter{figure}{2}
\vspace{-.295030cm}
\hspace*{.4811cm}
\includegraphics[angle=270,scale=0.60504]{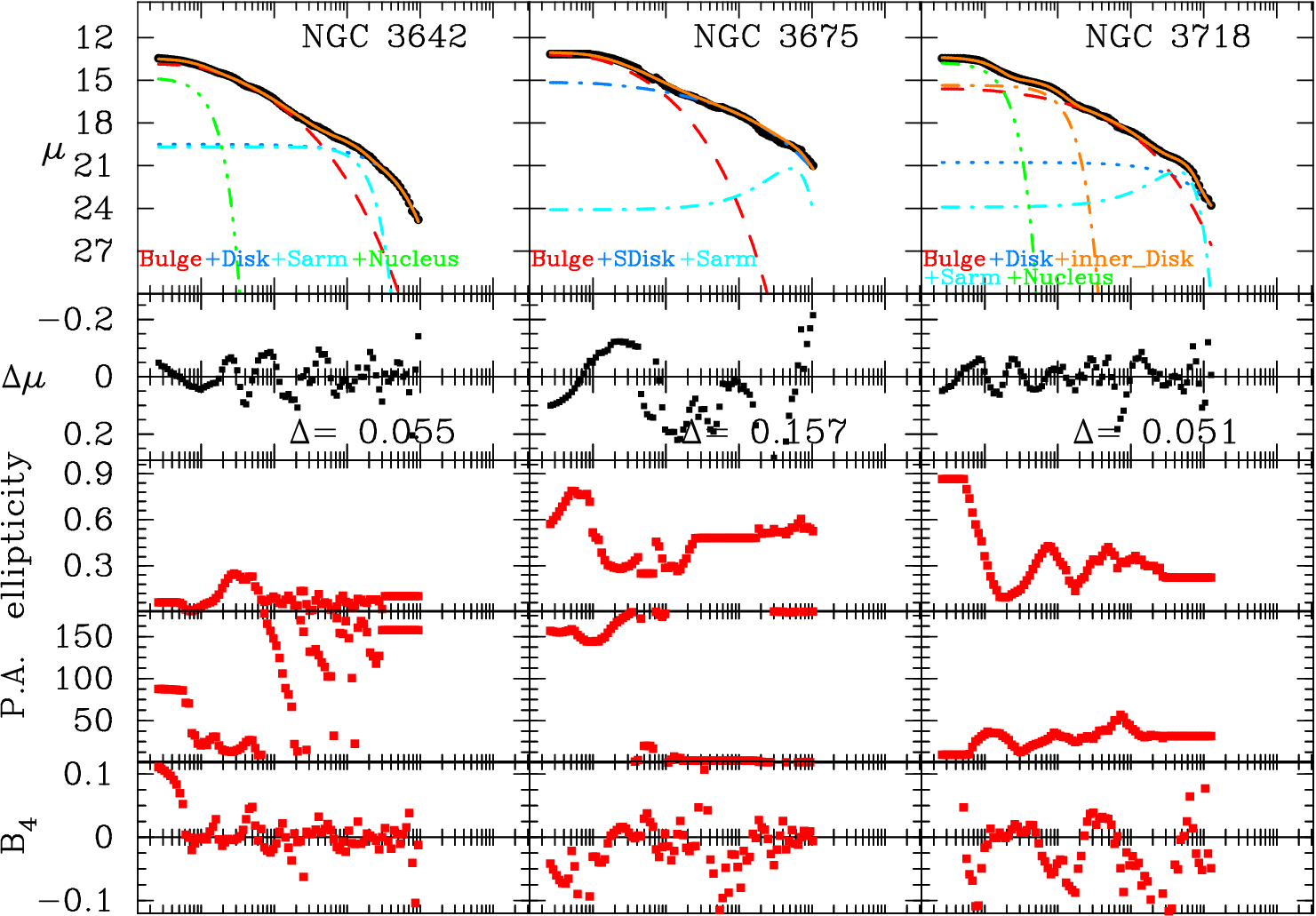}\\
\hspace*{.4811cm}
\includegraphics[angle=270,scale=0.60504]{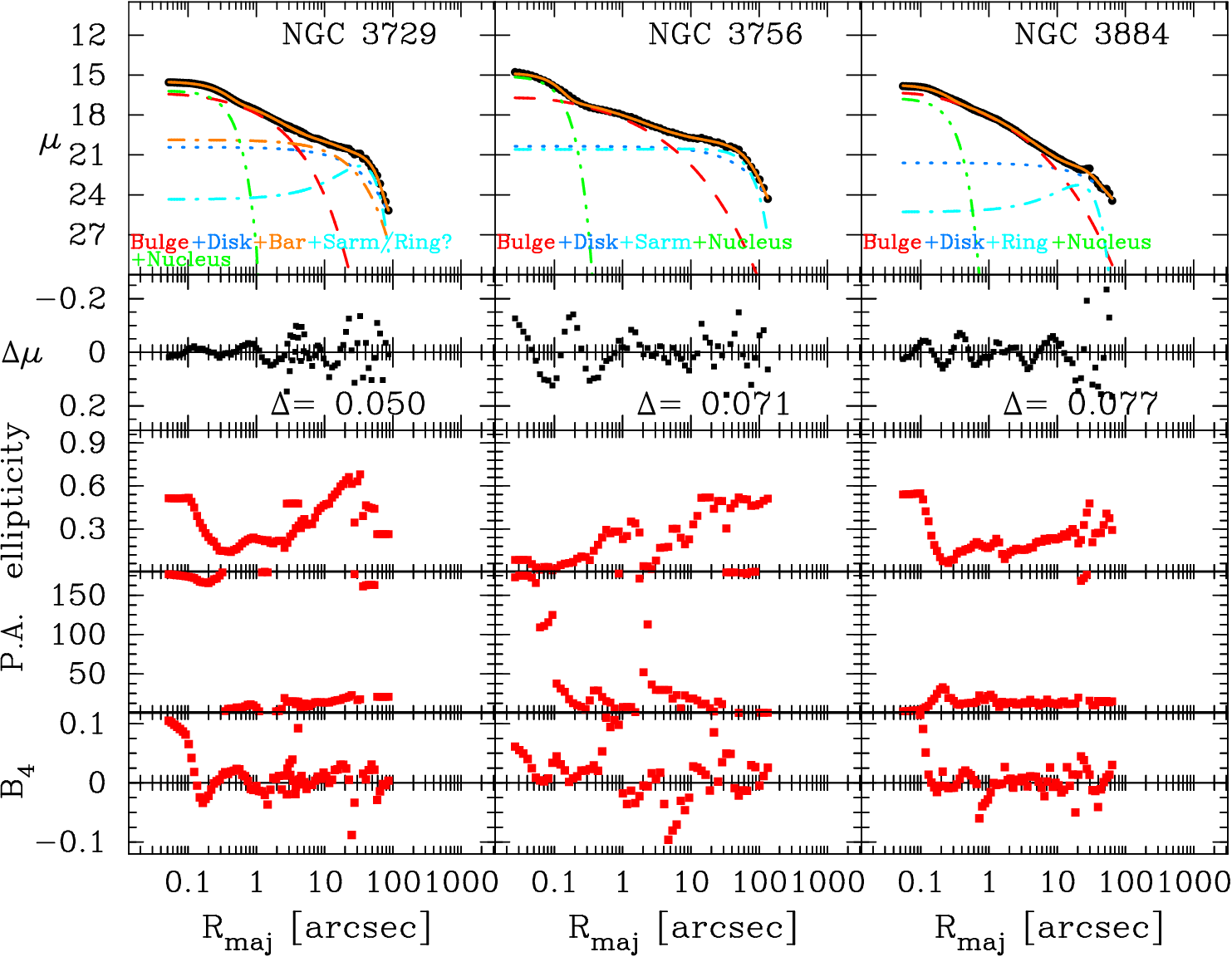}\\
\vspace{.2030cm}
\caption{\it continued.}
\end{figure*}

\begin{figure*}
\setcounter{figure}{2}
\vspace{-.295030cm}
\hspace*{.4811cm}
\includegraphics[angle=270,scale=0.60504]{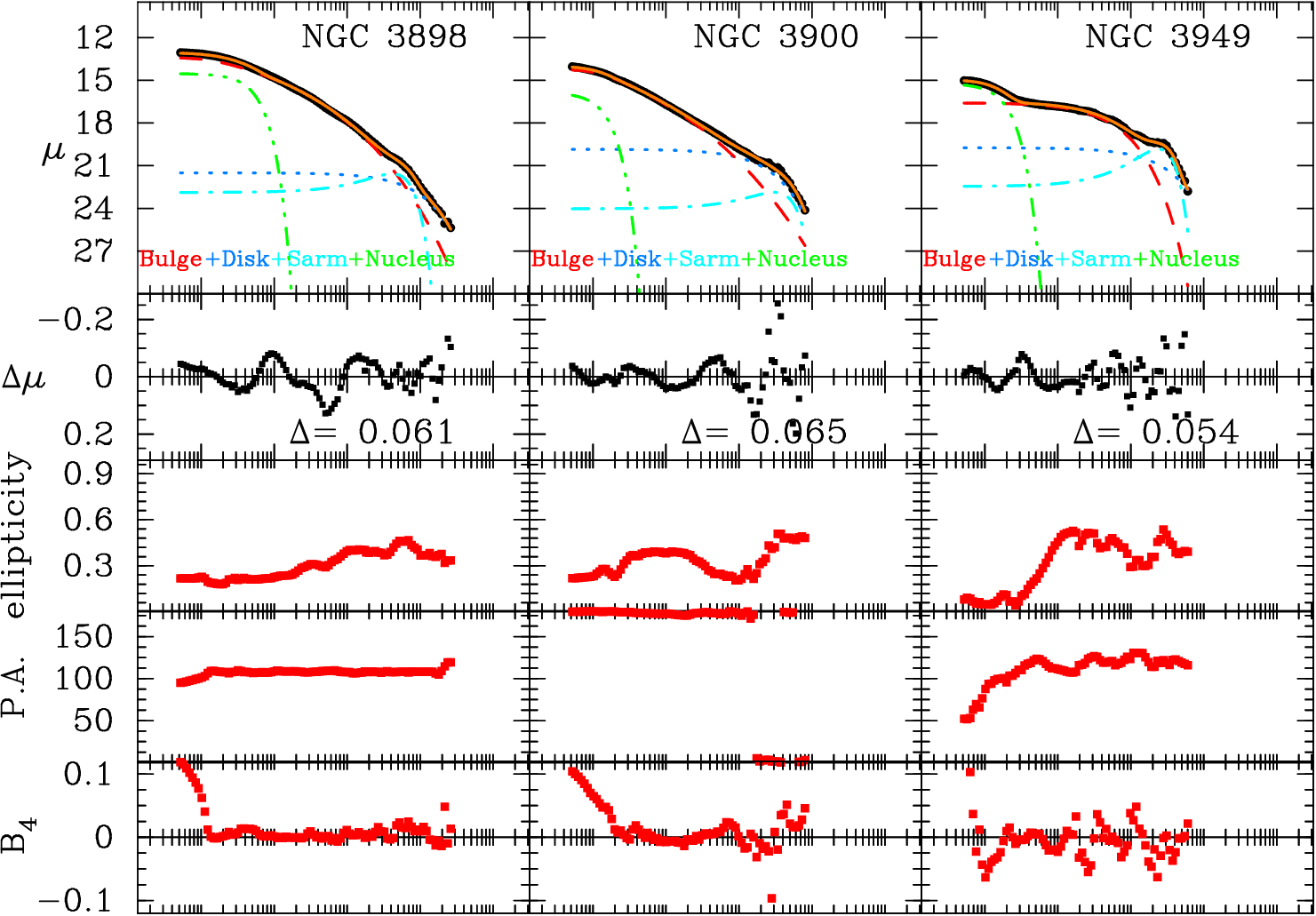}\\
\hspace*{.4811cm}
\includegraphics[angle=270,scale=0.60504]{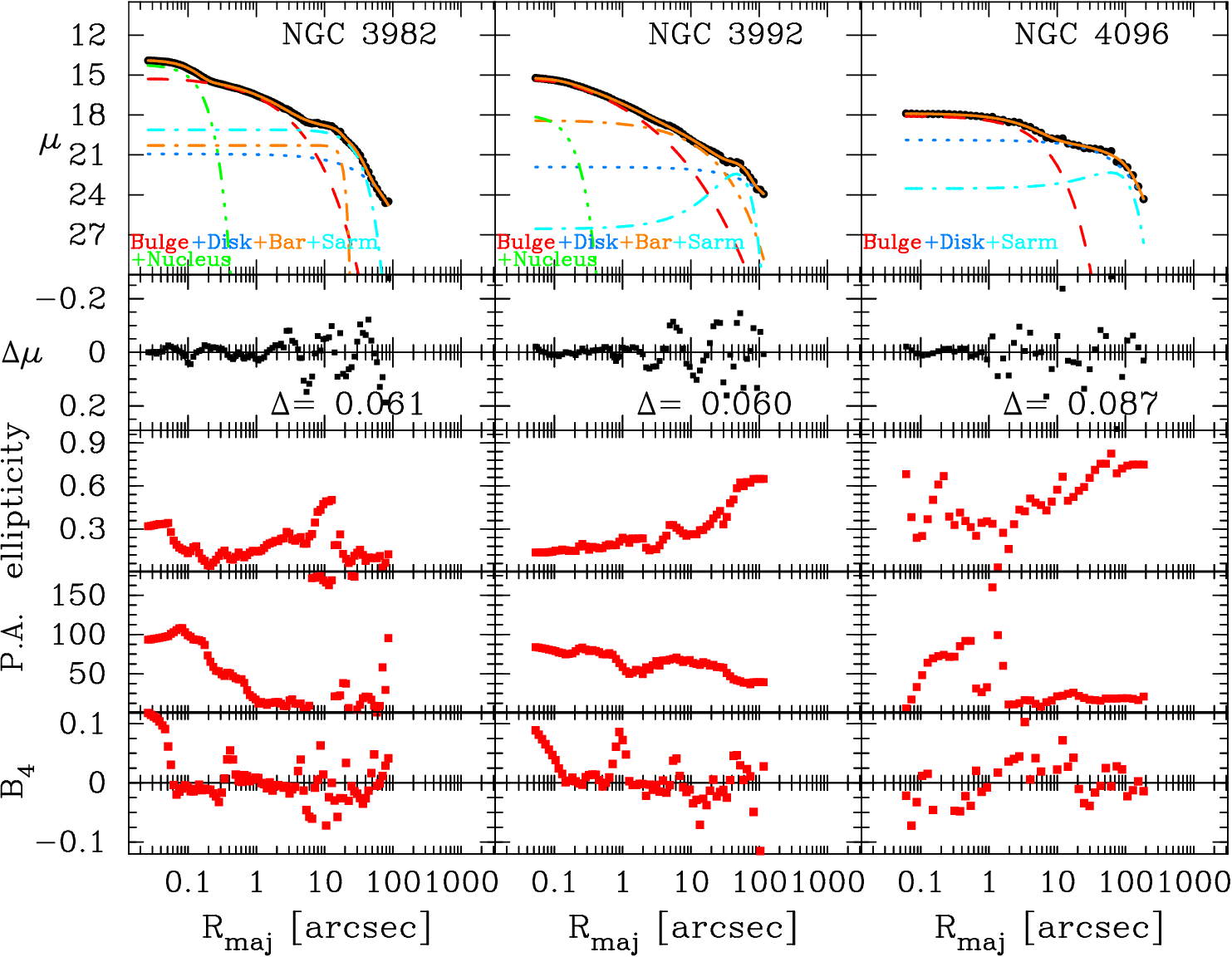}\\
\vspace{.2030cm}
\caption{\it continued.}
\end{figure*}

\begin{figure*}
\setcounter{figure}{2}
\vspace{-.295030cm}
\hspace*{.4811cm}
\includegraphics[angle=270,scale=0.60504]{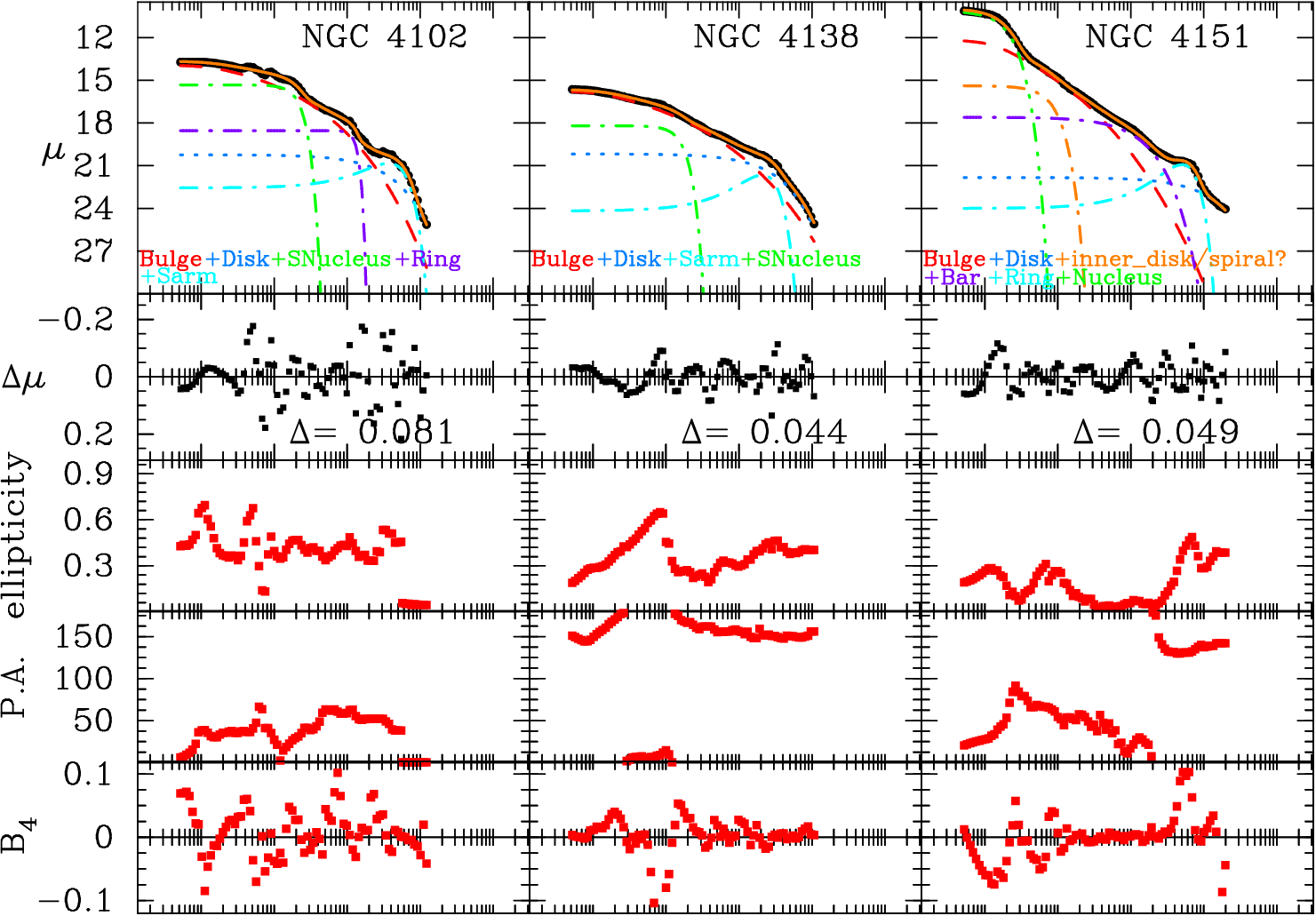}\\
\hspace*{.4811cm}
\includegraphics[angle=270,scale=0.60504]{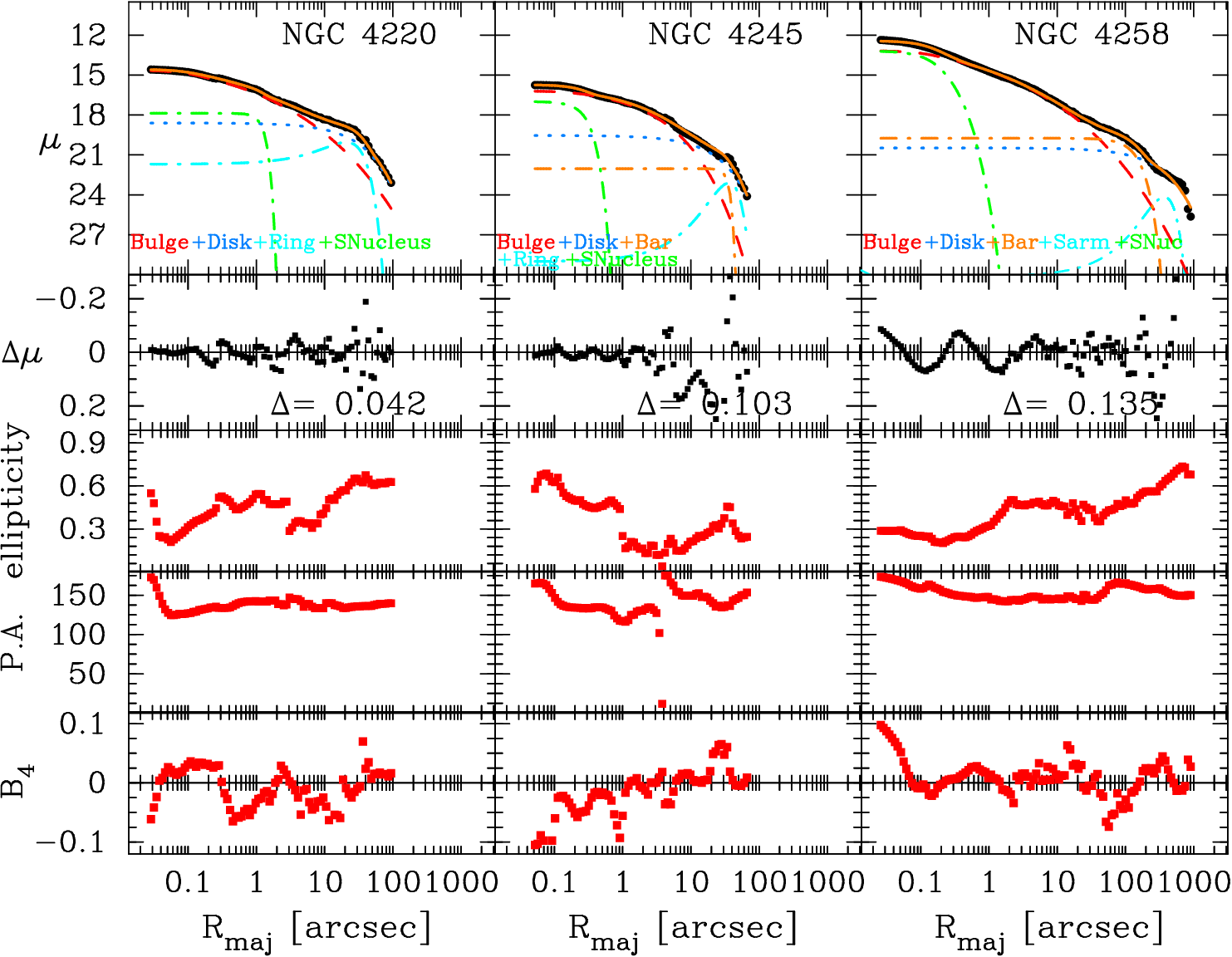}\\
\vspace{.2030cm}
\caption{\it continued.}
\end{figure*}

\begin{figure*}
\setcounter{figure}{2}
\vspace{-.295030cm}
\hspace*{.4811cm}
\includegraphics[angle=270,scale=0.60504]{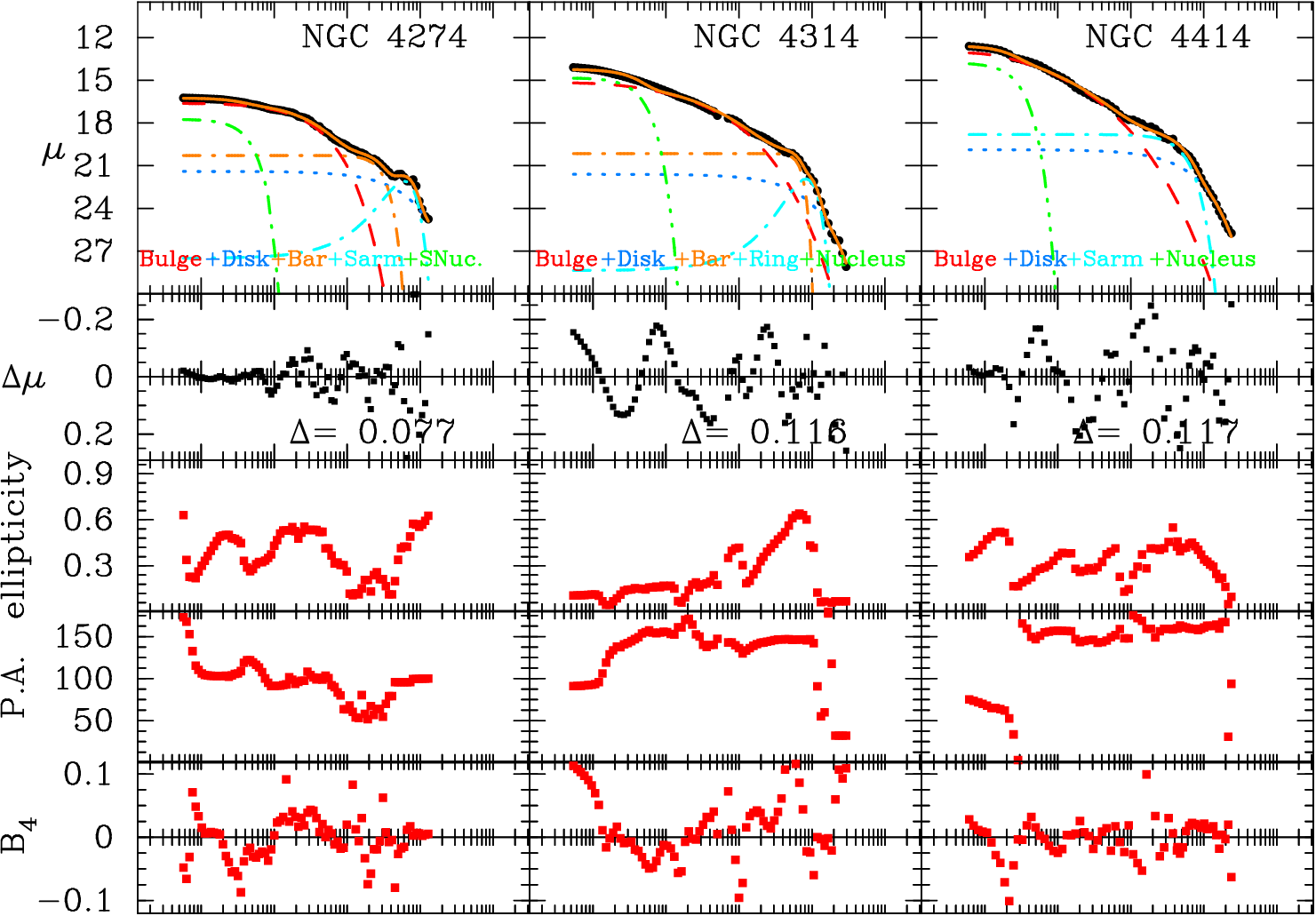}\\
\hspace*{.4811cm}
\includegraphics[angle=270,scale=0.60504]{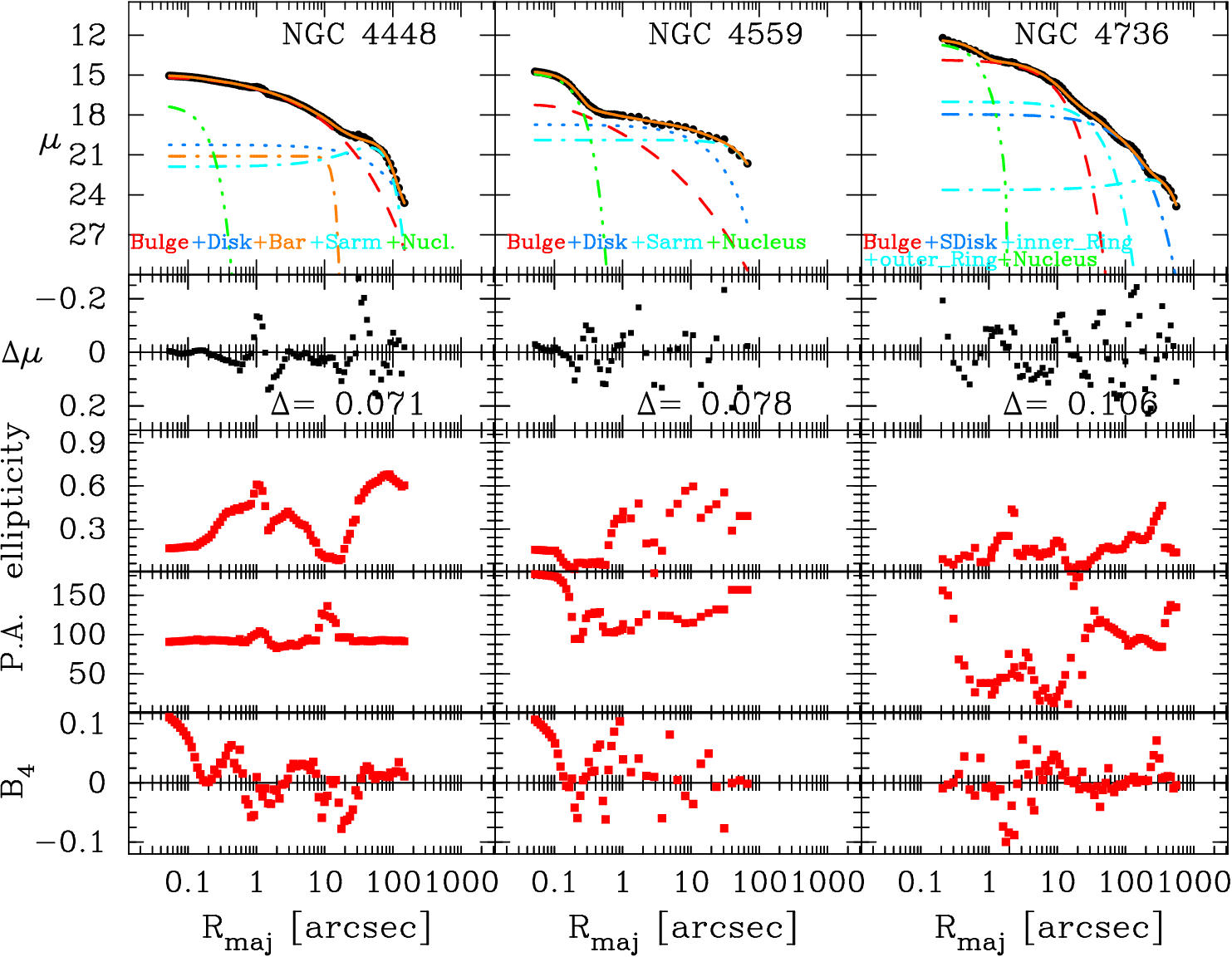}\\
\vspace{.2030cm}
\caption{\it continued.}
\end{figure*}

\begin{figure*}
\setcounter{figure}{2}
\vspace{-.295030cm}
\hspace*{.4811cm}
\includegraphics[angle=270,scale=0.60504]{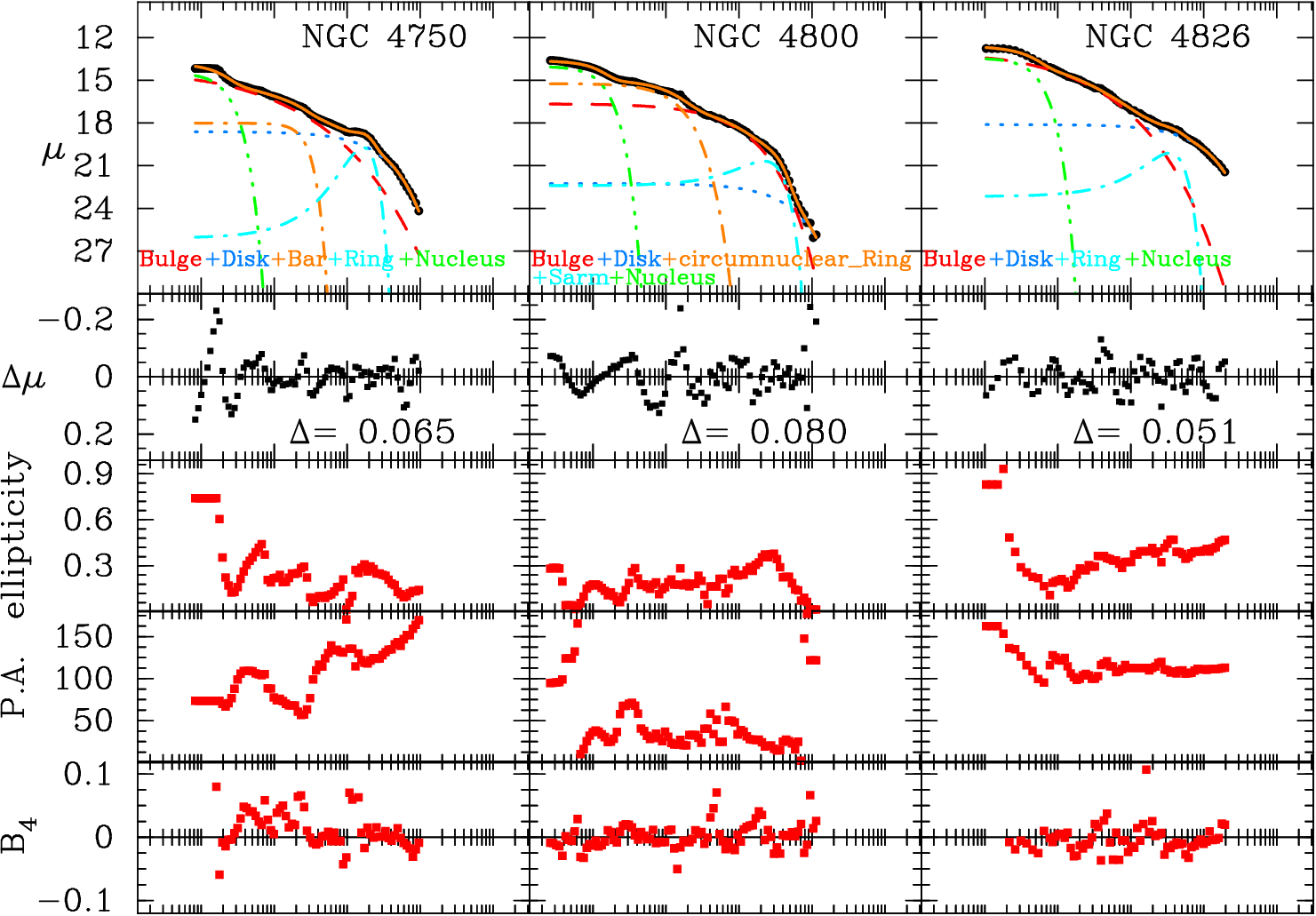}\\
\hspace*{.4811cm}
\includegraphics[angle=270,scale=0.60504]{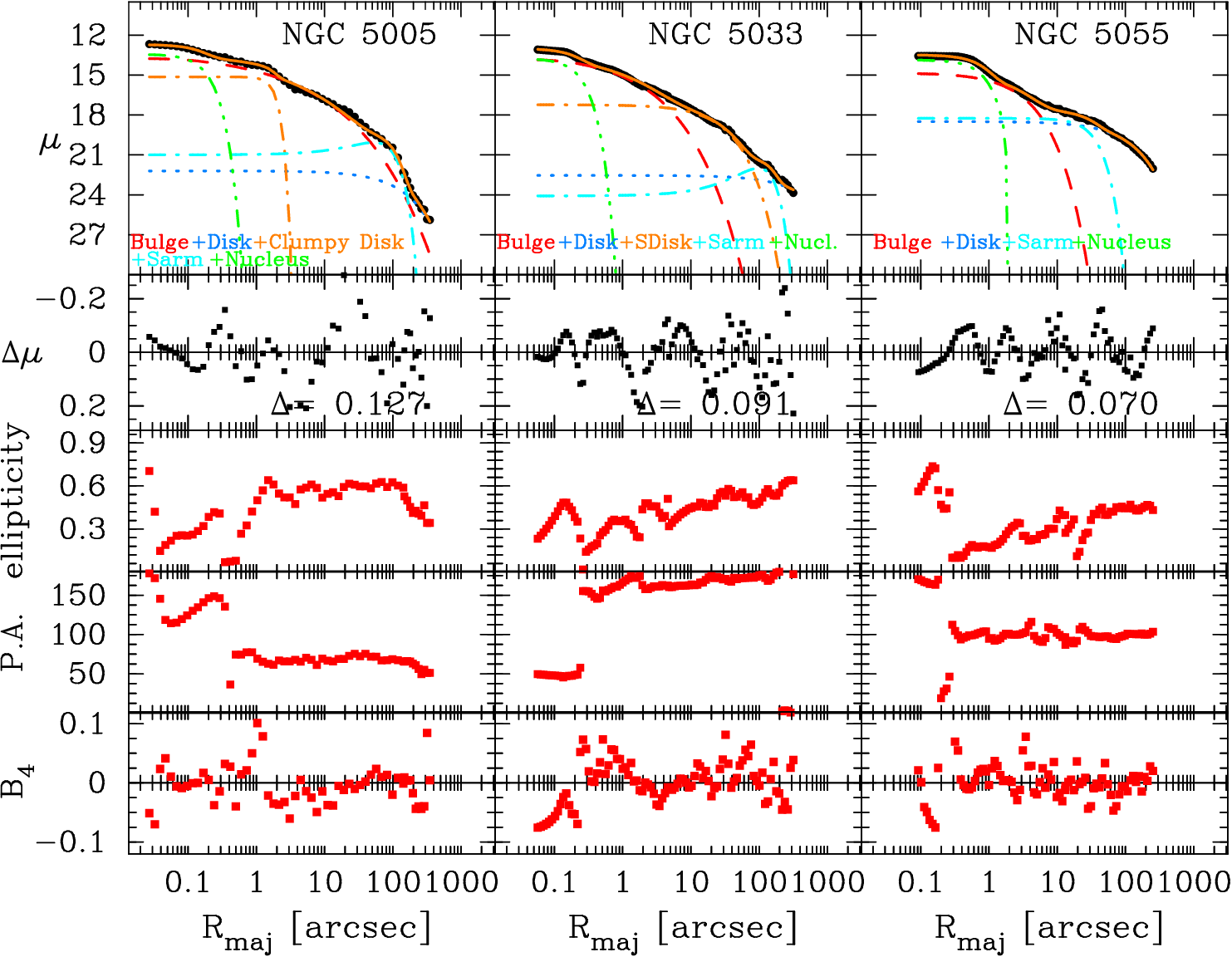}
\vspace{.2030cm}
\caption{\it continued.}
\end{figure*}

\begin{figure*}
\setcounter{figure}{2}
\vspace{-.295030cm}
\hspace*{.4811cm}
\includegraphics[angle=270,scale=0.60504]{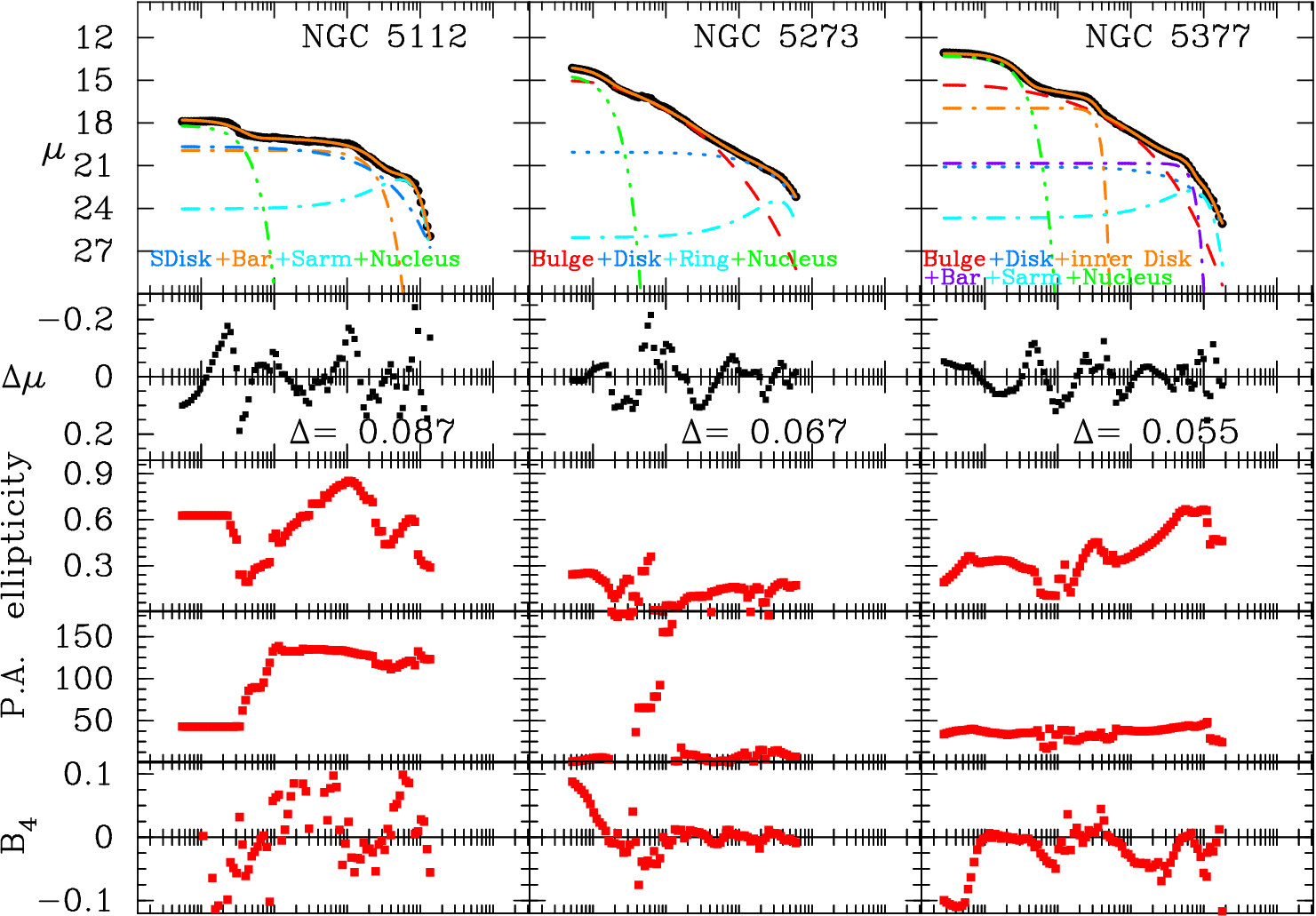}\\
\hspace*{.4811cm}
\includegraphics[angle=270,scale=0.60504]{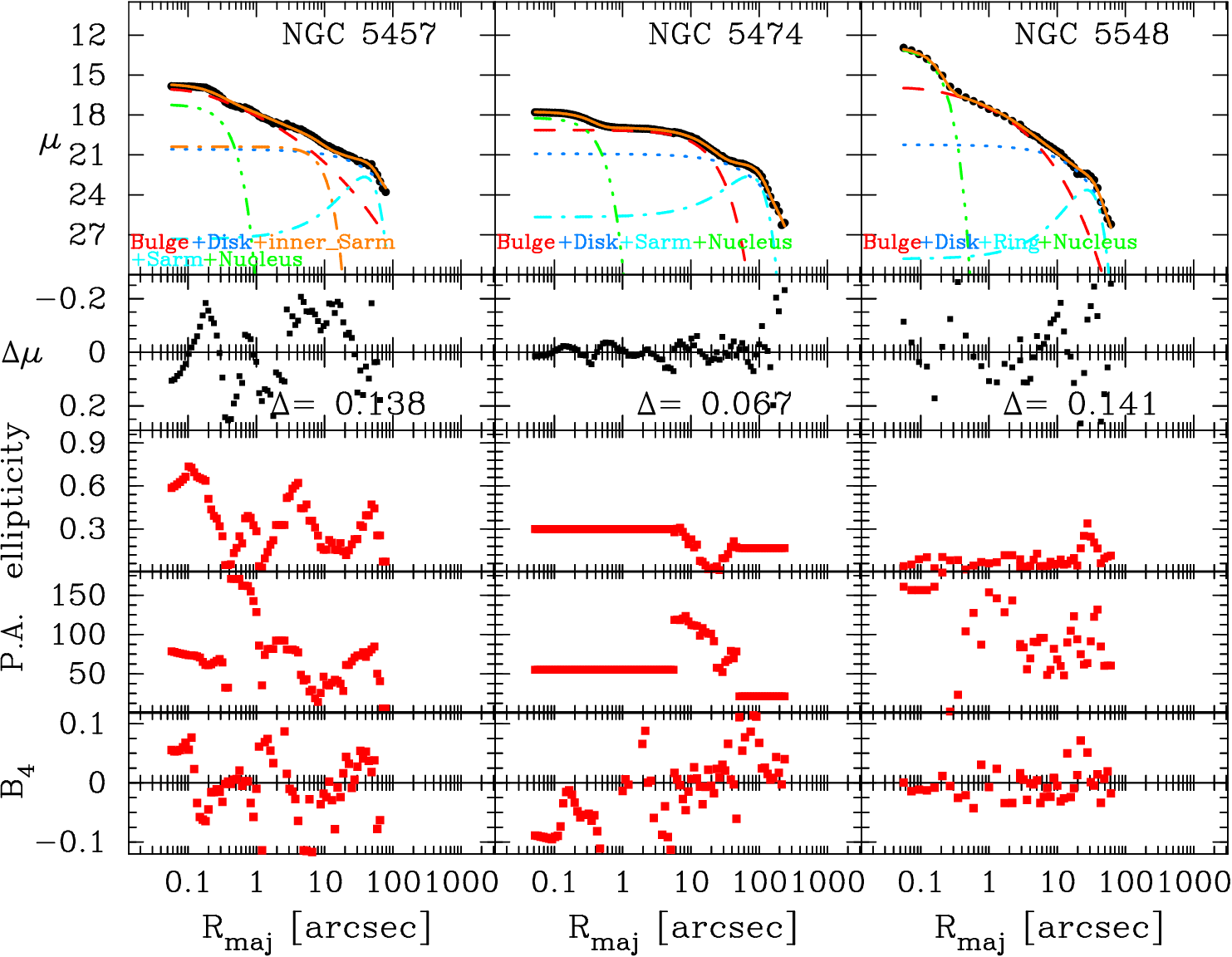}
\vspace{.2030cm}
\caption{\it continued.}
\end{figure*}

\begin{figure*}
\setcounter{figure}{2}
\vspace{-.295030cm}
\hspace*{.4811cm}
\includegraphics[angle=270,scale=0.60504]{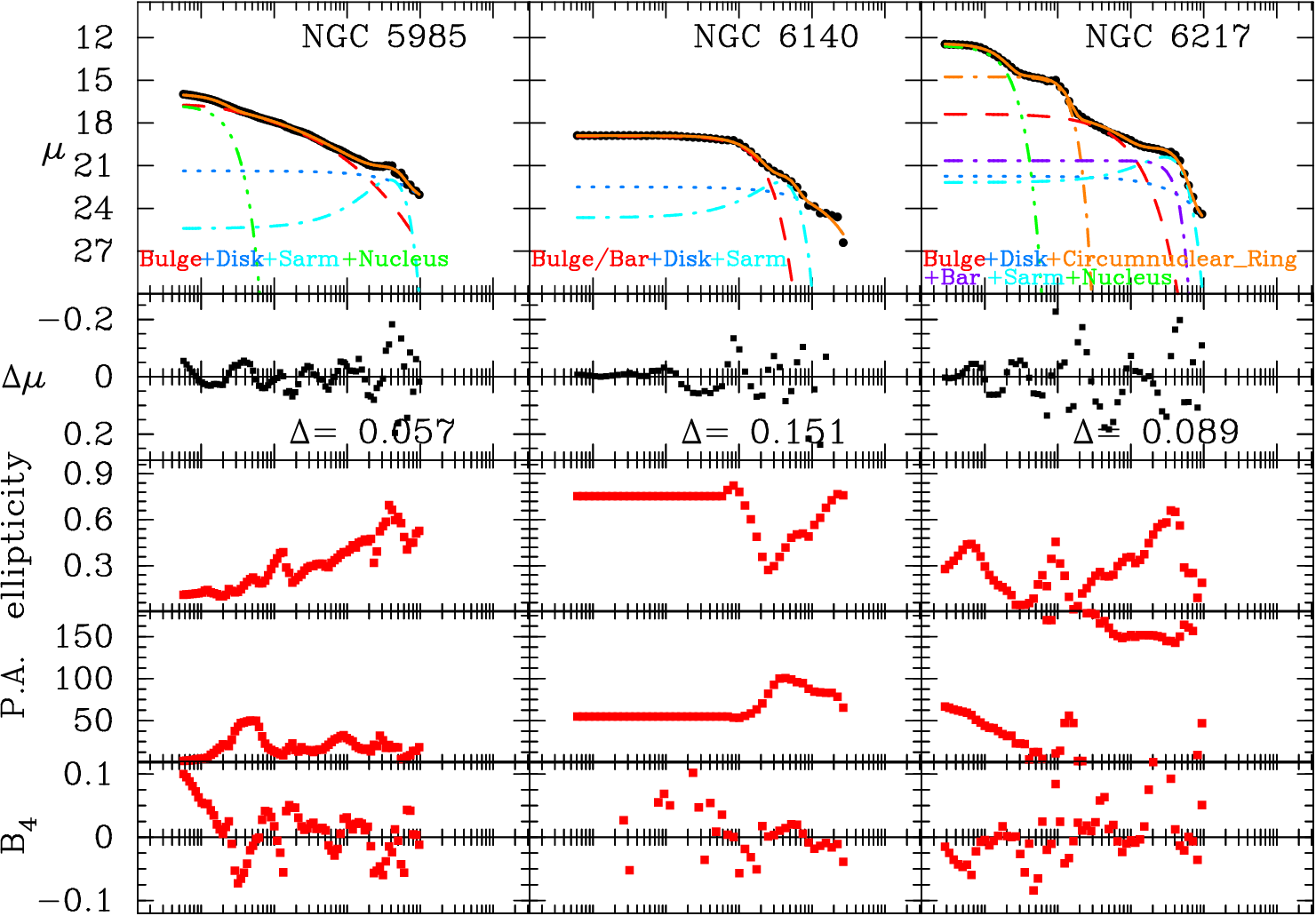}\\
\hspace*{.4811cm}
\includegraphics[angle=270,scale=0.60504]{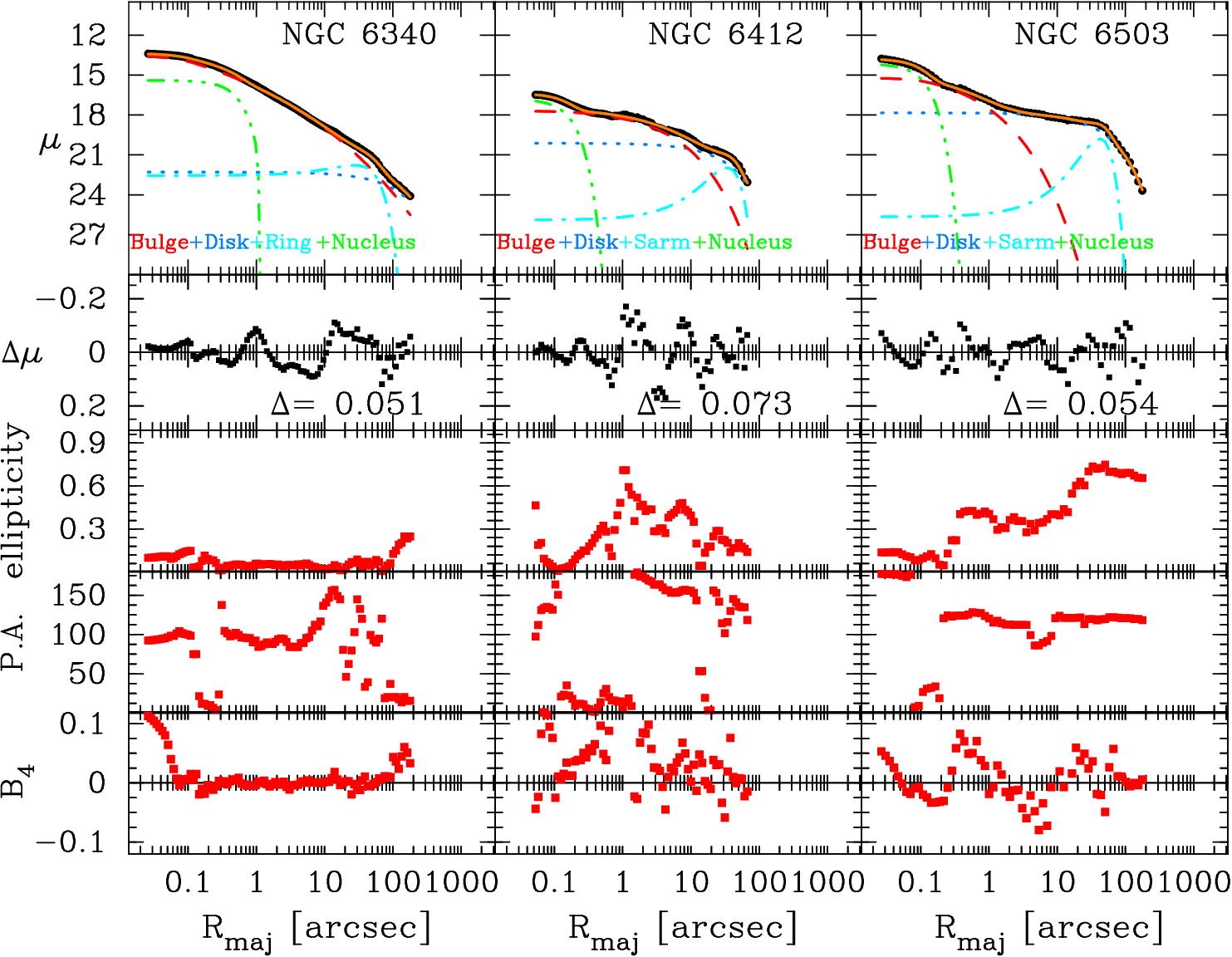}
\vspace{.2030cm}
\caption{\it continued.}
\end{figure*}

\begin{figure*}
\setcounter{figure}{2}
\vspace{-.295030cm}
\hspace*{.4811cm}
\includegraphics[angle=270,scale=0.60504]{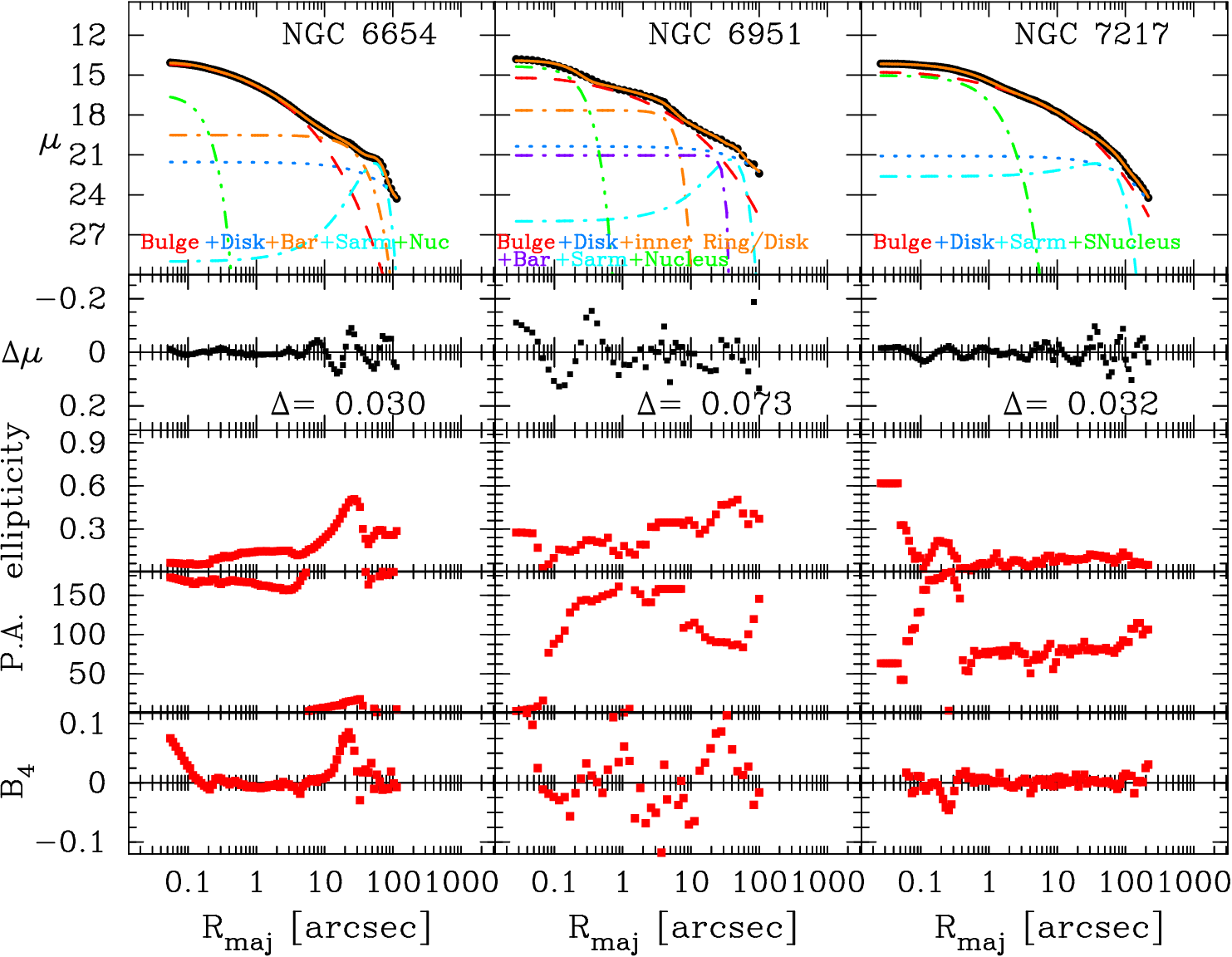}\\
\hspace*{.4811cm}
\vspace*{.409561624811cm}
\includegraphics[angle=270,scale=0.60504]{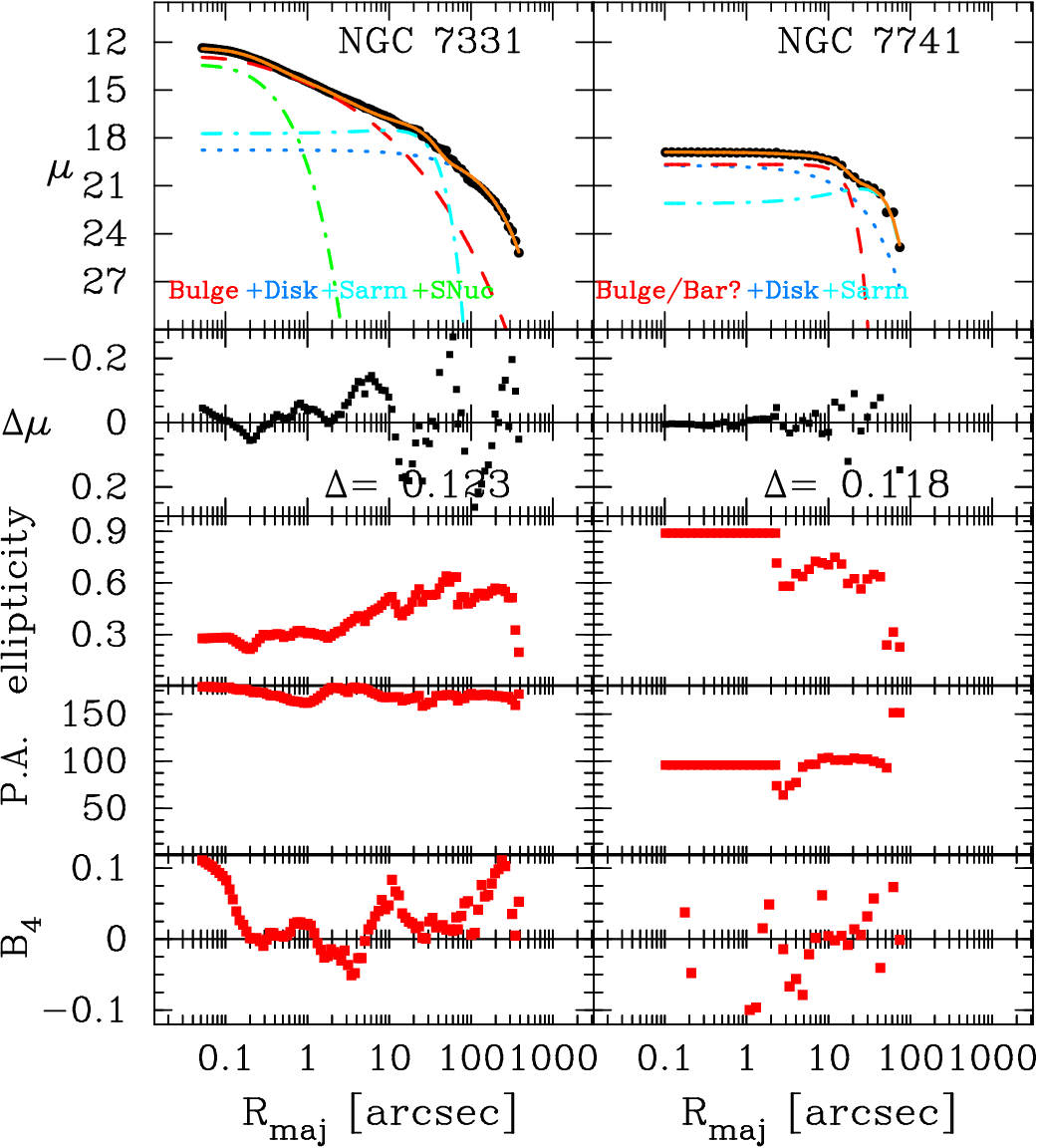}
\vspace{.2030cm}
\caption{\it continued.}
\end{figure*}

\begin{figure*}
\hspace*{.3930cm}
\vspace{.009830cm}
\includegraphics[angle=270,scale=0.60504]{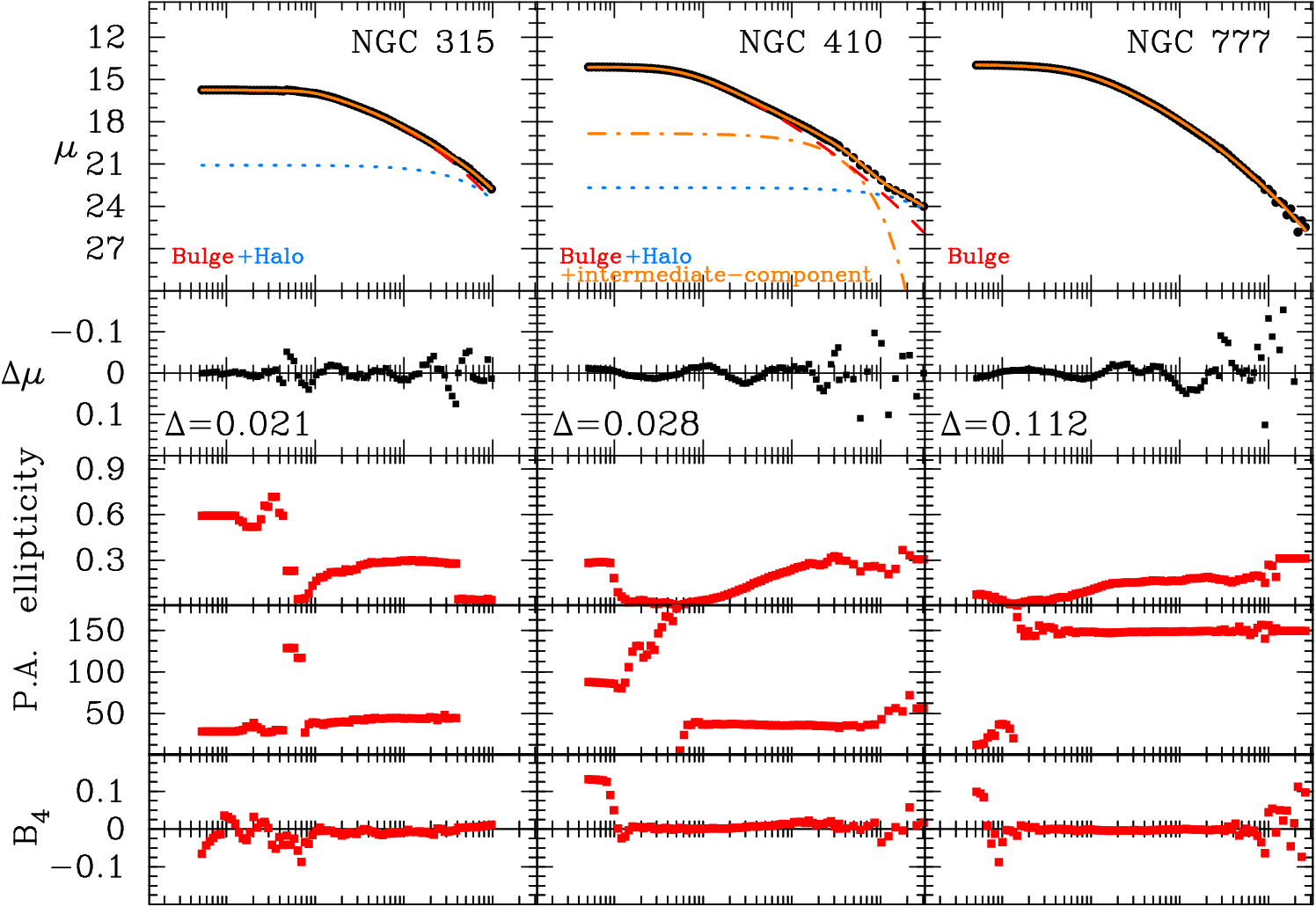}\\
\hspace*{.3930cm}
\vspace{.39830cm}
\includegraphics[angle=270,scale=0.60504]{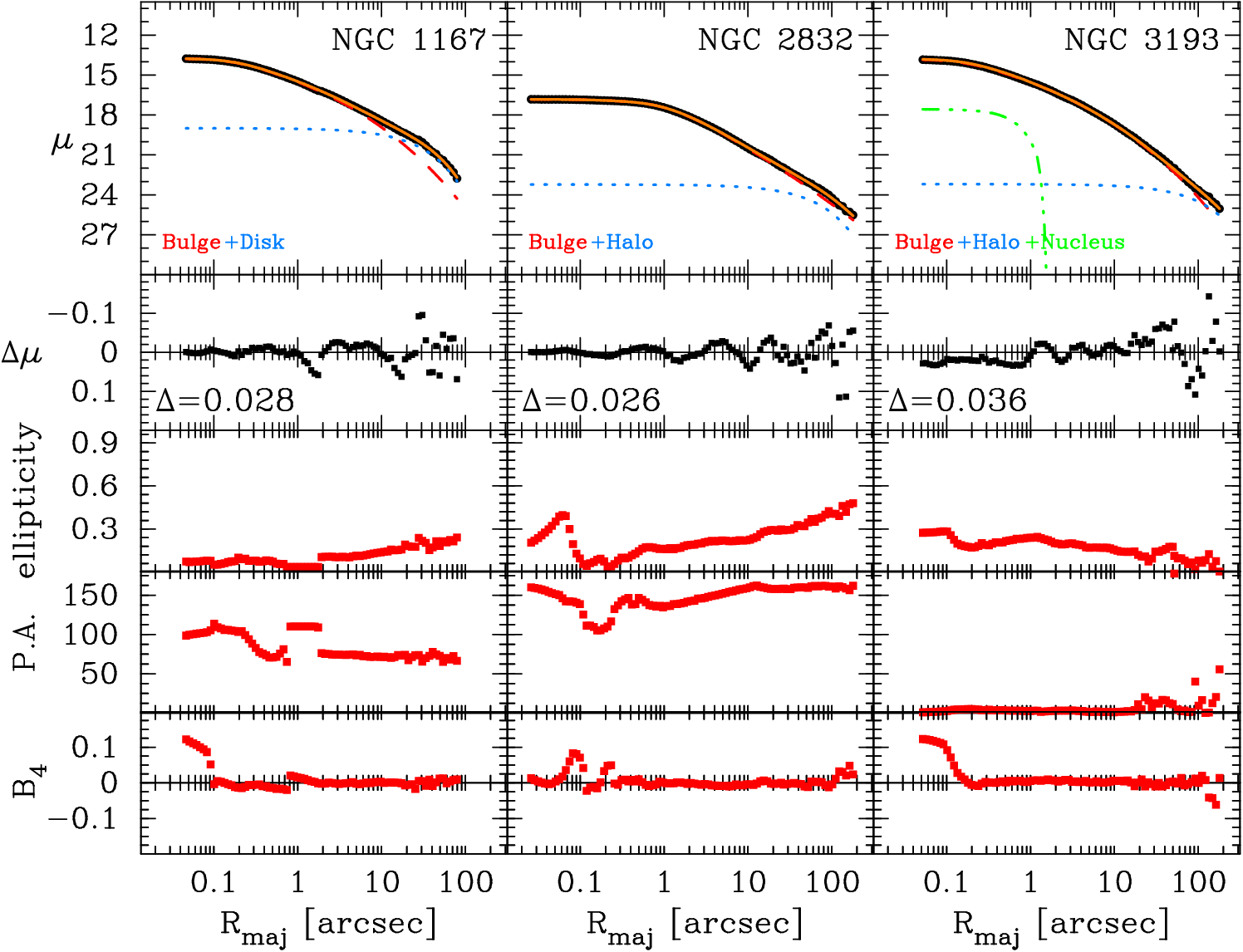}\\
\vspace{.2030cm}
\caption{Similar to Fig.~\ref{FigSer1} but showing here
  multi-component decompositions of core-S\'ersic LeMMINGs galaxies
  (see Table~\ref{Tab1cS}).  These galaxies have a central deficit of
  light relative to the inward extrapolation of their bulge's outer
  S\'ersic profile.  Of the 15 core-S\'ersic elliptical galaxies in
  the full sample, 6 (NGC~0315, NGC~0410, NGC~2832, NGC~3193, NGC~5322
  and NGC~6482) have an outer stellar halo component that we modelled
  with an exponential function.  }
 \label{FigCSer}
\end{figure*}

\begin{figure*}
\setcounter{figure}{3}
\hspace*{.3930cm}
\includegraphics[angle=270,scale=0.60504]{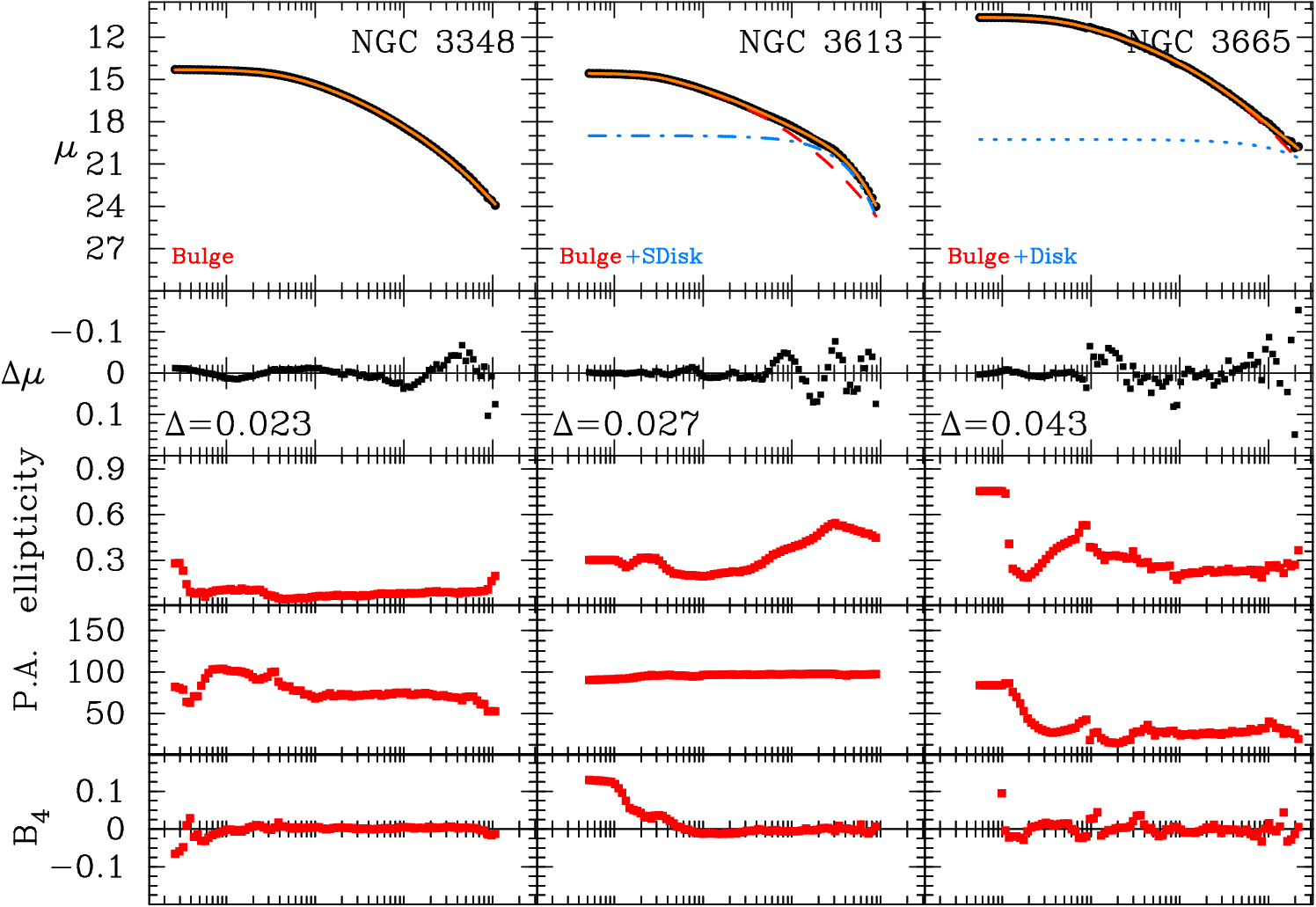}\\
\hspace*{.3930cm}
\includegraphics[angle=270,scale=0.60504]{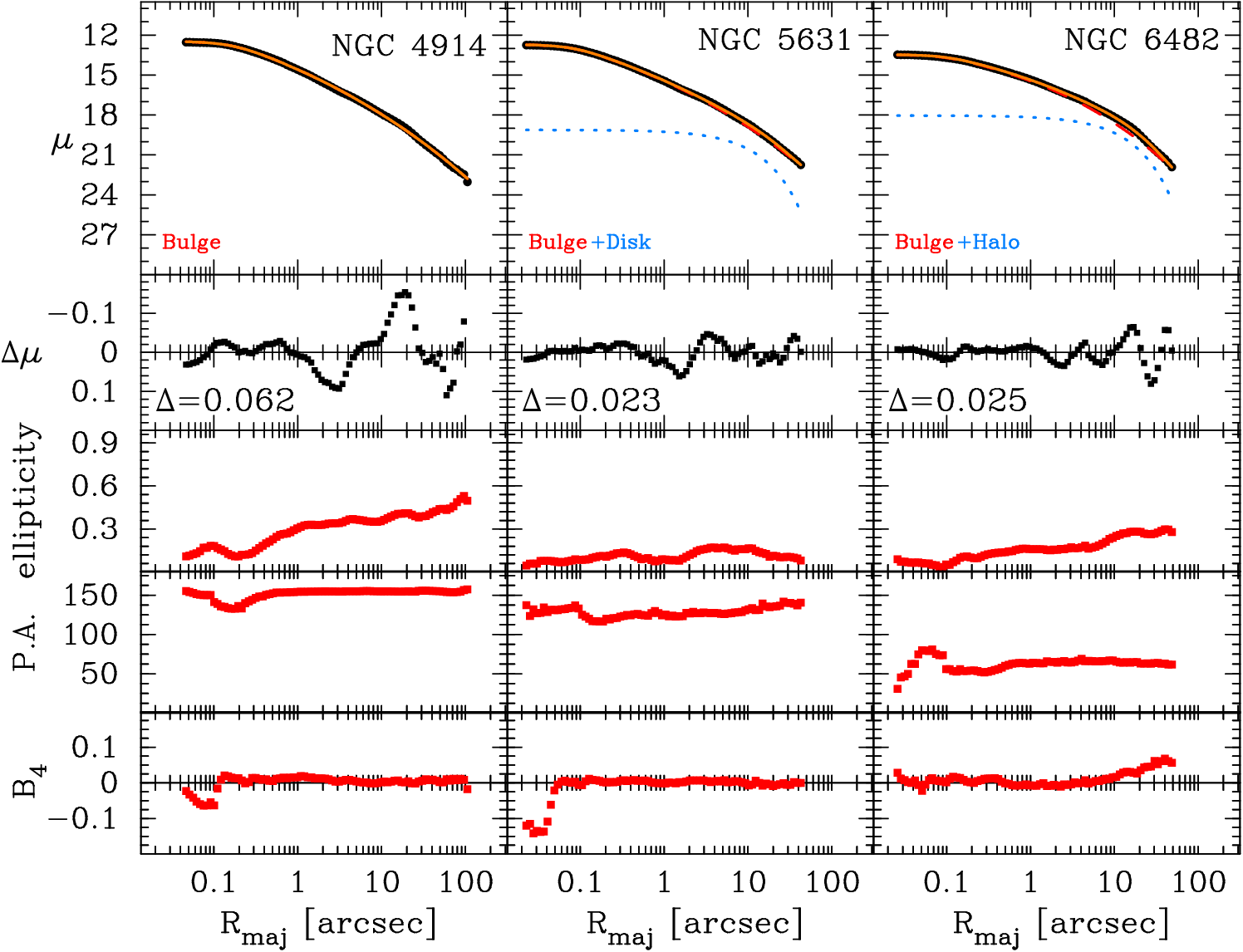}\\
\hspace*{-1.0cm}
\vspace{.2030cm}
\caption{\it continued.}
\end{figure*}

\begin{figure*}
\vspace{-.166095030cm}
\hspace{.31529993969008214cm}
\includegraphics[angle=270,scale=0.5261]{Dullo_GroupI_tullyR.ps}
\hspace{-.143947992997380cm}
\includegraphics[angle=270,scale=0.5261]{Dullo_GroupII_tullyR.ps}\\
\vspace{.0cm}
\hspace{.229926214cm}
\includegraphics[angle=270,scale=0.5261]{Dullo_GroupIII_tullyR.ps}
\hspace{-.143924799997380cm}
\includegraphics[angle=270,scale=0.5261]{Dullo_GroupIV_tullyT.ps}
\vspace{.14030cm}
\caption{Major-axis surface brightness profile decompositions of a
  dozen LeMMINGs galaxies using multiple S\'ersic models that are
  PSF-convolved with a Moffat function. There is strong agreement
  between the galaxies' fits given here and those convolved with a
  Gaussian PSF (see Figs.~\ref{FigSer1}, \ref{FigSerR} and
  ~\ref{Fig3}).}
 \label{FigMoff}
\end{figure*}

\begin{figure*}
	\includegraphics[trim={-1.1cm 0mm -4cm 1cm},clip, width=01.05099506\linewidth]{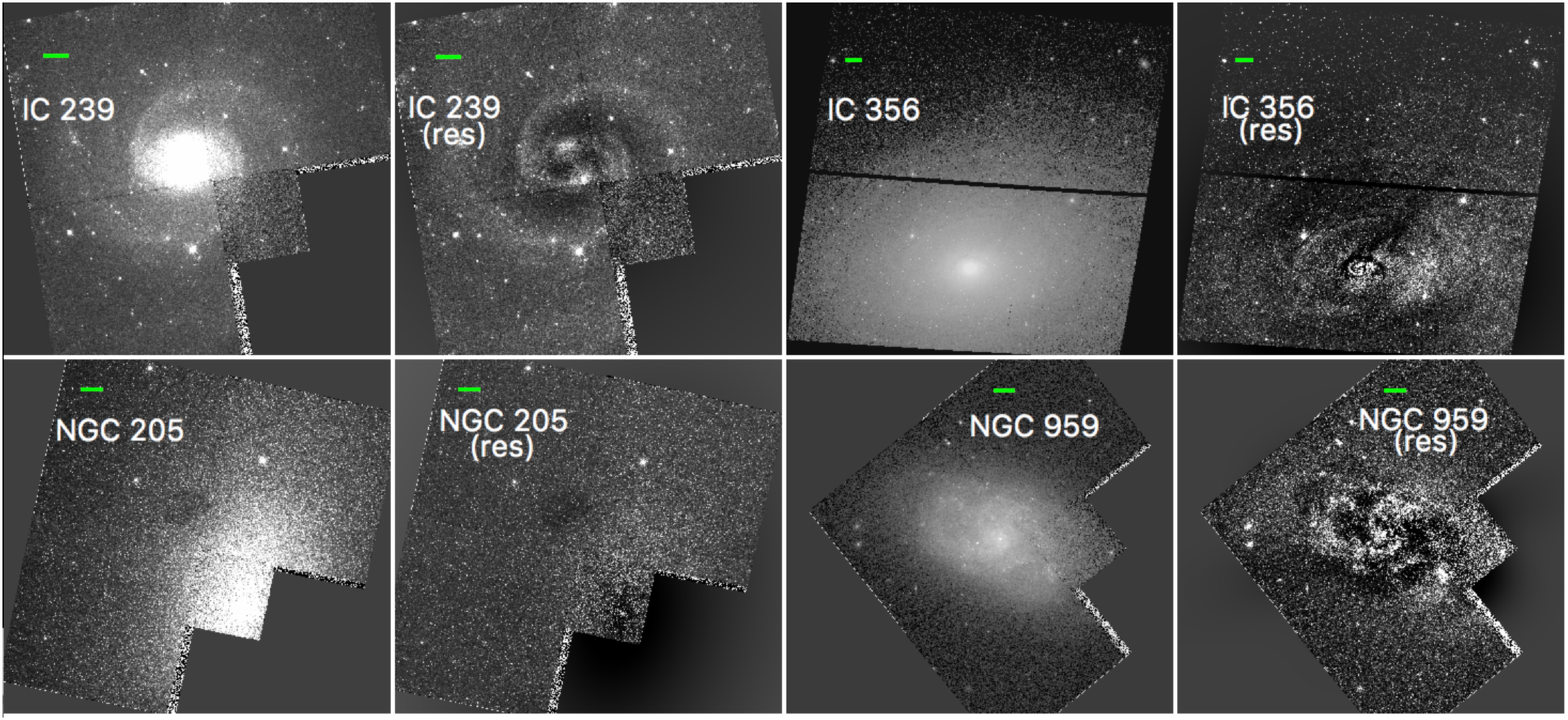}\\
	\hspace{4.997380cm}
	\includegraphics[trim={-1.1cm 0mm -4cm 0cm},clip, width=01.05099506\linewidth]{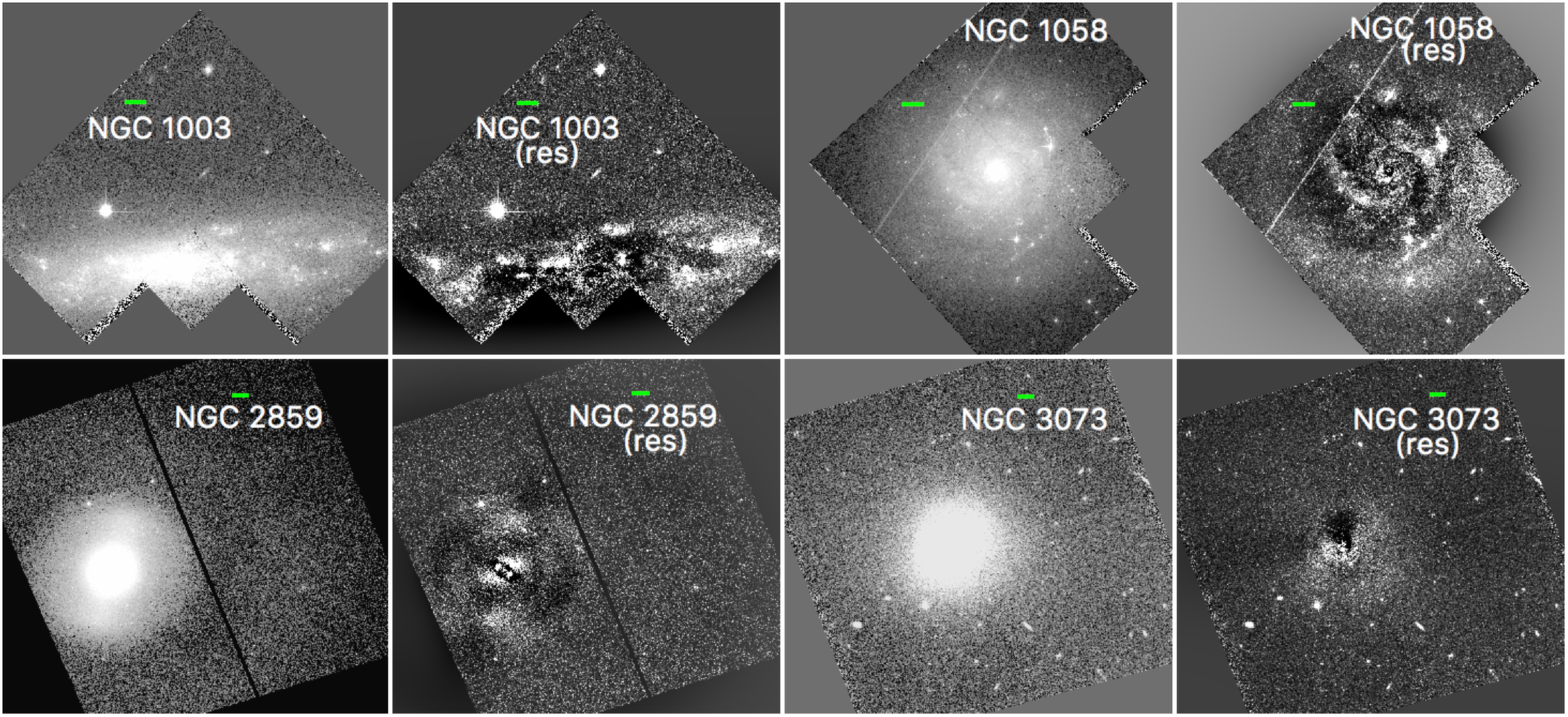}\\
	\includegraphics[trim={-1.1cm 0mm -4cm 1cm},clip, width=01.05099506\linewidth]{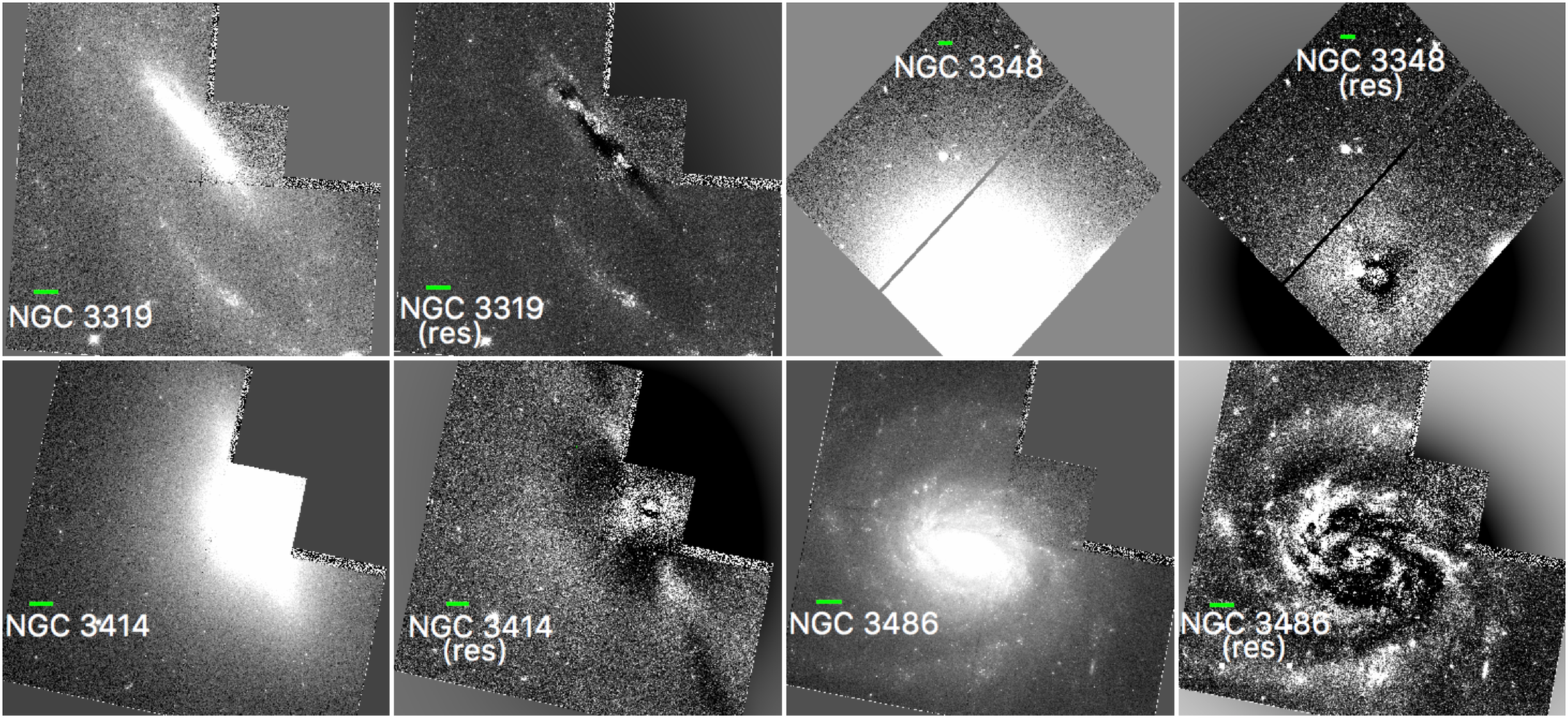}\\
\vspace{-0.328114cm}
\caption{  {\it HST} images and the corresponding residual images
    from our 2D multi-component decompositions for a sample of 28
    LeMMINGs galaxies, which are representative of the 65 sample
    galaxies with 2D decompositions in the paper, see
    Tables~\ref{NewTabA1}-\ref{Tab1cS}.  The green scale bar is
    10$\arcsec$. For the 2D fits, the 2D models generated using {\sc
      imfit} have the same type and number of galaxy structural
    components as the corresponding 1D modelling.  We find strong
    agreement between the 1D and 2D fits for all the 65 galaxies.}
 \label{2Dfits}
\end{figure*}

\begin{figure*}
\setcounter{figure}{5}
\includegraphics[trim={-1.1cm 0mm -4cm 1cm},clip, width=01.05099506\linewidth]{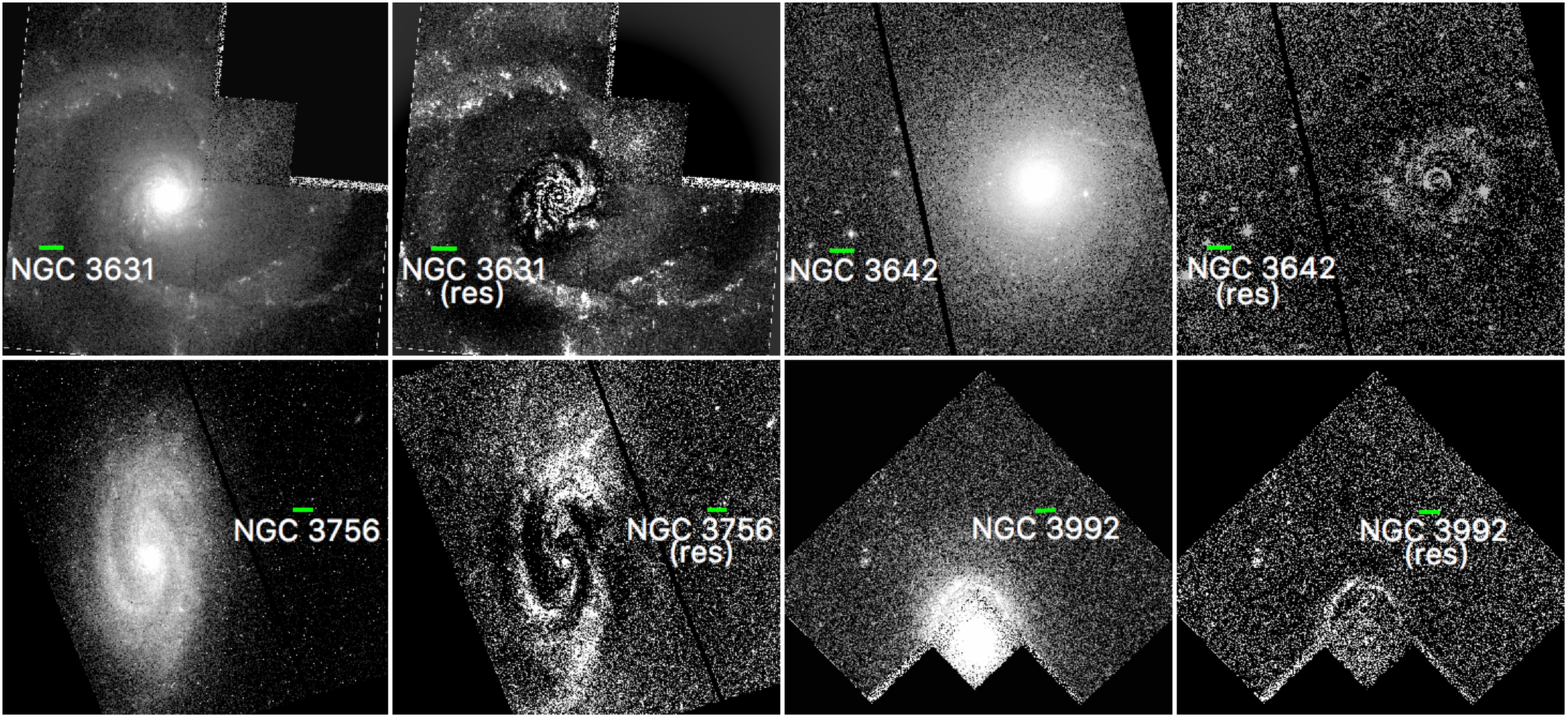}
\includegraphics[trim={-1.1cm 0mm -4cm 1cm},clip, width=01.05099506\linewidth]{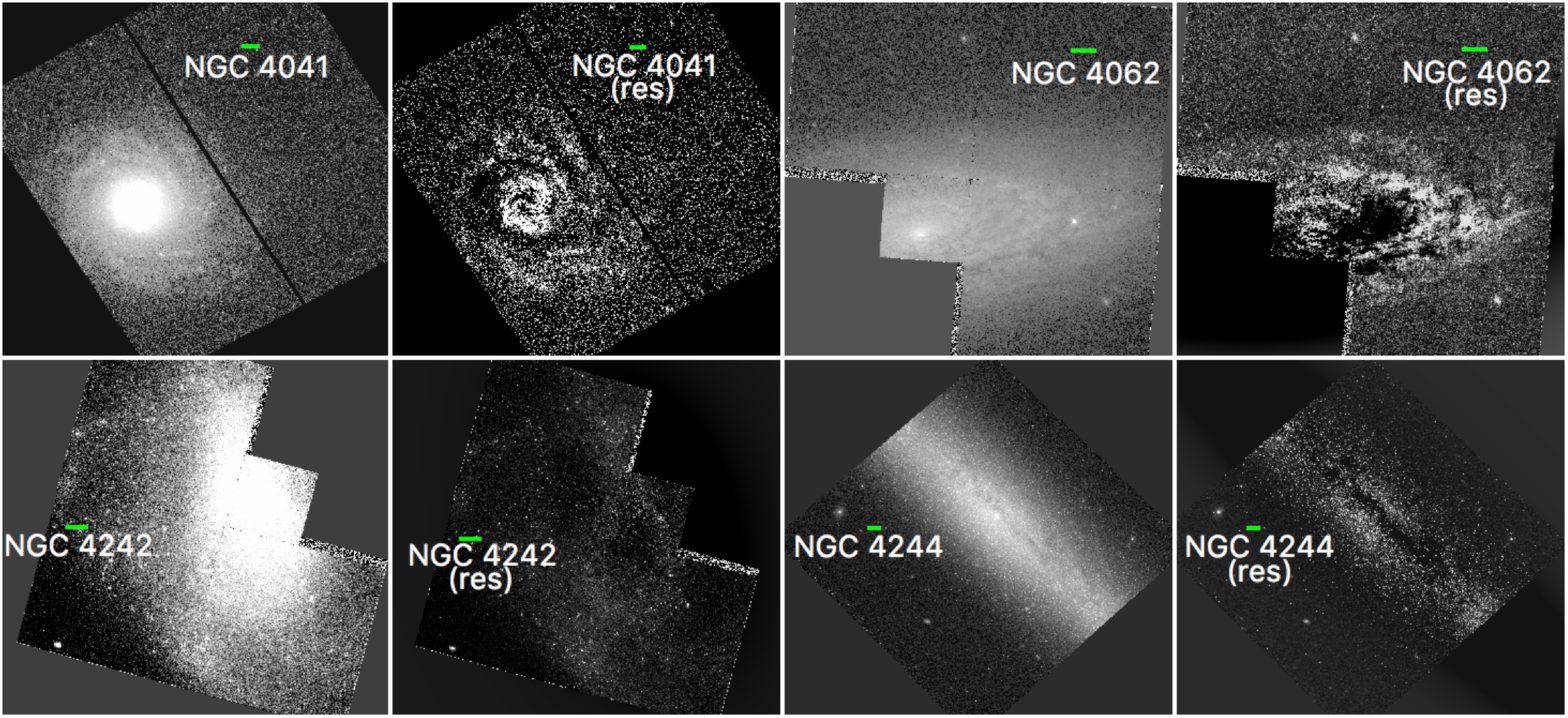}\\
\includegraphics[trim={-1.1cm 0mm -4cm 1cm},clip, width=01.05099506\linewidth]{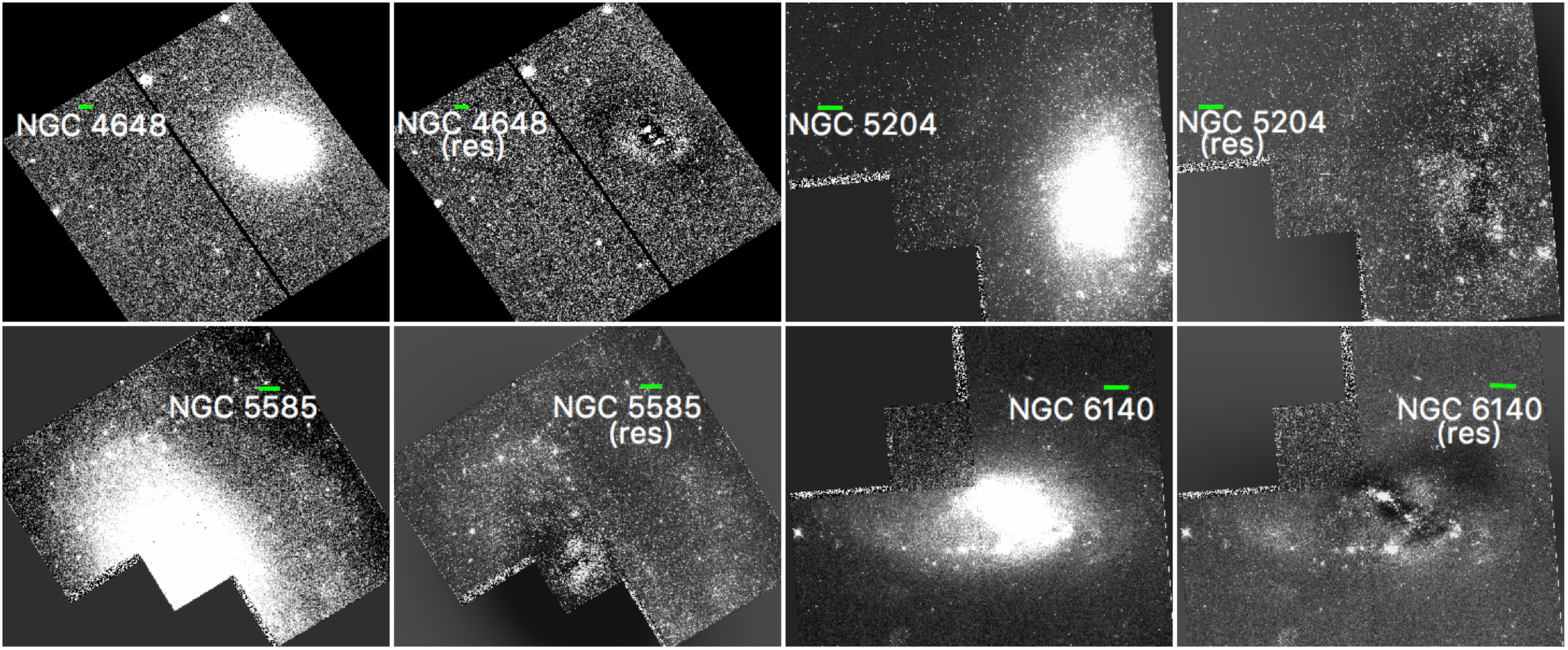}\\
\vspace{-0.20360cm}
\caption{\it continued.}
\end{figure*}

\begin{figure*}
\setcounter{figure}{5}
\includegraphics[trim={-1.1cm 0mm -4cm 1cm},clip, width=01.05099506\linewidth]{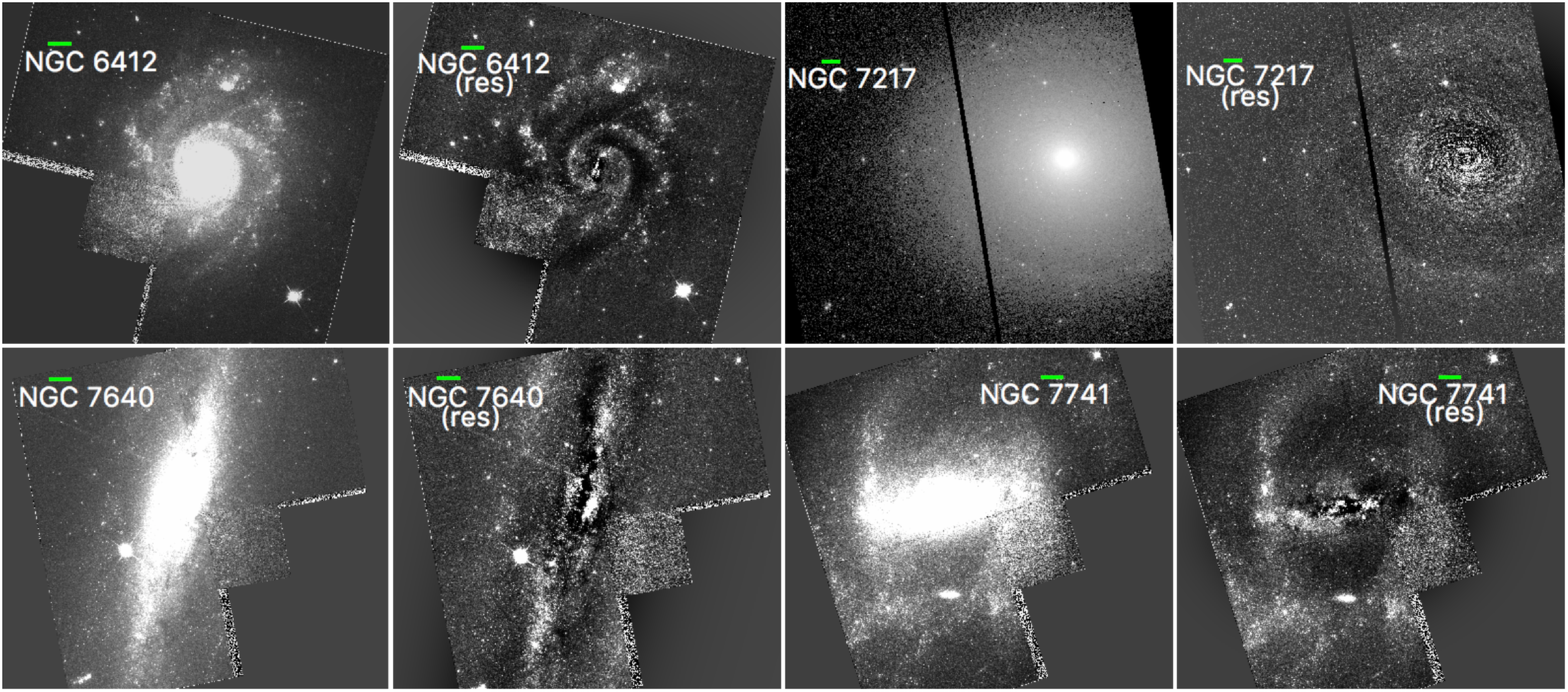}
\vspace{-0.20360cm}
\caption{\it continued.}
\end{figure*}

\section{Error Analysis}\label{AppendD}
This section describes our considerable endeavour dedicated to explore
the robustness of the decomposition of the surface brightness profiles
and the accuracy of the measured best-fitting structural parameters
using Monte Carlo (MC) simulated profiles. Doing so has a twofold
merit: it provides realistic error estimates for the best-fitting
structural parameters and allows us to derive a MC-based galaxy
component structural parameters. More than 100 realisations of each
galaxy's light profile were generated by running a series of MC
simulations. To achieve this, for each galaxy, we perturb the data
points of the galaxy light profile by sampling from a Gaussian with
sigma, where the sigma value is the mean of all the residuals obtained
from the actual galaxy light profile fitting (see Appendices
\ref{DataTables} and \ref{AppendC}) We then add a correlated noise (i.e., a
single, constant value) to the entire profile to account for potential
errors from inaccurate sky subtraction.  The standard deviation around
the mean of the median sky background values measured in
Section~\ref{SkYB} were consulted to roughly simulate potential sky
subtraction errors.  Each realisation was then decomposed following
the exact same fitting procedure as the modelling of the original  galaxy
light profile, including PSF convolution.

 In Fig.~\ref{Figerr}, we display the distributions of best-fitting
 parameters for the bulge, second galaxy component and the outer disc
 for 50 LeMMINGs galaxies together with Gaussian models fitted to the
 parameter distributions. The adopted errors for the best-fitting
 parameters (Tables~\ref{Tab1}$-$\ref{Tab1cS}) are computed using the
 standard deviations (i.e., $1\sigma$ values) from the fitted
 Gaussians to the 100 MC best-fit values (i.e., assuming a normal
 distribution) together with the difference between the mean values of
 the Gaussians and the best-fitting parameters adopted in this work.
 We note that the median $1\sigma$ values (see Fig.~\ref{Figerr})
 associated with the bulge's $\mu_{\rm e}$, $R_{\rm e}$, $n$ for the
 LeMMINGs galaxies are $\sim$ 0.20, 10\%, 10\%, respectively. We also
 remind the reader that the errors we derive here are consistent
 within the framework of the S\'ersic models. However the total errors
 could modestly exceed those estimated from the model fits as the
 fitting functions are not perfect and cannot possibly reproduce all structural  
 features in nearby objects such as those dealt with in this work. We note that the mean values of
 the Gaussians (see Fig.~\ref{Figerr}) are the MC-based best-fitting
 structural parameters for the galaxy components, which overall are in
 good agreement with those adopted in this work. The application of
 the galaxy decomposition methods on the simulated light profiles
 reveal that they are fairly robust.

To quantify the errors on the bulge magnitudes and stellar masses
(Table~\ref{Tab4}), the uncertainties on
the associated best fitting parameters were propagated.
Uncertainties in $M/L$ are also considered in the stellar mass error
budget (see Section~\ref{Sec3.4}).

\begin{figure*}
\vspace{-.1630cm}
\hspace*{.8130cm}
\includegraphics[angle=0,scale=0.35]{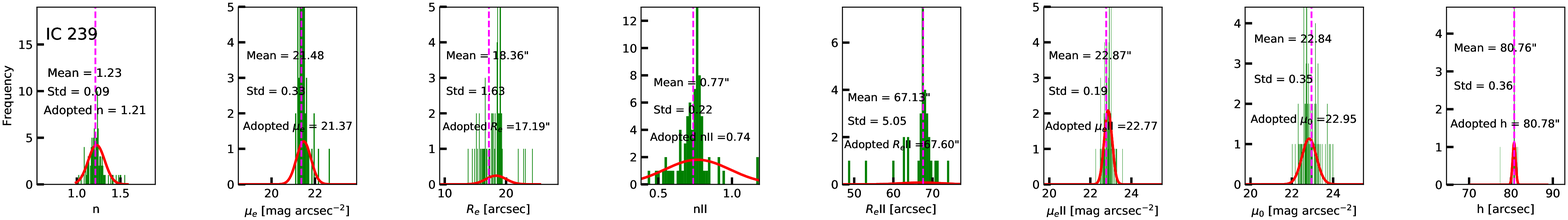}\\
\hspace*{.8130cm}
\includegraphics[angle=0,scale=0.35]{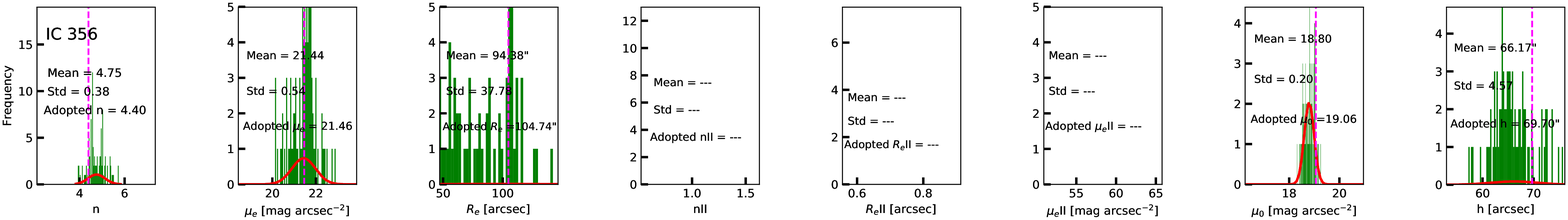}\\
\hspace*{.8130cm}
\includegraphics[angle=0,scale=0.35]{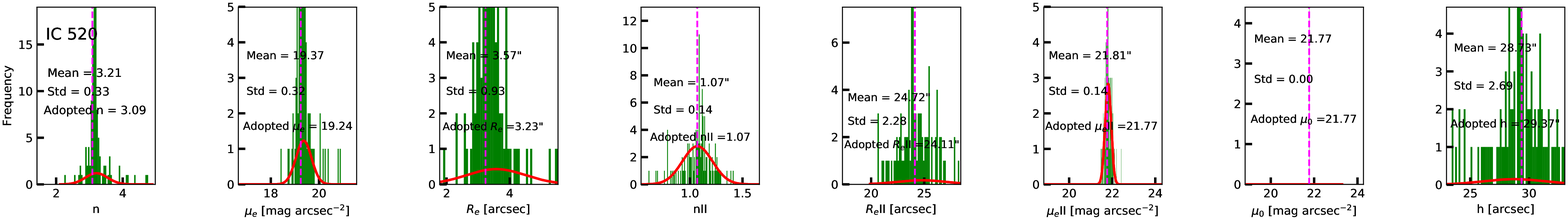}\\
\hspace*{.8130cm}
\includegraphics[angle=0,scale=0.35]{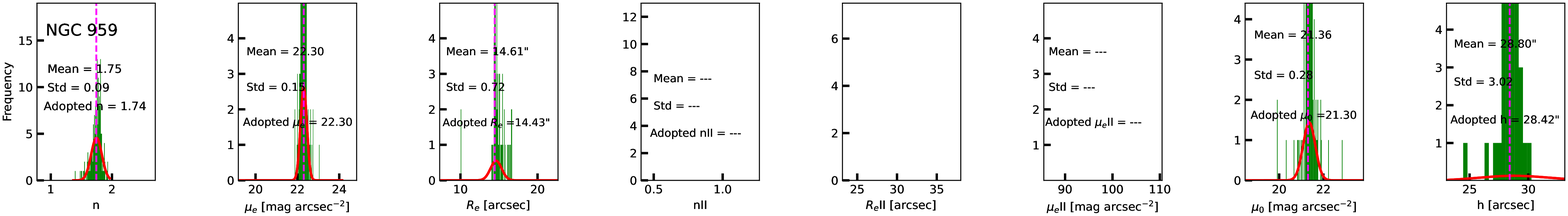}\\
\hspace*{.8130cm}
\includegraphics[angle=0,scale=0.35]{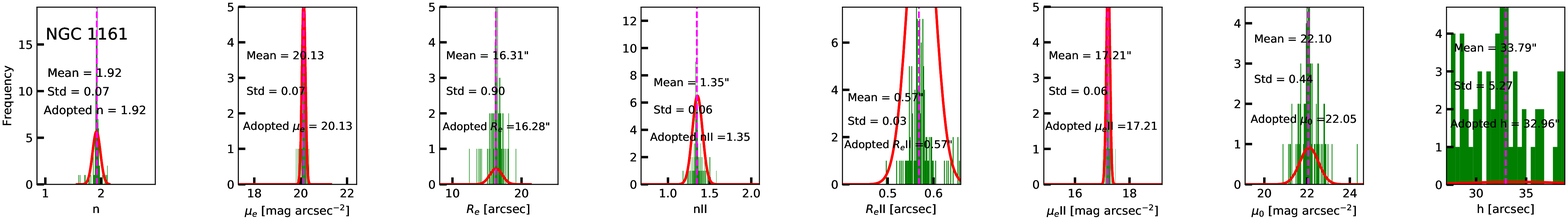}\\
\hspace*{.8130cm}
\includegraphics[angle=0,scale=0.35]{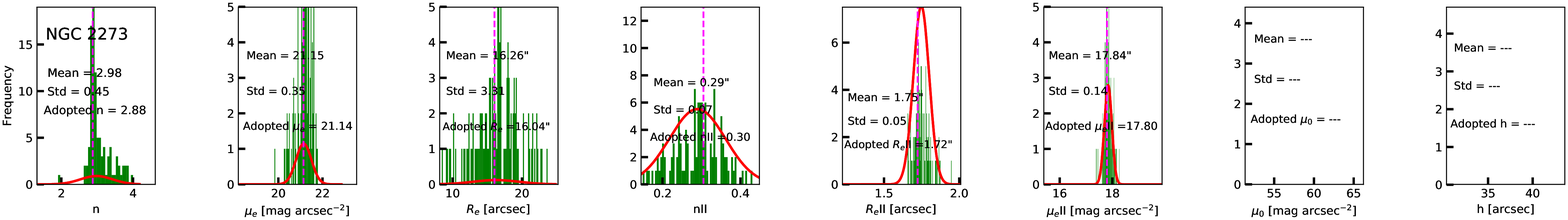}\\
\hspace*{.8130cm}
\includegraphics[angle=0,scale=0.35]{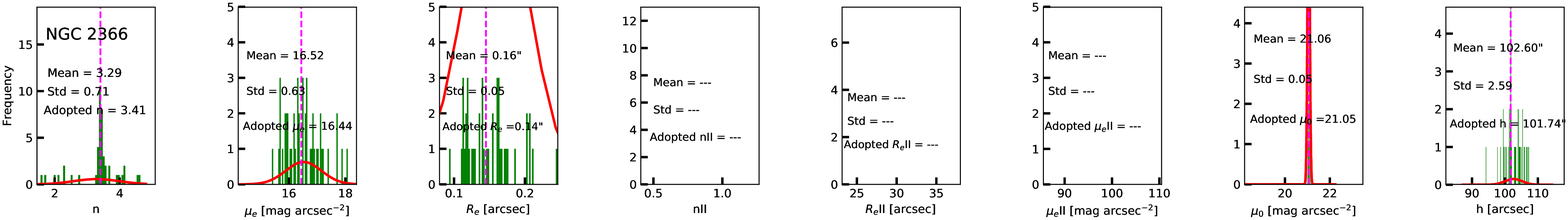}\\
\hspace*{-9.130cm}
\hspace{-19.630cm}
\vspace{-.298530cm}
\caption{For each sample galaxy, 100 realisations of the light profile
  were generated and then decomposed in the same way as the modelling
  of the actual galaxy light profile (see Fig.~\ref{FigSer1}).  This
  exercise has a twofold merit: to estimates of the uncertainties on
  the best-fitting structural parameters and to perform an MC-based
  estimates of the structural parameters for the galaxy components .
  Panels show the histograms of the best-fitting parameters for the
  bulge, disc and second S\'ersic component from fits to the
  realisations for 50 LeMMINGs galaxies (see the text for details).
  Gaussian model fits to the best-fitting parameter distributions (red
  curves) together with the associated mean and standard deviation
  values are shown. These mean values are MC-based measurements of the
  best-fitting structural parameters for the galaxy components, which
  overall are in good agreement with those adopted in this work
  indicated by dashed vertical line (see Tables~\ref{Tab1} and
  \ref{Tab1Ring}). }
 \label{Figerr}
\end{figure*}

\begin{figure*}
\setcounter{figure}{6}
\vspace{-.1630cm}
\hspace*{.8130cm}
\includegraphics[angle=0,scale=0.35]{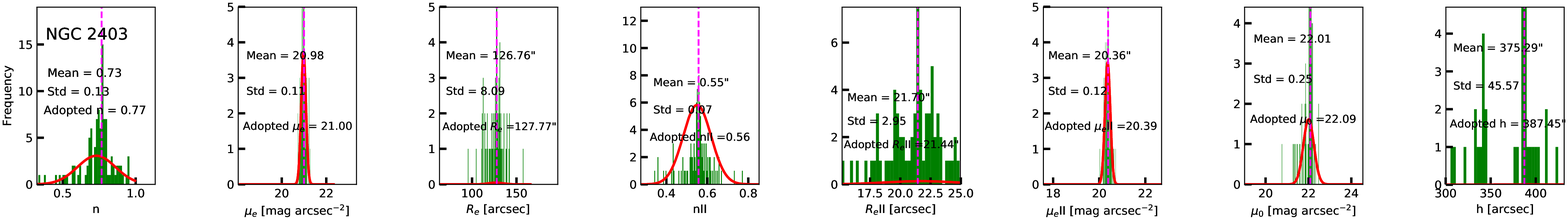}\\
\hspace*{.8130cm}
\includegraphics[angle=0,scale=0.35]{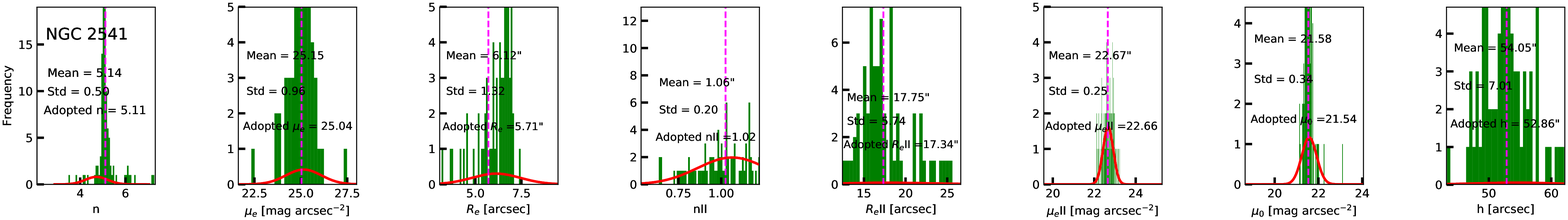}\\
\hspace*{.8130cm}
\includegraphics[angle=0,scale=0.35]{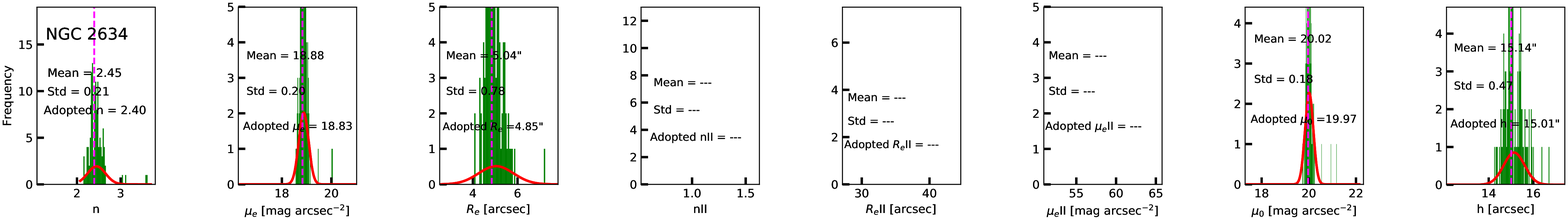}\\
\hspace*{.8130cm}
\includegraphics[angle=0,scale=0.35]{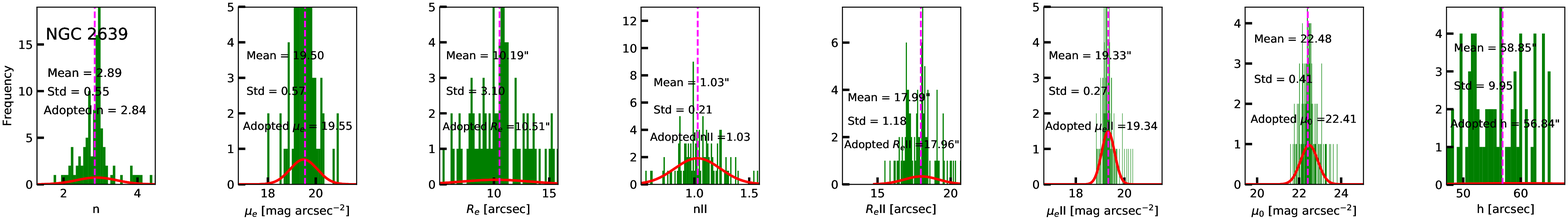}\\
\hspace*{.8130cm}
\includegraphics[angle=0,scale=0.35]{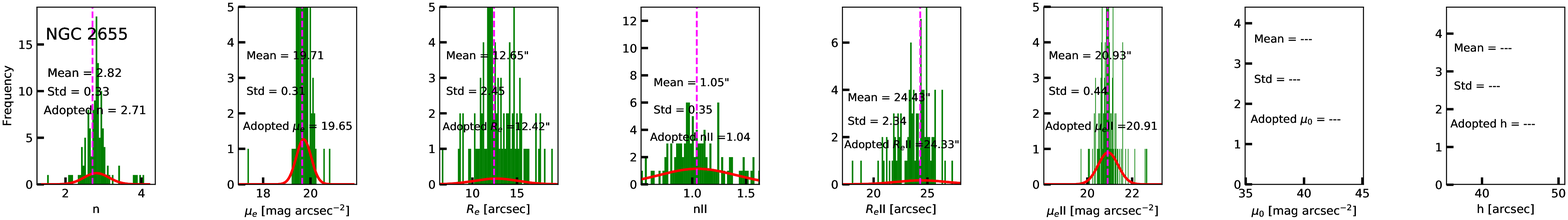}\\
\hspace*{.8130cm}
\includegraphics[angle=0,scale=0.35]{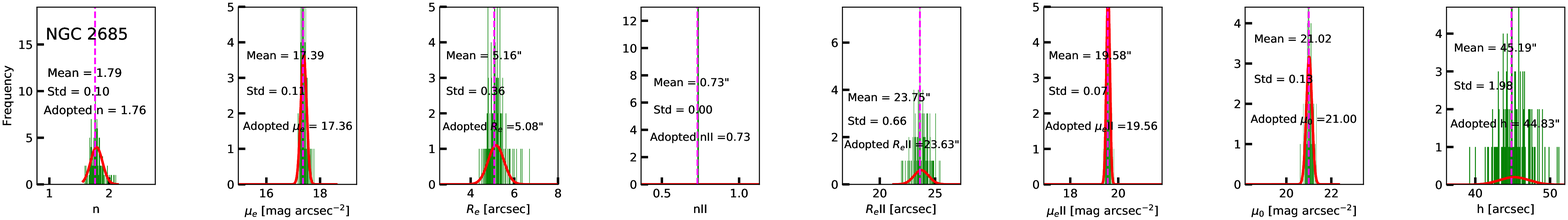}\\
\hspace*{.8130cm}
\includegraphics[angle=0,scale=0.35]{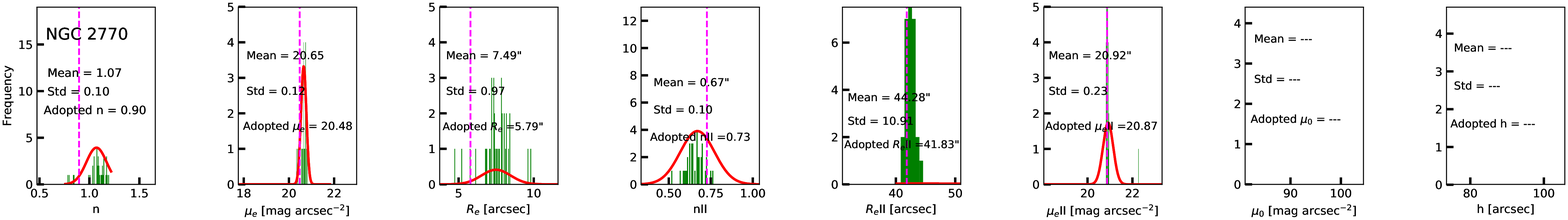}\\
\hspace*{.8130cm}
\includegraphics[angle=0,scale=0.35]{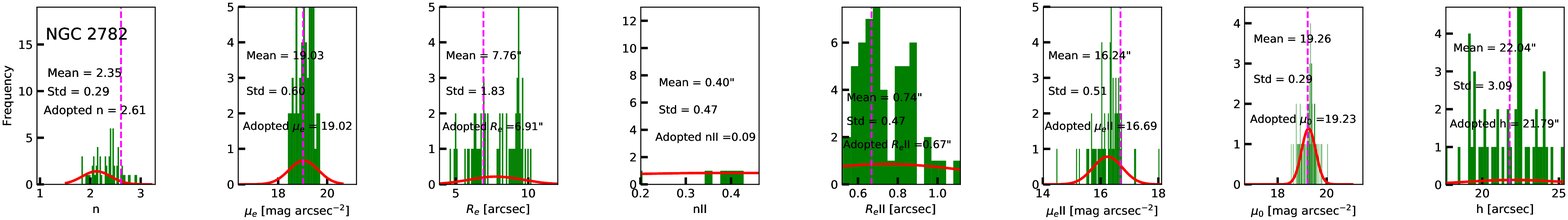}\\
\hspace{-1.630cm}
\vspace{-.298530cm}
\caption{\it continued.}
\end{figure*}

\begin{figure*}
\setcounter{figure}{6}
\vspace{-.1630cm}
\hspace*{.8130cm}
\includegraphics[angle=0,scale=0.35]{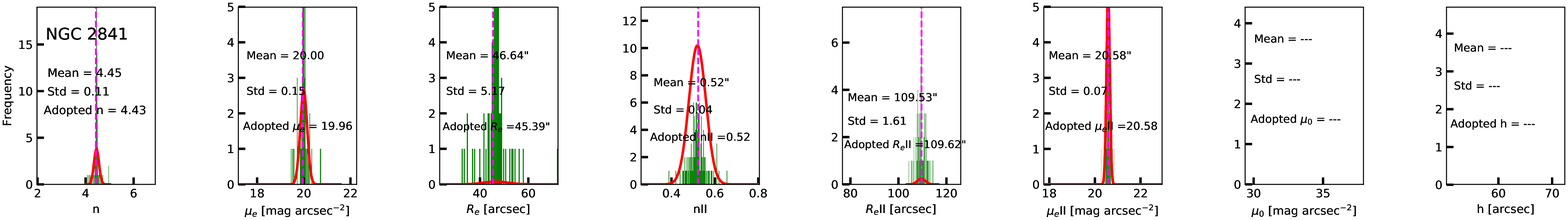}\\
\hspace*{.8130cm}
\includegraphics[angle=0,scale=0.35]{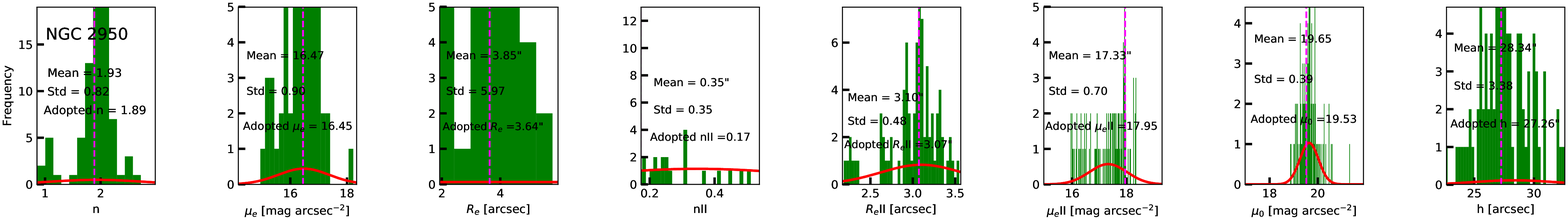}\\
\hspace*{.8130cm}
\includegraphics[angle=0,scale=0.35]{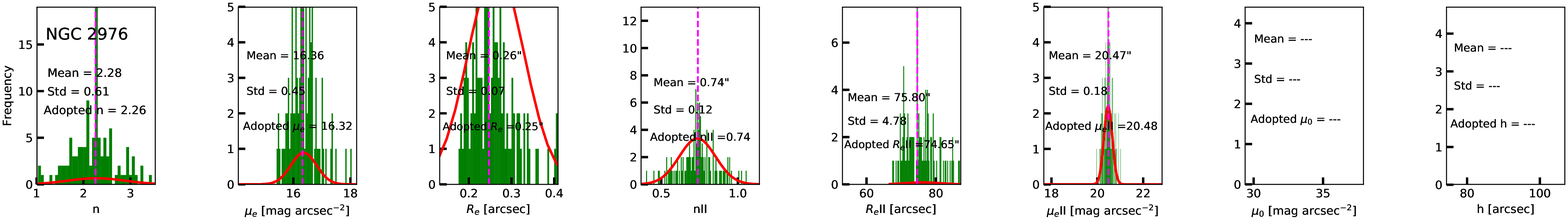}\\
\hspace*{.8130cm}
\includegraphics[angle=0,scale=0.35]{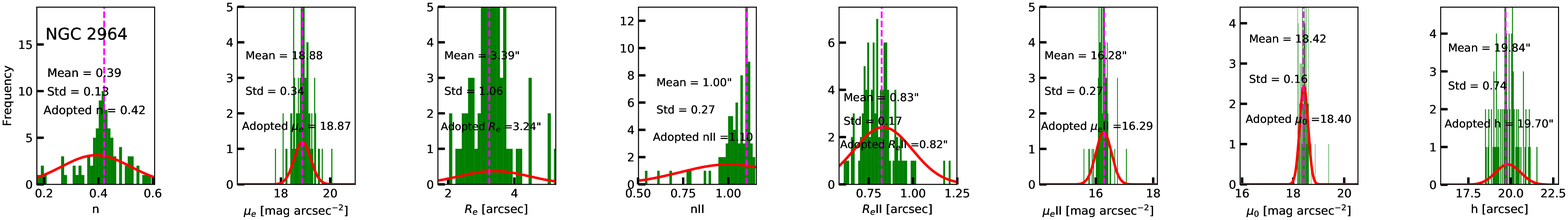}\\
\hspace*{.8130cm}
\includegraphics[angle=0,scale=0.35]{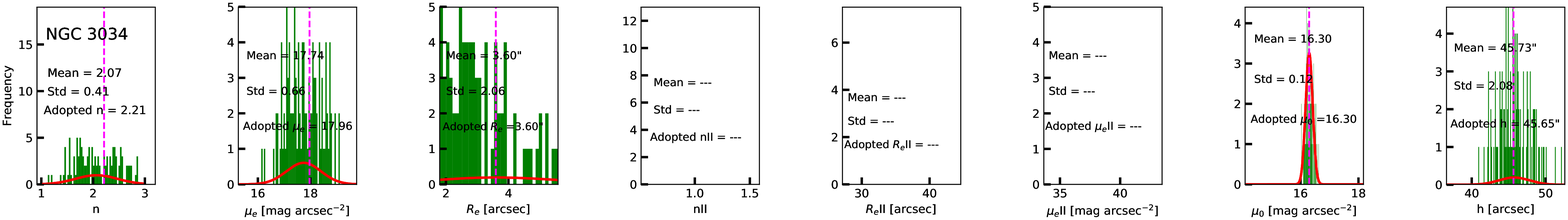}\\
\hspace*{.8130cm}
\includegraphics[angle=0,scale=0.35]{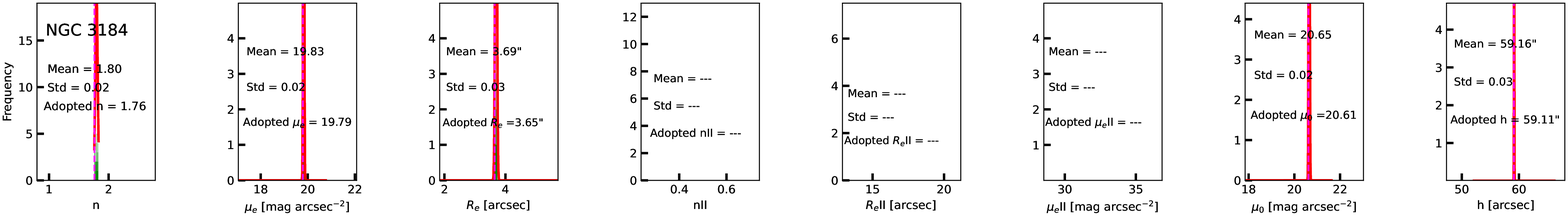}\\
\hspace*{.8130cm}
\includegraphics[angle=0,scale=0.35]{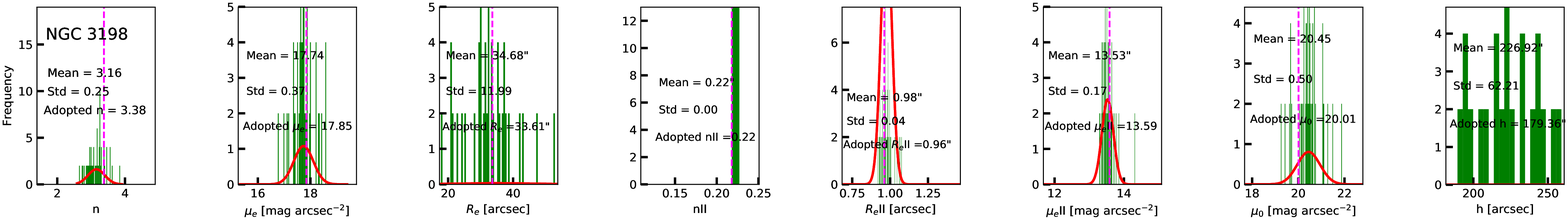}\\
\hspace*{.8130cm}
\includegraphics[angle=0,scale=0.35]{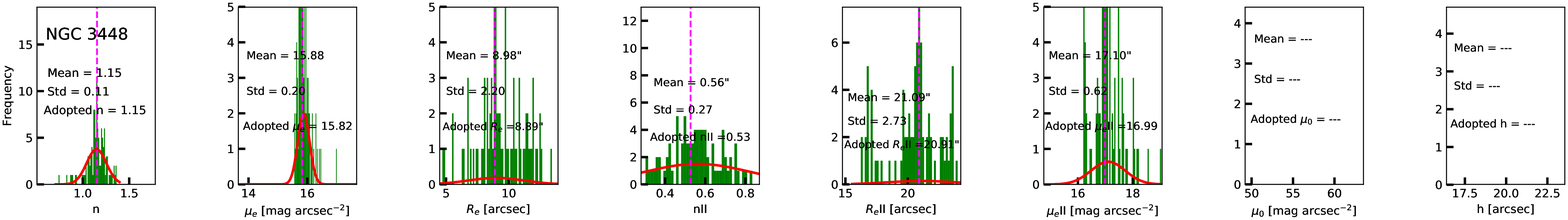}\\
\hspace*{.8130cm}
\includegraphics[angle=0,scale=0.35]{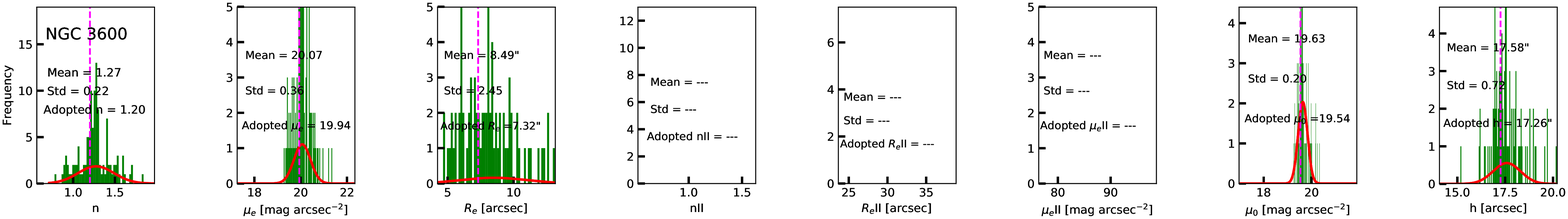}\\
\hspace{-15.630cm}
\vspace{-.298530cm}
\caption{\it continued.}
\end{figure*}

\begin{figure*}
\setcounter{figure}{6}
\vspace{-.1630cm}
\hspace*{.8130cm}
\includegraphics[angle=0,scale=0.35]{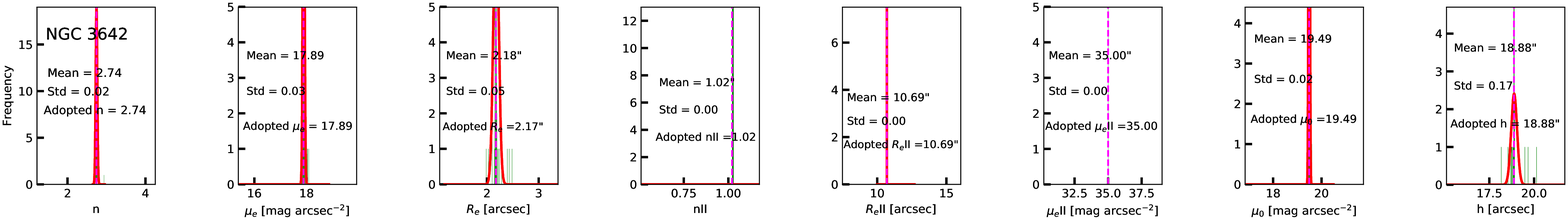}\\
\hspace*{.8130cm}
\includegraphics[angle=0,scale=0.35]{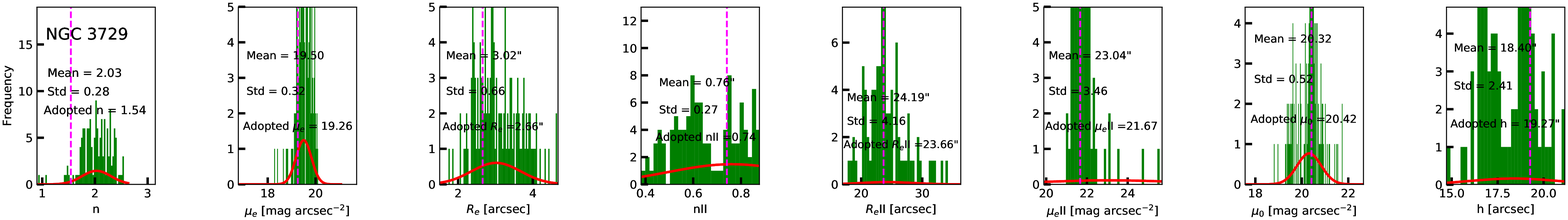}\\
\hspace*{.8130cm}
\includegraphics[angle=0,scale=0.35]{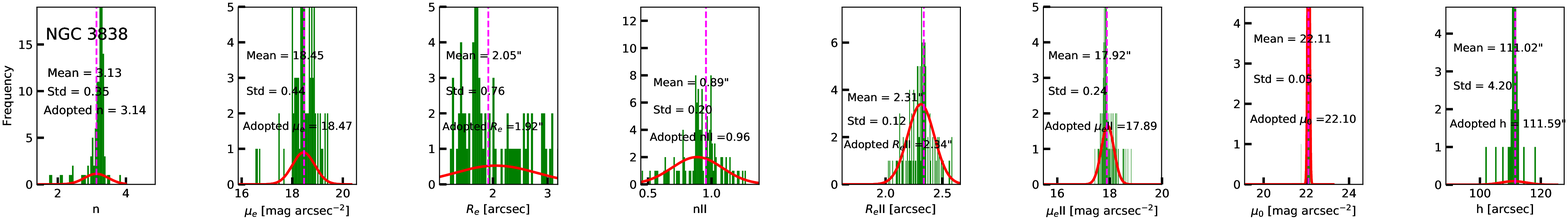}\\
\hspace*{.8130cm}
\includegraphics[angle=0,scale=0.35]{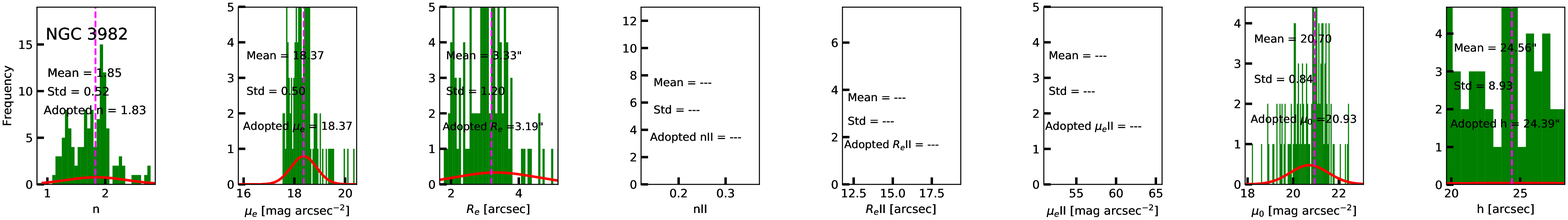}\\
\hspace*{.8130cm}
\includegraphics[angle=0,scale=0.35]{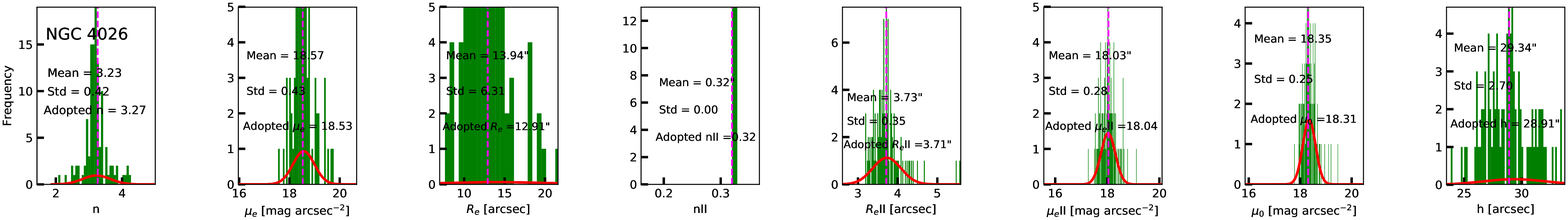}\\
\hspace*{.8130cm}
\includegraphics[angle=0,scale=0.35]{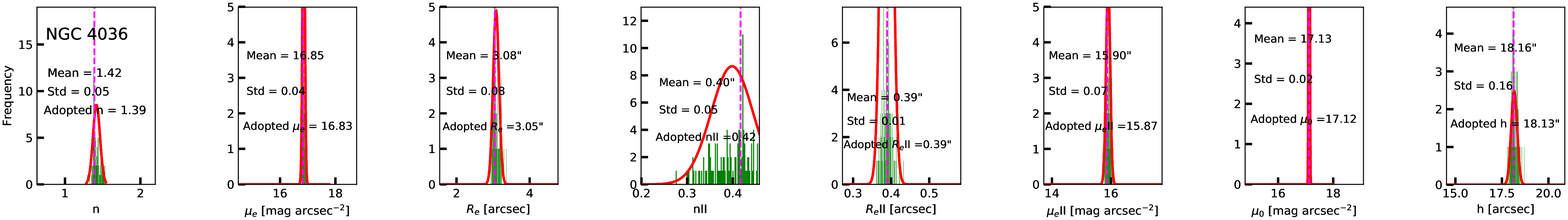}\\
\hspace*{.8130cm}
\includegraphics[angle=0,scale=0.35]{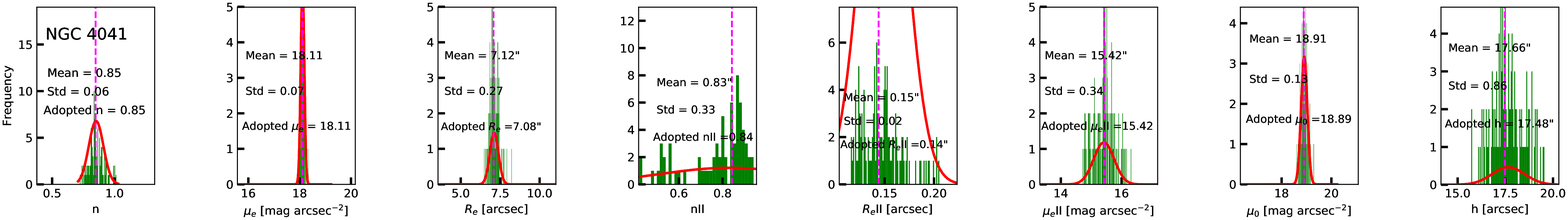}\\
\hspace*{.8130cm}
\includegraphics[angle=0,scale=0.35]{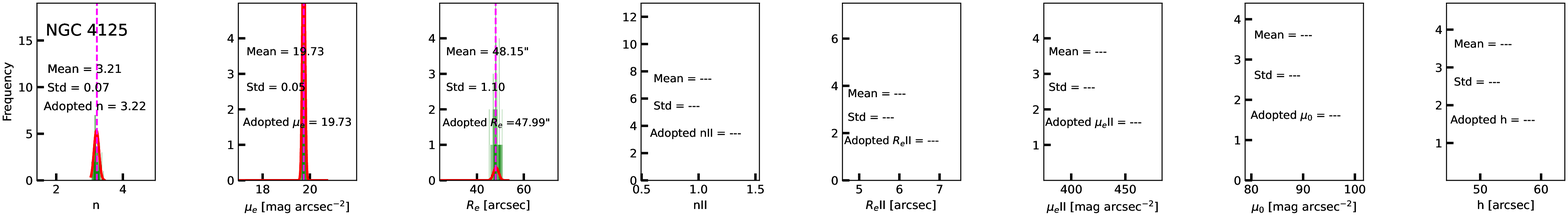}\\
\hspace*{.8130cm}
\includegraphics[angle=0,scale=0.35]{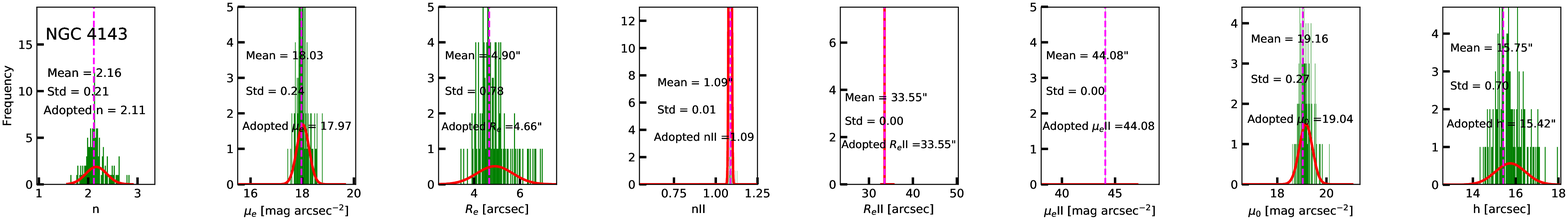}\\
\hspace{-15.630cm}
\vspace{-.298530cm}
\caption{\it continued.}
\end{figure*}

\begin{figure*}
\setcounter{figure}{6}
\vspace{-.1630cm}
\hspace*{.8130cm}
\includegraphics[angle=0,scale=0.35]{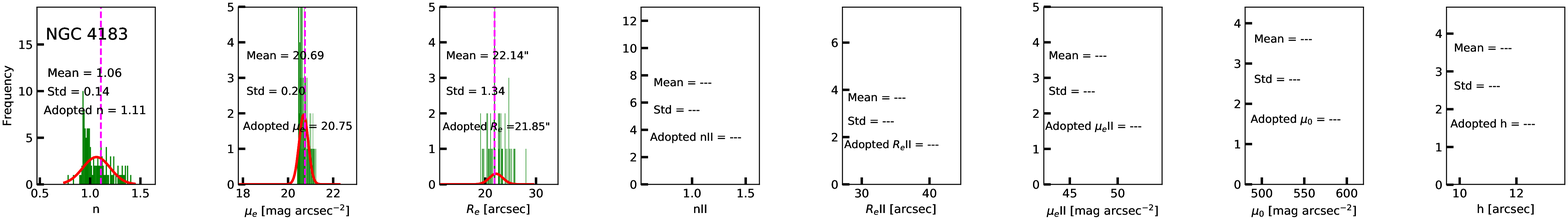}\\
\hspace*{.8130cm}
\includegraphics[angle=0,scale=0.35]{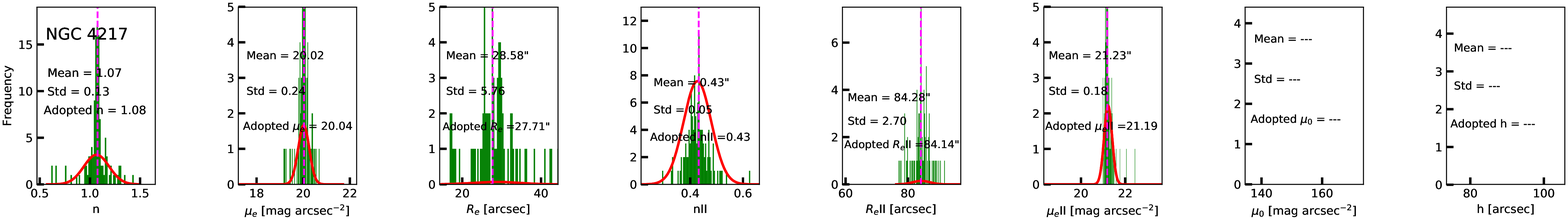}\\
\hspace*{.8130cm}
\includegraphics[angle=0,scale=0.35]{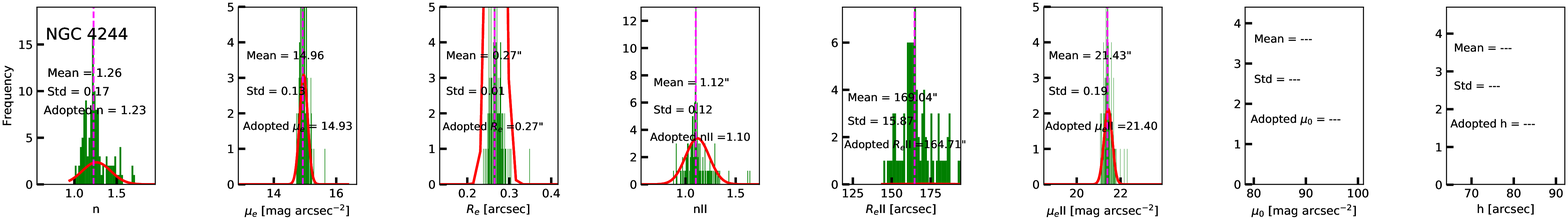}\\
\hspace*{.8130cm}
\includegraphics[angle=0,scale=0.35]{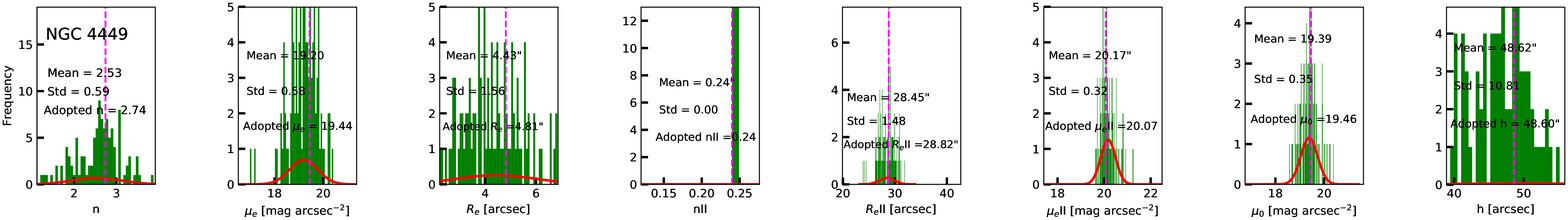}\\
\hspace*{.8130cm}
\includegraphics[angle=0,scale=0.35]{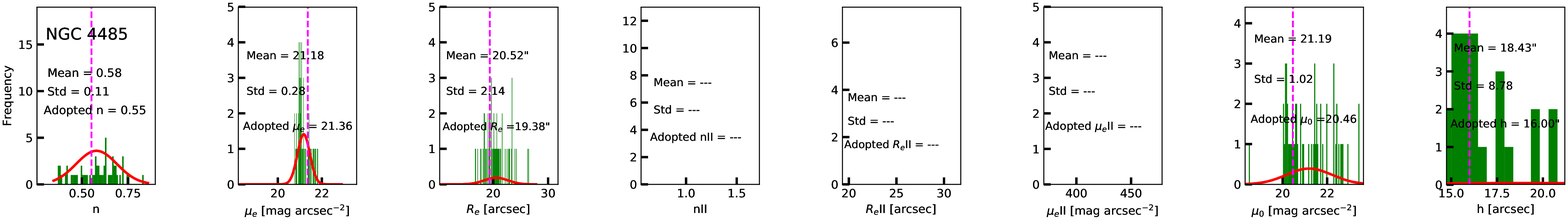}\\
\hspace*{.8130cm}
\includegraphics[angle=0,scale=0.35]{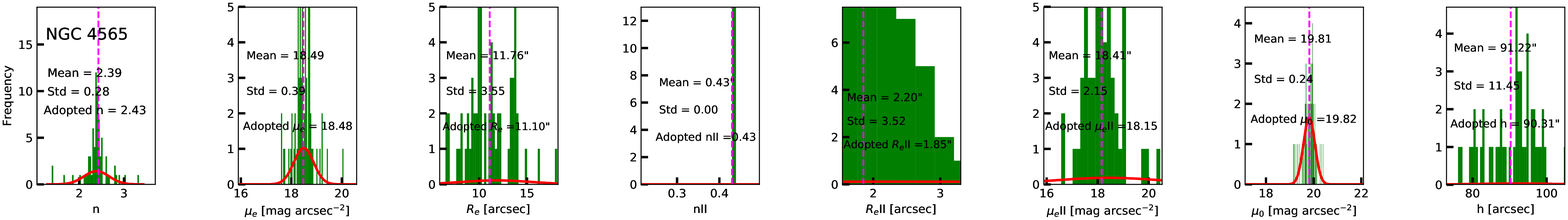}\\
\hspace*{.8130cm}
\includegraphics[angle=0,scale=0.35]{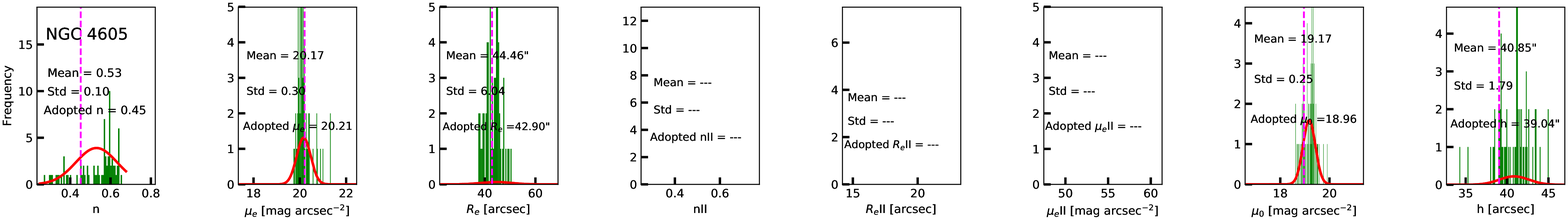}\\
\hspace*{.8130cm}
\includegraphics[angle=0,scale=0.35]{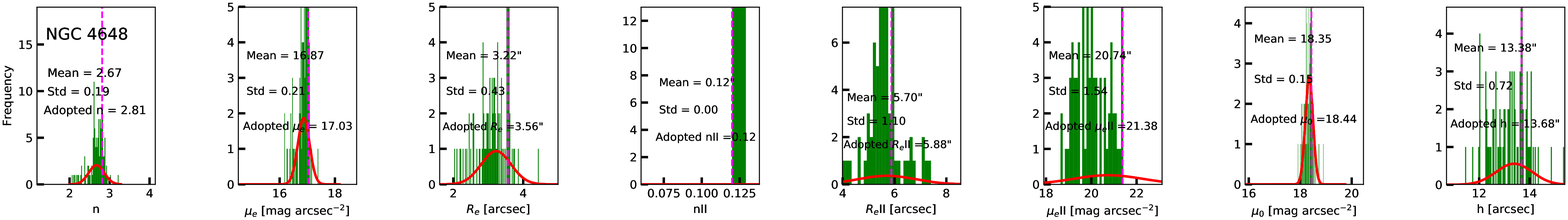}\\
\hspace*{.8130cm}
\includegraphics[angle=0,scale=0.35]{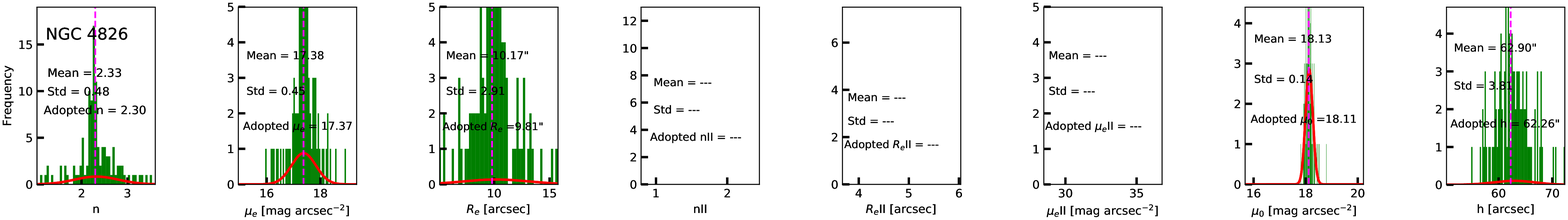}\\
\hspace{-15.630cm}
\vspace{-.298530cm}
\caption{\it continued.}
\end{figure*}

\begin{figure*}
\setcounter{figure}{6}
\vspace{-.1630cm}
\hspace*{.8130cm}
\includegraphics[angle=0,scale=0.35]{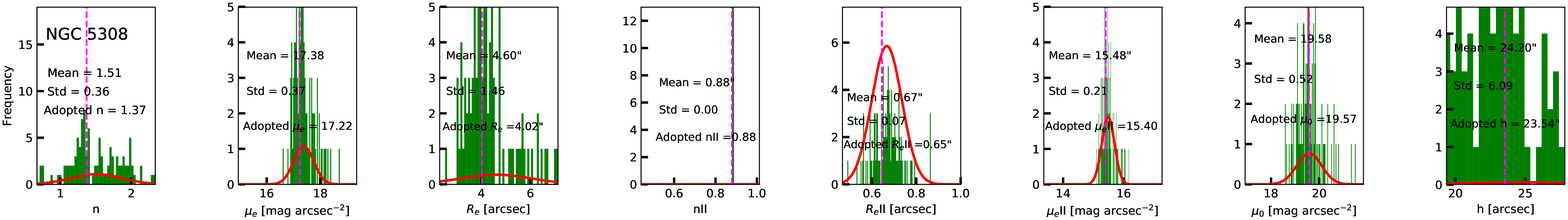}\\
\hspace*{.8130cm}
\includegraphics[angle=0,scale=0.35]{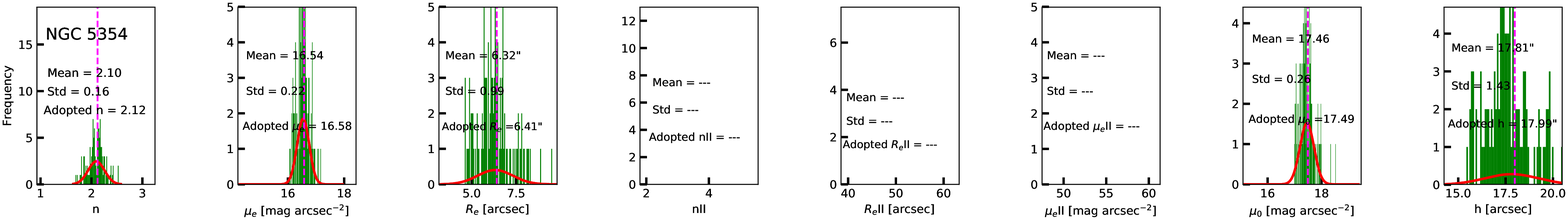}\\
\hspace*{.8130cm}
\includegraphics[angle=0,scale=0.35]{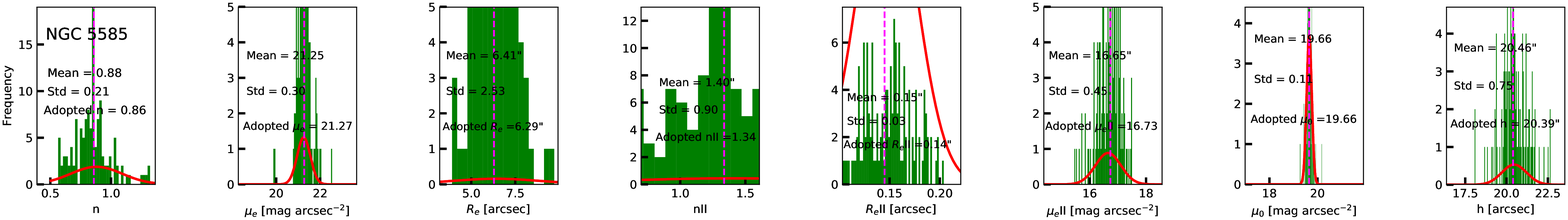}\\
\hspace*{.8130cm}
\includegraphics[angle=0,scale=0.35]{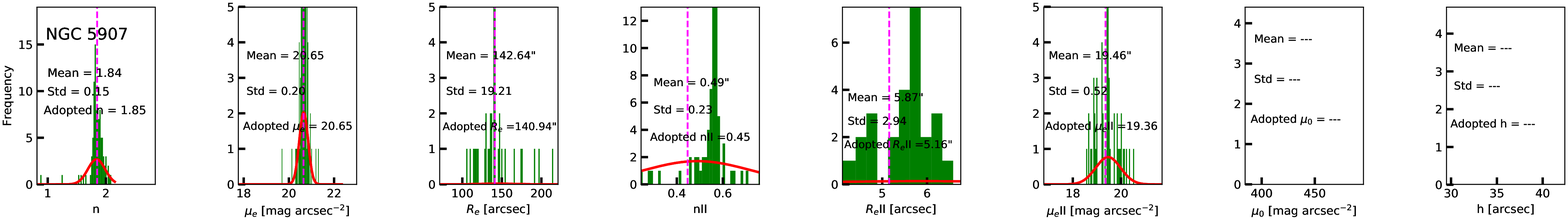}\\
\hspace*{.8130cm}
\includegraphics[angle=0,scale=0.35]{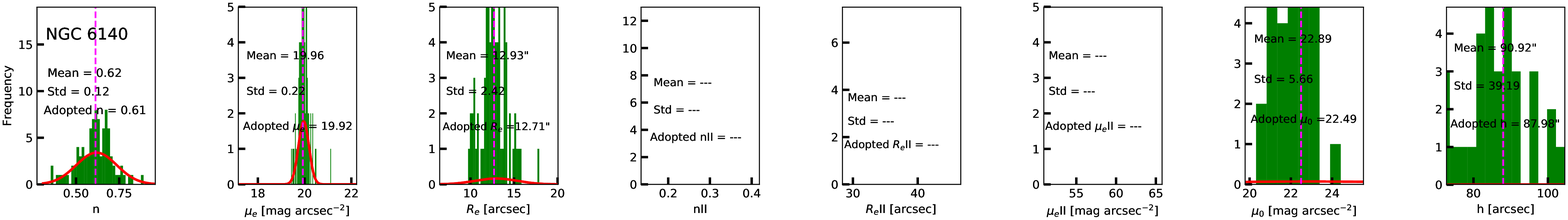}\\
\hspace*{.8130cm}
\includegraphics[angle=0,scale=0.35]{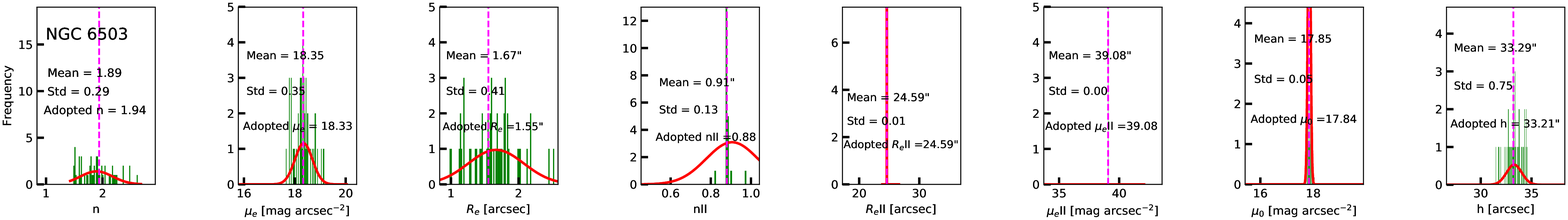}\\
\hspace*{.8130cm}
\includegraphics[angle=0,scale=0.35]{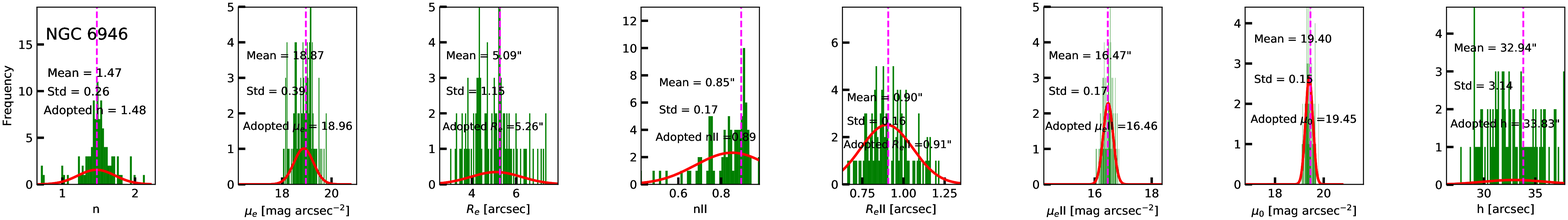}\\
\hspace*{.8130cm}
\includegraphics[angle=0,scale=0.35]{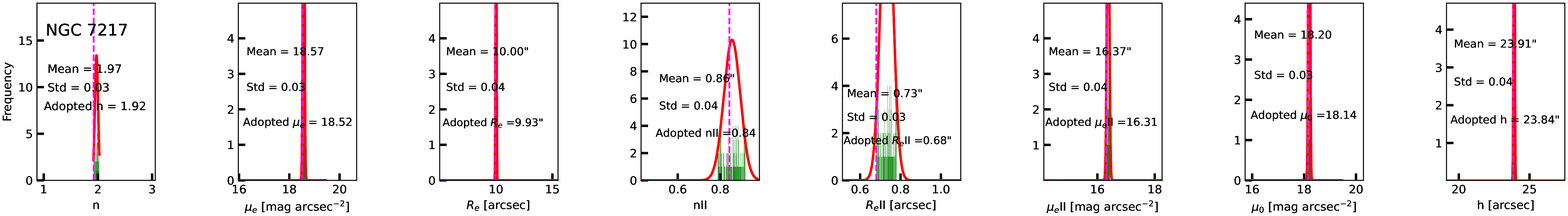}\\
\hspace{-25.630cm}
\vspace{-.298530cm}
\caption{\it continued.}
\end{figure*}

\label{lastpage}
\end{document}